\documentclass[notitlepage,nofootinbib,10pt,prx,aps, twocolumn]{revtex4-1}
\usepackage[maxfloats=256]{morefloats}
\usepackage[colorlinks=true]{hyperref}
\usepackage{amsmath}
\usepackage{amssymb}
\usepackage{color}
\usepackage{times}
\definecolor{deepblue}{rgb}{0,0,0.5}
\definecolor{deepred}{rgb}{0.6,0,0}
\definecolor{deepgreen}{rgb}{0,0.5,0}
\usepackage[mmddyy]{datetime}
\usepackage{moreverb}
\usepackage{graphicx}
\usepackage{hhline}
\usepackage{listings}
\usepackage{ctable}
\usepackage{qcircuit}
\usepackage[caption=false]{subfig}
\usepackage{blkarray}

\newcommand\pythonstyle{\lstset{
language=Python,
basicstyle=\ttm,
numbers=left,
stepnumber=1,
otherkeywords={self},             
keywordstyle=\ttb\color{deepblue},
emph={MyClass,__init__},          
emphstyle=\ttb\color{deepred},    
stringstyle=\color{deepgreen},
frame=tb,                         
showstringspaces=false            %
}}

\lstnewenvironment{python}[1][]
{
\pythonstyle
\lstset{basicstyle=\small,#1}
}
{}

\newcommand\pythoninline[1]{{\pythonstyle\lstinline!#1!}}

\begin{document}

\maxdeadcycles=1000

\newcommand{\comb}{\mathrm{comb}}
\newcommand{\rect}{\mathrm{rect}}
\newcommand{\tri}{\mathrm{tri}}
\newcommand{\sinc}{\mathrm{sinc}}
\newcommand{\Gaus}{\mathrm{Gaus}}
\newcommand{\somb}{\mathrm{somb}}
\newcommand{\fstop}{f/\#}
\newcommand{\warn}[1]{{\color{red} \textbf{#1}}}
\newcommand{\blue}[1]{{\color{blue} #1}}
\newcommand{\Ito}{It$\hat{\mathrm{o}}$}
\newcommand{\CNOT}[2]{\mathrm{CNOT}\left(#1,#2\right)}
\newcommand{\CROT}[4]{\mathrm{CROT}_{#1}\left(#2,#3;#4\right)}

\title{Generative machine learning with tensor networks: benchmarks on near-term quantum computers}
\author{Michael L. Wall}
\affiliation{The Johns Hopkins University Applied Physics Laboratory, Laurel, MD 20723}
\author{Matthew R.~Abernathy}
\address{The Johns Hopkins University Applied Physics Laboratory, Laurel, MD 20723}
\author{Gregory Quiroz}
\address{The Johns Hopkins University Applied Physics Laboratory, Laurel, MD 20723}

\begin{abstract}
Noisy, intermediate-scale quantum (NISQ) computing devices have become an industrial reality in the last few years, and cloud-based interfaces to these devices are enabling exploration of near-term quantum computing on a range of problems.  As NISQ devices are too noisy for many of the algorithms with a known quantum advantage, discovering impactful applications for near-term devices is the subject of intense research interest.  We explore quantum-assisted machine learning (QAML) on NISQ devices through the perspective of tensor networks (TNs), which offer a robust platform for designing resource-efficient and expressive machine learning models to be dispatched on quantum devices.  In particular, we lay out a framework for designing and optimizing TN-based QAML models using classical techniques, and then compiling these models to be run on quantum hardware, with demonstrations for generative matrix product state (MPS) models.  We put forth a generalized canonical form for MPS models that aids in compilation to quantum devices, and demonstrate greedy heuristics for compiling with a given topology and gate set that outperforms known generic methods in terms of the number of entangling gates, e.g., CNOTs, in some cases by an order of magnitude.  We present an exactly solvable benchmark problem for assessing the performance of MPS QAML models, and also present an application for the canonical MNIST handwritten digit dataset.  The impacts of hardware topology and day-to-day experimental noise fluctuations on model performance are explored by analyzing both raw experimental counts and statistical divergences of inferred distributions.  We also present parametric studies of depolarization and readout noise impacts on model performance using hardware simulators.
\end{abstract}

\maketitle

\section{Introduction}

In recent years, gate-based quantum computing has emerged as a relatively mature technology, with many platforms offering cloud-based interfaces to machines with a few to dozens of qubits~\cite{smith2016practical,steiger2018projectq,haner2018software,Qiskit,larose2019overview}, as well as classical emulators of quantum devices of this class~\cite{QCSims}.  Today's quantum computing resources remain a long way from the millions of qubits~\cite{campbell2017roads} required to perform canonical quantum computing tasks such as integer factorization with error correction~\cite{shor1999polynomial,gidney2019factor}, and present devices are either engineered with a specific demonstration goal or designed for general-purpose research-scale exploration~\cite{corcoles2019challenges}.  With the advent of noisy, intermediate-scale quantum (NISQ) devices~\cite{preskill2018quantum}, whose hardware noise and limited qubit connectivity and gate sets pose challenges for demonstrating scalable universal quantum computation, we are faced with a different form of quantum application discovery in which algorithms need to be robust to noise, limited qubit connectivity and gate sets, and highly resource-efficient.

Machine learning (ML) has been put forward as a possible application area for NISQ devices, with a range of recent proposals~\cite{biamonte2017quantum,perdomo2018opportunities,ciliberto2018quantum}.  ML may prove promising for NISQ applications because well-performing ML algorithms feature robustness against noise, quantum circuits can be designed for ML applications that are highly qubit-efficient~\cite{huggins2019towards}, and quantum models can be designed whose expressibility increases exponentially with qubit depth~\cite{huggins2019towards,glasser2019expressive}.  The most impactful near-term ML application likely lies in quantum-assisted machine learning (QAML), in which a quantum circuit's parameters are classically optimized based on measurement outcomes that may not be efficiently classically simulable~\cite{benedetti2019parameterized}; this also includes kernel-based learning schemes with a quantum kernel~\cite{havlivcek2019supervised}.  Tensor networks (TNs) provide a robust means of designing such parameterized quantum circuits that are quantum-resource efficient and can be implemented and optimized on classical or quantum hardware.  TN-based QAML algorithms hence leverage the significant research effort into optimization strategies for TNs~\cite{schollwock2011density,orus2014practical,orus2019tensor}, and also enable detailed benchmarking and design of QAML models classically, with a smooth transition to classically intractable models.

\begin{figure*}[t]
\centering
\includegraphics[width=1.69\columnwidth]{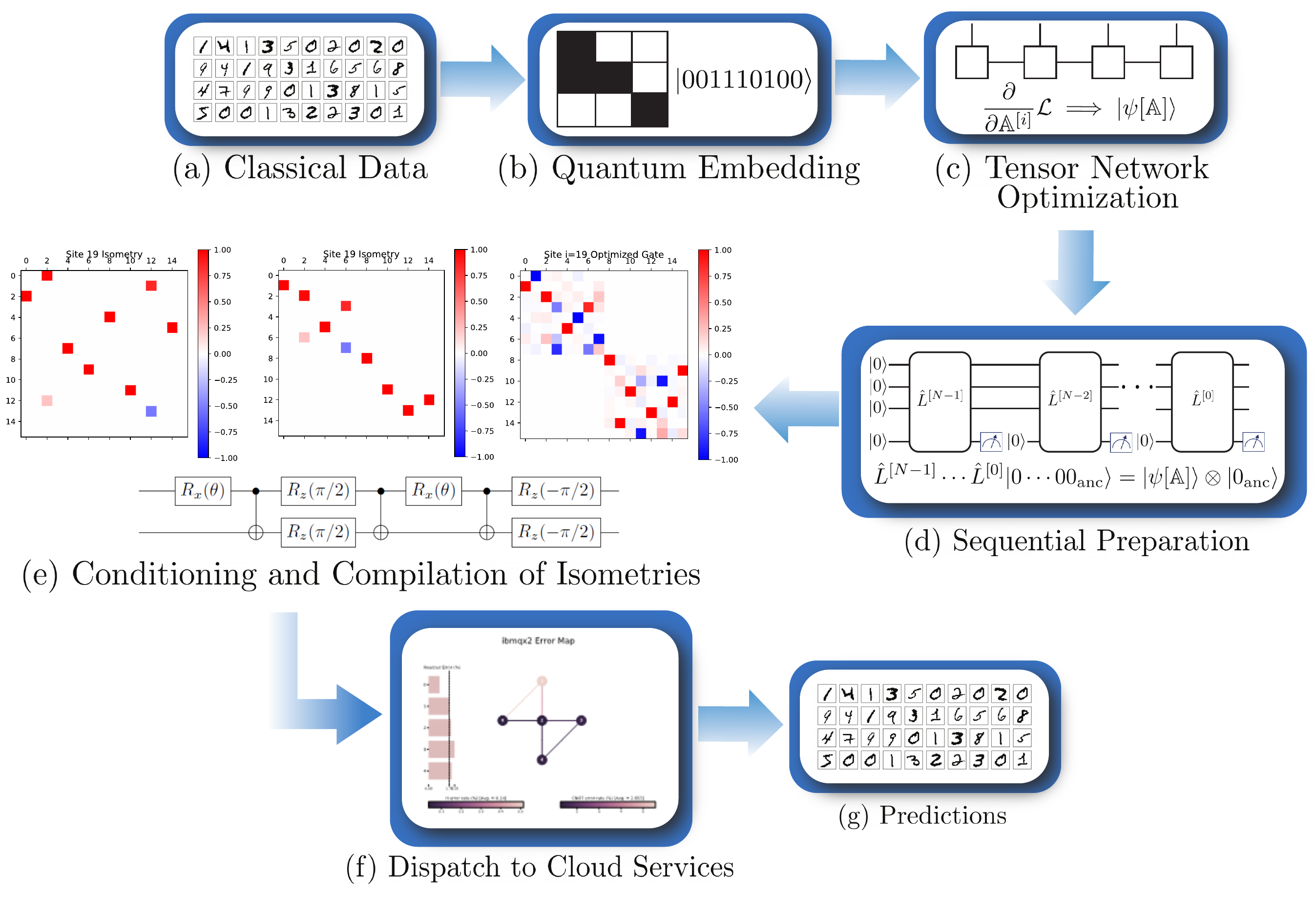}  
\caption{\label{fig:Overview} \emph{Overview of QAML workflow.} Classical data in (a) is pre-processed and transformed to quantum states embedded in an exponentially large Hilbert space in (b).  A TN model is learned from a collection of quantum training data in (c), which has the interpretation in (d) of a sequential preparation scheme involving a small number of readout qubits coupled to ancillary resources.  The isometries of the sequential preparation scheme in (d) are conditioned using inherent freedom in the TN representation in (e), and then converted into native gates for a target hardware architecture, displayed as the IBMQ-X2 processor for concreteness.  Running on cloud-based hardware in (f), we obtain measurements defining output predictions, as in (g).  For interpretation of graphical representations, see text.}
\end{figure*}

In this work, we explore the applicability of QAML with TN architectures on NISQ hardware and hardware simulators, exploring the effects of present-day and near-term hardware noise, qubit connectivity, and restrictions on gate sets.  We focus on fully generative unsupervised learning tasks, which have been identified as a promising avenue for QAML~\cite{perdomo2018opportunities}, and focus on the most resource-efficient matrix product state (MPS) TN topology.  We present a framework for QAML--outlined in Fig.~\ref{fig:Overview}--that includes translation of classical data into quantum states, optimization of an MPS model using classical techniques, the conversion of this classically-trained model into a sequence of isometric operations to be performed on quantum resources, and the optimization and compilation of these isometric operations into native operations for a given hardware topology and allowed gate set.  In particular, we develop several novel techniques for the compilation stage aimed at TN models for QAML on NISQ devices, such as the permutation of auxiliary quantum degrees of freedom in the TN to optimize mapping to hardware resources and heuristics for the translation of isometries into native operations using as few entangling operations (e.g., CNOTs) as possible.  

The tools developed herein enable the robust design and performance assessment of QAML models on NISQ devices in the regime where classical simulations are still possible, and will inform architectures and noise levels for scaling to the classically intractable regime.  Even in the classically intractable regime in which the model must be optimized using a quantum device in a hybrid quantum/classical loop~\cite{xiang2013hybrid,schuld2020circuit}, our techniques provide a means of obtaining an approximate, classically trained ``preconditioner" for the quantum models that can help avoid local minima and reduce optimization time.  We present exemplar results for our workflow for synthetic data that can be described by an exactly solvable two-qubit MPS QAML model, as well as on features extracted from the canonical MNIST handwritten digit dataset~\cite{lecun2010mnist}.

The remainder of this work is organized as follows: Sec.~\ref{sec:QAMLTN} discusses QAML with tensor networks (TNs) broadly, including embedding of classical data into quantum states, classical training of a TN model, and the conversion of TN models into resource-efficient sequential preparation schemes; Sec.~\ref{sec:compilation} discusses our approach for compiling TN-based QAML models for running on quantum hardware, including the utilization of ambiguity in the TN representation and greedy compilation heuristics for minimizing model gate depth; Sec.~\ref{sec:ExactMPS} presents an exactly solvable two-qubit QAML model and assesses the performance of our QAML workflow on quantum hardware; in Sec.~\ref{sec:MNIST} we give an example application to generative modeling of features extracted from the MNIST dataset and analyze the performance of our models as a function of hardware noise using a quantum hardware simulator; finally, in Sec.~\ref{sec:Concl} we conclude and give an outlook.  Details of the MNIST model studied in Sec.~\ref{sec:MNIST} are given in Appendix \ref{sec:MNISTIso}.

\section{Quantum-assisted machine learning with tensor networks}
\label{sec:QAMLTN}

Fig.~\ref{fig:Overview} broadly outlines the QAML workflow explored in the present work.  We begin with a collection of classical data vectors in a training set $\mathcal{T} = \{\mathbf{x}_j\}_{j=1}^{N_T}$, where each element $\mathbf{x}_j$ is an $N$-length vector.  The first step in our QAML workflow is to define a mapping of classical data vectors to vectors in a quantum Hilbert space.  Here, the only restriction we will place on the encoding of classical data in quantum states is that each classical data vector is encoded in an unentangled product state.  This is useful for several reasons.  For one, unentangled states are the simplest to prepare experimentally with high fidelity, and also enable us to use qubit-efficient sequential preparation schemes.  From a learning perspective, encoding individual data vectors in product states ensures that any entanglement that results in a quantum model comes from correlations in an ensemble of data and not from \emph{a priori} assumptions about pre-existing correlations for individual data vectors~\cite{stoudenmire2016supervised}.  For encoding of an $N$-dimensional classical data vector $\mathbf{x}$ into an ensemble of $N$ qubits, a convenient parameterization is
\begin{align}
|\Phi\left(\mathbf{x}\right)\rangle&=\bigotimes_{j=0}^{N-1} \left(\sum_{i_j=0}^{1}\phi_j\left(x_{i}\right)|i_j\rangle\right)\, ;
\end{align}
that is, in terms of local maps $\phi_j\left(x\right)$ mapping a single data element into a superposition of qubit states.  In order that the full map $\Phi\left(\mathbf{x}\right)$ maps each data instance into a normalized vector in Hilbert space, we require that 
\begin{align}
\label{eq:Nmap} \sum_j \left|\phi_j\left({x}\right)\right|^2=1\;\;\; \forall x\, .
\end{align}
When encoding data for use in generative applications it is also useful for the maps to have the \emph{orthonormality property}
\begin{align}
\label{eq:ONmap} \prod_{j=0}^{N-1}\int dx_j \phi_{i_j}^{\star}\left(x_j\right)\phi_{i_j'}\left(x_j\right)&=\prod_j \delta_{i_j,i_j'}\, ,
\end{align}
which ensures that the wavefunction encoding data 
\begin{align}
|\psi\rangle&=\sum_{i_0\dots i_{N-1}} c_{i_0\dots i_{N-1}}|i_0\dots i_{N-1}\rangle\, ,
\end{align}
is normalized whenever
\begin{align}
\sum_{i_0\dots i_{N-1}}\left|c_{i_0\dots i_{N-1}}\right|^2=1\, .
\end{align}
That is, maps satisfying Eq.~\eqref{eq:ONmap} map the data into an orthonormal Hilbert space.  

The simplest case occurs when the data is discrete, and so can be formulated as vectors $\mathbf{x}$ where $x_j\in\left\{0,1\right\}$.  We map each element to a qubit as~\cite{farhi2018classification,schuld2019quantum}
\begin{align}
\label{eq:BinaryMap}\phi_j\left(x\right)&=\delta_{j,x}\, .
\end{align}
This map clearly satisfies the properties Eqs.~\eqref{eq:Nmap} and \eqref{eq:ONmap} above, and so is suitable for either generative or discriminative applications.  In the case in which the data is continuous, $\mathbf{x}\in\mathbb{R}^N$, we now have freedom to choose how to encode it in Hilbert space.  The phase-like encoding
\begin{align}
\label{eq:phasemap}\phi_0\left(x\right)&=\cos\left(\frac{\pi}{2}\frac{x-x_{\mathrm{min}}}{x_{\mathrm{max}}-x_{\mathrm{min}}}\right)\, ,\\
\nonumber \phi_1\left(x\right)&=\sin\left(\frac{\pi}{2}\frac{x-x_{\mathrm{min}}}{x_{\mathrm{max}}-x_{\mathrm{min}}}\right)\, ,
\end{align}
has been used in Refs.~\cite{stoudenmire2016supervised} to encode data for quantum-inspired ML applications.  Eq.~\eqref{eq:phasemap} satisfies Eqs.~\eqref{eq:Nmap} but not Eq.~\eqref{eq:ONmap}.  A related map that satisfies both conditions is~\cite{stoudenmire2016supervised}
\begin{align}
\phi_0\left(x\right)&=e^{3\pi i x/2}\cos\left(\frac{\pi}{2}\frac{x-x_{\mathrm{min}}}{x_{\mathrm{max}}-x_{\mathrm{min}}}\right)\, ,\\
\nonumber \phi_1\left(x\right)&=e^{-3\pi i x/2}\sin\left(\frac{\pi}{2}\frac{x-x_{\mathrm{min}}}{x_{\mathrm{max}}-x_{\mathrm{min}}}\right)\, .
\end{align}
In the present work, we will focus on the case of binary data, and so utilize the map Eq.~\eqref{eq:BinaryMap}.

\subsection{Tensor networks and sequential preparation}
\label{sec:TNSP}
The next step in our QAML workflow outlined in Fig.~\ref{fig:Overview} is to learn a quantum model for the collection to quantum states $\{|\Phi\left(\mathbf{x}_j\right)\rangle\}_{j=1}^{N_T}$ resulting from applying the encoding map from the previous section to the training data.  Here, we define a quantum model as a collection of operations applied to quantum resources to produce a state that encodes the properties of the ensemble $\{|\Phi\left(\mathbf{x}_j\right)\rangle\}$.  In what follows, we specialize to the case of tensor network (TN) models, which provide a convenient parameterization of the structure of quantum operations and resources.  Generally speaking, TNs represent the high-rank tensor describing a quantum wavefunction in a specified basis as a contraction over low-rank tensors, and hence define families of low-rank approximations whose computational power can be expressed in terms of the maximum dimension of any contracted index $\chi$, known as the bond dimension.  

A wide variety of TN topologies have been considered which are able to efficiently capture certain classes of quantum states~\cite{schollwock2011density,orus2014practical,orus2019tensor}; in the present work we focus on matrix product states (MPSs).  MPSs use a one-dimensional TN topology, as shown using the Penrose graphical notation for tensors~\cite{schollwock2011density} in Fig.~\ref{fig:Overview}(b), and form the basis for the enormously successful density matrix renormalization group (DMRG) algorithm in quantum condensed matter physics~\cite{white1992density}.  MPSs have several properties that make them attractive for QAML.  For one, they are undoubtedly the most well-understood and mature of all tensor networks, which has led to robust optimization strategies that are widely used in the quantum many-body community.  In addition, MPSs are highly quantum resource efficient, in the sense that their associated wavefunctions can be sequentially prepared, and so qubits can be re-used in deployment on quantum hardware.  In fact, it can be shown that every state that can be sequentially prepared can be written as an MPS~\cite{schon2005sequential,schon2007sequential,perez2007matrix}.

In recent years, TNs have found applications outside of the condensed matter and quantum information domains.  The mathematical analysis community has proposed TN methods for data analysis, e.g., large-scale principle component analysis~\cite{cichocki2014tensor,cichocki2017tensor}.  In this community, MPSs are referred to as tensor trains~\cite{oseledets2011tensor}.  Using TN methods to design quantum-inspired ML models was first proposed by Stoudenmire and Schwab~\cite{stoudenmire2016supervised}, who put forth a scheme using a MPS network as a linear classifier in a Hilbert space whose dimension is exponentially large in the length of the raw data vector.  Since then, many other proposals for quantum-assisted or quantum-inspired TN ML models have appeared in the literature~\cite{stoudenmire2018learning,grant2018hierarchical,PhysRevE.98.042114,carrasquilla2019reconstructing,evenbly2019number,klus2019tensor,PhysRevB.99.155131,liu2019machine,glasser2020probabilistic,trenti2020quantum,bradley2020modeling,gillman2020tensor,miller2020tensor,selvan2020tensor,wang2020anomaly,reyes2020multi,PhysRevA.101.010301}, including generative modeling of binary data using MPSs in Ref.~\cite{PhysRevX.8.031012}.  In the majority of approaches, DMRG-inspired algorithms for optimization have been employed.  However the authors of Ref.~\cite{efthymiou2019tensornetwork} recently demonstrated an alternate strategy where a TN was implemented as a neural network using standard deep learning software, and the tensors of the TN were optimized using backpropagation strategies ubiquitous in classical ML.  While this strategy has shown good performance, it has also been shown to be suboptimal with respect to the DMRG-like approach~\cite{sun2020tangent}.  Nonetheless, the use of deep learning ``preconditioners" and the intersection of QAML and neural networks remains intriguing~\cite{levine2018bridging,chen2018equivalence,glasser2018neural}.

The fact that MPSs define a sequential preparation scheme means that MPSs define highly resource efficient schemes for learning~\cite{huggins2019towards} and quantum simulation~\cite{foss2020holographic}.  In particular, the qubit resource requirements for an MPS model are logarithmic in the bond dimension $\chi$, which encapsulates the expressivity of the model, and are independent of the length of the input data vector $N$.  In order to illustrate how this property comes about, consider that we have a register of $N$ qubits with states $|j_i\rangle$, $i=0,\dots,N-1$, $j_i=0,1$ in which we want to encode data and a $\chi$-level ancilla $|\alpha\rangle$, $\alpha=0,\dots,\chi-1$ that can be used to entangle the qubits.  Starting at the ``right" end of the system, we can initialize the $(N-1)^{\mathrm{\mathrm{st}}}$ qubit using an operator $\hat{L}_{N-1}$ defined as
\begin{align}
\label{eq:firstPrep}\hat{L}_{N-1}&=\sum_{\alpha,j_{N-1}}L_{\alpha}^{[N-1]j_{N-1}}|j_{N-1}\alpha\rangle\langle 0 0|\, ,
\end{align}
in which the coefficients $L^{[N-1]}$ satisfy the isometry condition
\begin{align}
\sum_{\alpha,j_{N-1}}L_{\alpha}^{j_{N-1}\star}L_{\alpha}^{j_{N-1}} = 1\, .
\end{align}
Clearly, if we start our qubit and ancilla system in the state $|00\rangle$, this operation transforms it into the (entangled) state $\sum_{\alpha j_{N-1}}L_{\alpha}^{j_{N-1}}|\alpha j_{N-1}\rangle$, and the isometry condition ensures that this state is normalized.  Moving to the next qubit, we now entangle it with the ancilla using the operator
\begin{align}
\label{eq:LNmo}\hat{L}_{N-2}&=\sum_{\alpha,j_{N-2},\beta}L_{\beta\alpha}^{[N-2]j_{N-2}}| j_{N-2}\beta\rangle\langle  0 \alpha|\, ,
\end{align}
which is subject to the isometry condition
\begin{align}
\label{eq:Lcanonical}\sum_j \mathbb{L}^{[N-2]j\dagger}\mathbb{L}^{[N-2]j}=\mathbb{I}_{\chi}\, ,
\end{align}
with $\mathbb{I}_{\chi}$ the $\chi\times\chi$ identity matrix.  This operation now puts the system in the state
\begin{align}
&\hat{L}_{N-2}\hat{L}_{N-1}|0_{N-2}0_{N-1}0_{\mathrm{ancilla}}\rangle\\
\nonumber &=\sum_{j_{N-2},j_{N-1},\alpha}\left[\mathbb{L}^{[N-2]j_{N-2}}\mathbb{L}^{[N-1]j_{N-1}}\right]_{\alpha} | j_{N-2}j_{N-1}\alpha\rangle\, .
\end{align}
We follow this same logic for all subsequent qubits, defining isometric operators that entangle them to the rest of the system using the ancilla, until we reach qubit 1, which is attached using the isometric operator
\begin{align}
\hat{L}_0&=\sum_{i_0,\beta} |i_0 0\rangle\langle  0 \beta|\, .
\end{align}
This operator puts the full system into the state
\begin{align}
&\hat{L}_0\dots \hat{L}_{N-1}| 0_0\dots 0_{N-1} 0_{\mathrm{ancilla}}\rangle\\
\nonumber &=\sum_{j_0\dots j_{N-1}}\mathbb{L}^{[0]j_0}\dots \mathbb{L}^{[N-1]j_{N-1}}| j_0\dots j_{N-1} 0_{\mathrm{ancilla}}\rangle\, .
\end{align}
Hence, in the last step, the qubit states decouple from the ancilla.  The qubit state takes the form of an MPS with the additional constraint that each of the MPS matrices $\mathbb{L}$ satisfies the left-orthogonal condition Eq.~\eqref{eq:Lcanonical}.  The above procedure can readily be read in reverse; given a general MPS QAML model with bond dimension $\chi$,
\begin{align}
\sum_{j_0\dots j_{N-1}}\mathbb{A}^{[0]j_0}\dots \mathbb{A}^{[N-1]j_{N-1}}|j_0\dots j_{N-1}\rangle\, ,
\end{align}
we can convert it into a sequential qubit preparation scheme with a $\chi$-dimensional ancilla by putting the MPS in left-canonical form.  This transformation to left-canonical form can be done without loss of generality using a well-known procedure involving an orthogonal decomposition, e.g. the singular value or QR decomposition~\cite{schollwock2011density}.  Thus, the tensors appearing in an MPS, which could result from a classical training optimization, can be formally (i.e., modulo compilation into native quantum operations for a given hardware architecture) translated into operations for deployment on quantum resources.

The above prescription assumed the presence of a register of $N$ qubits, but due to the sequential nature of the preparation this is unnecessary, and a single ``physical" qubit together with the $\chi$-level ancilla suffices, provided we are not measuring any multi-qubit properties of the state.  As an example, we will consider drawing a sample from an MPS wavefunction generative model with the binary map Eq.~\eqref{eq:BinaryMap}.  In this application, we first couple the qubit and ancilla as in Eq.~\eqref{eq:firstPrep} starting from both in the fiducial state $|0\rangle$.  We then measure the qubit in the computational basis, record its outcome as $x_{N-1}$, and then return it to the fiducial $|0\rangle$ state while leaving the ancilla unmeasured.  We note that the ability to re-initialize a single qubit independent of the others is not universally available in present-day hardware, but has been demonstrated in, e.g., trapped ion platforms~\cite{pino2020demonstration}.  We then re-entangle the ancilla and qubit using the operator $\hat{L}_{N-2}$ defined in Eq.~\eqref{eq:LNmo}, measure the qubit and record the outcome as $x_{N-2}$, and again return the qubit to the $|0\rangle$ state.  This procedure is repeated with the other operations $\hat{L}_j$ until a complete set of $N$ measurements $\mathbf{x}$ is made, which constitutes a data sample.  This procedure is denoted graphically in Fig.~\ref{fig:Overview} (d).  Clearly, this only requires a single ``physical"  or ``data" qubit (i.e., the one that is sampled) independent of the input data size $N$, and the construction of the $\chi$-level ancilla requires only $\log_2\chi$ qubits.  We stress that the scheme above is formal in the sense that it produces isometries acting on quantum resources without reference to their actual physical representation or other hardware constraints such as limited coherence time, connectivity, gate sets, etc..  The translation of these formal isometries into operations to be dispatched on a given target hardware are detailed in Sec.~\ref{sec:compilation}.

\subsection{Generative MPS models and classical training procedure}
\label{sec:ClassicalTraining}

We now further specialize to generative models, in which a collection of quantum data vectors are encoded into a wavefunction such that the probability distribution evaluated at data vector $\mathbf{x}$ is
\begin{align}
\label{eq:BornMachine}P\left(\mathbf{x}\right)&=\frac{\langle \psi |\Phi\left(\mathbf{x}\right)\rangle \langle \Phi\left(\mathbf{x}\right) |\psi\rangle}{Z}\, .
\end{align}
Here, $Z=\langle \psi|\psi\rangle=\int d\mathbf{x} \langle \psi |\Phi\left(\mathbf{x}\right)\rangle\langle \Phi\left(\mathbf{x}\right)|\psi\rangle$ is a normalization factor, and we assume the property Eq.~\eqref{eq:ONmap} holds for the Hilbert space encoding map.  As this corresponds to Born's rule for measurement outcomes, the resulting structure is referred to as a \emph{Born machine}~\cite{PhysRevX.8.031012,coyle2020born}.  

In order to discuss data representation using Born machines, we define the average log-likelihood of the data in the training set $\mathcal{T}$ as
\begin{align}
\mathcal{L}\left(\mathcal{T}\right)&=\frac{1}{N_T}\sum_{\mathbf{x}\in \mathcal{T}} \log \left[\frac{\langle \psi |\Phi\left(\mathbf{x}\right)\rangle\langle \Phi\left(\mathbf{x}\right)|\psi\rangle}{Z}\right]\, .
\end{align}
The minimization of the negative log-likelihood with respect to the parameters in our Born machine is equivalent to maximizing the probability that the data is generated by the Born machine.  We will parameterize the wavefunction to be trained as an MPS and assume that the data is encoded in terms of an orthonormal map as in Eq.~\eqref{eq:ONmap}, resulting in
\begin{align}
\nonumber &\mathcal{L}\left(\mathcal{T}\right)=\frac{1}{N_T}\sum_{\mathbf{x}\in \mathcal{T}} \log \Big[\sum_{i_0\dots i_{N-1}; i_0'\dots i_{N-1}'} \frac{\prod_j \phi_{i_j}^{\star}\left(x_j\right)\phi_{i_j'}\left(x_j\right)}{Z}\\
&\times \mathrm{Tr}\left[\mathbb{A}^{i_0\dagger}\dots \mathbb{A}^{i_{N-1}\dagger}\right]\mathrm{Tr}\left[\mathbb{A}^{i_0'}\dots \mathbb{A}^{i_{N-1}'}\right] \Big]\, ,
\end{align}
where the normalization factor (partition function) is
\begin{align}
Z&=\sum_{i_0\dots i_{N-1}}\mathrm{Tr}\left[\mathbb{A}^{i_0\dagger}\dots \mathbb{A}^{i_{N-1}\dagger}\right]\mathrm{Tr}\left[\mathbb{A}^{i_0}\dots \mathbb{A}^{i_{N-1}}\right]\, .
\end{align}

We will optimize the Born machine by a DMRG-style procedure using gradient descent, where the gradient is taken with respect to the tensors of the MPS.  Namely, we will consider the gradient with respect to a group of $s$ neighboring tensors $\Theta=\mathbb{A}^{i_l}\dots \mathbb{A}^{i_{l+s}}$, with $s$ typically being one or two, noting that the gradient of an object with respect to a tensor is a tensor whose elements are the partial derivatives with respect to the individual tensor elements.  We take the gradient with respect to the conjugates of the tensors $\mathbb{A}^{i_j}$, formally considering these conjugates independent of the tensors themselves.  This gradient may be written as
\begin{align}
\nabla_{\Theta^{\star}} \mathcal{L}\left(\mathcal{T}\right)=&\frac{1}{N_T}\sum_{\mathbf{x}\in\mathcal{T}}\frac{\nabla_{\Theta^{\star}} \langle \psi |\Phi\left(\mathbf{x}\right)\rangle\langle\Phi\left(\mathbf{x}\right)|\psi\rangle}{\langle \psi |\Phi\left(\mathbf{x}\right)\rangle\langle\Phi\left(\mathbf{x}\right)|\psi\rangle}-\frac{\nabla_{\Theta^{\star}} Z}{Z}\, ,\\
\nonumber =&\frac{1}{N_T}\sum_{\mathbf{x}\in\mathcal{T}}\frac{\nabla_{\Theta^{\star}} \langle \psi |\Phi\left(\mathbf{x}\right)\rangle\langle\Phi\left(\mathbf{x}\right)|\psi\rangle}{\langle \psi |\Phi\left(\mathbf{x}\right)\rangle\langle\Phi\left(\mathbf{x}\right)|\psi\rangle}\\
&-\sum_{\mathbf{x}}\frac{\nabla_{\Theta^{\star}} \langle \psi |\Phi\left(\mathbf{x}\right)\rangle\langle\Phi\left(\mathbf{x}\right)|\psi\rangle}{Z}\, .
\end{align}
With this gradient in hand, we update the local block of tensors as
\begin{align}
\Theta&\to \Theta +\eta \nabla_{\Theta^{\star}} \mathcal{L}\left(\mathcal{T}\right)\, ,
\end{align}
in which $\eta$ is a learning rate (note that this is equivalent to minimizing the negative log likelihood).  For the single-site algorithm ($s=1$), this update does not change the bond dimension or canonical form of the MPS.  For the two-site algorithm ($s=2$), we can now split the updated tensor $\Theta$ into its component MPS tensors as
\begin{align}
\label{eq:ThetaSplit} \Theta_{\alpha \beta}^{ij}&=\sum_{\mu} A_{\alpha\mu}^{i}A_{\mu \beta}^j\, ,
\end{align}
using, e.g., the SVD.  Hence, the addition of the gradient can increase the bond dimension, and thus the representation power, adaptively based on the data.  The bond dimension can also be set implicitly by requiring that the $L_2$-norm of the tensor $\Theta$ is represented with a bounded relative error $\varepsilon$.  The above update has affected only a small group of the tensors with all others held fixed.  We now shift the orthogonality center to a neighboring tensor, and perform the same local optimization procedure.  For the two-site case, the shift of orthogonality center can be accomplished simultaneously with the splitting of the tensor $\Theta$ in Eq.~\eqref{eq:ThetaSplit}.  In the one-site case, the orthogonality center is moved to the next tensor in the optimization cycle using either the SVD or the QR decomposition.  A complete optimization cycle, or ``sweep," occurs when we have updated all tensors twice, moving in a back-and-forth motion over the MPS.  The sweeping process is converged once the negative log-likelihood no longer decreases substantially.  Example convergence behavior will be given later in Sec.~\ref{sec:MNIST}.

\section{Compilation of MPS models for quantum hardware}
\label{sec:compilation}
In this section, we address how to take an MPS model resulting from the classical optimization procedure outlined in Sec.~\ref{sec:ClassicalTraining} and convert it into a sequence of operations to be performed on a quantum device.  We will refer to this operation as \emph{quantum compilation}.  Many modern NISQ software ecosystems, for example Qiskit~\cite{Qiskit} and Forest~\cite{smith2016practical}, have routines for compiling quantum instructions, usually required to be supplied in the form of an abstract quantum circuit model.  These compilers typically perform multiple passes through the abstract circuit to map virtual qubits from the abstract model onto the hardware qubits of the device, route operations between the virtual qubits to hardware qubits, e.g., by placing SWAP gates, and optimization to minimize some property of the circuit, such as entangling gate count.  We note that quantum compilation remains an active area of research, and currently available \emph{generic} methods for quantum compilation tend to produce ``deep" circuits with significant numbers of entangling gates.

There are several unique properties of our particular quantum computing use case --compiling isometries encoding TN models for QAML-- that make them unique compared to traditional quantum computing use cases.  For one, our isometries are defined on the Hilbert space of a physical qubit and a formal $\chi$-level ancilla, and so may not uniquely describe an isometric operation on a set of virtual qubits, e.g., when $\chi$ is not a power of 2.  Further, since the ancilla degrees of freedom are never directly measured, there is no preferred basis or state ordering for these states.  Both of these properties give freedom that can be utilized to simplify compilation.  In addition, the isometries are the result of an optimization procedure that has a finite tolerance (see Sec.~\ref{sec:ClassicalTraining}), and so do not need to be compiled exactly to meet some fine-tuned property.  That is to say, model predictions are not more accurate when using a compiled unitary that matches the isometry better than the optimization tolerance.  For NISQ devices in particular, fine-tuning of isometry properties through the introduction of additional entangling gates may in fact produce worse results due to the increased noise in the circuit compared to a shallower representation.  These properties have motivated us to pursue optimizations of the tensor network structure as well as a set of greedy compilation heuristics, inspired by Ref.~\cite{davis2019heuristics}, that we outline in what follows.

\begin{figure}[h]
  \begin{center}
\includegraphics[width=0.7\columnwidth]{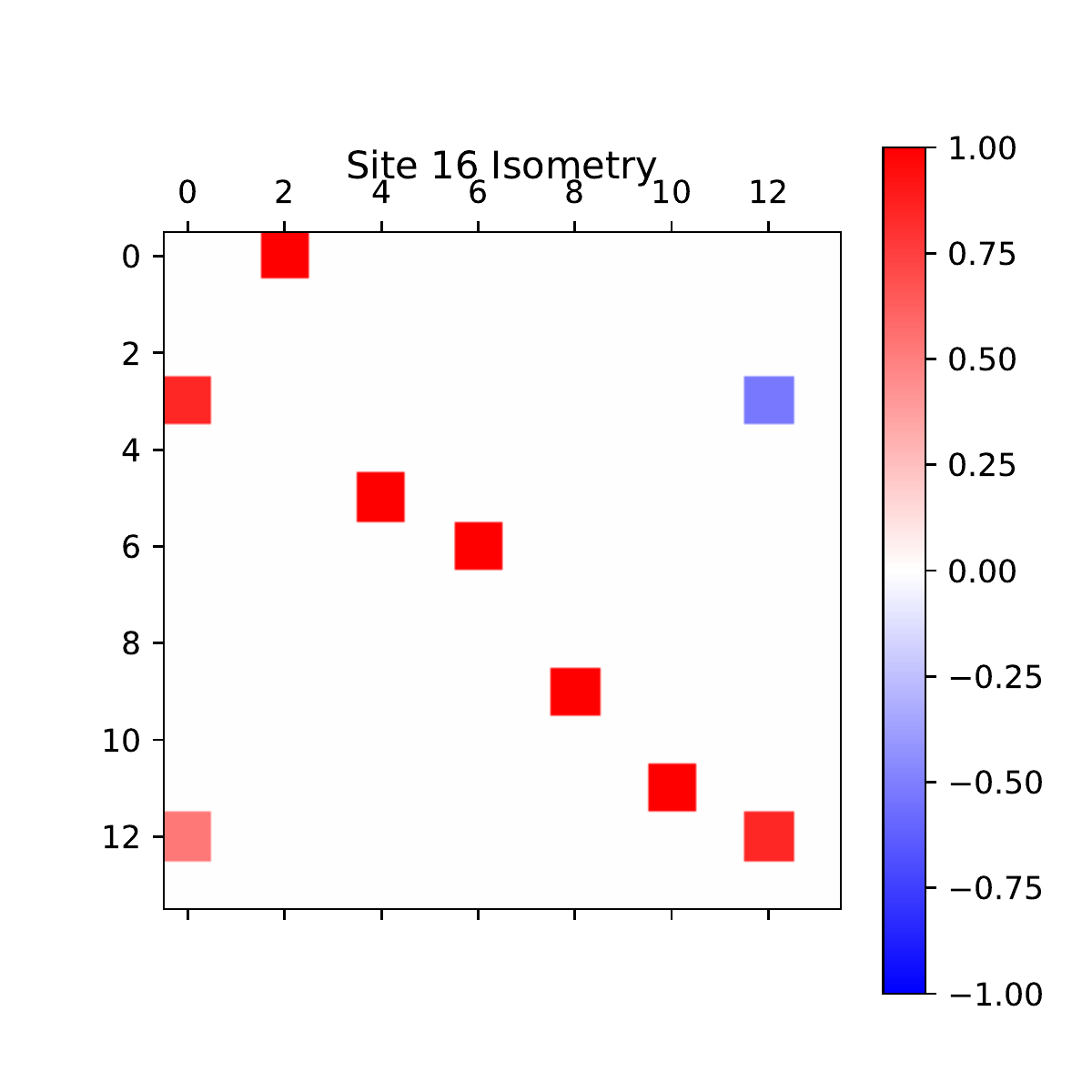}   
\caption{\label{fig:IsoExample} \emph{Example isometry for optimization}.   An example isometry acting on a single physical qubit in the state $|0\rangle$ and a $\chi=7$-level ancilla, taken from the MNIST example in Sec.~\ref{sec:MNIST}.  The isometry has been cleaned to remove small numerical values resulting from classical optimization, but no further optimization has been applied.}
\end{center}
\end{figure}

The key objects that we want to optimize in this section are the isometries $\hat{L}^{[i]}$ defined by the elements of the MPS in left-canonical form, see Sec.~\ref{sec:TNSP}.  Given that the binary encoding map used in this work, Eq.~\eqref{eq:BinaryMap}, is real-valued, all MPS tensors are real-valued, and this extends to the isometries.  We will display the isometries using plots of their matrix representations in a fixed basis, as in Fig.~\ref{fig:IsoExample}.  In this and similar plots, the basis ordering is defined with the physical qubit (i.e., the qubit that begins in the $|0\rangle$ state and is read out after each isometric operation) as the least significant qubit such that an isometry acting on a $\chi$-dimensional ancilla $\alpha\in\{0,\dots,\chi-1\}$ and a physical qubit $q\in\{0,1\}$ has state indices
\begin{align}
\mathrm{index}\left(|\alpha q\rangle\right)&=2\alpha+q\, .
\end{align}
For isometries that have their ancilla states decomposed into qubits, we order those qubits $a_i\in\{0,1\}$ such that significance increases with label index $i$, i.e.
\begin{align}
\mathrm{index}\left(|a_{n_{\mathrm{anc}}}\dots a_{1} q\rangle\right)&=\sum_{i=1}^{n_{\mathrm{anc}}} 2^{i} a_i+q\, ,
\end{align}
The isometry in Fig.~\ref{fig:IsoExample} acts on a physical qubit and a $\chi=7$ dimensional ancilla, transforming the state $|00\rangle$ into a superposition of $|11\rangle$ and $|60\rangle$, the state $|10\rangle$ into $|00\rangle$, and so on.  We note that the isometry in Fig.~\ref{fig:IsoExample} is undefined when acting on states with $|q=1\rangle$ in accordance with the sequential preparation scheme, but takes arbitrary ancilla states as inputs.  Because of the isometry property, we only need to account for the nonzero elements of the operation when matching to a unitary, and so do not need to distinguish between zero elements and undefined elements.

As a first step in compilation, we will want to ``clean" the isometries from the classical model in order to remove noise at the level of the classical optimization tolerance, otherwise we will expend effort attempting to compile this noise into quantum operations that will not improve the fidelity of the calculation.  This amounts to implementing a filter on the MPS to remove elements below some tolerance level $\varepsilon$, which can be accomplished by using MPS compression to find the MPS with specified resources (e.g., restricted bond dimension $\chi$) $|\phi\rangle$ that is closest in the $L_2$-norm to a target MPS $|\psi\rangle$ that has higher resource requirements ($\chi'$).  While this is optimally done variationally~\cite{schollwock2011density}, a simple and practical method for performing this operation is to use local SVD compression, in which the MPS tensor of the orthogonality center $A^{[i]}$ is decomposed by the SVD as
\begin{align}
\label{eq:LCFrecur} A^{[i]j_i}_{\alpha\beta}&\to \sum_{\mu} U_{\left(\alpha j_i\right)\mu}S_{\mu} V_{\mu \beta}\, ,\\
\label{eq:RCFrecur} A^{[i]j_i}_{\alpha\beta}&\to \sum_{\mu} U_{\alpha\mu}S_{\mu} V_{\mu \left(j_i \beta\right)}\, ,
\end{align}
where the upper expression is for a right-moving update and the lower for a left-moving update.  We can truncate the bond dimension by keeping only the $\chi$ largest singular values, or determine the new bond dimension implicitly through a singular value cutoff $\varepsilon$ as
\begin{align}
1-\sum_{\mu=1}^{\chi} S_{\mu}^2/\sum_{\mu}S_{\mu}^2<\varepsilon\, .
\end{align}
When the MPS tensor is the orthogonality center, this condition is equivalent to a $L_2$-norm optimization of the full wavefunction.  Replacing $A^{[i]}$ by the truncated $U$ for a right-moving update or by $V$ for a left-moving update and contracting the truncated $SV$ or $US$ into the neighboring tensor completes the local optimization.  Sweeping the optimization across all tensors completes the filtering step.  Since the optimization only deals with the parameters of a single MPS tensor at a time, it is not guaranteed to be globally optimal, but this simple procedure works well in practice.  As a side benefit, ending the optimization by applying the update Eq.~\eqref{eq:LCFrecur} and replacing the MPS tensor $A^{[i]}$ with $U$ for each tensor places the MPS in left-canonical form, from which the isometries for sequential preparation can be constructed from the tensor elements.

\subsection{Ancilla permutation and the diagonal gauge}
\label{sec:DG}
The conversion of an MPS into left canonical form uses the gauge freedom inherent in MPSs, namely that any invertible matrix $\mathbb{X}$ and its inverse can be placed between any two tensors of the MPS, i.e.
\begin{align}
\tilde{\mathbb{A}}^{[i]j_i}&=\mathbb{A}^{[i]j_i}\mathbb{X}\, ,\\
\tilde{\mathbb{A}}^{[i+1]j_{i+1}}&=\mathbb{X}^{-1}\mathbb{A}^{[i+1]j_{i+1}}\, ,
\end{align}
such that each of the tensors in the left-canonical MPS satisfies the isometry constraint
\begin{align}
\label{eq:Isoconstraint}\sum_{\alpha j}L^{[i]j}_{\alpha\beta}L^{[i]j\star}_{\alpha\beta'}&=\delta_{\beta\beta'}\, ,
\end{align}
without changing the overall quantum state.  However, we note that the constraint Eq.~\eqref{eq:Isoconstraint} still allows for the insertion of any \emph{unitary} matrix and its inverse on either the left or right bond basis of an MPS tensor $\mathbb{L}^{[i]j}$ without changing the state or the isometry conditions.  This freedom stems from the fact that the bond degrees of freedom are only used to mediate correlations between the physical degrees of freedom and are not directly measured, and so have no preferred basis for representation.  We can attempt to exploit this freedom to produce MPS models that are more amenable to compilation on a given target hardware.  We note that, just as with the ordinary gauge freedom of MPSs, a change of gauge affects two neighboring MPS tensors at a time, and so an operation that may benefit one tensor also affects its neighbors and so on down the network. Thus, the optimal choice of gauge requires a global optimization across all tensors.

\begin{figure}[t]
  \begin{center}
\includegraphics[width=0.49\columnwidth]{Site16NPIsometry.pdf} \includegraphics[width=0.49\columnwidth]{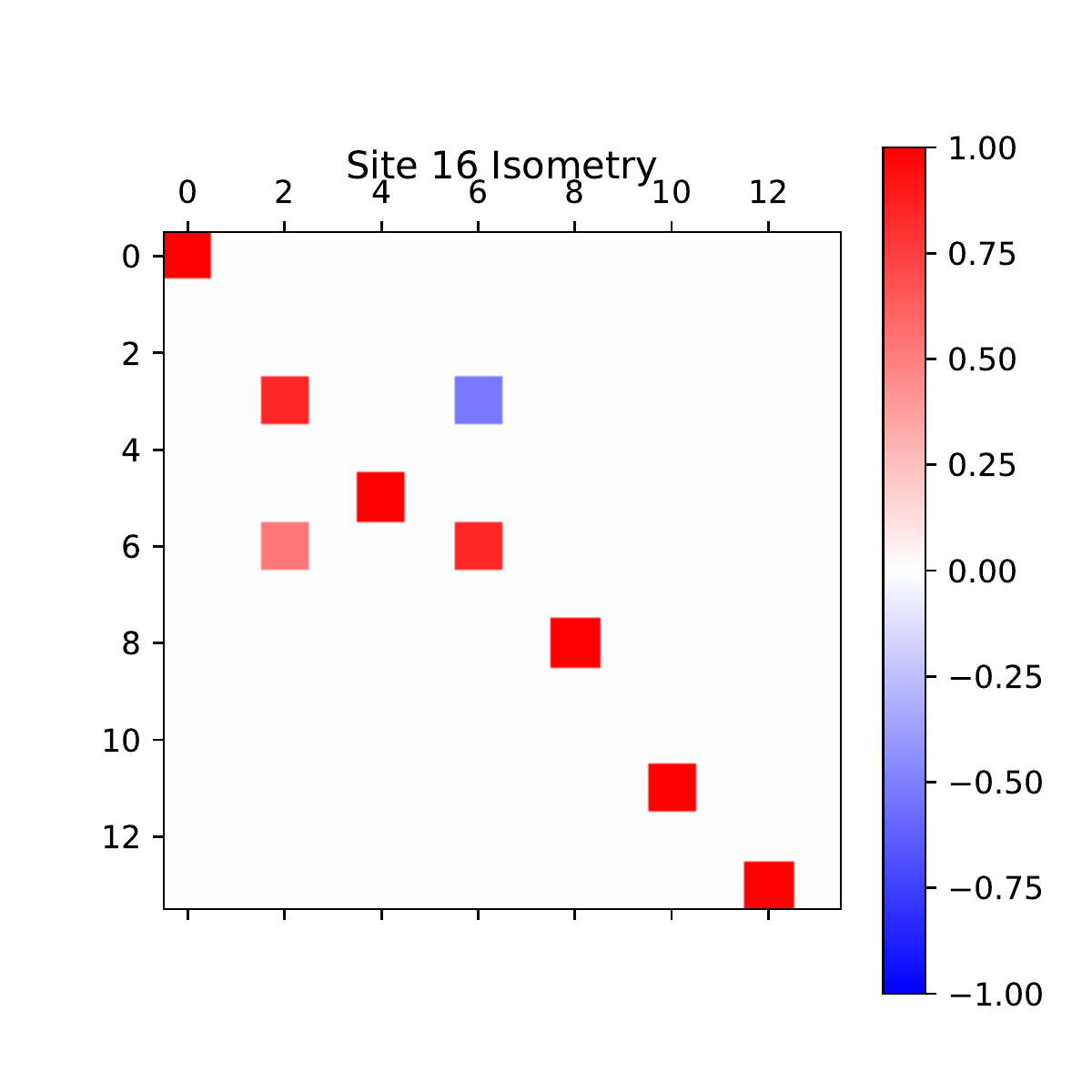}  
\caption{\label{fig:DiagonalGauge} \emph{Application of diagonal gauge.} Example isometries for an operations with a $\chi=7$ dimensional ancilla before (left panel-same isometry as Fig.~\ref{fig:IsoExample}) and after (right panel) applying the diagonal gauge transformation Eq.~\eqref{eq:Lperm} to the right ancilla basis states.}
\end{center}
\end{figure}

To utilize the ambiguity in the basis representation of the ancilla states, we have devised a simple procedure that we have found to aid in compiling isometries for QAML models.  The heuristic guiding our scheme is to ensure that operations are as ``diagonal" as possible, in the sense that qubits preferentially remain in their same state rather than being swapped or mixed with other ancilla qubits.  Operationally, in order to work only within the ancilla basis where we have freedom of representation, we define a matrix of overlaps
\begin{align}
M_{\alpha\beta}^{[i]}&=\sum_{j} L^{[i]j\star}_{\alpha \beta} L^{ [i]j}_{\alpha\beta}\, ,
\end{align}
which ``integrates out" the physical qubit from the isometry used for sequential preparation, and so acts only in the ancilla space.  A diagonal $\mathbb{M}$ is desired, as this would perfectly preserve the individual ancilla basis states and so reduce the number of quantum operations required.  Recalling that we are only changing either the left or right basis of $\mathbb{M}$ at a time, one possible option to increase its diagonal dominance through transformation of either the left or right basis is to use the polar decomposition $\mathbb{M}\to\mathbb{U}\mathbb{P}$ or $\mathbb{M}\to\mathbb{P}\mathbb{U}$ with $\mathbb{U}$ unitary and $\mathbb{P}$ Hermitian and positive semidefinite.  Using $\mathbb{U}^{1/2}$ to transform the basis of $L$ would transform $\mathbb{M}$ into $\mathbb{P}$; however, this transformation does not preserve sparsity in $L$, and we have found that it often leads to more complex operators in practice.  Instead, we use the values of $\mathbb{U}$ from the polar decomposition to define a permutation of the ancilla basis states as, e.g.,
\begin{align}
\label{eq:Lperm}\tilde{L}^{[i]j}_{\alpha, \mathrm{argmax}|\mathbb{U}_{:\beta}|}&=L^{[i]j}_{\alpha\beta}\, .
\end{align}
This operation does preserve sparsity, and results in more diagonal operations in the ancilla degrees of freedom.  An example of the isometries for a QAML model with and without this permutation procedure are shown in the right and left panels of Fig.~\ref{fig:DiagonalGauge}, respectively.  We see that the permutation of the basis states does result in a more diagonal isometry operator, as desired.

The permutation operation Eq.~\eqref{eq:Lperm} is ambiguous whenever multiple elements of a column of $\mathbb{U}$ have the same absolute value.  Recalling that our sequential MPS preparation scheme requires that the ancilla start and end in the vacuum state, we see that this occurs for tensors near the extremal values of the representation when an ancilla qubit is first utilized or an ancilla qubit is decoupled from the remaining qubits.  In such cases, we use the following alternate procedure to decide between permutations.  First, we enumerate all basis permutations resulting from these ambiguities for a given tensor $\mathbb{L}^{[i]j_i}$ and construct their associated isometries $\hat{\tilde{L}}^{(\zeta)}$, in which $\zeta$ indexes permutations.  To decide between these permutations, we again would like to make this operator as ``diagonal" as possible, in the sense of minimizing the number of qubit operations being applied.  We construct a simple cost function as follows: for each state indexed by the ancilla state $\alpha$ and the physical qubit $q$ as above, we convert the state index into its binary representation $\mathbf{b}$, which effectively maps the ancilla state onto a collection of $\log_2\chi$ qubits.  As an example, the states of a four-dimensional ancilla and a single physical qubit give the representations
\begin{align}
\mathrm{index}\left(|0,0\rangle \right)&=0\to \left(0,0,0\right)\, ,\\
\mathrm{index}\left(|0,1\rangle \right)&=1\to \left(0,0,1\right)\, ,\\
\mathrm{index}\left(|1,0\rangle \right)&=2\to \left(0,1,0\right)\, ,\\
\mathrm{index}\left(|1,1\rangle \right)&=3\to \left(0,1,1\right)\, ,\\
\vdots &\, , \\
\mathrm{index}\left(|3,1\rangle \right)&=7\to \left(1,1,1\right)\, .
\end{align}
We now calculate a distance between two basis states $\left(\alpha,j\right)$ and $\left(\alpha',j'\right)$ with respective binary representations $\mathbf{b}$ and $\mathbf{b}'$ as $\mathcal{D}\left[\left(\alpha,j\right),\left(\alpha',j'\right)\right]=\left(\sum_{\mu}\left|b_{\mu}-b_{\mu}'\right|\right)^2$.  The term in parentheses counts the number of individual qubit ``flips" required to convert one of the states into the other, and the square strongly penalizes multi-qubit coordinated flips.  We then use the cost function 
\begin{align}
\label{eq:Czeta} \mathcal{C}_{\zeta}&=\mathrm{Tr}\left(\left|\mathbb{L}^{(\zeta)}\right|\mathbb{D}\right)\, ,
\end{align}
in which $\mathbb{D}$ is the matrix with $\mathcal{D}\left[\bullet,\bullet\right]$ as elements and $\left|\mathbb{L}^{(\zeta)}\right|$ is the matrix of absolute values of $\mathbb{L}^{(\zeta)}$, to choose from between the $\mathbb{L}^{(\zeta)}$.

As with the usual transformation of MPS gauge to mixed canonical form~\cite{schollwock2011density}, there is a ``right-moving" update that permutes the right bond basis of a tensor $\mathbb{A}^{[i]}$ and the left bond basis of $\mathbb{A}^{[i+1]}$ and a ``left-moving" update that permutes the left bond basis of $\mathbb{A}^{[i]}$ and the right bond basis of $\mathbb{A}^{[i-1]}$.  When applied to all tensors, we say that the MPS is in the \emph{diagonal gauge}, as it is the gauge which enforces the isometries for state preparation to be as diagonal as possible (according to our particular cost functions).  We stress that the MPS is still in left-canonical form, and so the sequential preparation scheme still holds; the diagonal gauge merely uses the unitary freedom remaining in the left-canonical form to further optimize the state preparation procedure while maintaining sparsity.  There is a single tensor that is not optimized at a certain location $k$ in the transformation to the diagonal gauge that we call the \emph{diagonality center}, analogous to the orthogonality center of mixed canonical form.  While the location of the diagonality center can again be used as an optimization parameter, we have found it convenient to set the diagonality center to an isometry that is initially an identity matrix.  Such an isometry can always be introduced by padding the classical data vectors with a zero at location $k$.  The reason for our choice is that the permutation to diagonal gauge will transform this identity isometry into a permutation matrix, which is likely to be easier to compile with high fidelity than a general, non-sparse isometry.  Specific techniques for compilation will be presented in a later section.

In addition to the permutation ambiguity, there is also a sign ambiguity on each of the bond states of the isometry.  We again use diagonal dominance in fixing this sign ambiguity by reversing the sign of a column (row) if the element with magnitude above a certain threshold closest to the diagonal is negative during a right-moving (left-moving) update of the diagonal gauge, with the sign also being absorbed into the tensor to the right (left) of the one being optimized.  Following transformation to diagonal gauge, we fix the signs of all elements of the diagonality center (chosen, as above, to be a permutation operator) to be positive by absorbing any negative signs into the nearest tensor that has elements of mixed sign in the chosen bond direction.

\subsection{Greedy compilation heuristics}
\label{sec:Compilation}
\begin{figure}[b]
  \begin{center}
\includegraphics[width=0.75\columnwidth]{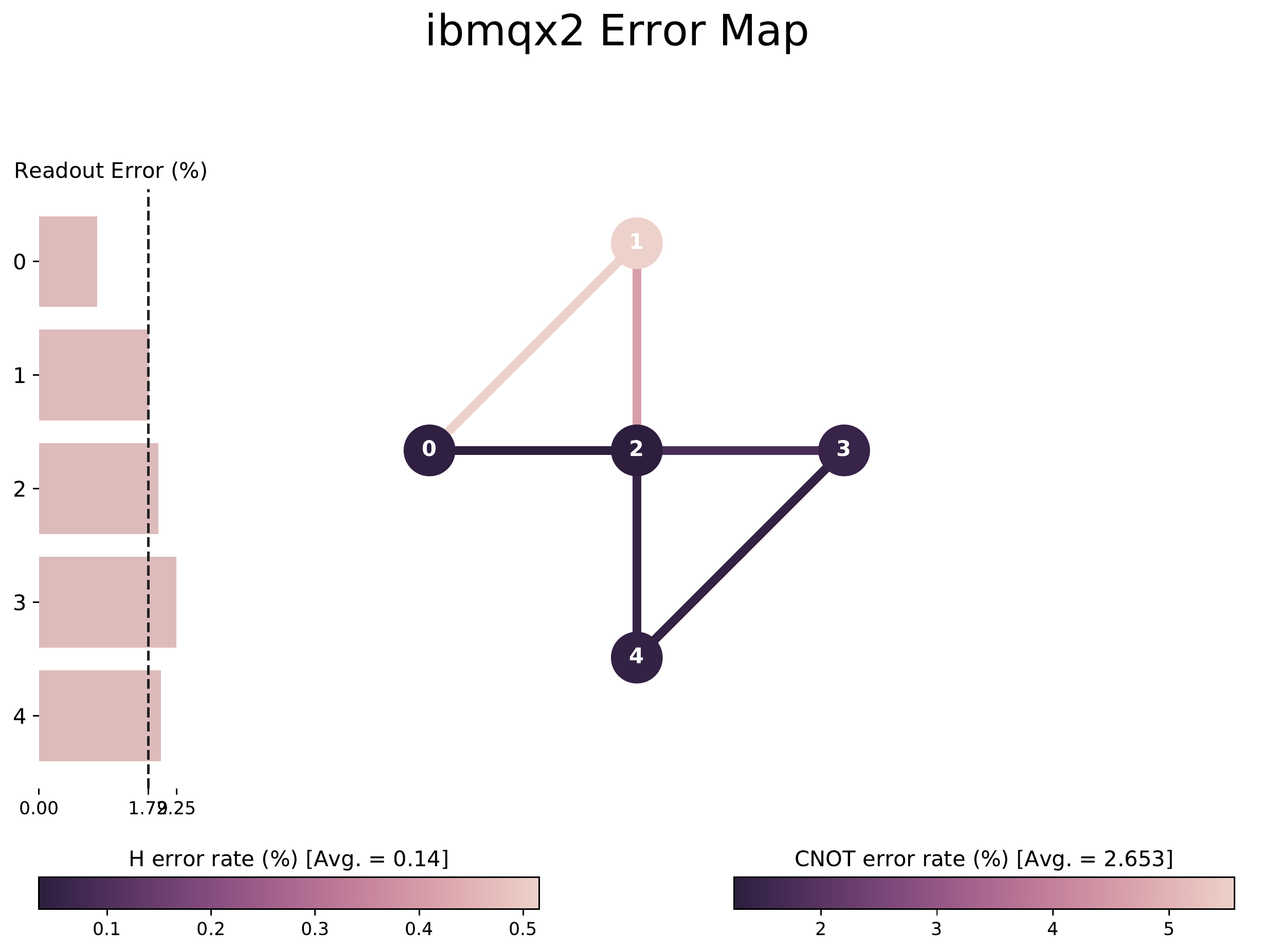}  
\caption{\label{fig:IBMQX2Err}  \emph{Exemplar NISQ hardware architecture.} The qubit layout (circles), CNOT coupling topology (lines), and error sources of the 5-qubit IBMQ-X2 device~\cite{Qiskit} as an exemplar NISQ machine.  We note that these error rates are a snapshot, and are subject to fluctuations.}
\end{center}
\end{figure}

Following the fixing of gauge outlined in the last subsection, we are in a position where we now want to transform the isometries $\hat{L}_i$ into operations to be performed on quantum hardware.  The target hardware will have a collection of qubits laid out with a given topology and an allowed gate set of single-qubit rotations and entangling gates between pairs of qubits.  Generally speaking, two-qubit gates are subject to higher degrees of noise than the single-qubit gates, and so higher-fidelity operations will be obtained by using as few two-qubit gates as possible.  As an example, the error map and qubit/gate topology for the IBMQ-X2 machine is shown in Fig.~\ref{fig:IBMQX2Err}.  For this device, the single-qubit gates are defined by~\cite{Qiskit}
\begin{align}
\label{eq:U3} \hat{U}_3\left(\theta,\phi,\lambda\right)&=\left(\begin{array}{cc} \cos\frac{\theta}{2} &-e^{i\lambda}\sin\frac{\theta}{2} \\ e^{i\phi}\sin\frac{\theta}{2}&e^{i(\lambda+\phi)}\cos\frac{\theta}{2}\end{array}\right)\, ,
\end{align}
and the two-qubit gates are controlled-NOT (CNOT) gates, which are allowed only between qubits designated with a solid line in Fig.~\ref{fig:IBMQX2Err}.  As shown in the figure, the average error of the CNOT gates at the time of this measurement was $\sim 2.6\%$, while the error of the single-qubit gates was $\sim 0.15\%$.  Hence, a goal in compiling our isometries is to use as few gates as possible, and especially to minimize the number of two-qubit gates.

In our compilation heuristic, we enumerate possible unitaries by constructing a tree of potential circuit structures with continuous parameters to be optimized.  The root node of our tree is comprised of a single-qubit gate (such as the $\hat{U}_3$ gate in Eq.~\eqref{eq:U3}) for each qubit.  Each node in the tree has a child node corresponding to the placement of an entangling gate in one of its allowed positions, and then adding single-qubit gates to the qubits acted on by the entangling gate.  Any circuit that can be constructed using the allowed entangling gates and single-qubit rotations corresponds to a node in this tree, as proved in Ref.~\cite{davis2019heuristics}.  In order to select between nodes in this tree, we define a cost function
\begin{align}
\label{eq:AStarCostFunction} \mathcal{C}\left(\hat{U},\hat{L}\right)&=\sum_{\left(i,j\right)\in\mathcal{S}}\left|{U}_{i,j}-{L}_{i,j}\right|^2\, ,
\end{align}
in which $\mathcal{S}$ denotes the set of indices such that the elements of the matrix representation of the isometry are greater than some tolerance $|{L}_{i,j}|>\delta$.  Because of the isometry property of $\hat{L}$ and the unitarity of the candidate gates $\hat{U}$, we can optimize only over the elements in $\mathcal{S}$, which reduces the computational complexity of the cost function.  Our optimization will select a particular unitary $\hat{U}$ as being acceptable when the cost function drops below a specified tolerance $\varepsilon$.  

The optimization procedure begins by optimizing the root node (single-qubit gates) over its parameters and checking the cost function if an acceptable gate is found.  If no acceptable gate is found, a queue of gates corresponding to adding an entangling gate and a pair of single-qubit gates to the root node in all allowed locations as outlined above is formed, and these gates are optimized and their cost functions recorded.  If no gate from this queue is acceptable, a priority queue is formed by sorting the gates from this set according to their cost functions and then appending entangling gates and single-qubit rotations as above.  In order to avoid an exponential growth of the number of search considerations, we limit the number of gates forming the starting point of the priority queue (i.e., before appending new entangling gate and single-qubit rotations) to a fixed number.  This number is used as a convergence parameter, and can vary between optimization cycles; we find that it is useful to allow more gates in early optimization cycles where the operations involve fewer parameters and so optimization is fast, and then to decrease the number of kept gates as the circuits become deeper.  Also, we note that it may be useful to add a gate-dependent heuristic function $h$ to the cost function when sorting gates to add to the priority queue, as advocated in Ref.~\cite{davis2019heuristics}.   This can be used to account for, e.g., hardware-dependent noise~\cite{cincio2020machine}; we will return to our choice of this function shortly.

Here, we briefly note details of our implementation of the above procedure, along with some problem-specific optimizations.  Our subroutine for the cost function takes as input a vector of parameters $\boldsymbol{\theta}$, constructs a matrix representation of the parameterized gate sequence
\begin{align}
\hat{U}\left(\boldsymbol{\theta}\right)&=\hat{M}_{N_G}\left(\boldsymbol{\theta}_{N_G}\right)\dots\hat{M}_{1}\left(\boldsymbol{\theta}_{1}\right)\, ,
\end{align}
in which $\boldsymbol{\theta}_{i}$ is the vector of parameters used by gate $i$, and then evaluates the cost function Eq.~\eqref{eq:AStarCostFunction}.  This enables us to obtain analytic gradients of the cost function also as elements of products of matrices.  We optimize the cost function using the BFGS method, and allow for multiple batches of input parameters with random variations added to avoid local minima.  Additionally, as noted above, all of the isometries that result from the use of a real-valued quantum embedding map will be real, and so we can restrict our attention to real-valued gates.  Hence, in our implementation, we parameterize single-qubit gates as $y$-rotations
\begin{align}
\hat{R}_y\left(\theta\right)&\equiv \left(\begin{array}{cc} \cos\frac{\theta}{2} &-\sin\frac{\theta}{2} \\ \sin\frac{\theta}{2}&\cos\frac{\theta}{2}\end{array}\right)\, ,
\end{align}
which relate to the gates in Eq.~\eqref{eq:U3} as $\hat{R}_y\left(\theta\right)=\hat{U}_3\left(\theta,0,0\right)$, and CNOTs for the entangling gates.  While we have made the above gate choices for use in this paper, we stress that our methods apply to any other choice of single-qubit and entangling gates.  While there is no guarantee that there are not operations with fewer entangling gates that could be found using complex-valued gates, we find that the reduction in the number of parameters when using real gates significantly improves the optimization time.

\begin{figure*}
  \begin{center}
\includegraphics[width=1.99\columnwidth]{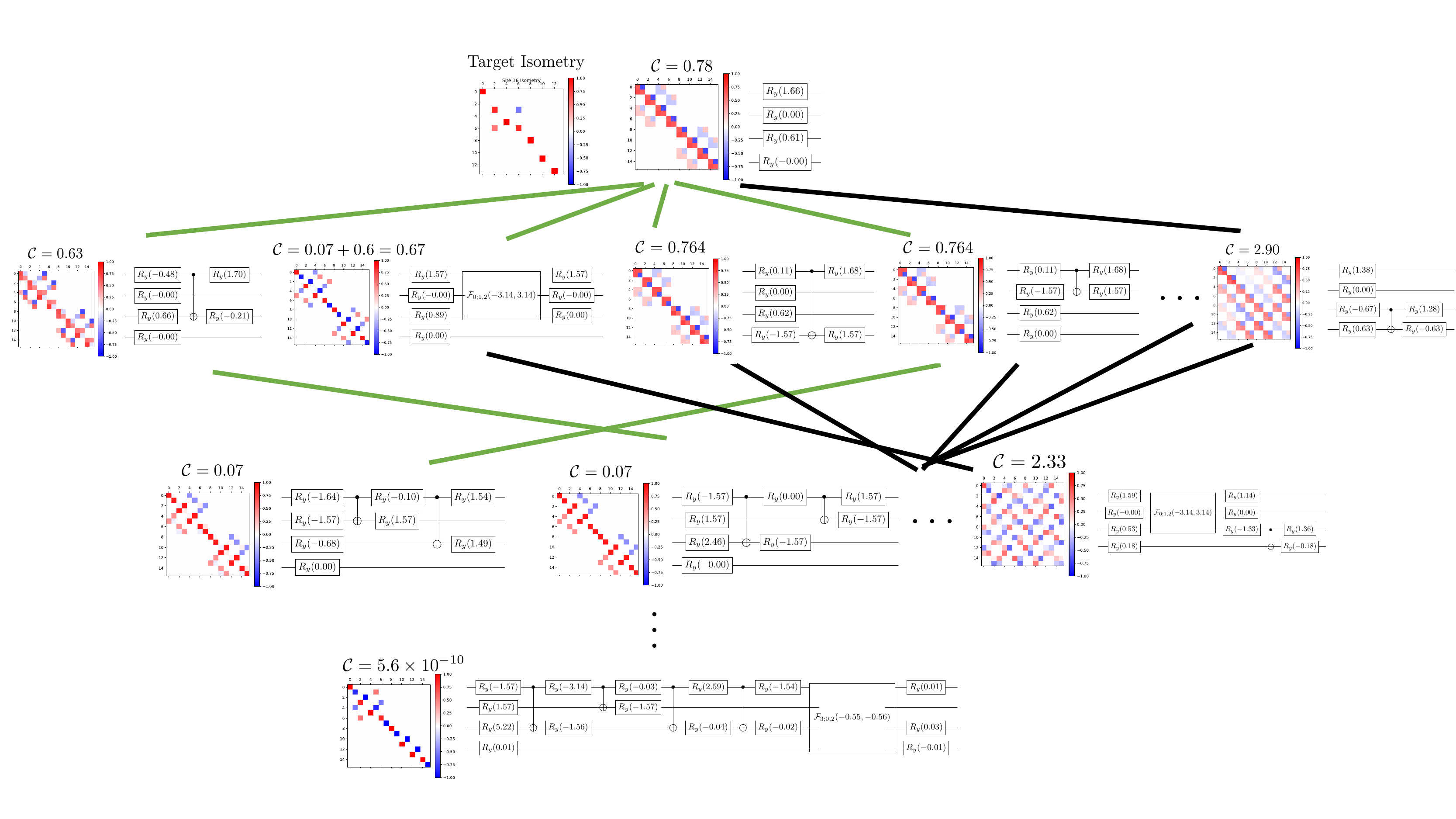}  
\caption{\label{fig:AStarFigure}  \emph{Example of greedy compilation procedure.}  Example gates represented as circuits and matrix plots resulting from applying the greedy compilation procedure to the isometry shown in the upper left (same isometry as in the right panel of Fig.~\ref{fig:DiagonalGauge}).  The starting ansatz is a single-qubit rotation on each qubit, given in the top center of the figure.  The next row down shows the gates resulting from adding a single entangling gate to this ansatz, ordered left to right by their cost functions $\mathcal{C}$.  A constant penalty $0.6$ is added to the cost function for use of a $\hat{\mathcal{F}}$ gate in ordering the priority queue, resulting in the given ordering.  The gates indicated by green lines denote those passed to the next level of optimization.  This procedure terminates in the gate shown at the bottom of the figure with the given cost function tolerance of $5\times 10^{-4}$.}
\end{center}
\end{figure*}

The final optimization we have included is to introduce longer gate sequence ``motifs" into the optimization alongside the native entangling gates.  In particular, the two motifs we have utilized in our work are a two-qubit rotation gate
\begin{align}
\label{eq:Sdef}&\hat{\mathcal{S}}\left(\theta,\theta'\right)=\\
\nonumber &\left(\begin{array}{cccc} \cos\left(\frac{\theta-\theta'}{2}\right)&0&0&\sin\left(\frac{\theta-\theta'}{2}\right)\\ 0&\cos\left(\frac{\theta+\theta'}{2}\right)&\sin\left(\frac{\theta+\theta'}{2}\right)&0 \\ 0&-\sin\left(\frac{\theta+\theta'}{2}\right)&\cos\left(\frac{\theta+\theta'}{2}\right)&0 \\ -\sin\left(\frac{\theta-\theta'}{2}\right)&0&0&\cos\left(\frac{\theta-\theta'}{2}\right)\end{array}\right)\, ,
\end{align}
which is allowed between any two qubits that have CNOT connectivity, and a version of the $\hat{\mathcal{S}}$ gate we call $\hat{\mathcal{F}}$ that is controlled on a third qubit.  We find that the former gate can be compiled using two CNOTs using the ansatz sequence shown in Eq.~\eqref{eq:Scomp}
\begin{equation}
\label{eq:Scomp} 
\Qcircuit @C=1em @R=.7em {&\gate{R_y\left(\phi_0\right)}&\ctrl{1}&\gate{R_y(\phi_1)}&\ctrl{1}&\gate{R_y\left(\phi_2\right)}&\qw\\&\gate{R_y(\phi_0')}&\targ&\gate{R_y(\phi_1')}&\targ&\gate{R_y(\phi_2')}&\qw\\}\, ,
\end{equation}
and the latter gate with control on qubit $c$ and the operation $\hat{\mathcal{S}}$ applied to qubits $q_1$ and $q_2$ can be constructed using 
\begin{align}
\nonumber &\hat{\mathcal{F}}_{c;q_1q_2}\left(\theta,\theta'\right)=\mathrm{CNOT}\left(c,q_2\right)\mathrm{CNOT}\left(c,q_1\right)\hat{\mathcal{S}}_{q_1q_2}\left(-\frac{\theta}{2},-\frac{\theta'}{2}\right)\\
\label{eq:Fcomp}&\times \mathrm{CNOT}\left(c,q_2\right)\mathrm{CNOT}\left(c,q_1\right)\hat{\mathcal{S}}_{q_1q_2}\left(\frac{\theta}{2},\frac{\theta'}{2}\right)\, .
\end{align}
Hence, $\hat{\mathcal{S}}$ gates require 2 CNOTs for compilation and $\hat{\mathcal{F}}$ gates require 8 CNOTs for compilation.  Both gates were identified from experiments with the greedy optimization procedure outlined above using only CNOTs, and their direct inclusion into the optimization enables more rapid convergence.  As these gates require multiple entangling gates, it is useful to introduce a heuristic penalty function $h$ into the cost function for ordering the next priority queue to ensure that they are not chosen over shorter gates with a similar cost function.  The choice of this penalty function will be problem-specific, and finding ways for optimizing it in a data-driven fashion for problems of interest is an intriguing area for further research.  We also note that the use of multi-qubit controlled gates is penalized through the choice of the cost function Eq.~\eqref{eq:Czeta} for choosing the permutation to diagonal gauge; the choice of a cost function of 4 or 8 for a gate requiring two and three bit flips is in rough accordance with the number of CNOTs required for $\hat{\mathcal{S}}$ and $\hat{\mathcal{F}}$, respectively.

An example application of this procedure to the isometry shown in the right panel of Fig.~\ref{fig:DiagonalGauge} is given in Fig.~\ref{fig:AStarFigure}.  Here, we give cost function penalties of $0.6$ and $0.2$ for $\hat{\mathcal{F}}$ and $\hat{\mathcal{S}}$ gates, respectively, use a cost function tolerance of $5\times 10^{-4}$, and keep the 4 lowest cost gates to generate the priority queue from the first optimization and the 2 lowest-cost gates on subsequent optimizations.  The successive rows show the optimized gates resulting from adding a single entangling gate to the ansatz resulting from the last round of optimization, starting with a single-qubit rotation on each qubit (top center).  The green lines show the gates which are kept to form the new priority queue.  Here and throughout, the quantum circuits are ordered with the physical (i.e. readout) qubit on the top line and the ancilla qubits in increasing order on lower lines.  Following an optimization in which $\hat{\mathcal{S}}$ or $\hat{\mathcal{F}}$ gates may be used, the ``raw" circuit containing these parameterized gates is then compiled into CNOTs using Eqs.~\eqref{eq:Scomp} and \eqref{eq:Fcomp}, products of single-qubit rotations are collected together, and then optimization passes are run to determine if single-qubit gates with rotations smaller than a certain threshold can be removed without affecting the cost function.  We note that no cost function penalty is applied when an $\hat{\mathcal{S}}$ or $\hat{\mathcal{F}}$ gate brings the cost function below its desired tolerance, as in the last step of the optimization shown in Fig.~\ref{fig:AStarFigure}, but is only used for ordering the priority queue when no gates meet the cost function tolerance.

Several ``generic" methods for the compilation of isometries exist, as reviewed in, e.g., Ref.~\cite{PhysRevA.93.032318}.  These algorithms also underlie the implementation in Qiskit~\cite{Qiskit}.  In the generic approach, the matrix representation of the isometry is decomposed, e.g., a single column at a time or by the cosine-sine decomposition, and the resulting decompositions expressed in terms of multi-qubit controlled operations, which are themselves decomposed into a target gate set using known representations.  These approaches are constructive, and so will find decompositions of any isometry in principle, but they are not designed to find the most efficient representation by some metric, e.g., the number of entangling gates.  Further, as noted above, the use of such generic algorithms requires an ``isometric completion" in the case that the bond dimension $\chi$ is not a power of 2, and may expend additional resources in exactly compiling noise in the isometries.  Special purpose methods have also been developed for compiling permutation gates in Ref.~\cite{soeken2019compiling}, which have been shown to outperform the {generic algorithms} in some cases.  This method uses a reversible logic synthesis to map the permutation into a reversible circuit comprised of single-target gates, and these single-target gates are then compiled into networks of CNOTs, Hadamard gates, and $\hat{R}_z\left(\theta\right)=|0\rangle\langle0|+e^{i\theta}|1\rangle\langle 1|$ rotations.

In order to compare our methods with the generic, constructive method for compiling isometries, we again consider the isometry in Figs.~\ref{fig:DiagonalGauge} and \ref{fig:AStarFigure}.  As noted above, in order to utilize the generic methods we have to map this isometry into a complete isometry over a set of qubits, which requires us to define the action of the isometry on the state in which the ancilla qubits are all in the state $|1\rangle$, which was left unconstrained by the optimization procedure.  For simplicity, we use the ``isometric completion" in which the operator takes this state to itself without modifying the state of the physical qubit.  Using the \verb#iso# method of the \verb#QuantumCircuit# class from Qiskit~\cite{Qiskit} implementing the generic methods of Ref.~\cite{PhysRevA.93.032318} on the unconstrained \verb#ibmq_qasm_simulator# hardware topology produces a gate representation with 122 CNOTs at \verb#optimization_level# 0, and 120 CNOTs at \verb#optimization_level# 3.  The greedy compilation procedure presented in this work achieves a representation with a cost function error of $5.6\times 10^{-10}$ with an order of magnitude fewer entangling gates for this particular isometry.  An explicit circuit representation is given in Fig.~\ref{fig:site16}(d) of the appendix.

\begin{figure}[t]
  \begin{center}
 \subfloat[Isometry]{
\includegraphics[width=0.32\columnwidth]{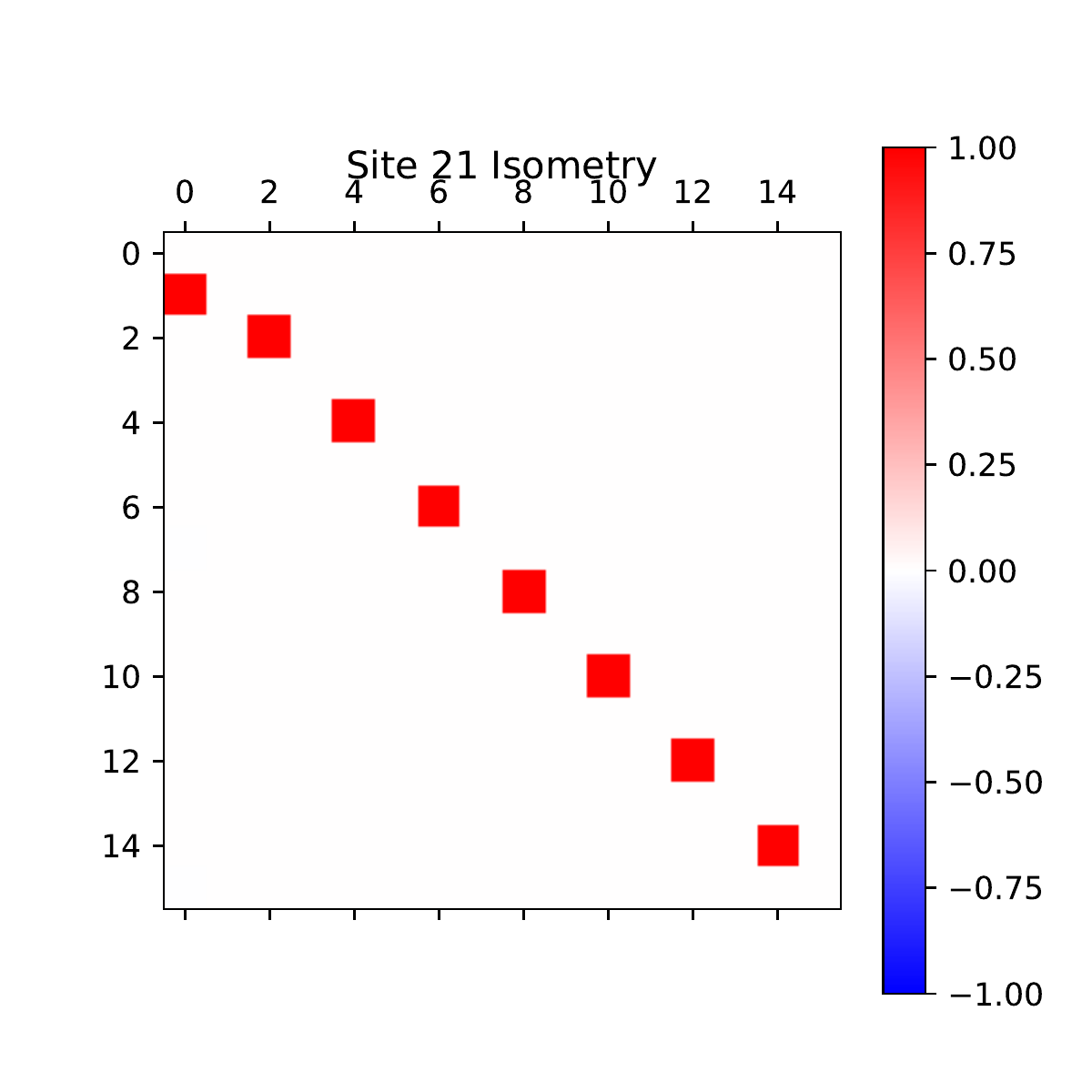}
}
\subfloat[Optimized gate]{
\includegraphics[width=0.32\columnwidth]{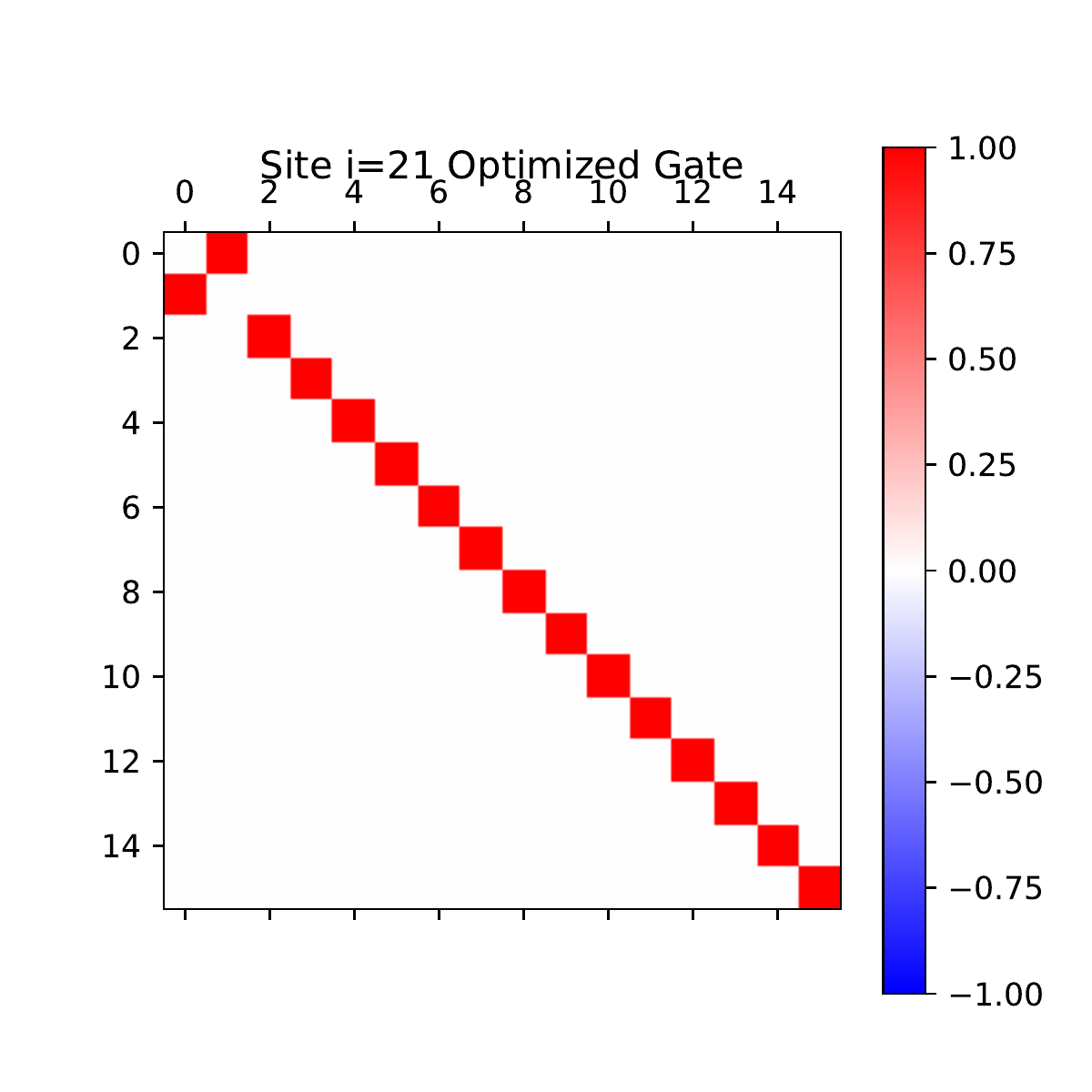}
}
\subfloat[Gate from Ref.~\cite{soeken2019compiling} method]{
 \includegraphics[width=0.32\columnwidth]{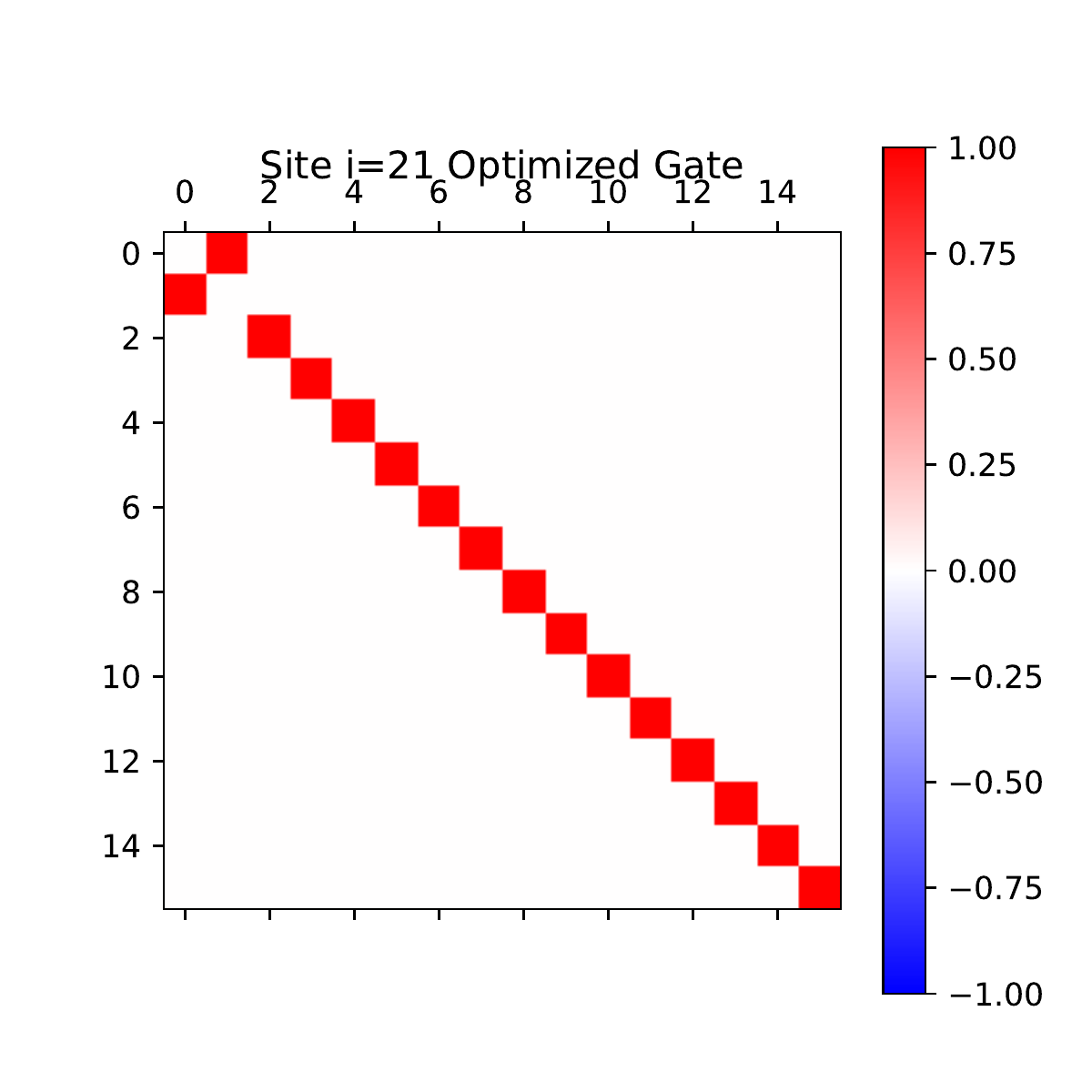} 
}\\
\subfloat[Expanded and cleaned circuit from optimization]{
   \includegraphics[width=0.9\columnwidth]{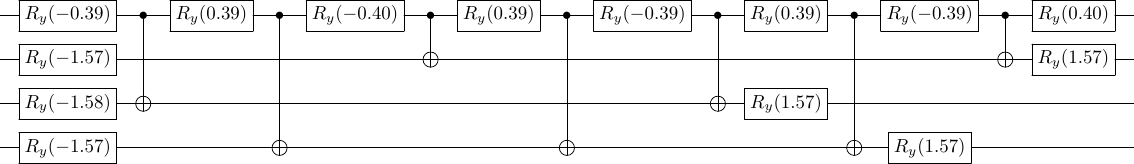} 
}\\
\subfloat[Circuit from method of Ref.~\cite{soeken2019compiling}.]{
 \includegraphics[width=0.9\columnwidth]{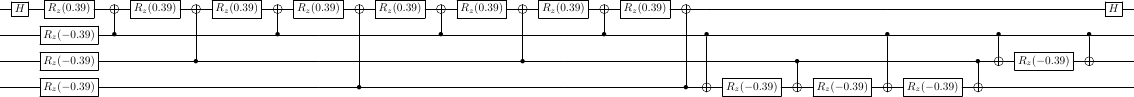} 
}\\

\caption{\label{fig:PermComp} \emph{Comparison of greedy gate compilation procedure with methods of Ref.~\cite{soeken2019compiling}}. (a) Target isometry, which can be completed to a permutation operator.  (b) Matrix plot of result from greedy compilation procedure (cost function $\sim 2\times 10^{-15}$).  (c) Matrix plot of result from the methods of Ref.~\cite{soeken2019compiling}.  (d) Quantum circuit representation of greedy compilation procedure result.  (e) Quantum circuit representation of Ref.~\cite{soeken2019compiling} result.}
\end{center}
\end{figure}

As a point of comparison for the specialized methods for permutation gates studied in Ref.~\cite{soeken2019compiling}, we consider the isometry shown in Fig.~\ref{fig:PermComp}(a).  This is indeed a permutation on the space acted upon, and so can be represented by a family of ``unitary completions."  We take the straightforward choice of unitary completion in which we leave the ancilla qubits unchanged by the permutation, as shown in Fig.~\ref{fig:PermComp}(c).  The result of applying our greedy compilation procedure is given in matrix form in panel (b), and in quantum circuit form in panel (d).  This gate requires 7 CNOTs, and has a cost function error of $\sim 2\times 10^{-15}$.  The result of applying the methods of Ref.~\cite{soeken2019compiling} are shown in panels (c) and (e); the gate here requires 14 CNOT operators.  Generally speaking, we find that our greedy compilation procedure finds comparable or better gates for isometries corresponding to near-diagonal permutations compared to using the methods of Ref.~\cite{soeken2019compiling} with the straightforward unitary completion given above.  However, it is also worth noting that our procedure is designed for isometries and so generally does not produce permutation operators on the entire space at the end of optimization.  That is to say, the optimized gate is a permutation in the space spanned by the isometry, but the full unitary is not a permutation, see, e.g., Fig.~\ref{fig:site15} of the appendix.  It is also worth noting that for complex, highly non-diagonal permutations, as can occur for the diagonality center when transforming to the diagonal gauge, the methods of Ref.~\cite{soeken2019compiling} can produce more efficient representations.

\section{Exactly solvable benchmark model}
\label{sec:ExactMPS}
\begin{figure}[h]
  \begin{center}
\includegraphics[width=0.99\columnwidth]{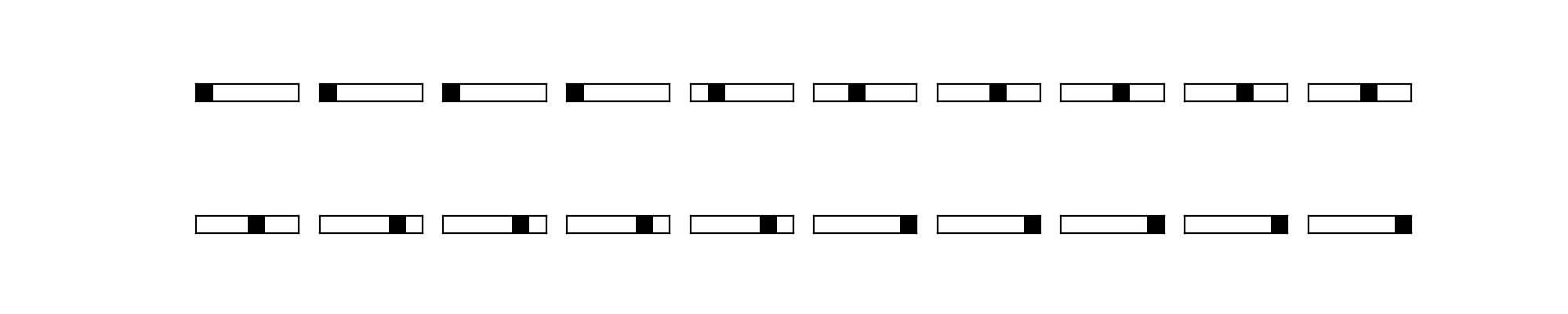}  
\caption{\label{fig:OneHot} \emph{Example dataset for an exactly solvable MPS generative model} with a six-dimensional probability vector $\mathbf{p}=(1/5,1/20,1/20,1/4,1/5,1/4)$.}
\end{center}
\end{figure}

As an exactly solvable benchmark, we consider an MPS Born machine encoding the probability distribution of classical discrete data vectors $\mathbf{x}$, $x_i\in\left\{0,1\right\}$ $\forall i$.  The simplest nontrivial situation is when the data vectors consist of all zeros except for a single 1, closely related to the canonical bars and stripes (BAS) dataset.  Let us denote the probability that the 1 resides at location i as $p_i$, with $\sum_{i=0}^{N-1}p_i=1$.  It can be shown that this data can be represented exactly as a bond dimension 2 MPS Born machine with tensors
\begin{align}
\label{eq:exMPS1}A^{[0] 0}_{00}&=1\, ,\; A^{[0] 1}_{01}=e^{i\phi_0}\sqrt{p_0}\, ,\\
A^{[j] 0}_{00}&=1\, ,\;\; A^{[j] 1}_{01}=e^{i\phi_j}\sqrt{p_j}\, ,\;\; A^{[j] 0}_{11}=1\, ,\\
\label{eq:exMPS3}A^{[N-1] 0}_{10}&=1\, ,\;\; A^{[N-1] 1}_{00}=e^{i\phi_{N-1}}\sqrt{p_{N-1}}\, ,
\end{align}
with the $\{\phi_j\}$ denoting arbitrary phases.  The presence of a large number of arbitrary phases is a generic feature of TN models for generative applications: since the square of the wavefunction is used to generate classical data samples, the phase structure of the wavefunction is generally underconstrained.  This in turn implies that TN models can have some flexibility over the particular gate set used to entangle the physical qubits to the ancillae without affecting the sampling outcomes.  The exactly solvable model encapsulated by Eqs.~\eqref{eq:exMPS1}-\eqref{eq:exMPS3} is a useful benchmark both because it is the simplest nontrivial example of a sequentially preparable QAML model, involving a single ancilla qubit, and because it can be exactly solved for any classical data vector length and probabilities $\mathbf{p}$.  An example dataset for $\mathbf{p}=(1/5,1/20,1/20,1/4,1/5,1/4)$ is given in Fig.~\ref{fig:OneHot}.

The construction in Eqs.~\eqref{eq:exMPS1}-\eqref{eq:exMPS3} is reminiscent of the well-known MPS representation of the W state~\cite{perez2007matrix}.  In order to convert this generic MPS into a sequential qubit preparation scheme we should place the MPS into left-canonical form.  Since the bond dimension is known, we can do so in terms of the QR decomposition.  For simplicity of exposition, we will take all phases $\phi_j=0$, though we will relax this condition shortly.  Performing the QR decomposition on the first tensor, we find
\begin{align}
A^{[0]}&=\left(\begin{array}{cc} 1&0\\ 0&\sqrt{p_0}\end{array}\right)\\
\to &QR=\left(\begin{array}{cc} 1&0 \\ 0 &\sqrt{\frac{p_0}{\left|p_0\right|}}\end{array}\right)\left(\begin{array}{cc} 1&0 \\ 0 &\sqrt{p_0}\end{array}\right)\, ,\\
\Rightarrow L^{[0]}&=\left(\begin{array}{cc} 1&0 \\ 0 &\sqrt{\frac{p_0}{\left|p_0\right|}}\end{array}\right)\, ,\\
A^{[1] 0}_{00}&=1\, ,\;\; A^{[1] 1}_{01}=\sqrt{p_1}\, ,\;\; A^{[2] 0}_{11}=\sqrt{p_0}\, .
\end{align}
Reshaping the second tensor and decomposing, we find
\begin{align}
A^{[1]}_{\left(\alpha i\right)\beta}&=\left(\begin{array}{cccc} 1&0\\ 0&\sqrt{p_1}\\ 0&\sqrt{p_0} \\ 0&0\end{array}\right)\\
\to &QR=\left(\begin{array}{cccc} 1&0\\ 0&\sqrt{\frac{p_1}{p_0+p_1}}\\ 0&\sqrt{\frac{p_0}{p_0+p_1}} \\ 0&0\end{array}\right)\left(\begin{array}{cc} 1&0\\ 0 &\sqrt{p_0+p_1}\end{array}\right)\, ,\\
\Rightarrow L^{[1] 0}_{00}&=1\, ,\;\; L^{[1] 1}_{01}=\sqrt{\frac{p_1}{p_0+p_1}}\, ,\;\; L^{[1] 0}_{11}=\sqrt{\frac{p_0}{p_0+p_1}}\, .
\end{align}
This generalizes to
\begin{align}
L^{[j] 0}_{00}&=1\, ,\;\; L^{[j] 1}_{01}=\sqrt{\frac{p_j}{\sum_{i\le j}p_i}}\, ,\;\; L^{[j] 0}_{11}=\sqrt{\frac{\sum_{i<j}p_i}{\sum_{i\le j}p_i}}\, ,
\end{align}
with the last tensor being
\begin{align}
L^{[N-1] 0}_{10}&=\sqrt{\sum_{i< N-1}p_i}\, ,\;\; L^{[N-1] 1}_{00}=\sqrt{p_{N-1}}\, .
\end{align}

We now take these left-canonical tensors and reshape them to correspond to isometries acting on a single physical qubit $|i_q\rangle$ and an ancilla qubit $|\alpha_a\rangle$.  We start from the $N^{\mathrm{th}}$ tensor, where both the qubit and ancilla are in the state $0$.  The isometry is
\begin{align}
\hat{L}_{N-1}&=\left(\sqrt{1-p_{N-1}}|1_a0_q\rangle+\sqrt{p_{N-1}}|0_a1_q\rangle\right)\langle 0_a0_q|\, .
\end{align}
Following this, the physical qubit can be measured in the computational basis and its outcome (classically) stored, and then the physical qubit is returned to the state $|0_q\rangle$.  We then repeat this procedure of acting with isometries, measuring the physical qubit, classically recording its output, and returning the physical qubit to 0, using the isometries
\begin{align}
\label{eq:LjQR}\hat{L}_{j}&=|0_a0_q\rangle\langle 0_a0_q|\\
\nonumber &+\left(\sqrt{\frac{p_j}{\sum_{i\le j}p_i}}|0_a1_q\rangle+\sqrt{\frac{\sum_{i<j}p_i}{\sum_{i\le j}p_i}}|1_a0_q\rangle\right)\langle 1_a0_q|\, .
\end{align}
We note that Eq.~\eqref{eq:LjQR} also holds for the final site, $j=0$, and produces an unentangled ancilla in the state $|0_a\rangle$.  With these operators in hand, we can re-insert the arbitrary phases on the elements resulting in the state $|1_q\rangle$, yielding
\begin{align}
&\hat{L}_{N-1}=\left(\sqrt{1-p_{N-1}}|1_a0_q\rangle+e^{i\phi_{N-1}}\sqrt{p_{N-1}}|0_a1_q\rangle\right)\langle 0_a0_q|\, , \\
\label{eq:LjQRp} &\hat{L}_{j}=|0_a0_q\rangle\langle 0_a0_q|\\
\nonumber &+\left(e^{i\phi_j}\sqrt{\frac{p_j}{\sum_{i\le j}p_i}}|0_a1_q\rangle+\sqrt{\frac{\sum_{i<j}p_i}{\sum_{i\le j}p_i}}|1_a0_q\rangle\right)\langle 1_a0_q|\, .
\end{align}
We note that there is a ``natural" unitary completion of the operators in Eq.~\eqref{eq:LjQRp} given by
\begin{align}
\label{eq:LjQRpp}  &\hat{U}_{j}=|0_a0_q\rangle\langle 0_a0_q|+|1_a1_q\rangle\langle 1_a1_q|\\
\nonumber &+\left(e^{i\phi_j}\sqrt{\frac{p_j}{\sum_{i\le j}p_i}}|0_a1_q\rangle+\sqrt{\frac{\sum_{i<j}p_i}{\sum_{i\le j}p_i}}|1_a0_q\rangle\right)\langle 1_a0_q|\\
\nonumber &+\left(\sqrt{\frac{\sum_{i<j}p_i}{\sum_{i\le j}p_i}}|0_a1_q\rangle-e^{-i\phi_j}\sqrt{\frac{p_j}{\sum_{i\le j}p_i}}|1_a0_q\rangle\right)\langle 0_a1_q|\, ,
\end{align}
in which the state $|1_a1_q\rangle$--that is never populated under ideal operation--is left unchanged and the action on the--also ideally unpopulated--state $|0_a1_q\rangle$ is determined by orthogonality.  Written in the basis representation $\{|0_a0_q\rangle , |0_a1_q\rangle, |1_a0_q\rangle, |1_a1_q\rangle\}$, we find
\begin{align}
\label{eq:LjMatRep}\left[\hat{U}_{j}\right]&=\left(\begin{array}{cccc} 1&0&0&0\\ 0&\cos\theta_j&e^{i\phi_j}\sin\theta_j&0 \\ 0&-e^{-i\phi_j}\sin\theta_j&\cos\theta_j&0\\ 0&0&0&1\end{array}\right)\, ,
\end{align}
in which $\cos\theta_j=\sqrt{\frac{\sum_{i<j}p_i}{\sum_{i\le j}p_i}}$ and so $\sin\theta_j=\sqrt{\frac{p_j}{\sum_{i\le j}p_i}}$.  This gate has a natural interpretation as a rotation within the subspace of a single quantum of excitation shared between the qubit and ancilla, with the rotation angle set by the classical data vector probabilities (for $p_j\to 0$, $\theta_j\to 0$ and Eq.~\eqref{eq:LjMatRep} becomes the identity).  An analogous unitary completion for the isometry $\hat{L}_{N-1}$ is given by
\begin{align}
\nonumber &\hat{U}_{N-1}=\left(\cos\theta_{N-1}|1_a0_q\rangle+e^{i\phi_{N-1}}\sin\theta_{N-1}|0_a1_q\rangle\right)\langle 0_a0_q|\\
\nonumber &+\left(-e^{-i\phi_{N-1}}\sin\theta_{N-1}|1_a0_q\rangle+\cos\theta_{N-1}|0_a1_q\rangle\right)\langle 0_a1_q|\\
\nonumber &+\left(e^{i\phi_{N-1}}\sin\theta_{N-1}|1_a1_q\rangle+\cos\theta_{N-1}|0_a0_q\rangle\right)\langle 1_a0_q|\\
\label{eq:UN}&+\left(-e^{-i\phi_{N-1}}\sin\theta_{N-1}|0_a0_q\rangle+\cos\theta_{N-1}|1_a1_q\rangle\right)\langle 1_a0_q|\, .
\end{align}

From a gate-based perspective, the operators in Eqs.~\eqref{eq:LjMatRep} with $\phi=-\pi/2$ are described by the Fermionic Simulation, or fSim$(\theta,\varphi)$ gate~\cite{PhysRevLett.120.110501}, with $\varphi=0$ and $\theta=\theta_j$; this gate has been recently been demonstrated in gmon qubits~\cite{foxen2020demonstrating}.  Alternatively, they are a one-parameter generalization of the iSWAP gate~\cite{schuch2003natural}.  We note that the unitary completion $\hat{U}_j$ at $\phi_j=0$ is given by $\hat{\mathcal{S}}\left(\theta_j,\theta_j\right)$ in the notation of Eq.~\eqref{eq:Sdef}, and so for the gate set employed by the IBMQ processors, the shortest decomposition for $U_j$ is given by Eq.~\ref{eq:Scomp}.  While in some alternative hardware platforms, such those employing tunable qubits~\cite{kjaergaard2020superconducting}, iSWAP gates can be implemented natively, partial iSWAPs still require decomposition.  We also note that the operation Eq.~\eqref{eq:LjMatRep} is generated by the effective Hamiltonian
\begin{align}
\hat{\mathbb{H}}_j&=\theta_j\left(\hat{\sigma}^+_q\hat{\sigma}^-_a+\hat{\sigma}^-_q\hat{\sigma}^+_a\right)\, ,
\end{align}
for ``unit time" in the sense that
\begin{align}
\hat{U}_{j}&=\exp\left(-i \hat{\mathbb{H}}_j\right)\, ,
\end{align}
when $\phi_j=\pi/2$.  This gate is readily achieved in trapped ion-based quantum computers using an equally weighted combination of XX and YY M{\o}lmer-S{\o}renson gates~\cite{sorensen2000entanglement}, as well as a variety of other platforms implementing XY effective spin-spin interactions~\cite{PhysRevLett.102.100501}.  It is also interesting that the ``data angle" $\theta_j$ has a natural interpretation as an ersatz ``evolution time" in this perspective.  

\begin{figure}[b]
\centering

\subfloat[Circuit decomposition for Eq.~\eqref{eq:LjQRpp}.]{
\Qcircuit @C=.7em @R=.7em {
& \qw  &\ctrl{1}& \gate{R_y(\theta_j)} & \ctrl{1} & \gate{R_z(\pi/2)} & \ctrl{1} &  \gate{R_y(\theta_j)} & \ctrl{1} & \gate{R_z(-\pi/2)} & \qw\\
& \qw & \targ & \qw & \targ & \gate{R_z(\pi/2)} & \targ & \qw & \targ & \gate{R_z(-\pi/2)} & \qw
}
}

\subfloat[Circuit decomposition for Eq.~\eqref{eq:UN}.]{
\Qcircuit @C=1em @R=.7em {
& \gate{R_y(2\theta_{N-1})}&\gate{X} & \ctrl{1} &\gate{X}& \qw \\
& \qw&\qw & \targ & \qw & \qw
}
}

\caption{Circuit decompositions for $U_j$ [Panel (a)] and $U_{N-1}$ [Panel (b)] based on a gateset of single qubit rotations and CNOTs.  In both diagrams, the upper line is the physical (sampled) qubit and the lower line is the ancilla.}
\label{fig:HandCompiled}
\end{figure}
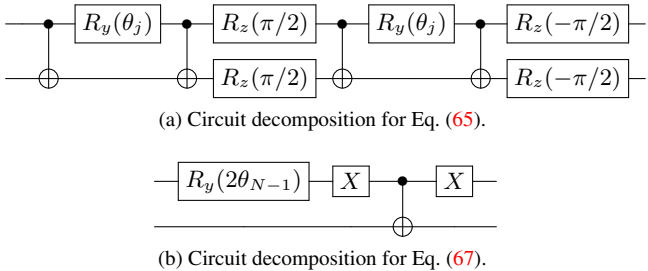

Before moving on from this exactly solvable example, we would like to point out how the freedom in representation of the bond basis manifests itself in this exactly solvable example.  Namely, the predictions of the model Eq.~\eqref{eq:exMPS1}-\eqref{eq:exMPS3} are unchanged if we reverse the roles of the $|0_a\rangle$ and $|1_a\rangle$ ancilla states in all but the first and last steps of preparation (using the unitary freedom exploited in the transformation to diagonal gauge discussed in Sec.~\ref{sec:DG}).  In this case, we have the isometries
\begin{align}
\hat{\tilde{L}}_{N-1}&=\left(\sqrt{1-p_{N-1}}|0_a0_q\rangle+e^{i\phi_{N-1}}\sqrt{p_{N-1}}|1_a1_q\rangle\right)\langle 0_a0_q|\, , \\ 
\hat{\tilde{L}}_j&=|1_a0_q\rangle\langle 1_a0_q|\\
&+\left(e^{i\phi_j}\sqrt{\frac{p_j}{\sum_{i\le j}p_i}}|1_a1_q\rangle+\sqrt{\frac{\sum_{i<j}p_i}{\sum_{i\le j}p_i}}|0_a0_q\rangle\right)\langle 0_a0_q|\, .
\end{align}
The natural unitary completions of these isometries take the matrix representation 
\begin{align}
\left[\hat{\tilde{U}}_j\right]&=\left(\begin{array}{cccc} \cos\theta_j&0&0&-e^{-i\phi_j}\sin\theta_j\\ 0&1&0&0 \\ 0&0&1&0\\ e^{i\phi_j}\sin\theta_j&0&0&\cos\theta_j\end{array}\right)\, ,
\end{align}
and so are described by $\hat{\mathcal{S}}\left(-\theta_j,\theta_j\right)$ at $\phi_j=0$, and are generated by the effective Hamiltonian
\begin{align}
\hat{\tilde{\mathbb{H}}}&=\theta_j\left(\hat{\sigma}^+_q\hat{\sigma}^+_a+\hat{\sigma}^-_q\hat{\sigma}^-_a\right)\, ,
\end{align}
at $\phi=\pi/2$.

\subsection{Training and compilation}
\label{sec:TandC}
In this section, we detail the application of the methods outlined in this paper to the exactly solvable benchmark in Sec.~\ref{sec:ExactMPS} using the probabilities $\mathbf{p}=\left(8/31,18/31,5/31\right)$.  As a point of comparison, we consider a ``hand compiled" version of the unitary completed isometries Eqs.~\eqref{eq:LjQRpp} and \eqref{eq:UN}.  Taking $\phi=0$, we can compile these gates using the circuits shown in Fig.~\ref{fig:HandCompiled}.  However, we additionally note that with the assumption that the physical qubit starts in the state $|0_q\rangle$, the first CNOT in Fig.~\ref{fig:HandCompiled}(a) is the identity, and so can be neglected, leading to a circuit with three CNOTs.

\begin{figure}[t!]
\centering
\subfloat[Site 0 isometry]{
\includegraphics[width=0.4\columnwidth]{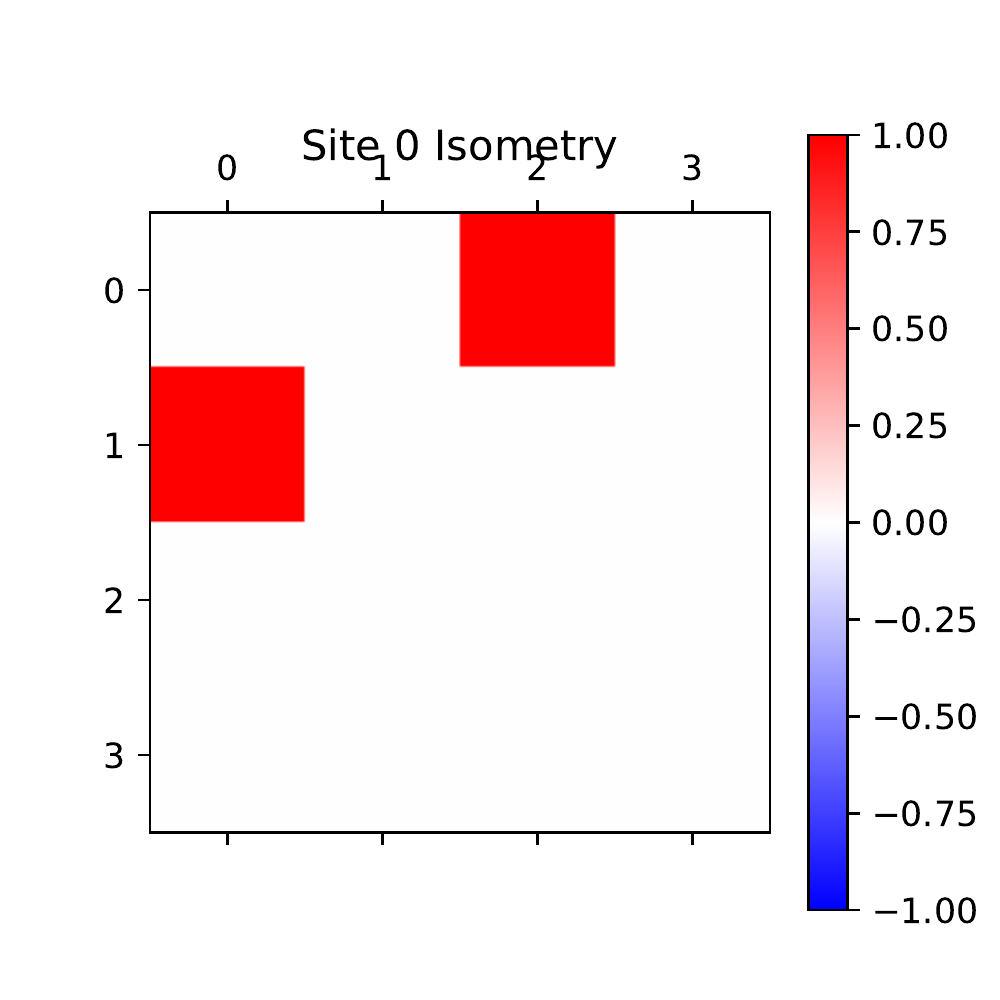}
}
\subfloat[Site 0 optimized gate]{
\includegraphics[width=0.4\columnwidth]{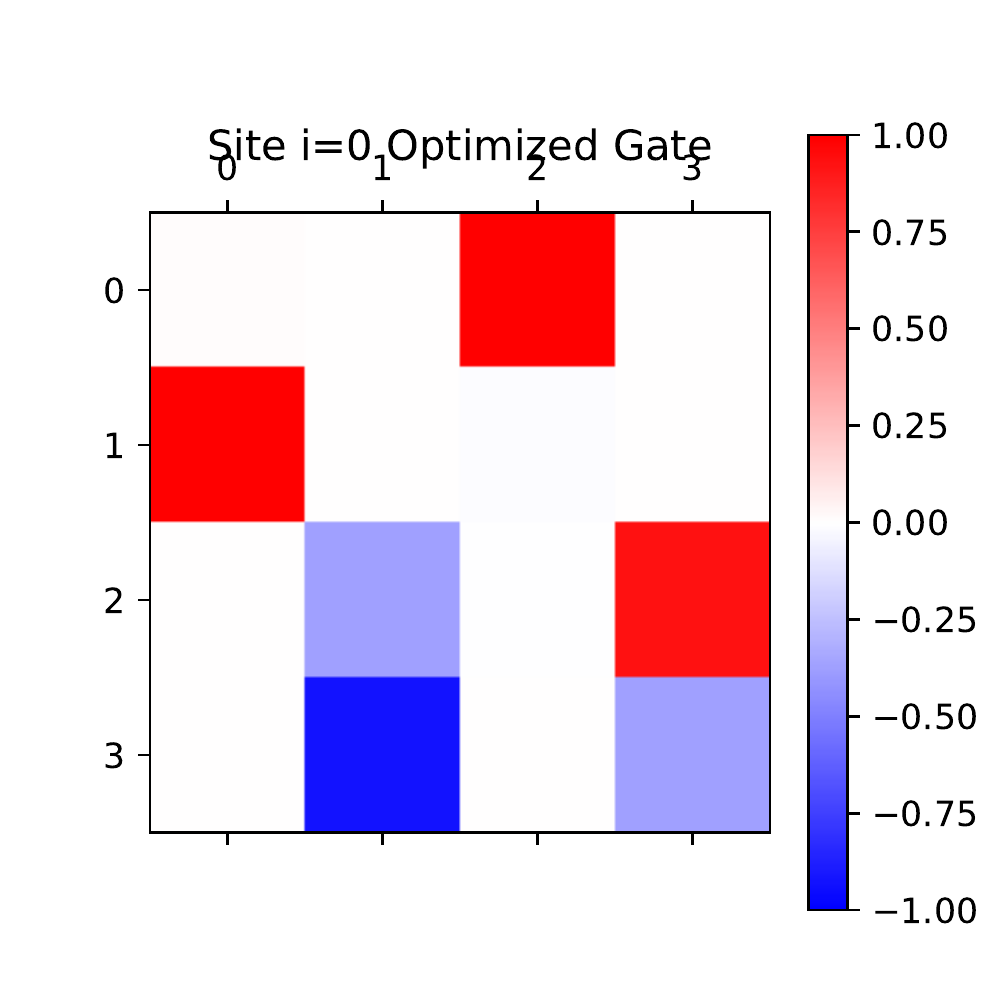}
}\\

\subfloat[Site 0 circuit from optimization]{
 \includegraphics[width=0.5\columnwidth]{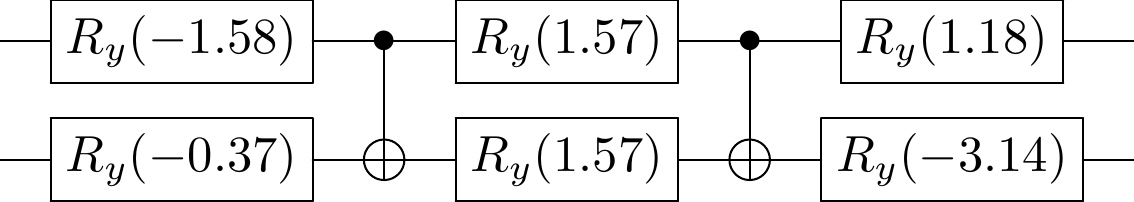} 
}

\subfloat[Site 1 optimized gate]{
\includegraphics[width=0.4\columnwidth]{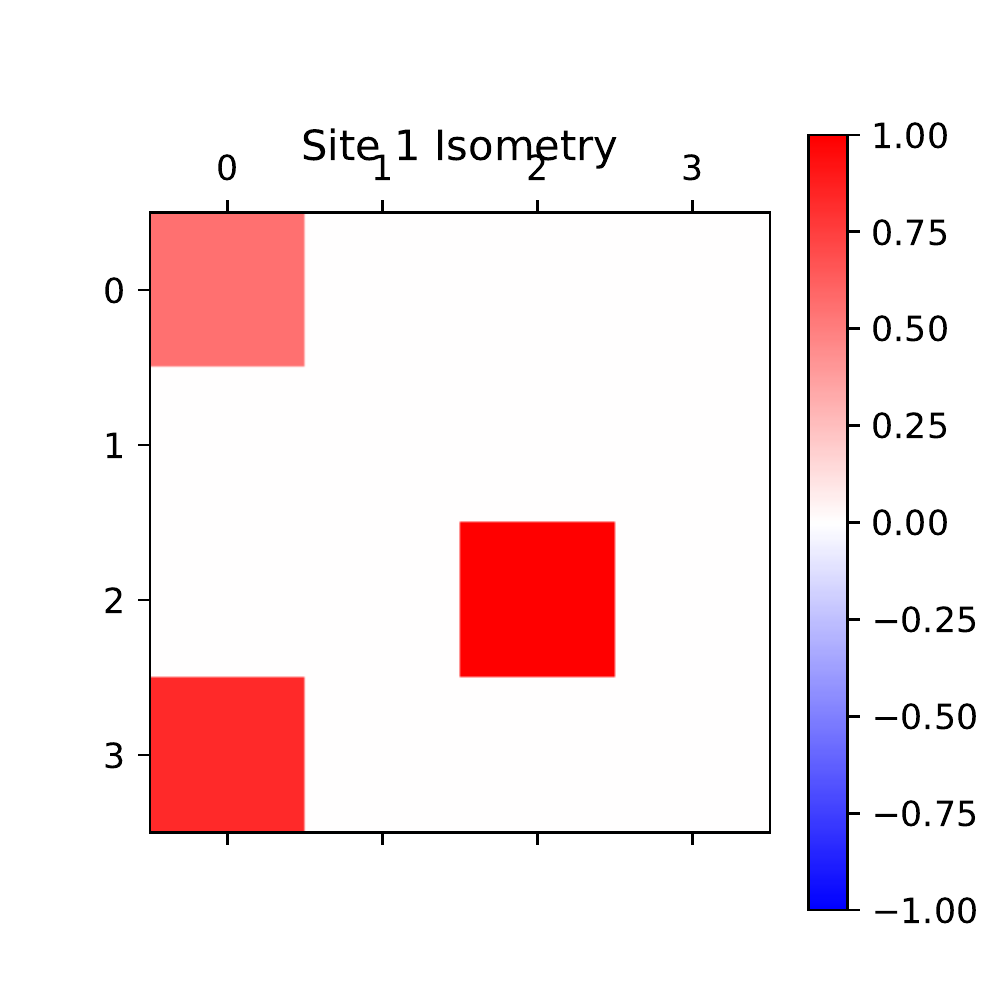}
}
\subfloat[Site 1 optimized gate]{
\includegraphics[width=0.4\columnwidth]{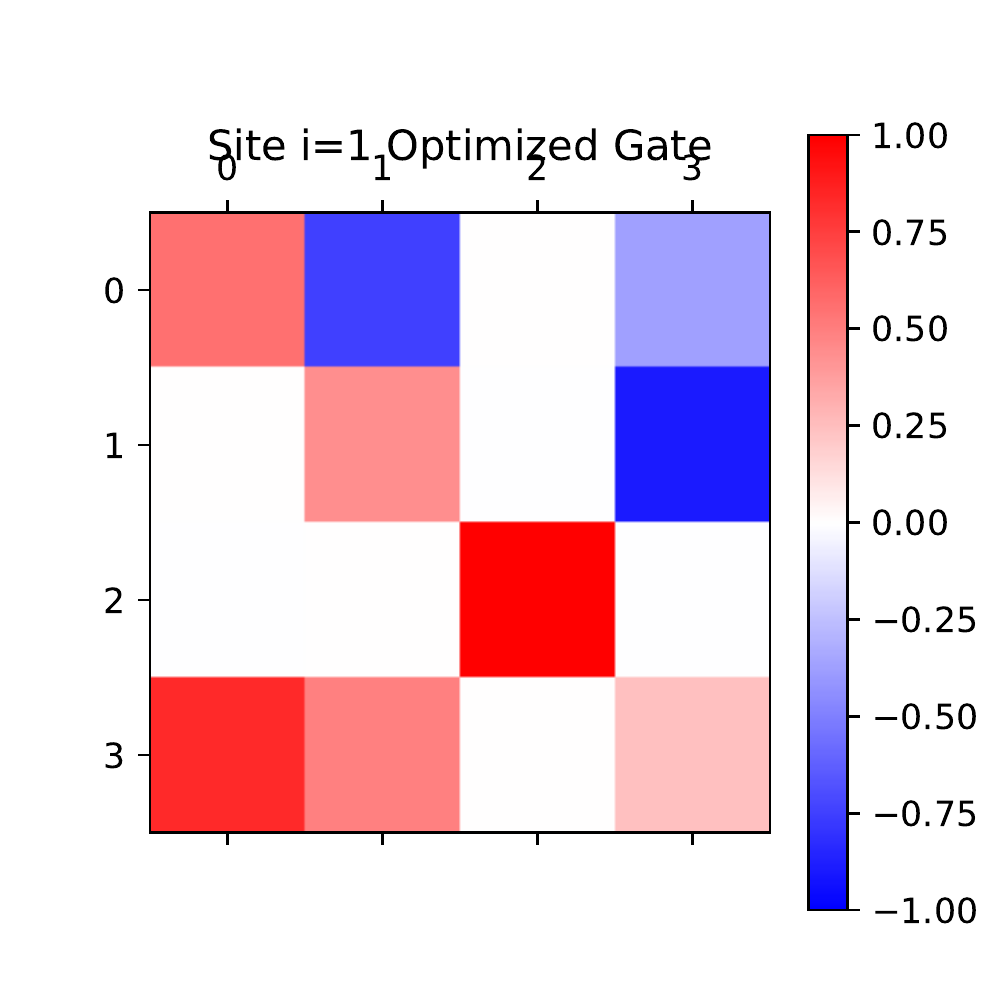}
}\\

\subfloat[Site 1 optimized gate]{
\includegraphics[width=0.5\columnwidth]{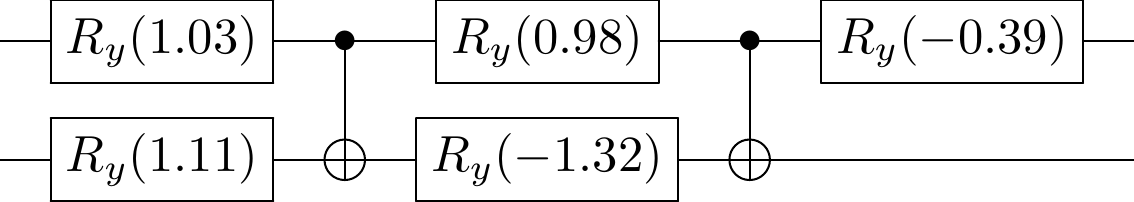}
}\\

\subfloat[Site 2 isometry]{
\includegraphics[width=0.4\columnwidth]{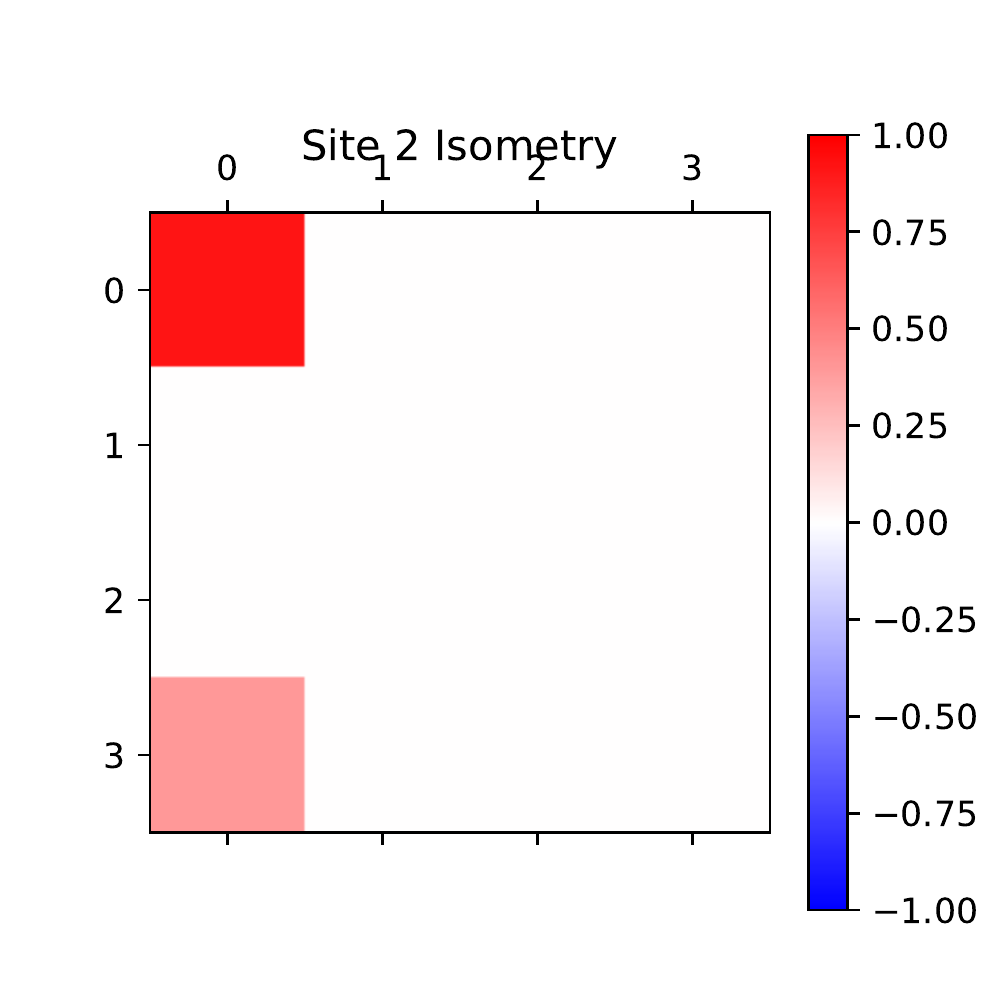}
}
\subfloat[Site 2 optimized gate]{
\includegraphics[width=0.4\columnwidth]{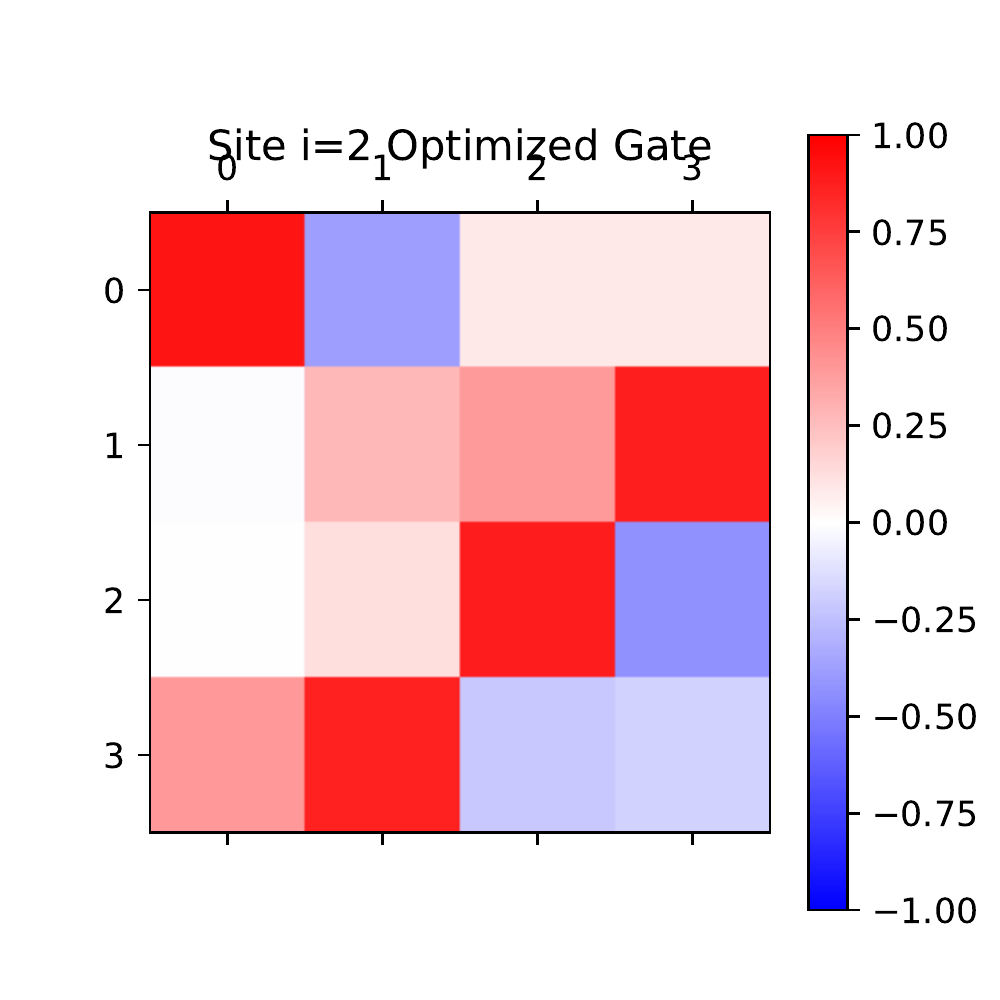}
}\\

\subfloat[Site 2 circuit from optimization]{
\includegraphics[width=0.5\columnwidth]{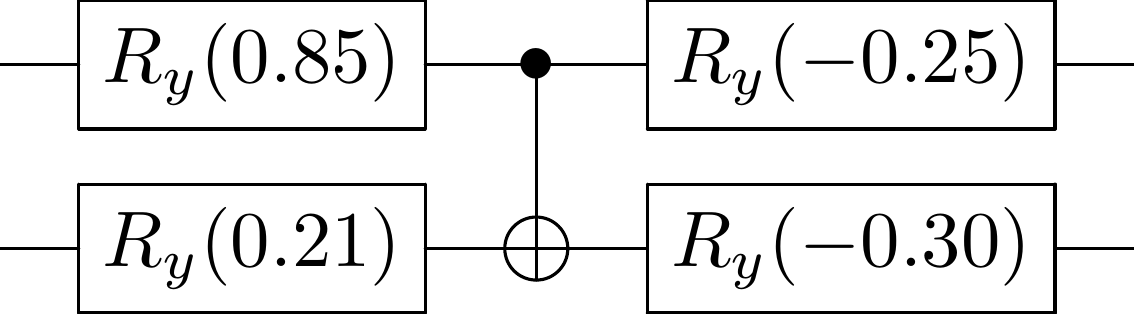}
}
\caption{\label{fig:ESBM} \emph{Exactly solvable Born Machine benchmark isometries.}  Isometries and optimized gates for the three-site exactly solvable Born Machine benchmark.   Panels (a), (d), and (g) are the isometries output from the classically trained model, panels (b), (e), and (h) are matrix plots of the unitaries output by our greedy compilation procedure, and panels (c), (f), and (i) are circuit representations of the optimized unitaries.}
\end{figure}

\begin{figure*}

\subfloat[Hand-compiled gate]{
\includegraphics[width=0.95\textwidth]{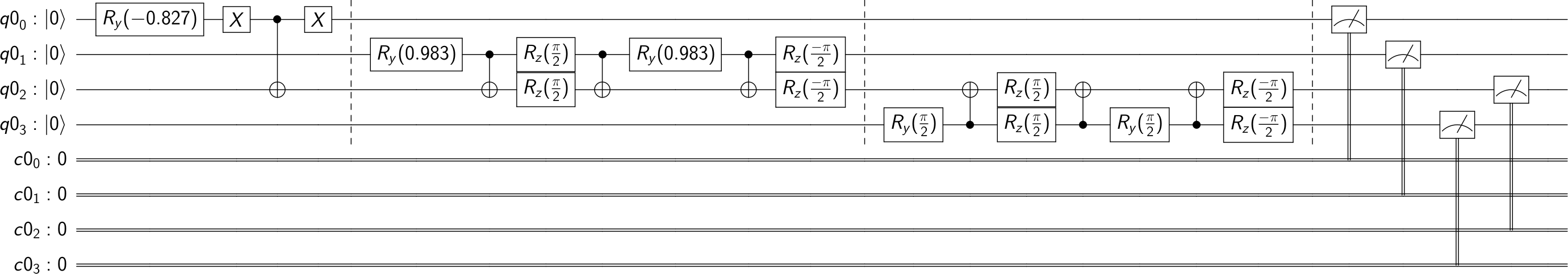}
}\\
\subfloat[Gate from QAML optimization workflow]{
\includegraphics[width=0.95\textwidth]{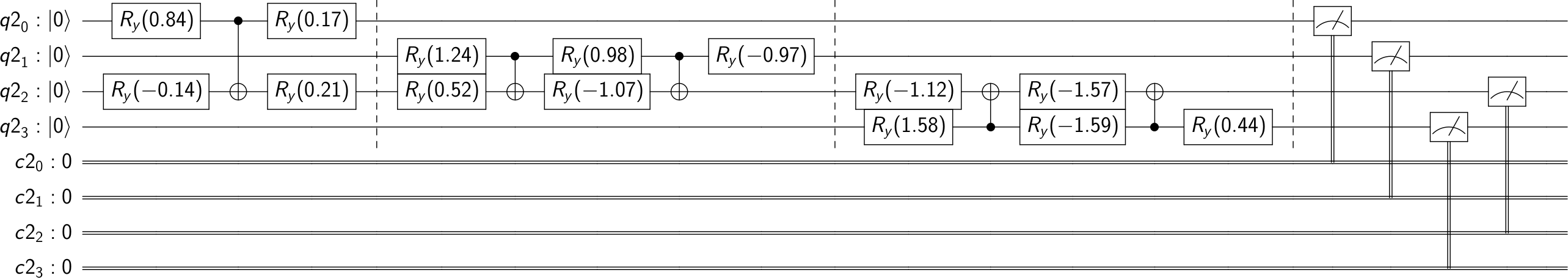}
}

\caption{\label{fig:HandVsAuto}  \emph{Comparison of hand-compiled and auto-compiled circuits for exactly solvable test case.}  The exactly solvable benchmark with three physical qubits (0, 1, and 3) and a single ancilla qubit (2) implemented as a quantum circuit using the hand-compiled circuits in Fig.~\ref{fig:HandCompiled} (upper panel) or the auto-compiled gates in Fig.~\ref{fig:ESBM} (lower panel).}
\end{figure*}

\subsection{Performance of benchmark on cloud-based hardware}

In this section, we present results for the exactly solvable benchmark model running on cloud-based NISQ hardware, using IBM devices as an example.  We note that the current IBM hardware does not allow measurement and re-initialization during an experimental run, and so our sequential preparation schemes cannot be directly implemented on these devices.  However, we can still test our generative models by implementing the gates $\hat{U}_j$ of the sequential preparation scheme on a register of $(N+1)$ qubits prepared in the $|0\dots 0\rangle$ state, coupling each physical qubit to the same ancilla in order from $(N-1)$ down to 0.  This procedure is limited in practice by the number of available qubits and their connectivity to a single ancilla qubit.  However, for devices with a cross-shaped topology, such as the IBMQ-X2, we can readily couple up to 4 qubits to a central ancilla qubit, and for devices with a T-shaped topology, such as the Vigo, we can couple up to 3 qubits to a single ancilla.

To demonstrate our methods to this benchmark case, we train a $\chi=2$ Born machine using the single-site gradient descent described in Sec.~\ref{sec:ClassicalTraining} and compile it into gates using the procedures of Sec.~\ref{sec:compilation} with the diagonality center at 1 and a greedy optimization tolerance of $5\times 10^{-4}$.  Following the usual parlance of MPSs from condensed matter physics, we will refer to the physical indices of the MPS tensors as sites.  The results of this procedure are shown in Fig.~\ref{fig:ESBM}, with panels (a)-(c) for site 0, panels (d)-(f) for site 1, and panels (g)-(i) for site 2.  The final cost functions for sites 0, 1, and 2 are $6.7\times 10^{-9}$, $7.3\times 10^{-10}$, and $2.0\times 10^{-9}$, respectively.  We see that the obtained quantum circuits are substantially different than those obtained by hand-compilation of the ``natural" unitary completion, but are still of very high fidelity in the space spanned by the isometry.  In addition, the gates for sites 0 and 1 are shallower than the hand-compiled gate, which may be anticipated based on known optimality results for two-qubit gates~\cite{vatan2004optimal}.

Utilizing this approach for the exactly solvable Born Machine model with the probability vector given in Sec.~\ref{sec:TandC}, we find the circuits shown in Fig.~\ref{fig:HandVsAuto}.  Here, the physical qubits (those that are sampled to obtain output classical data vectors) are assigned to be qubits 0, 1, and 3, and the ancilla is qubit 2.  The upper panel is for the hand-compiled circuits from Fig.~\ref{fig:HandCompiled}, and the lower panel is the circuit from Fig.~\ref{fig:ESBM} using the workflow put forth in this work.  The dashed vertical lines demarcate the circuits corresponding to the individual sites of the Born machine, but are inessential and neighboring single-qubit rotations can be joined for increased efficiency.

\begin{figure*}[t]
\subfloat[Hand-compiled results for IBMQ-X2]{
\includegraphics[width=0.95\columnwidth]{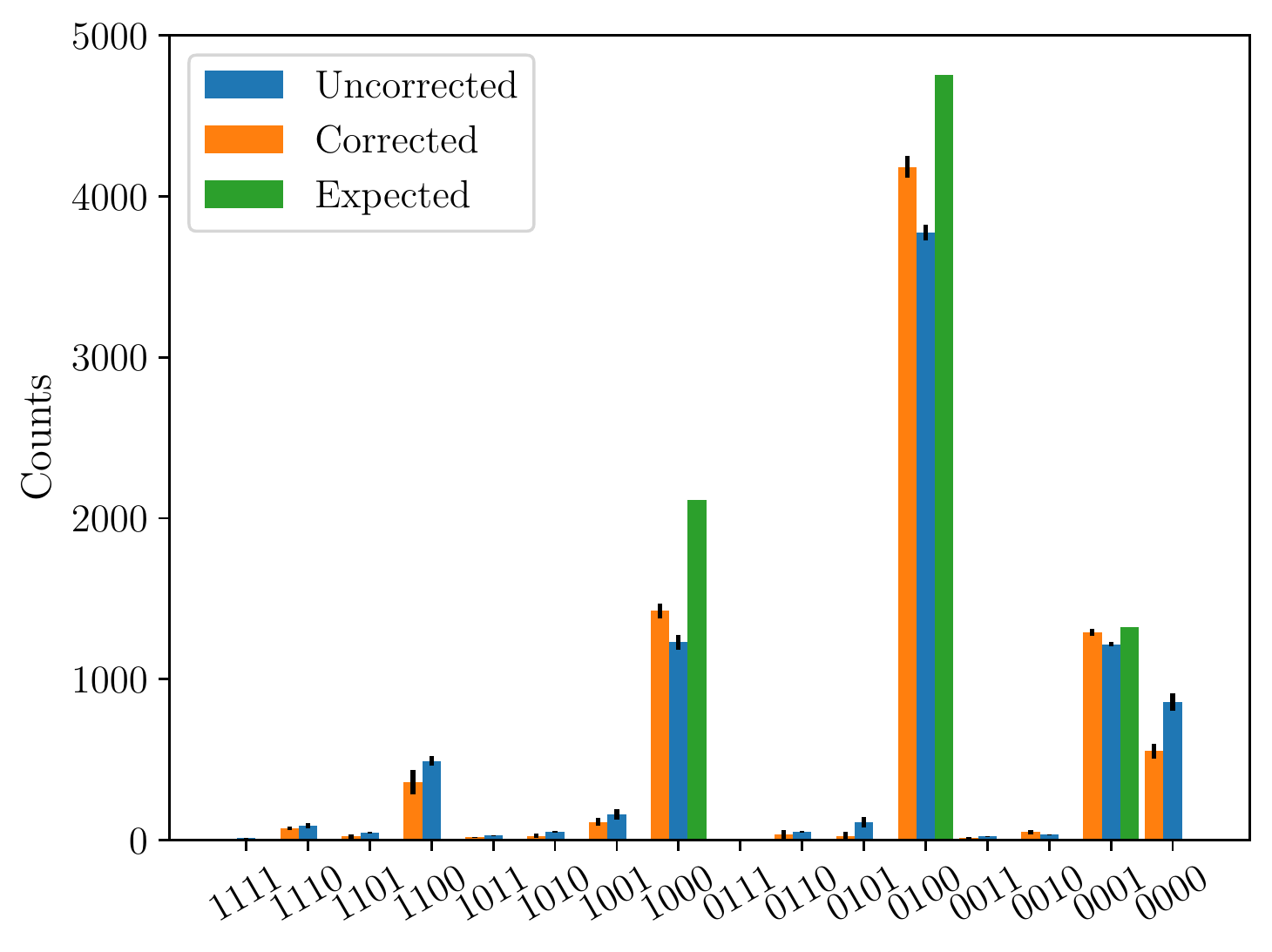}
}
\subfloat[Auto-compiled results for IBMQ-X2]{
\includegraphics[width=0.95\columnwidth]{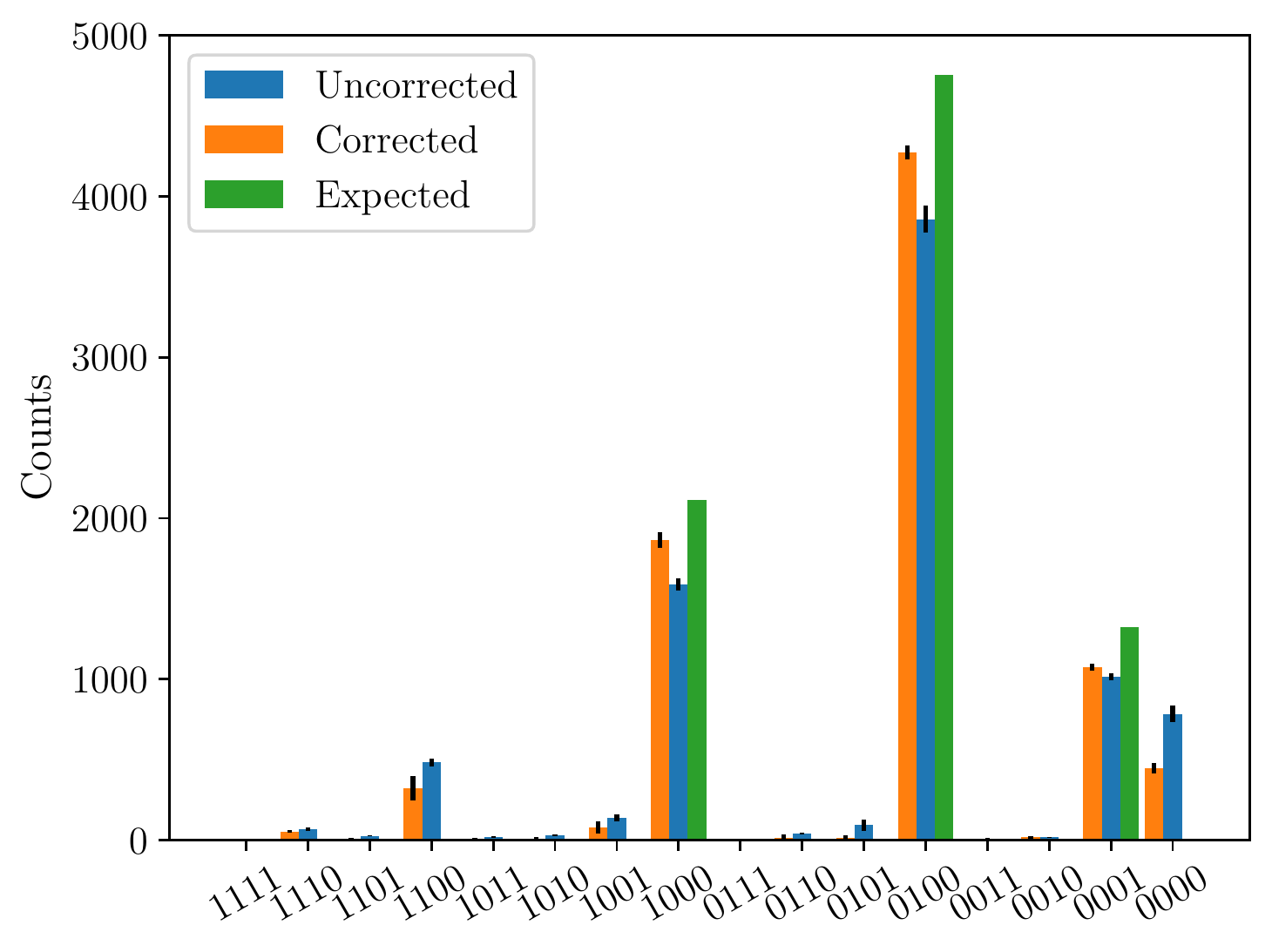}
}\\

\subfloat[Hand-compiled results for IBMQ-Vigo]{
\includegraphics[width=0.95\columnwidth]{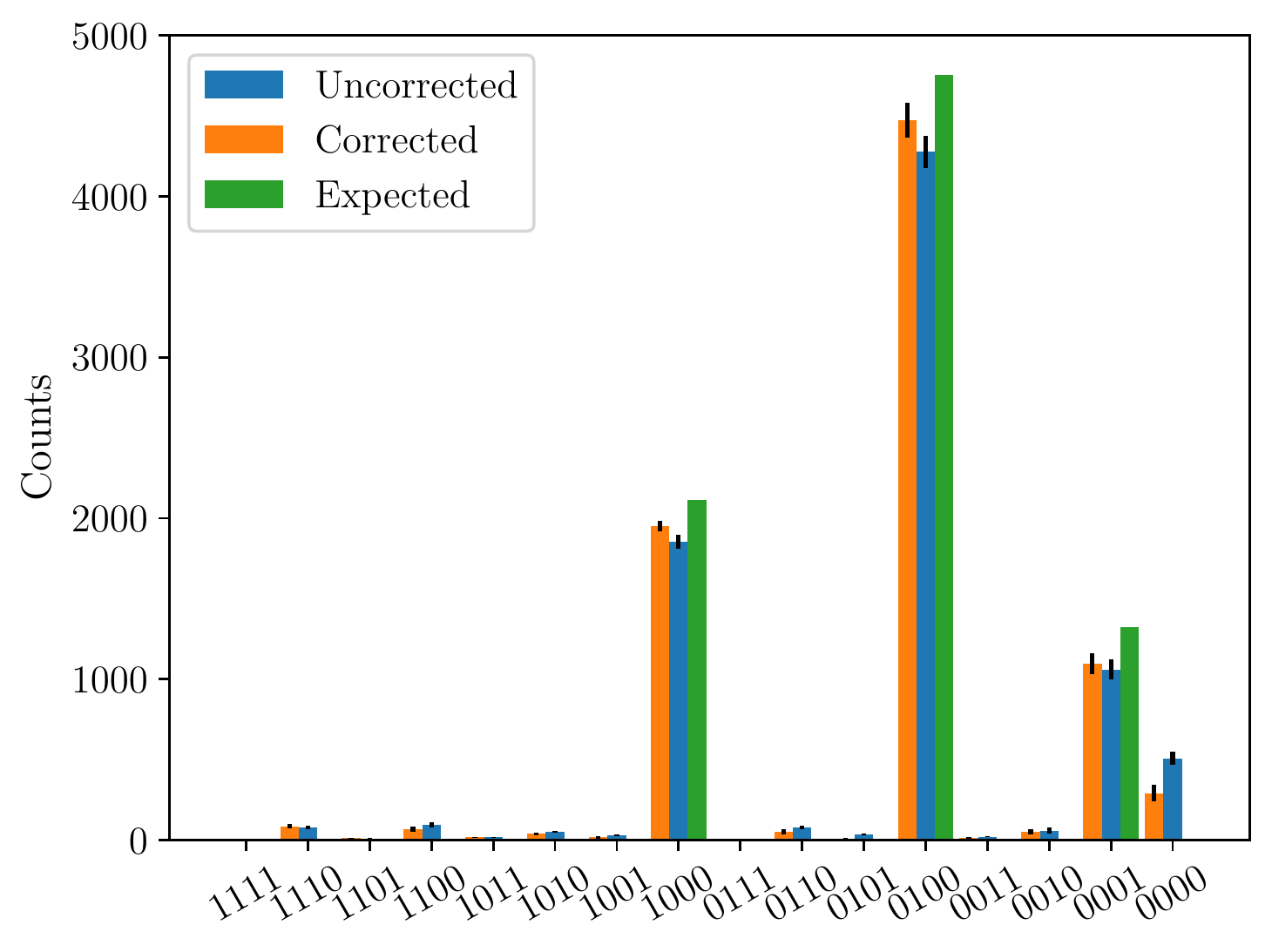}
}
\subfloat[Auto-compiled results for IBMQ-Vigo]{
\includegraphics[width=0.95\columnwidth]{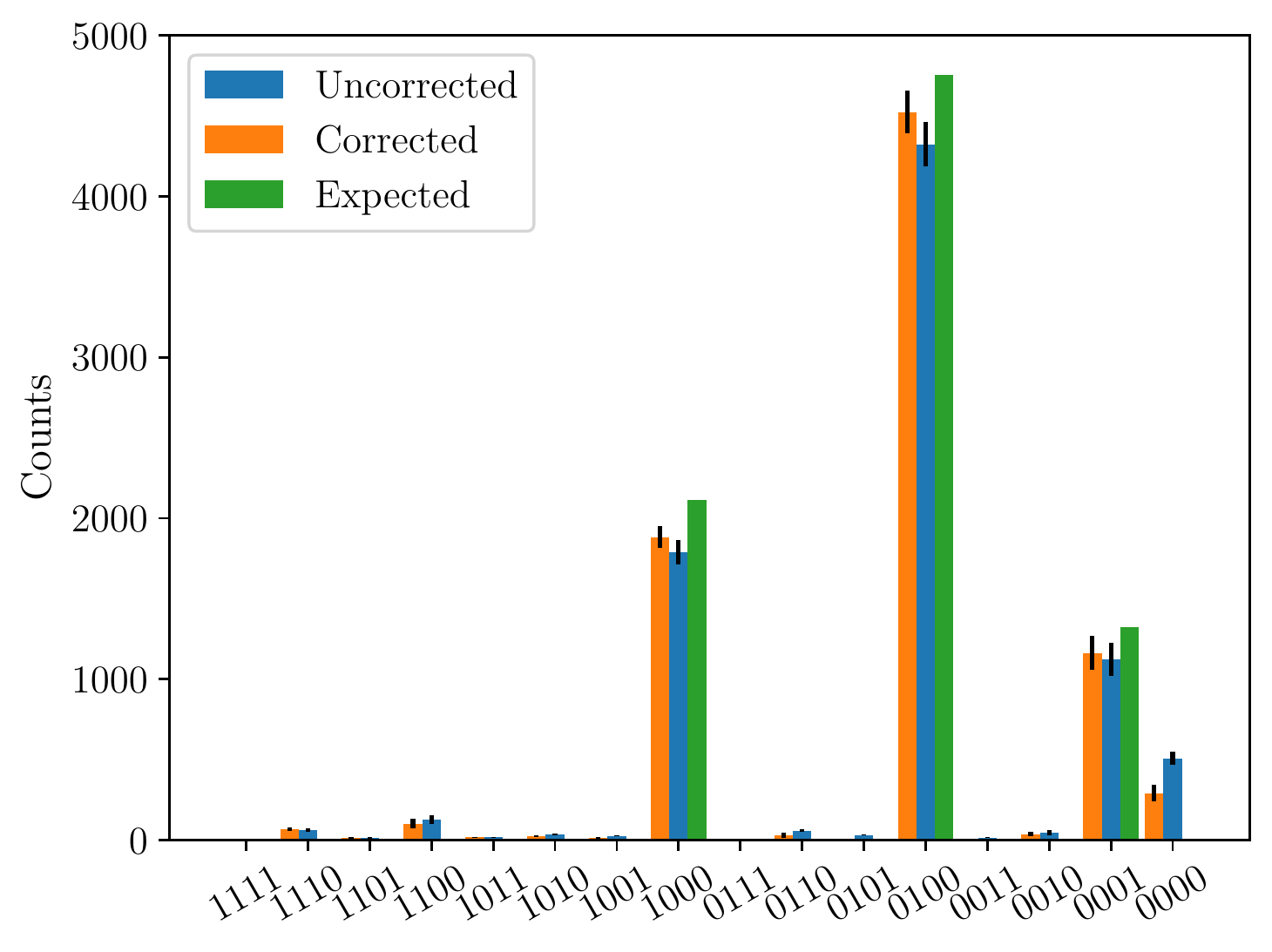}
}

\caption{\label{fig:ExBMProb} \emph{Comparison of ideal, measured, and noise-corrected measurement outcomes on quantum hardware.}  The results of the hand-compiled benchmark model shown in Fig.~\ref{fig:HandVsAuto}(a) are jackknifed over several days of independent experimental runs on the IBMQ-X2 [panel (a)] and IBMQ-Vigo [panel (c)] hardware.  Similarly, the jackknifed results for auto-compiled benchmark model shown in Fig.~\ref{fig:HandVsAuto}(b) are shown for the IBMQ-X2 [panel (b)] and IBMQ-Vigo [panel (d)] hardware.  The rightmost green bars are the noiseless expectations, the center blue bars are the raw measurements, and the left orange bars have a measurement filter applied.  Black lines indicate $1\sigma$ confidence intervals.}
\end{figure*}

\begin{figure}[t]
\subfloat[Convex KL divergences for IBMQ-X2]{
\includegraphics[width=0.95\columnwidth]{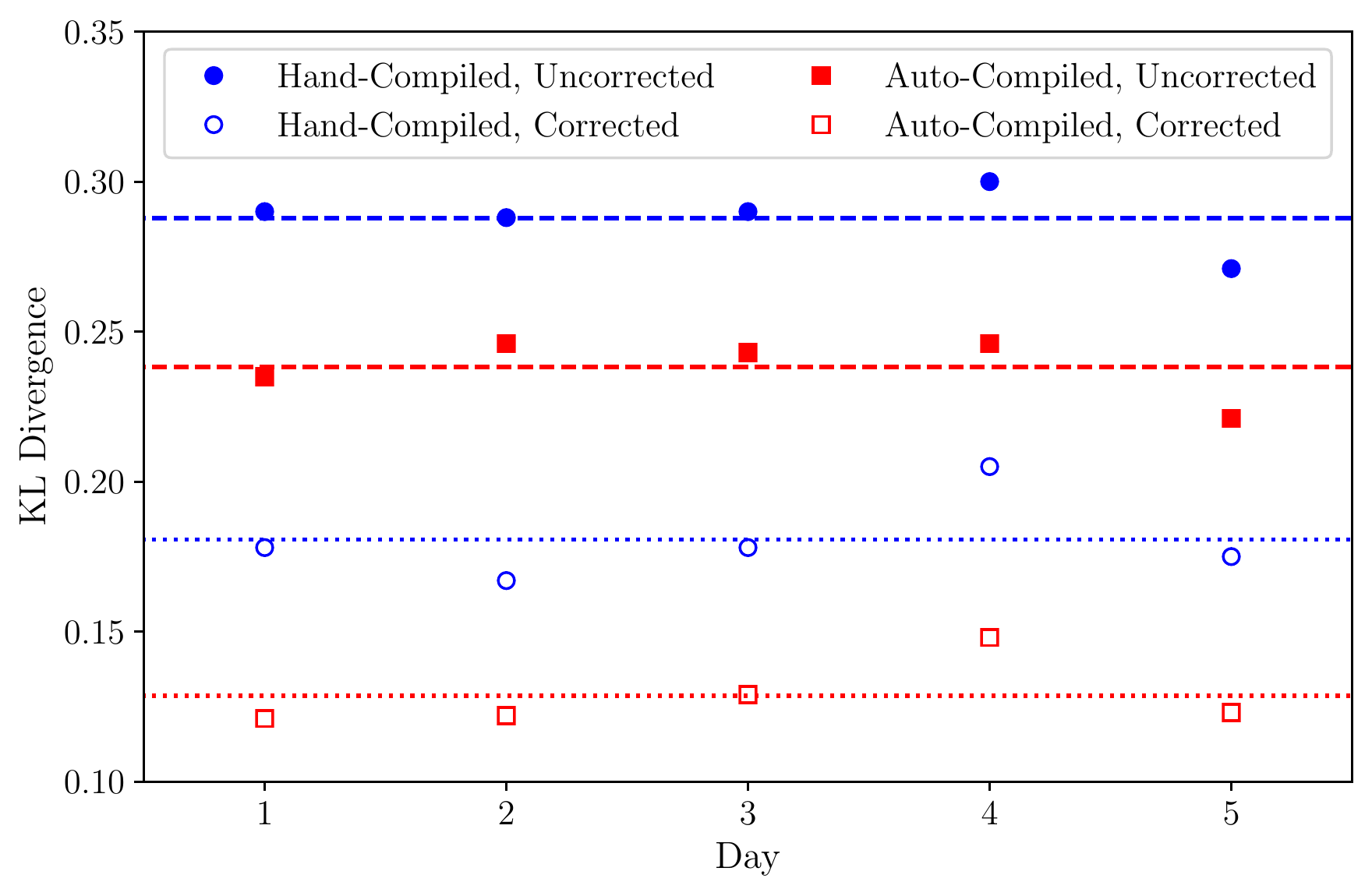}
}\\
\subfloat[Convex KL divergences for IBMQ-Vigo]{
\includegraphics[width=0.95\columnwidth]{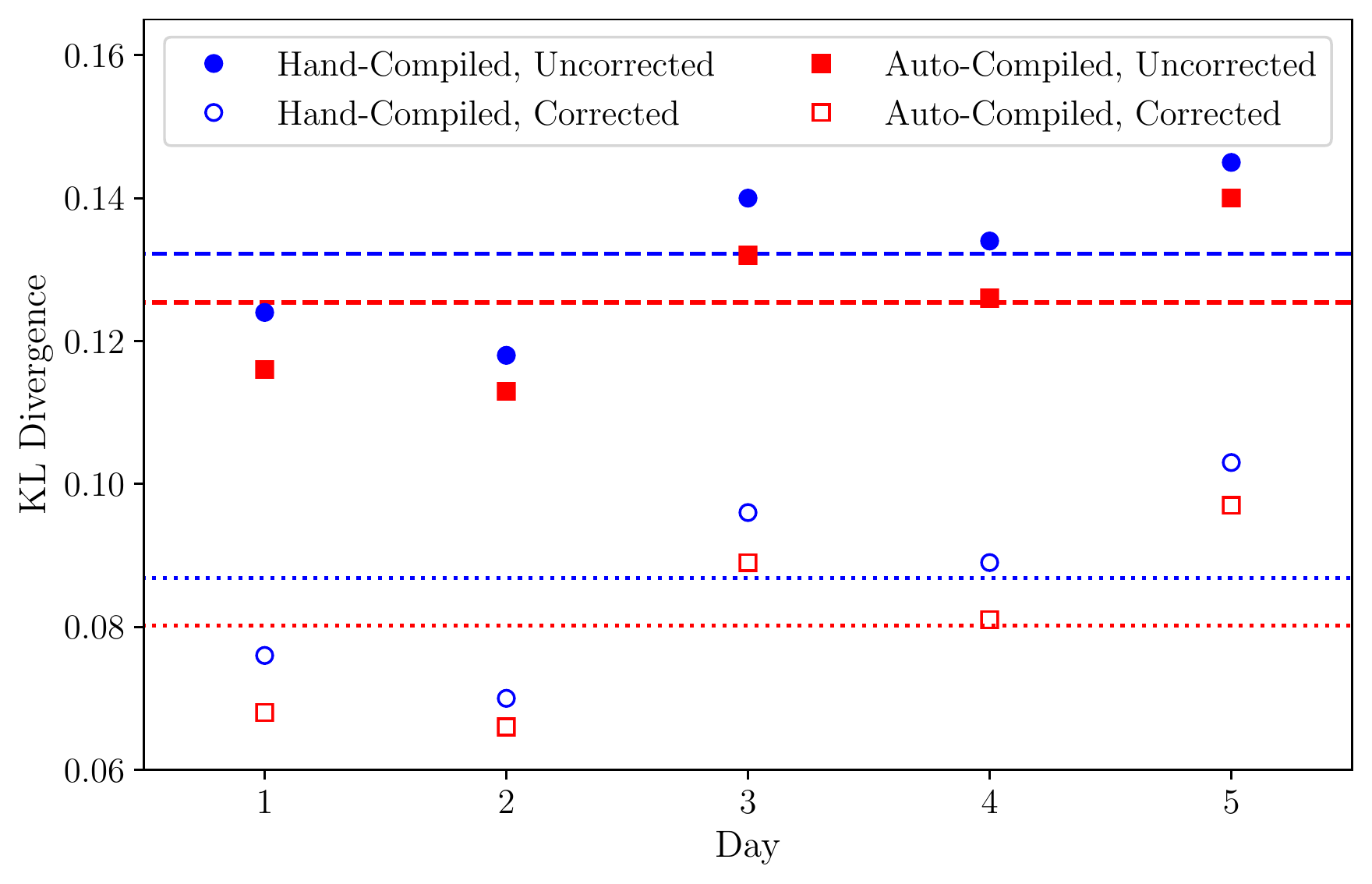}
}

\caption{\label{fig:KLBench} \emph{Convex KL divergence between ideal and measured QAML outcomes with time.} The convex KL divergence Eq.~\eqref{eq:KLd} between the ideal, noiseless measurement probabilities and the measurement probabilities inferred from 25 experiments of $2^{13}$ shots is shown as a function of experimental run day for the (a) IBMQ-X2 and (b) IBMQ-Vigo devices.  Filled symbols use the raw experimental counts (center blue bars in Fig.~\ref{fig:ExBMProb}) and empty symbols use the counts with measurement noise filter applied (left orange bars in Fig.~\ref{fig:ExBMProb}).  Lines indicate the KL divergences computed using all measurements from all days.}
\end{figure}

As metrics for assessing the performance of our QAML models, we utilize both the raw experimental counts used to infer measurement probability distributions and a convex version of the Kullback-Leibler (KL) divergence between the ideal ($p_T$) and estimated ($p_N$) distributions, as implemented in SciPy~\cite{2020SciPy-NMeth},
\begin{align}
\label{eq:KLd}\mathrm{KL}\left(p_T,p_N\right)&=\left\{\begin{array}{cc} p_T\left[\log \left(\frac{p_T}{p_N}\right)-1\right]+p_N& p_T>0,p_N>0\\ p_N&p_T=0, p_N\ge 0\\ \infty&\mbox{otherwise}\end{array}\right.\, .
\end{align}
The noise levels of NISQ devices fluctuate over time, and so to account for these statistical variations we implemented a jackknife procedure~\cite{efron1982jackknife} for the mean and variance including bias correction, utilizing 25 experimental runs per day of $2^{13}=8192$ shots each across 5 days.  We further refine each experimental run using the measurement noise filter implemented in Qiskit~\cite{Qiskit}, which produces a measurement noise correction map from a collection of calibration measurements which are performed immediately before the experimental shots.  

The results of our jackknife analysis on the experimental measurement counts per state are shown in Fig.~\ref{fig:ExBMProb}.  Here, panels (a) and (c) are the results for the hand-compiled model circuit in Fig.~\ref{fig:HandVsAuto}(a) and panels (b) and (d) are for the auto-compiled circuit in Fig.~\ref{fig:HandVsAuto}(b).  Panels (a) and (b) are run on the IBMQ-X2 device, and panels (c) and (d) are on the IBMQ-Vigo device.  In all panels, the rightmost green bar represents the ideal counts given by the model probability vector $\mathbf{p}=\left(8/31,18/31,5/31\right)$, the center blue bars are the raw experimental measurements without noise calibration applied, and the leftmost orange bars are the experimental measurements with the noise calibration applied.  The black lines centered on the tops of the bars indicate the $1\sigma$ confidence intervals from the jackknife procedure.  As noted above, qubits 0, 1, and 3 map to the probabilities $p_0$, $p_1$, and $p_2$, respectively, and qubit 2 is the ancilla.  Clearly, the application of the measurement noise filter improves the fidelity of the results.  Also, generally speaking, the results for the Vigo device (lower panels) are closer to the ideal results than for the X2 (upper panels).  The largest probability state resulting from errors is the state $|0000\rangle$ with no ``hot" physical bits, followed by $|1100\rangle$, with the two highest probability physical bits ``hot."  We note that the outcomes involving the ancilla qubit in the $|1\rangle$ state can be removed in postselection by virtue of the fact that the sequential preparation scheme should end with the ancilla in the $|0\rangle$ state (see Sec.~\ref{sec:TNSP}), but this results in small corrections for the present case.  Finally, we see that the auto-compiled results using the approach of Sec.~\ref{sec:compilation} (right panels) are generally closer to the ideal results than the hand-compiled circuits (left panels), though this is not true for each state individually.

In Fig.~\ref{fig:KLBench} we display the KL divergence between the ideal, noiseless probabilities of measuring each individual quantum state and the measurement probabilities estimated from 25 experiments of $2^{13}$ shots without (filled symbols) and with (empty symbols) the measurement noise calibration filter applied.  The x axis denotes consecutive experimental days, and the horizontal lines indicate the KL divergence resulting from the distributions averaged over all days.  Clearly, the application of the measurement noise filter improves the estimation of probabilities, as indicated by a lower KL divergence with respect to the ideal results.  In addition, the auto-compiled circuits (squares) show a lower KL divergence than the hand-compiled circuits, likely due to their shallower circuits.  Finally, we find that the Vigo results in panel (b) have lower KL divergence than the X2 results in panel (a), indicating an overall lower noise level for these days, in spite of the day-to-day fluctuations in the KL divergence being comparable in magnitude between the two machines.

\section{Example using the MNIST dataset}
\label{sec:MNIST}

\begin{figure}[t]
  \begin{center}
\includegraphics[width=0.95\columnwidth]{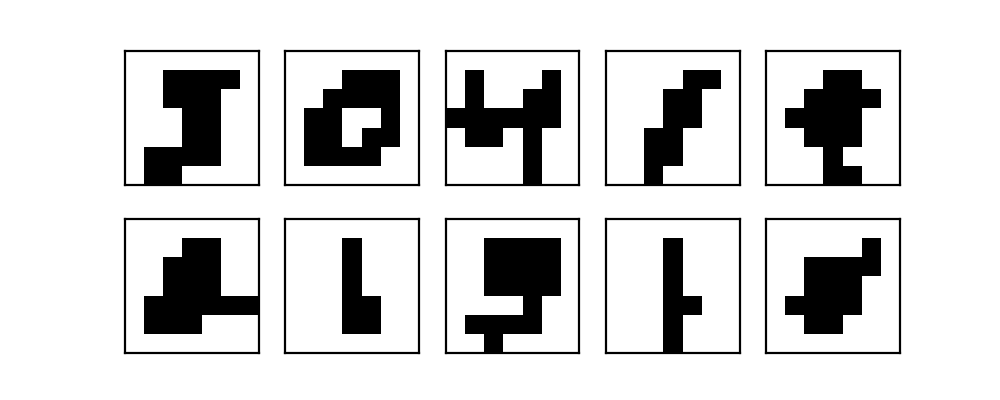}  
\caption{\label{fig:MNISTProcessed} \emph{Example processed MNIST data} produced by downsampling through a max filter to 7$\times$7 pixels and binarization.  Clockwise from top left, the truth labels are 5,0,4,1,9,4,1,3,1,2.}
\end{center}
\end{figure}

The exactly solvable benchmark presented in Sec.~\ref{sec:ExactMPS} provided data that was simple and well-structured enough that it could be exactly memorized using a single qubit to mediate bistring correlations.  In this section, we consider a QAML benchmark that is again a generative MPS Born machine, analogous to Sec.~\ref{sec:ExactMPS}, but using data from the MNIST handwritten digit dataset~\cite{lecun2010mnist}, a canonical ML test case.  The MNIST dataset consists of greyscale images, each consisting of 28$\times$28 pixels, of the numbers 0 through 9.  We process the data by passing through a filter that returns the max value from every contiguous 2$\times$2 pixel block twice, resulting in images of size $7\times7$.  While this is not necessary, and produces less raw data available for learning, it reduces the number of isometries in the sequential preparation scheme to compile, allowing for both a more detailed case-by-case analysis and reducing the overall gate depth for the ancilla in the sequential preparation scheme.  Our next step in processing this data is to binarize the greyscale images, such that we can use the binary qubit encoding for simplicity.  Examples of the processed data are shown in Fig.~\ref{fig:MNISTProcessed}.

\begin{figure}[b]
  \begin{center}
\includegraphics[width=0.95\columnwidth]{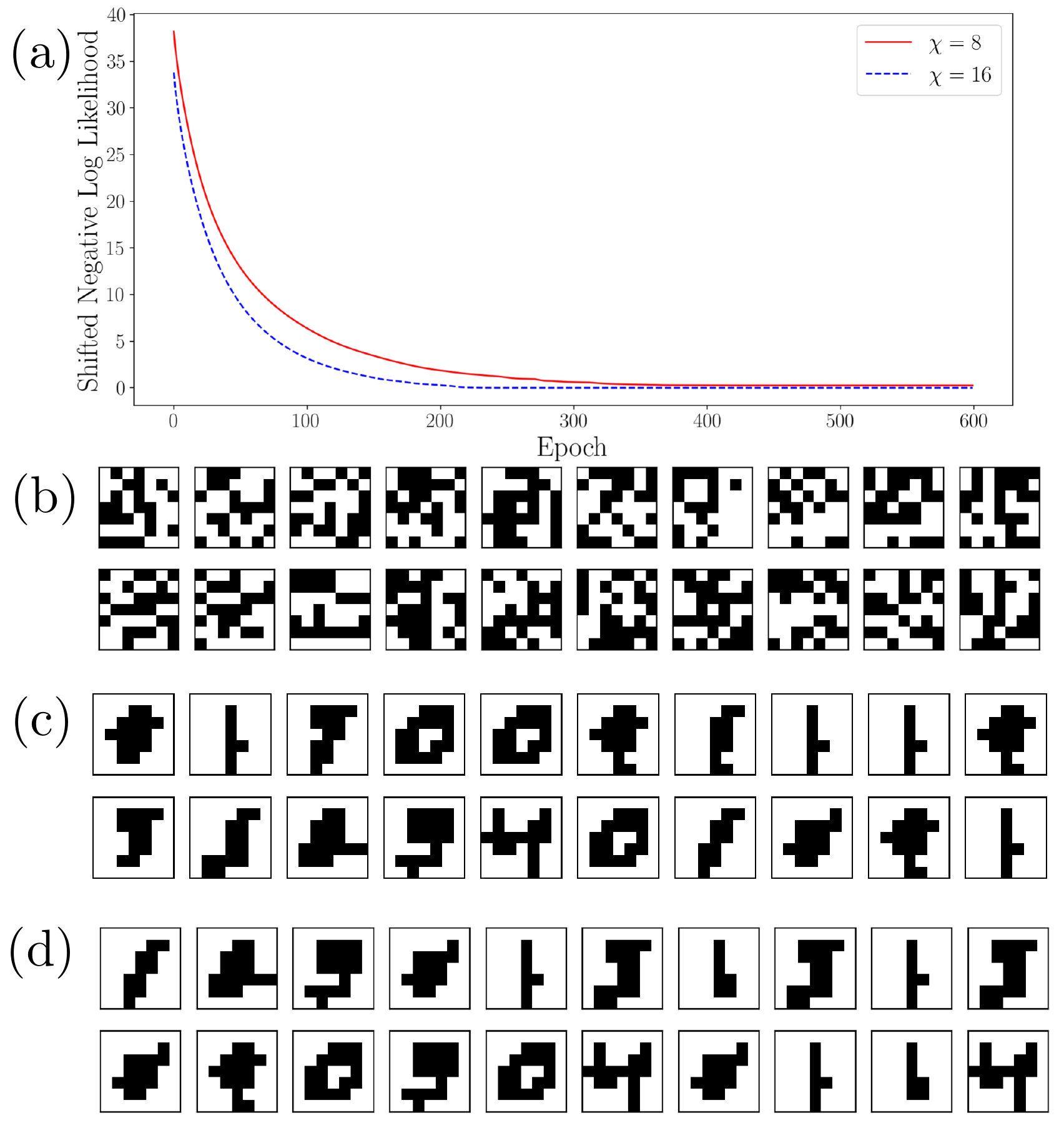}  
\caption{\label{fig:MNISTConvergence} \emph{Convergence of classical TN optimization.} The convergence behavior of the negative log-likelihood is shown for $\chi=8$ (red solid lines) and $\chi=16$ (blue dashed lines) in panel (a).  Both results use single-site gradient descent with a learning rate of $\eta=10^{-4}$.  Panel (b) displays sample data drawn from the initial, random $\chi=16$ MPS model before optimization, panel (c) shows samples drawn from the $\chi=8$ MPS model after optimization, and panel (d) shows samples drawn from the $\chi=16$ MPS model after optimization.  Because of the small size of the dataset, the $\chi=16$ model is able to reach the theoretical minimum and so memorize the full dataset.}
\end{center}
\end{figure}

In this work, we explore this dataset in the small-data regime where MPS models of modest bond dimension can memorize all patterns, analogous to the benchmark in Sec.~\ref{sec:ExactMPS}, but the bitstrings demonstrate more complex correlations that require more quantum resources.  The bitstrings are specified by mapping the $7\times7$ pixel arrays into binary vectors of length 49.  As an example of the classical optimization procedure, an MPS Born machine with $\chi=8$ (three ancilla qubits) converges to a negative log-likelihood of 2.563 following roughly 400 iterations of single-site gradient descent with a learning rate of $\eta=10^{-4}$ (see Sec.~\ref{sec:ClassicalTraining}) on the $N_T=10$ item dataset shown in Fig.~\ref{fig:MNISTProcessed}.  Using $\chi=16$ (four ancilla qubits), we reach the theoretical minimum value of the negative log-likelihood of $\log N_T$ after roughly 200 iterations.  The convergence behavior of these log-likelihoods is shown in Fig.~\ref{fig:MNISTConvergence}, together with samples drawn from the model before and after optimization.  Because the $\chi=8$ model does not reach the theoretical minimum, data elements not seen in the training data are present in the samples.  In contrast, the $\chi=16$ model reaches the theoretical minimum, and so only produces data samples from the training data.

Following the classical optimization of the MPS tensors, we clean the MPSs to remove small numerical values from the classical optimization procedure, place it into left-canonical form to describe a sequential preparation scheme, and then fix the remaining permutation ambiguity in the bond degrees of freedom using the transformation to diagonal gauge described in Sec.~\ref{sec:DG}.  For the case in which the diagonality center is at site 35, we find the isometries collected in Figs.~\ref{fig:site0}-\ref{fig:site48} of the appendix.  We compile these isometries using our greedy compilation procedure with a cost function (Eq.~\eqref{eq:AStarCostFunction}) tolerance of $\varepsilon=5\times 10^{-4}$ except for the diagonality center, which we compile using the methods of Ref.~\cite{soeken2019compiling} following the straightforward unitary completion procedure defined in Sec.~\ref{sec:Compilation}; the results are again shown in Figs.~\ref{fig:site0}-\ref{fig:site48} of the appendix.  In these figures, the ``raw" circuits may include non-native parameterized gates such as $\hat{\mathcal{S}}$ and single-qubit rotations with rotation angles near zero, while the ``expanded and cleaned" circuits compile the non-native gates into single-qubit gates and CNOTs using Eqs.~\eqref{eq:Scomp} and \eqref{eq:Fcomp}, collect adjacent single-qubit rotations, and then remove small single-qubit rotations with additional optimization passes.  We note that $\varepsilon=5\times 10^{-4}$ translates into roughly $\sqrt{\varepsilon}\sim 2\%$ error in the elements of the compiled unitary, which is comparable to the entangling gate error rates of current cloud-based machines.

\begin{figure}[t]
  \begin{center}
\includegraphics[width=0.65\columnwidth]{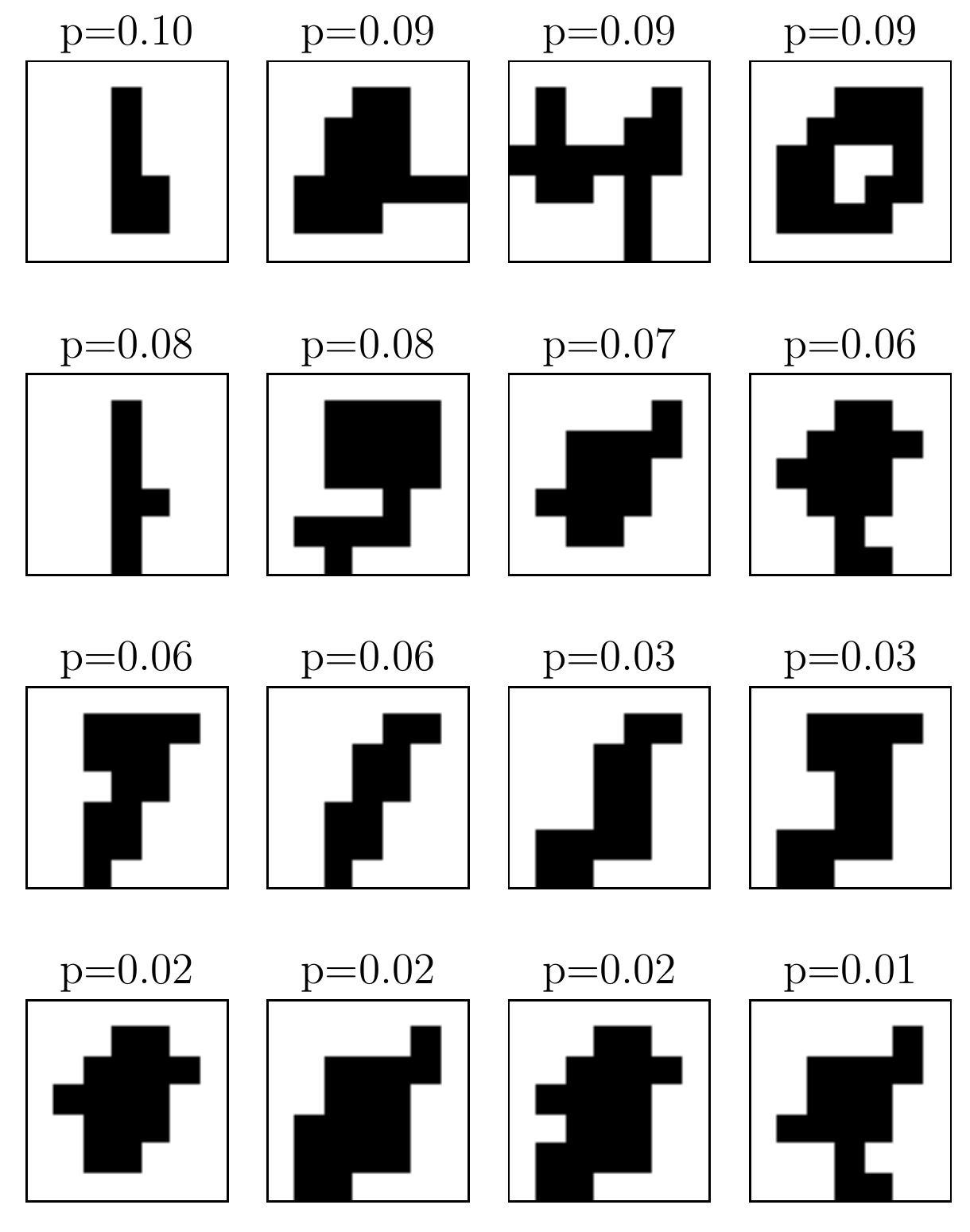}  
\caption{\label{fig:NoiselessMNIST} \emph{Example data sampled from MNIST model on simulated noiseless quantum hardware.} The data samples with probability $\ge 1\%$ obtained from the sequential preparation procedure defined by the isometries in Figs.~\ref{fig:site0}-\ref{fig:site48} are displayed together with their probability of occurrence estimated from $2^{13}$ shots.}
\end{center}
\end{figure}

To investigate the fidelity of the compiled model, we implemented the sequential preparation procedure in which a single data qubit is coupled to three ancilla qubits using the isometries in Figs.~\ref{fig:site0}-\ref{fig:site48} on the IBM \verb#qasm# hardware simulator.  As described in Sec.~\ref{sec:TNSP}, the isometries are applied from site 48 down to site 0 with data qubit measurement in the $z$ basis and reinitialization in the $|0\rangle$ state between the application of isometries.  The outcomes of these measurements constitute a data sample of the model.  While measurement and re-initialization in the midst of an experimental run are not supported on the current IBM quantum hardware, this operation is supported in the hardware simulators.  Example data generated from $2^{13}$ runs on ideal, noiseless hardware is shown in Fig.~\ref{fig:NoiselessMNIST}.  Here, the data samples with probability $\ge 1\%$ are shown together with their probabilities.  The training data from Fig.~\ref{fig:MNISTProcessed} are clearly recognized as the elements with highest probability, save for the digit 5, whose highest probability data sample involves confusion with the 1 digit.  We recall that deviations from the ideal training digits and their ideal occurrence probabilities of $10\%$ are a result of the restriction of the model to $\chi=8$ as well as the finite optimization tolerance $\varepsilon$.  Comparing with Fig.~\ref{fig:MNISTConvergence}(c), which shows samples taken from the classically trained model, it appears that the restriction to $\chi=8$ has a greater influence on the probabilities and sample variations than the finite compilation tolerance.

\begin{figure}[t]
  \begin{center}
\includegraphics[width=0.95\columnwidth]{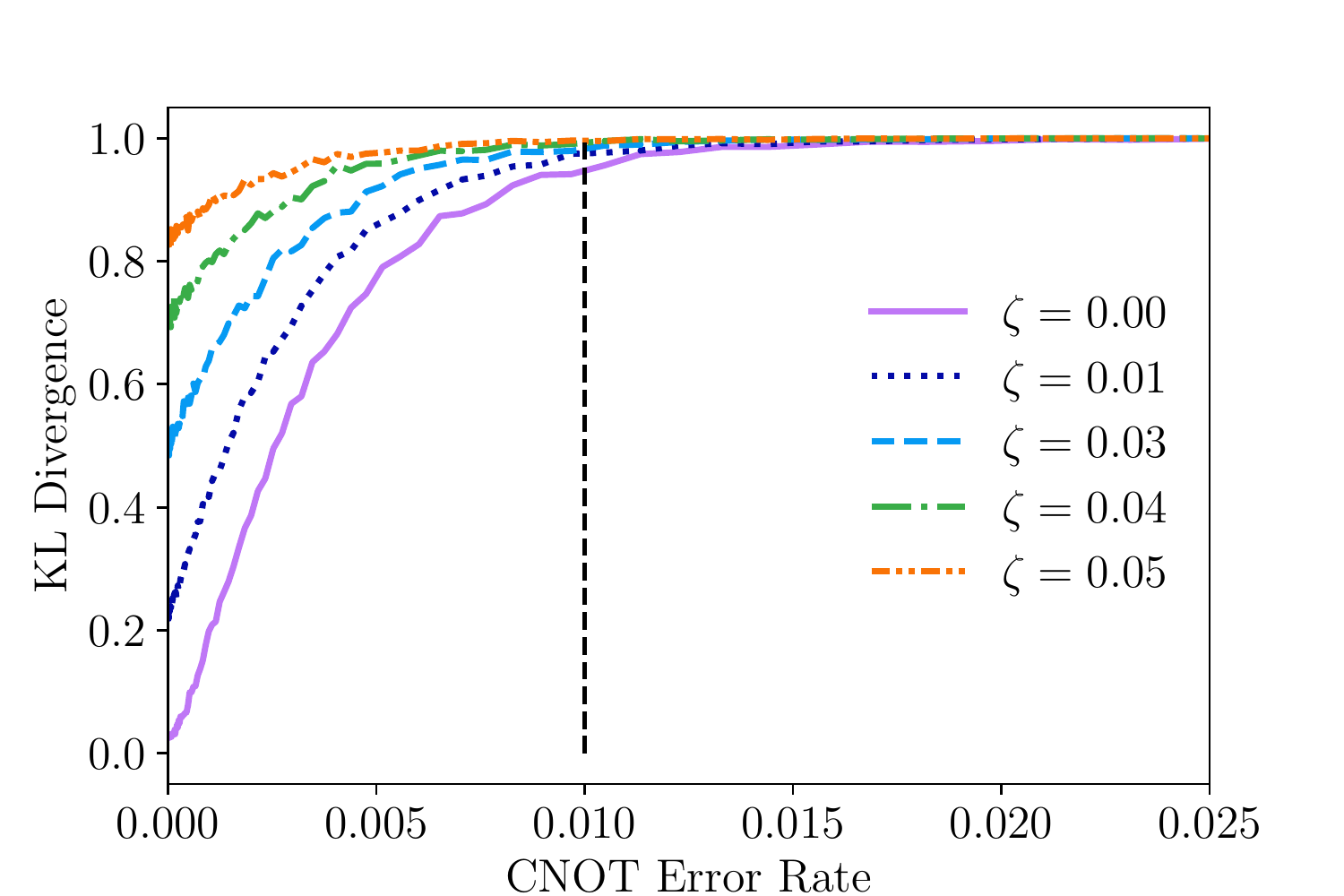}  
\caption{\label{fig:KLdiv} \emph{Convex KL divergence between sampled data without and with hardware noise.} The convex KL divergence Eq.~\eqref{eq:KLd} between a noiseless and noisy simulation is shown as functions of the CNOT error rate $\xi_2$ and the readout error $\zeta$.  The probability distributions of both the noiseless and noisy models are estimated from $2^{13}$ shots using a hardware simulator.  The vertical dashed line indicates 1\% CNOT error rate, a rough measure of the current state-of-the-art for NISQ devices.}
\end{center}
\end{figure}

To investigate the effects of hardware noise, we use a simple model of depolarization noise in which we take $\xi=1-F$ to be the average gate error, with $F$ the average gate fidelity.  With this, the depolarizing channel is represented by the operator
\begin{align}
\hat{\mathcal{E}}_{\mathrm{dep}}&=\left(1-p\right)\hat{I}+p\hat{D}\, ,
\end{align}
in which $p=2^{N_q}\xi/(2^{N_q}-1)$ with $N_q$ the number of qubits and the Kraus representation of the depolarizing channel is given by the operators
\begin{align}
\hat{\mathcal{E}}_{\mathrm{dep}}&=\left\{\sqrt{1-\left(4^{N_q}-1\right)p/4^{N_q}}\hat{I}^{\otimes N_q},\sqrt{p/4^{N_q}}\hat{\mathcal{P}}\right\}
\end{align}
in which 
\begin{align}
\hat{\mathcal{P}}=\left\{\hat{I},\hat{\sigma}_x,\hat{\sigma}_y,\hat{\sigma}_z\right\}^{\otimes N_q}\backslash \hat{I}^{\otimes N_q}\, ,
\end{align}
is the set of $N_q$-qubit products of Pauli operators without the $N_q$-qubit identity matrix.  In our model, we assign the same error $\xi_2$ to all CNOT gates and an error $\xi_1=\xi_2\times 10^{-2}$ to all single-qubit gates in accordance with typical IBM hardware characteristics.  In addition to the depolarizing error, we also include an uncorrelated-qubit (tensor product) readout noise model parameterized by~\cite{bravyi2020mitigating,nachman2019unfolding}
\begin{align}
P\left(0|0\right)&=1-\zeta\, ,P\left(1|0\right)=\zeta\, ,\\
P\left(0|1\right)&=2\zeta\, ,P\left(1|1\right)=1-2\zeta\, ,
\end{align}
in which $P\left(a|b\right)$ denotes the probability of obtaining measurement outcome $a$ from a preparation of the quantum state $|b\rangle$, assumed identical across all qubits for simplicity.

\begin{figure}[t]
  \begin{center}
\includegraphics[width=0.65\columnwidth]{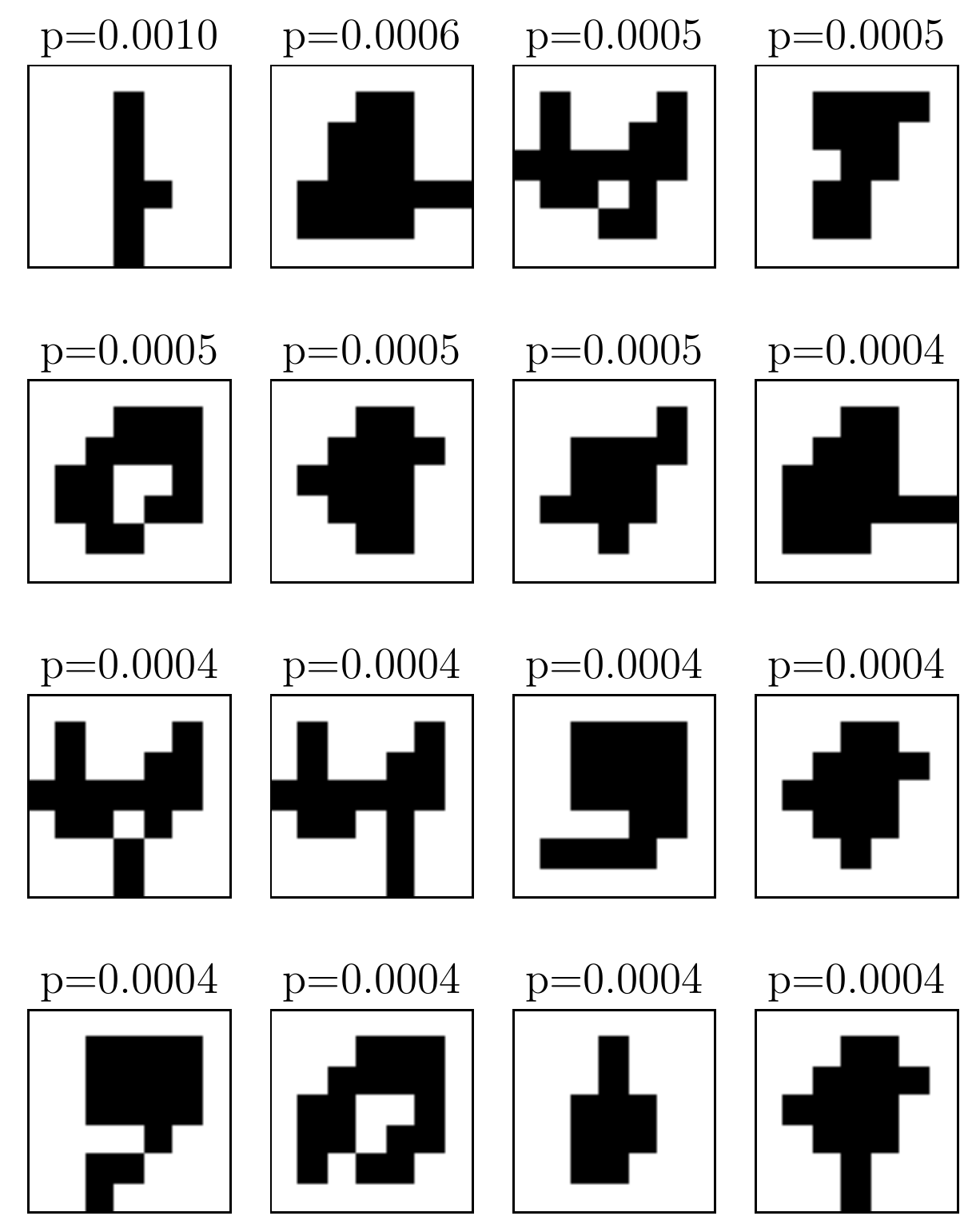}  
\caption{\label{fig:NoisyMNIST} \emph{Example data sampled from MNIST model on simulated noisy quantum hardware.} The highest probability data samples from the sequential preparation procedure defined by the isometries in Figs.~\ref{fig:site0}-\ref{fig:site48} on a simulated machine with depolarization error rate $\xi_2=0.01$ (dashed line in Fig.~\ref{fig:KLdiv}) are displayed together with their probability of occurrence estimated from $2^{13}$ shots.}
\end{center}
\end{figure}

\begin{figure}[t]
  \begin{center}
\includegraphics[width=0.95\columnwidth]{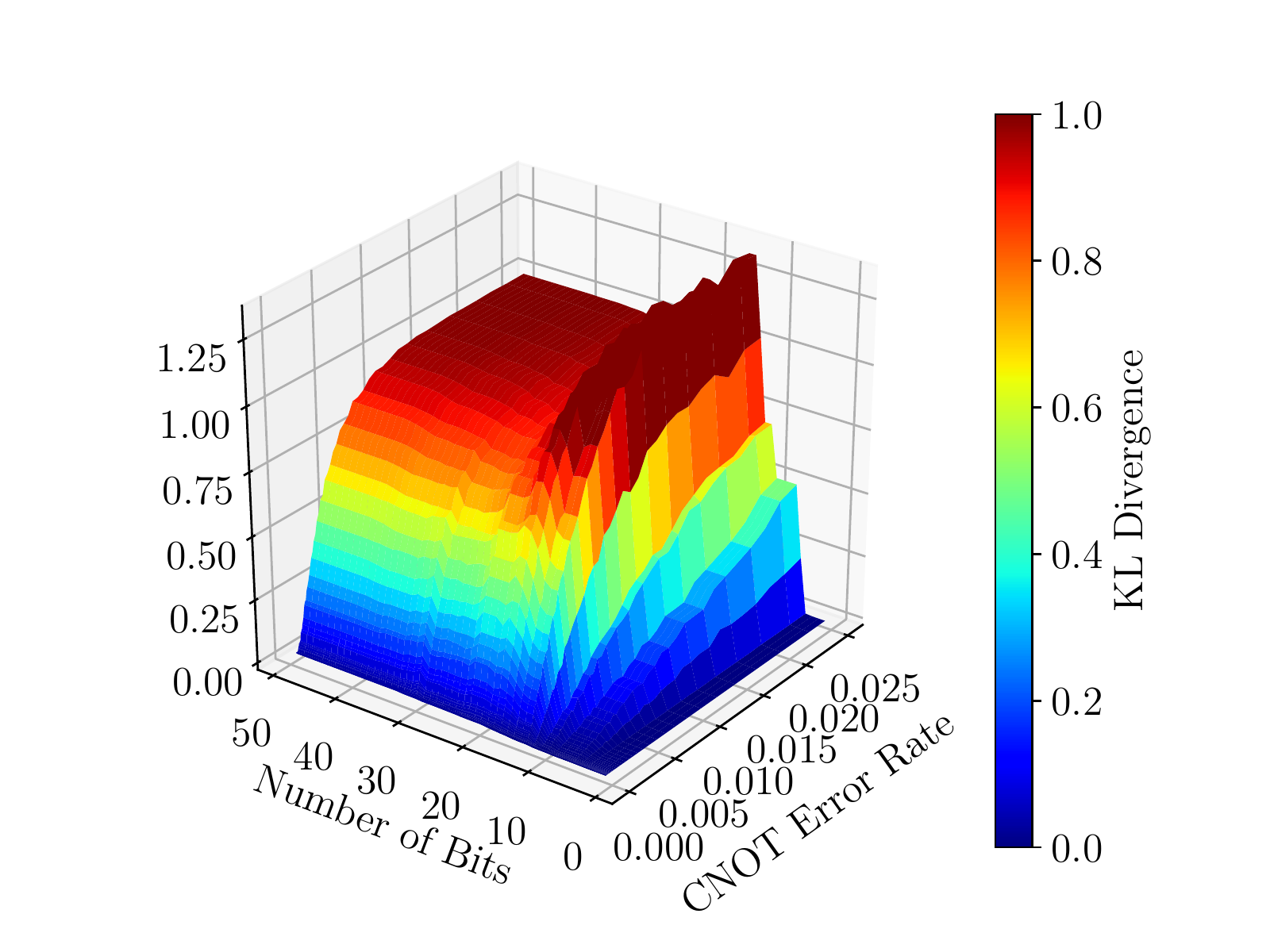}  
\caption{\label{fig:KLSurf} \emph{Convex KL divergence variation with noise and bitstring length.} The convex KL divergence Eq.~\eqref{eq:KLd} between a noiseless and noisy simulation is shown as a function of the CNOT error rate $\xi_2$ and the sampled bitstring length at zero readout error ($\zeta=0$).  The probability distributions of both the noiseless and noisy models are estimated from $2^{13}$ shots using a hardware simulator.}
\end{center}
\end{figure}

We again characterize the difference in predictions between the ideal and noisy outcomes using the convex KL divergence defined in Eq.~\eqref{eq:KLd}.  As before, we estimate the probabilities from ensembles of $N_s=2^{13}$ shots calculated using a hardware simulator for both the noiseless ``truth" model and the noisy models.  This divergence is shown in Fig.~\ref{fig:KLdiv} as a function of the error parameterizations $\xi_2$ and $\zeta$.  The CNOT error rate $\xi_2$ parameterizes both the two-qubit and single-qubit depolarization errors and $\zeta$ parameterizes the readout error.  The divergence shows a rapid rise driven by the appearance of data samples not present in the noiseless model.  The samples with greatest occurrence drawn from the model evaluated at the CNOT error rate $\xi_2\sim0.01$, indicated by the dashed vertical line in Fig.~\ref{fig:KLdiv}, and $\zeta=0$ are shown in Fig.~\ref{fig:NoisyMNIST}.  We can recognize many of the digits from the training set in this model, but they occur with significantly lower probabilities due to the appearance of additional noise-driven patterns.  Because of the sequential preparation, we can expect that the bits near the end of the bitstring (i.e.~for sites near 48) are produced at higher fidelity than those of lower site indices because of errors present in manipulating the ancilla qubits.  We see that this is the case in Fig.~\ref{fig:KLSurf}, which displays the convex KL divergence as functions of the CNOT error rate $\xi_2$ and the number of bits sampled in the bitstring at zero measurement error $\zeta=0$.  Non-monotonic behavior is due to the significant differences in complexity of the gate sequences to produce individual bits, see Figs.~\ref{fig:site0}-\ref{fig:site48} of the appendix.  A rise in error is seen as the number of bits increases as the total gate depth increases, peaking around ten bits.  After this point the KL divergence levels off as additional bits may require shallower gate sequences, bringing the overall agreement between the noisy and true probability distribution closer.

\section{Conclusions and outlook}
\label{sec:Concl}

We have presented a complete workflow for generative quantum-assisted machine learning (QAML) using Born machines with a matrix product state (MPS) tensor network (TN) structure.    In our workflow, classical data is encoded into quantum states using an embedding map, the ensemble of quantum states is learned as a TN Born machine using a classical DMRG-like procedure with gradient descent of the negative log-likelihood, and the model is compiled into operations for target quantum hardware to obtain data samples as measurement outcomes.  Using MPS-based models enables the use of highly quantum resource-efficient sequential preparation schemes requiring $\mathcal{O}\left(1\right)$ qubits for a classical data vector length $N$ and $\mathcal{O}\left(\log\chi\right)$ qubits for bond dimension $\chi$, which encapsulates the model expressivity.  We presented several optimizations in the compilation stage of the workflow, such as the introduction of the diagonal gauge of the MPS model that utilizes inherent freedom in the model representation to reduce the complexity of the compiled model, as well as greedy heuristics for finding shallow gate sequences matching a target isometry to a specified tolerance given hardware topology and allowed gate constraints.  We presented an exactly solvable benchmark model requiring two qubits, and assessed its performance on currently available quantum hardware.  We also presented an example application modeling features extracted from the MNIST dataset parametrically with depolarizing and readout hardware noise using a hardware simulator.

Our results lay the groundwork for utilizing TN models in practical QAML applications, and leave several avenues for future research.  First, the QAML demonstrations given in this work consist of overfit models, and so do not constitute ``true" machine learning models which should be able to appropriately generalize from data.  This is a result of either using data with very simple structure, as in our exactly solvable model, or using a very small sample size of training data, as in our MNIST application.  Small sample sizes were used in the present work to enable detailed analysis of model performance with limited quantum resources.  In future work, the generalization power of TN-based QAML models on NISQ hardware will be explored moving towards the large-data regime.  We also note that other studies have indicated that TN models with current training strategies generally have a tendency towards overfitting~\cite{Martyn2020Entanglement}.  Second, we have focused on the applications of MPSs to generative modeling, but other TN structures, such as tree tensor networks~\cite{PhysRevA.74.022320,stoudenmire2018learning} may also be useful for QAML applications, as well as other tasks such as feature extraction and classification.   The procedures outlined in this paper can be readily adapted to compiling the isometries appearing in models for other TNs and other applications.  Finally, the procedure outlined in this paper wherein a model is trained classically before being compiled to a quantum device cannot by itself yield a quantum advantage, as it requires the model to be both classically and quantumly simulable.  However, our procedures will be useful in designing and analyzing TN-inspired model structures for scaling towards the classically intractable regime, and can also serve as ``preconditioners" where a model trained using optimal classical strategies is augmented with additional quantum resources and then trained directly on the quantum device or in a hybrid quantum/classical optimization loop, potentially avoiding local minima and speeding up optimization times.

\section{Acknowledgements}

We would like to thank Dave Clader, Giuseppe D'Aguanno, and Colin Trout for useful discussions, and would like to acknowledge funding from the Internal Research and Development program of the Johns Hopkins University Applied Physics Laboratory.

\bibliography{Refs}
\bibliographystyle{apsrev4-1}

\clearpage

\appendix
\section{Isometries and optimized gates for MNIST dataset; $\chi=8$}
\label{sec:MNISTIso}
\begin{figure}[h]
%

\subfloat[Isometry]{
\includegraphics[width=0.3\columnwidth]{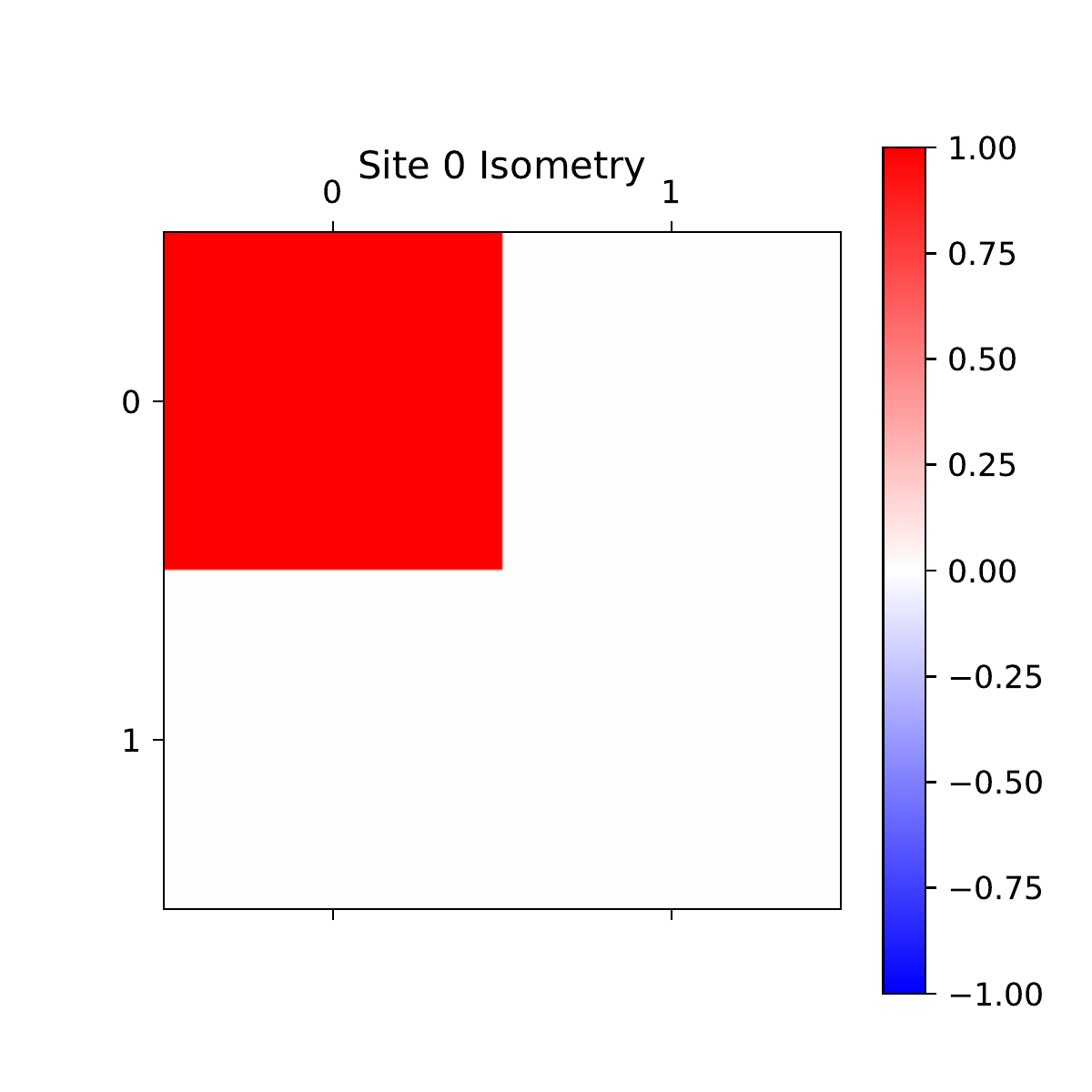}
}
\subfloat[Optimized gate]{
\includegraphics[width=0.3\columnwidth]{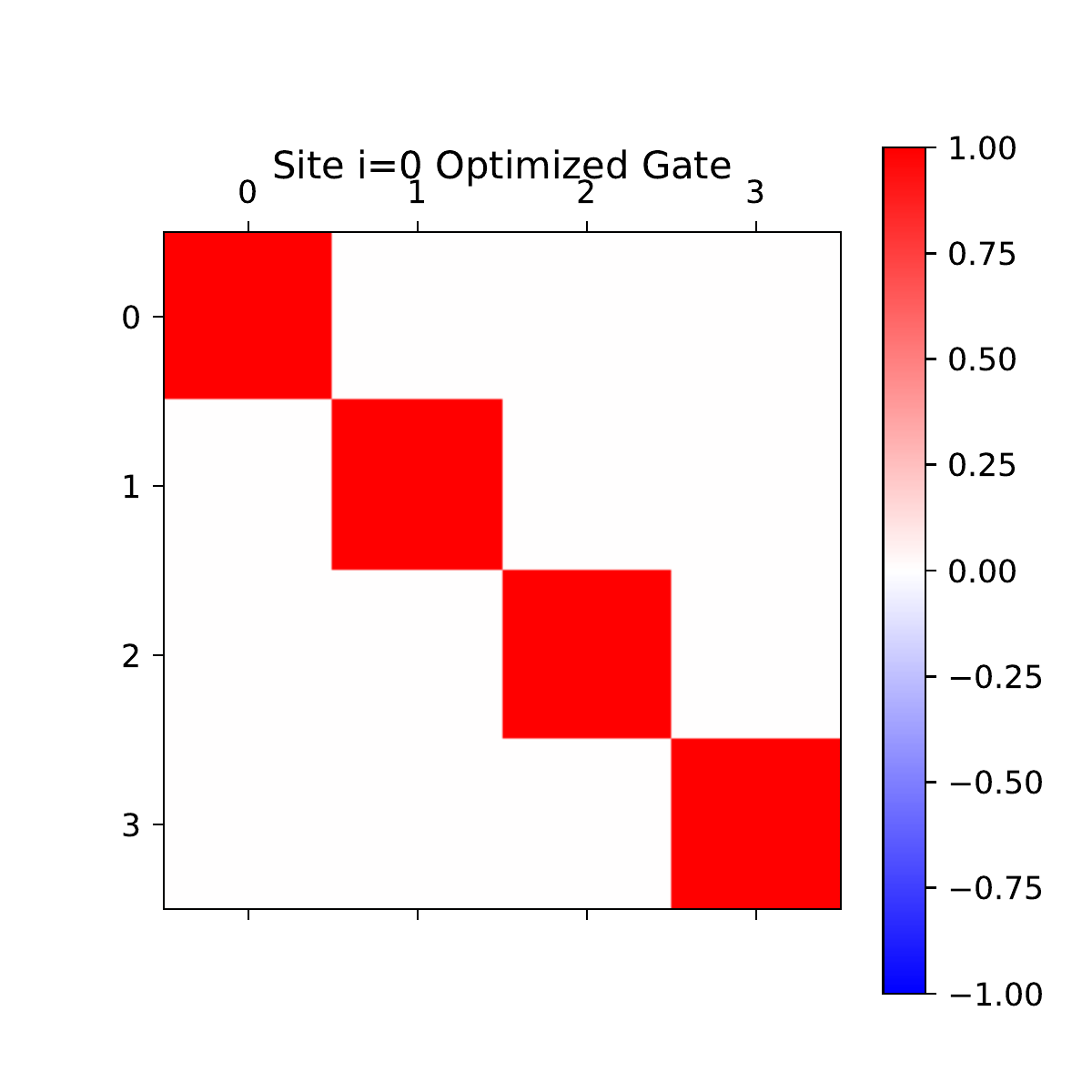}
}
\subfloat[Circuit from optimization]{
\includegraphics[width=0.3\columnwidth]{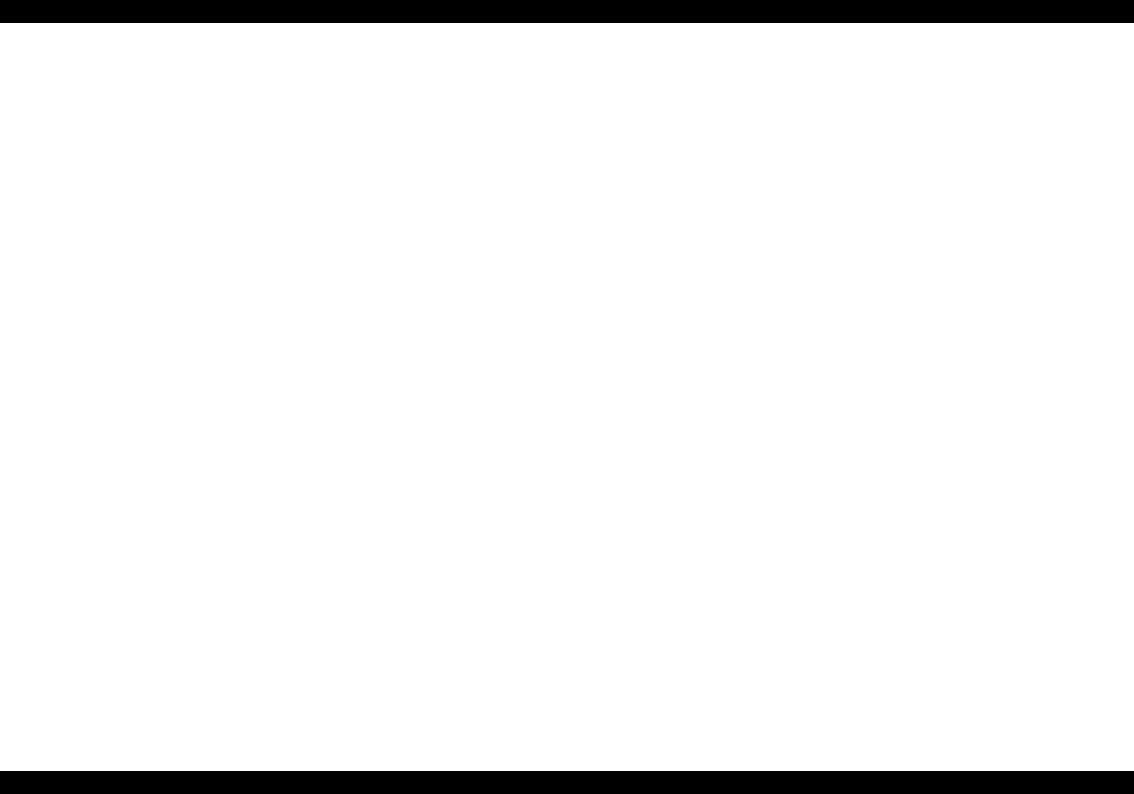}
}
\caption{\label{fig:site0} Optimization for site 0}
\end{figure}
%
\begin{figure}[h]
%

\subfloat[Isometry]{
\includegraphics[width=0.3\columnwidth]{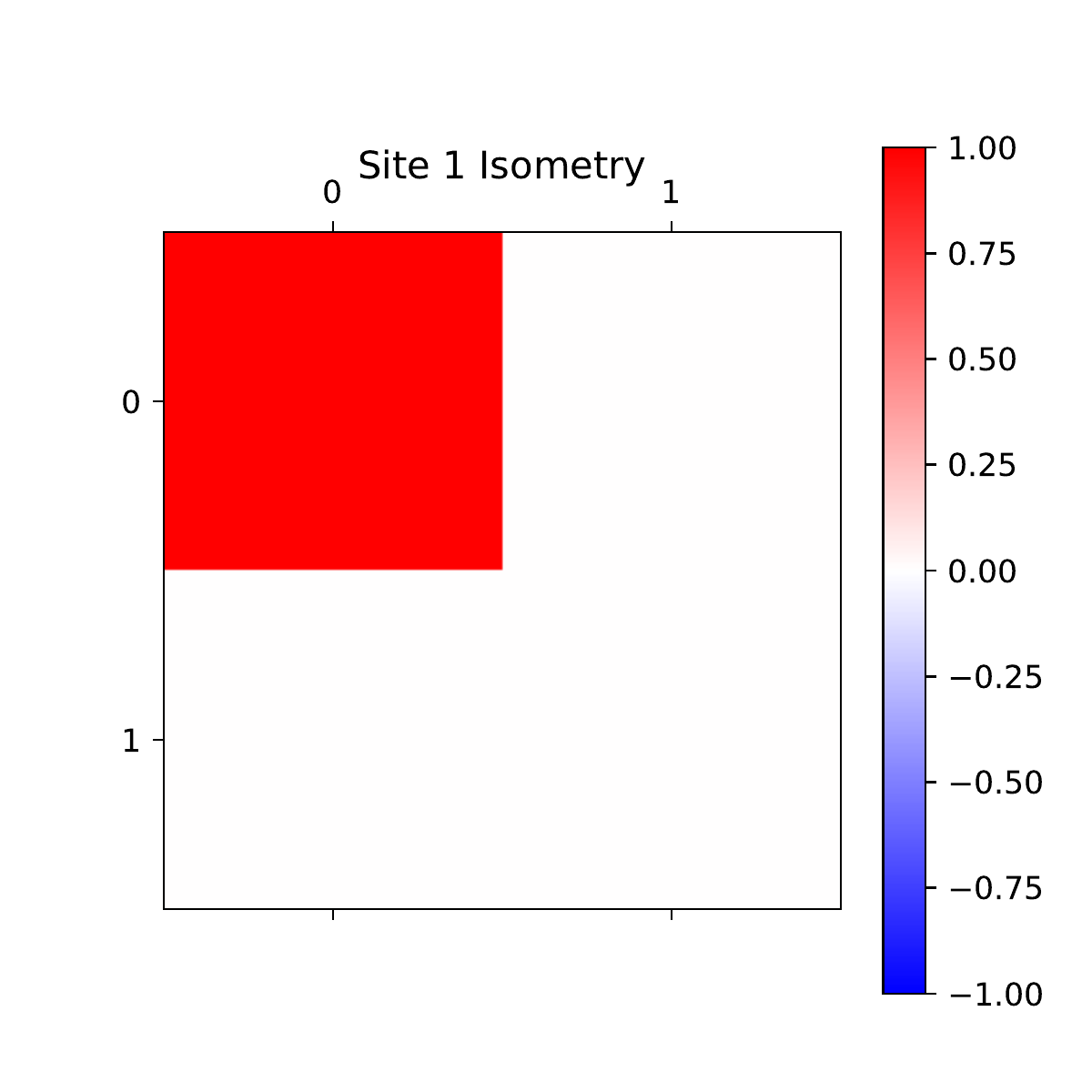}
}
\subfloat[Optimized gate]{
\includegraphics[width=0.3\columnwidth]{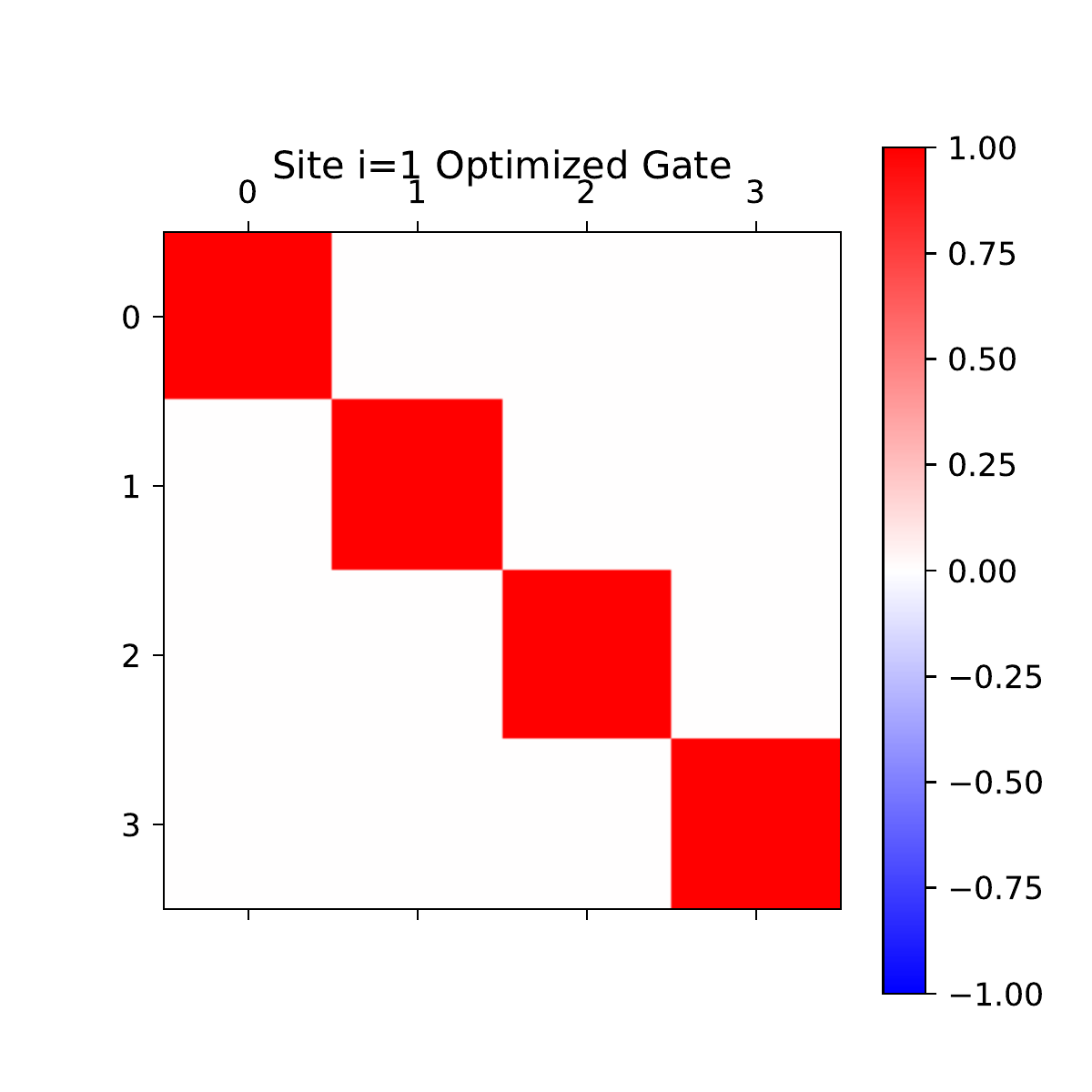}
}
\subfloat[Circuit from optimization]{
\includegraphics[width=0.3\columnwidth]{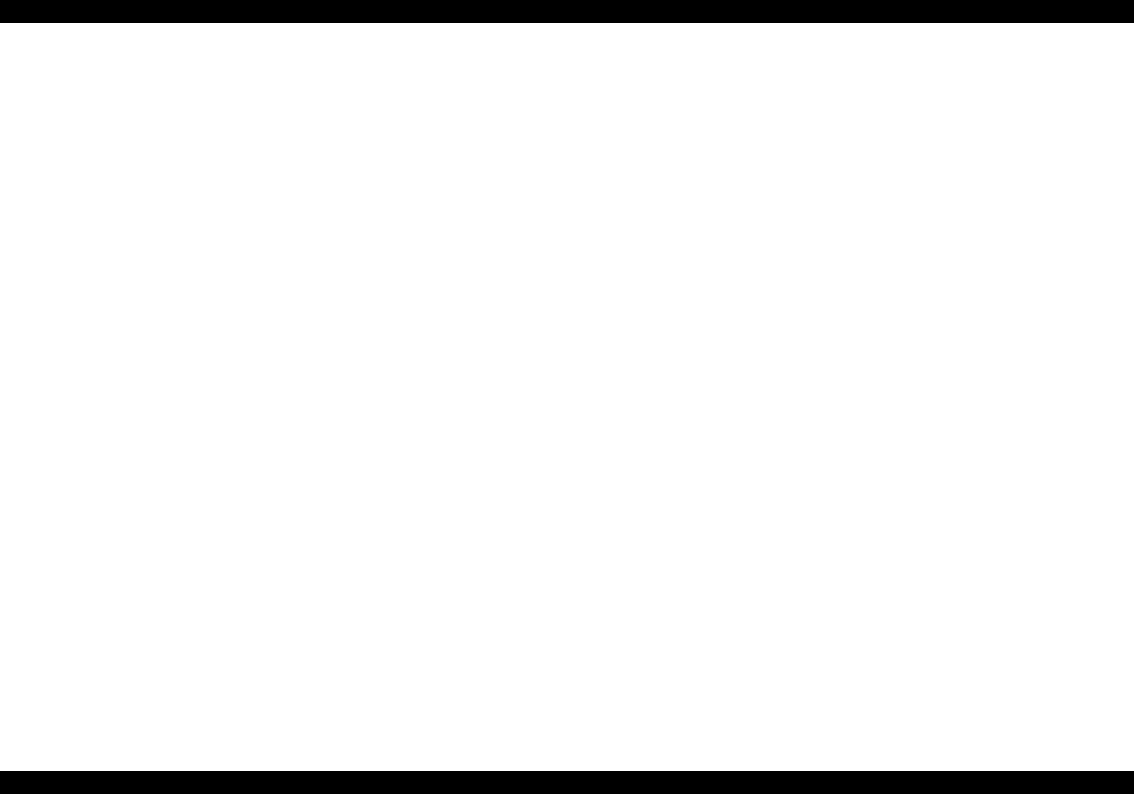}
}
\caption{Optimization for site 1}
\end{figure}
%
\begin{figure}[h]
%

\subfloat[Isometry]{
\includegraphics[width=0.3\columnwidth]{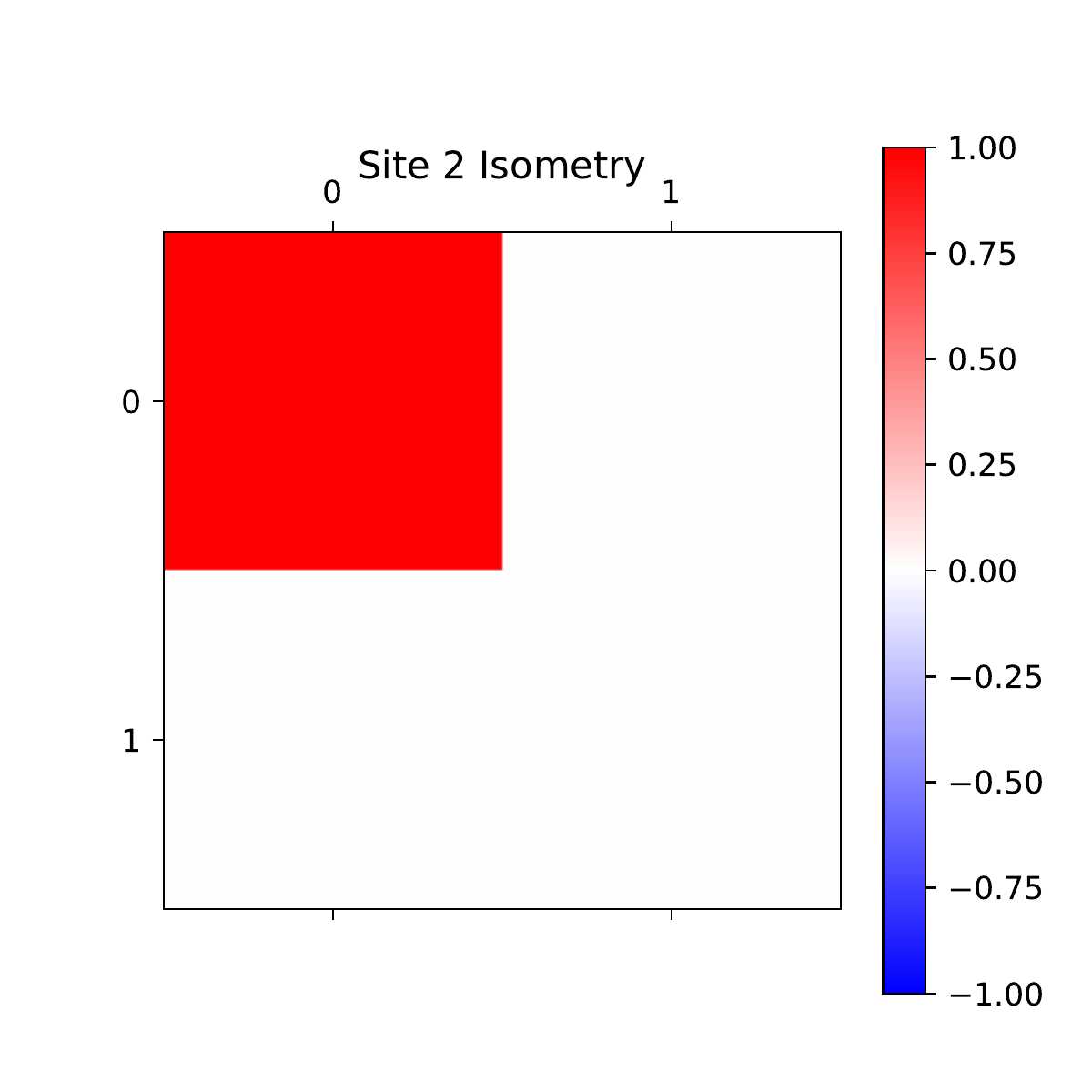}
}
\subfloat[Optimized gate]{
\includegraphics[width=0.3\columnwidth]{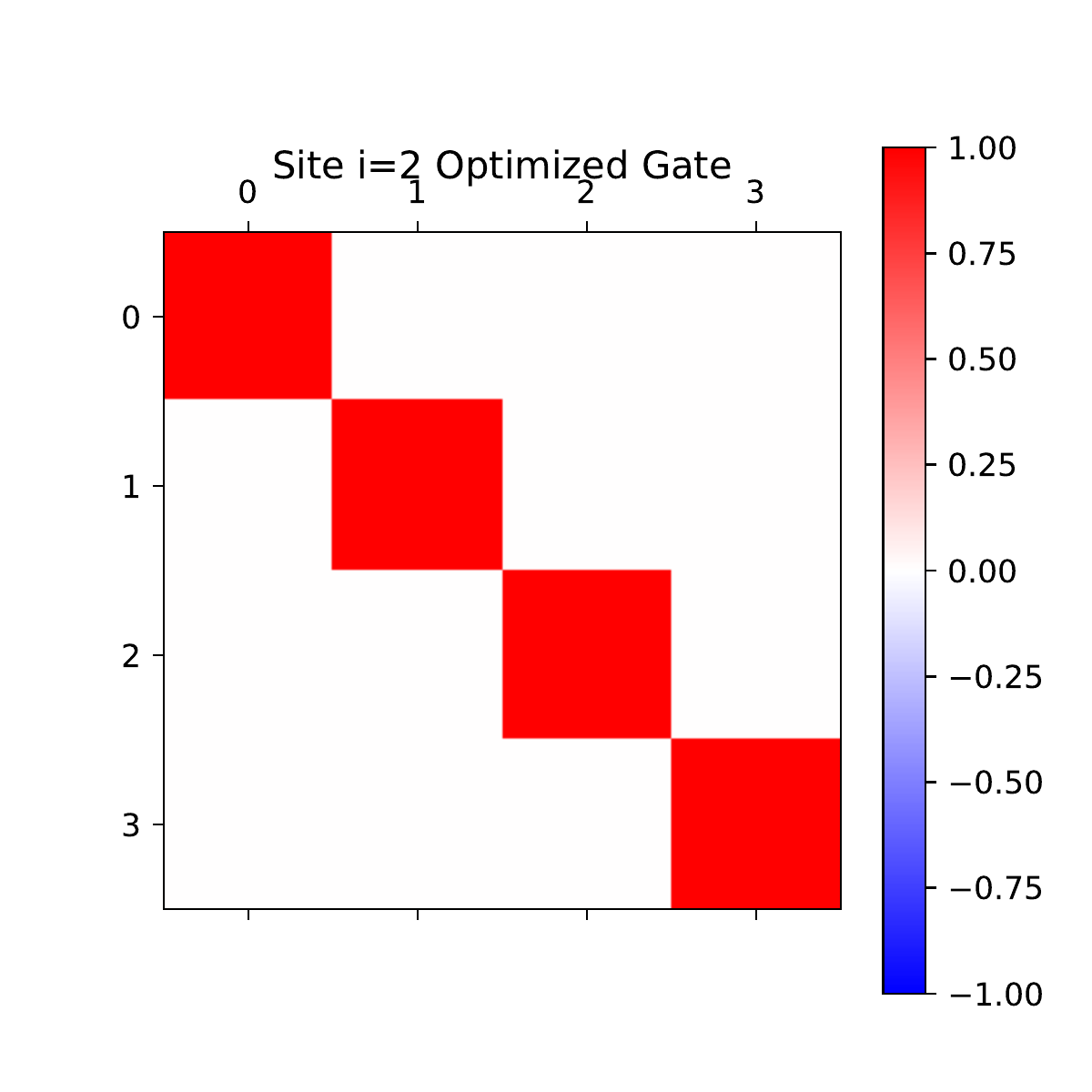}
}
\subfloat[Circuit from optimization]{
\includegraphics[width=0.3\columnwidth]{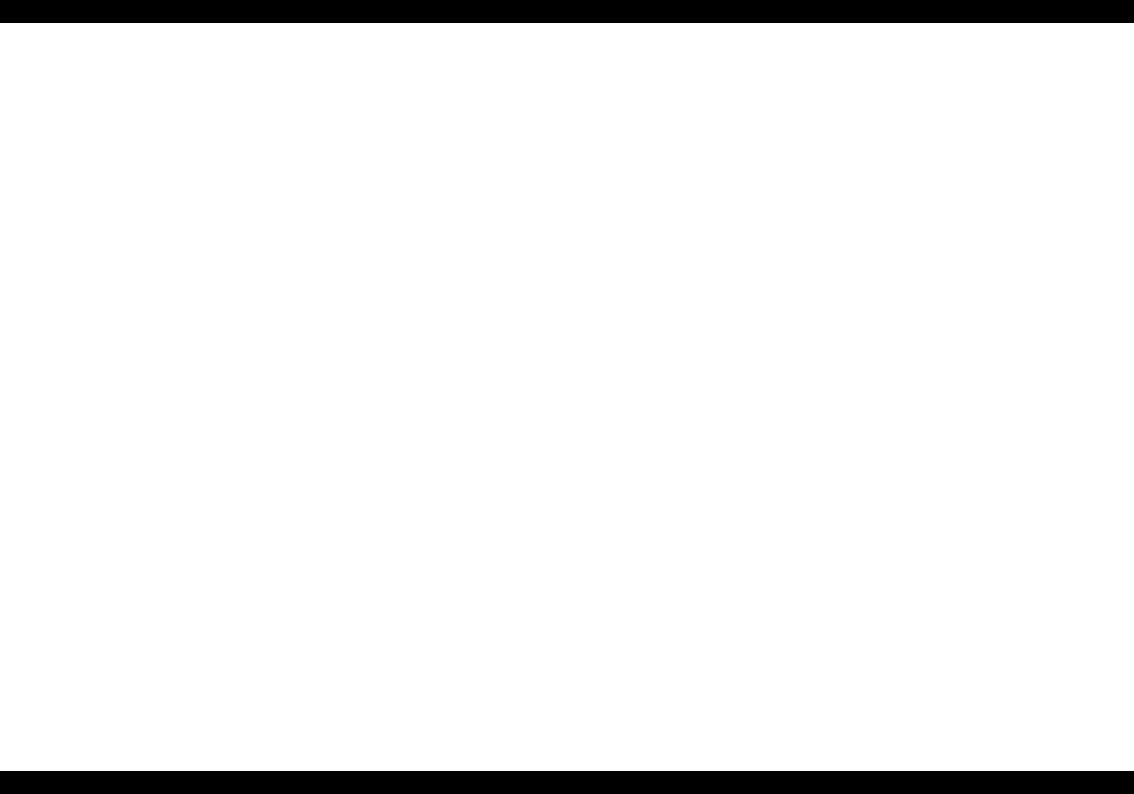}
}
\caption{Optimization for site 2}
\end{figure}
%
\begin{figure}[h]
%

\subfloat[Isometry]{
\includegraphics[width=0.3\columnwidth]{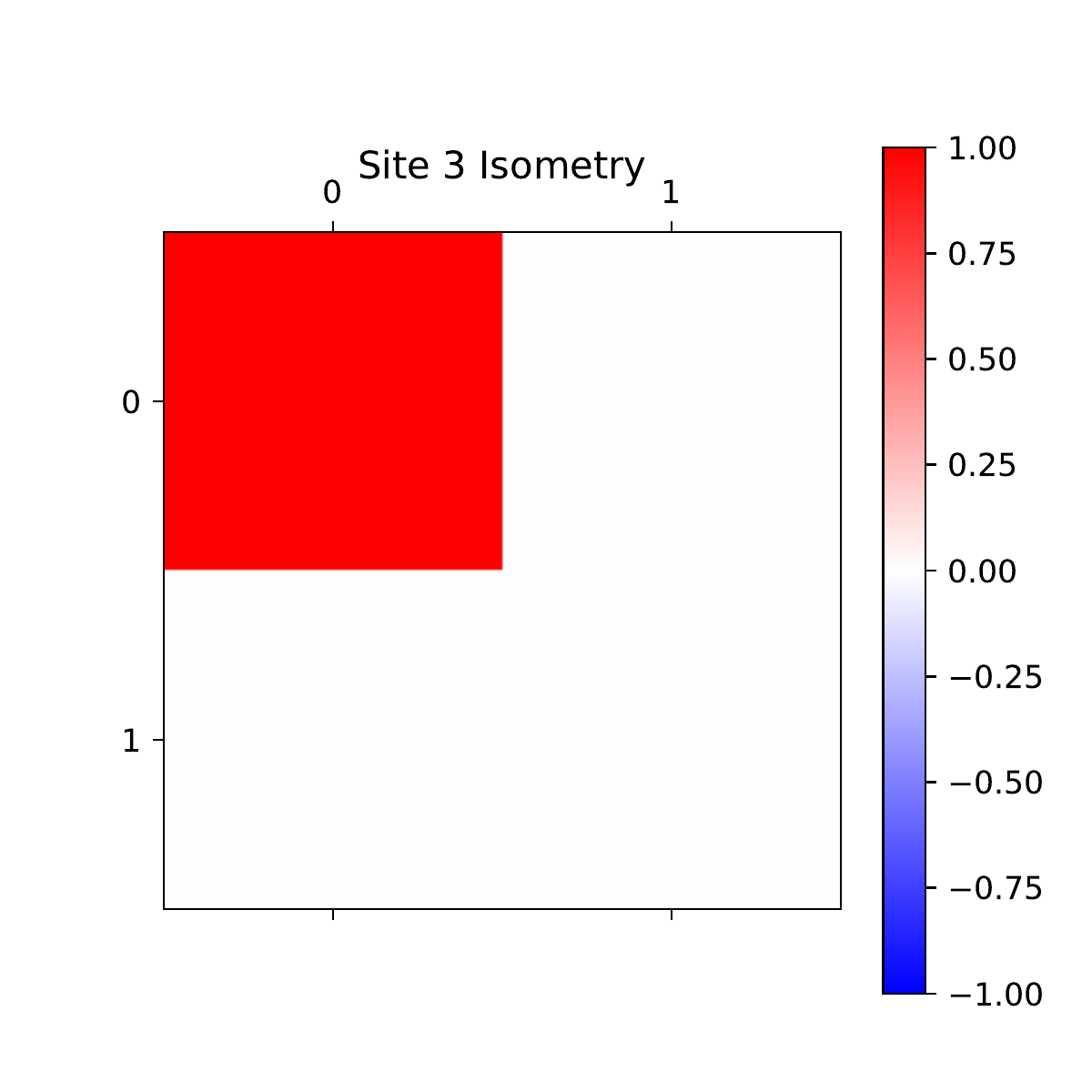}
}
\subfloat[Optimized gate]{
\includegraphics[width=0.3\columnwidth]{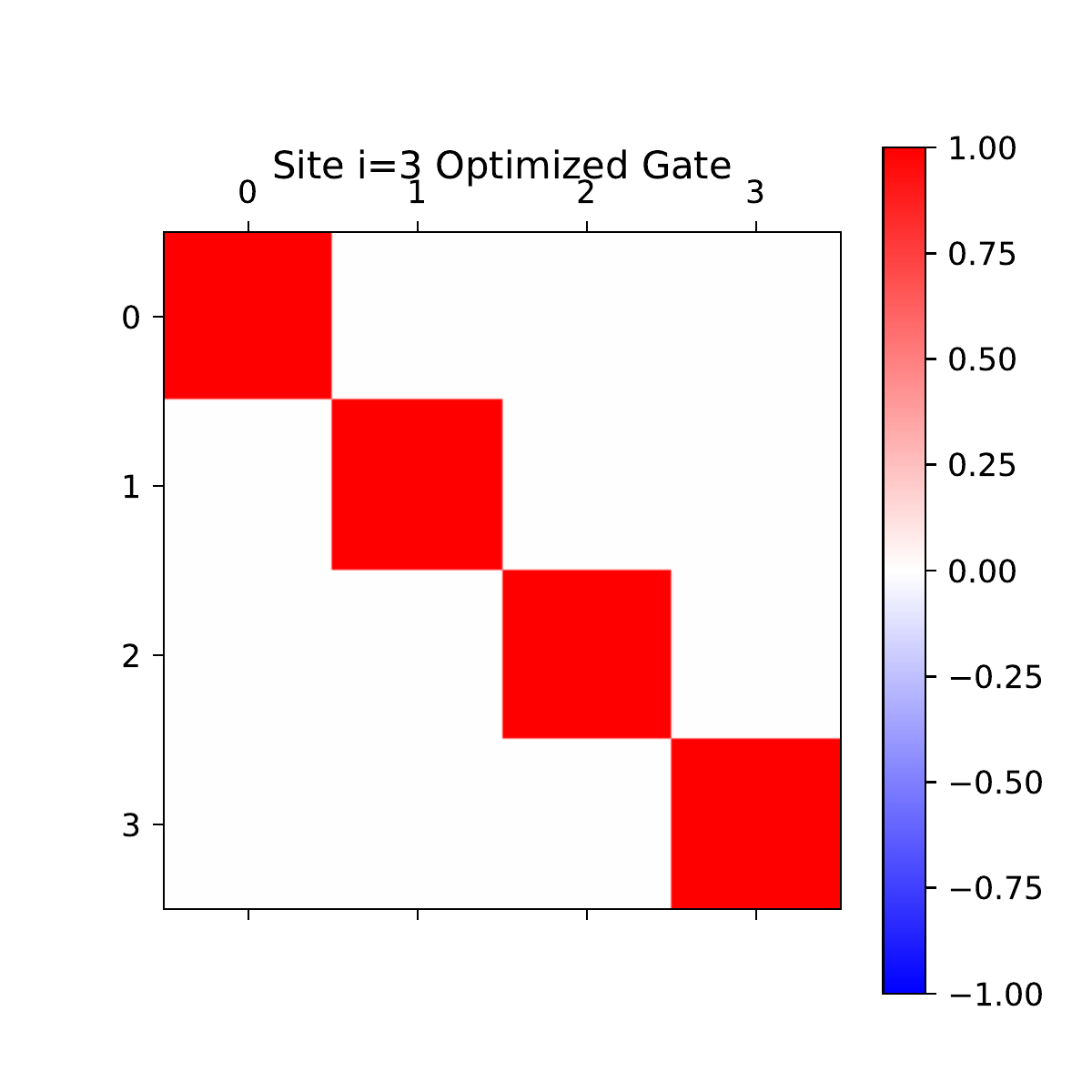}
}
\subfloat[Circuit from optimization]{
\includegraphics[width=0.3\columnwidth]{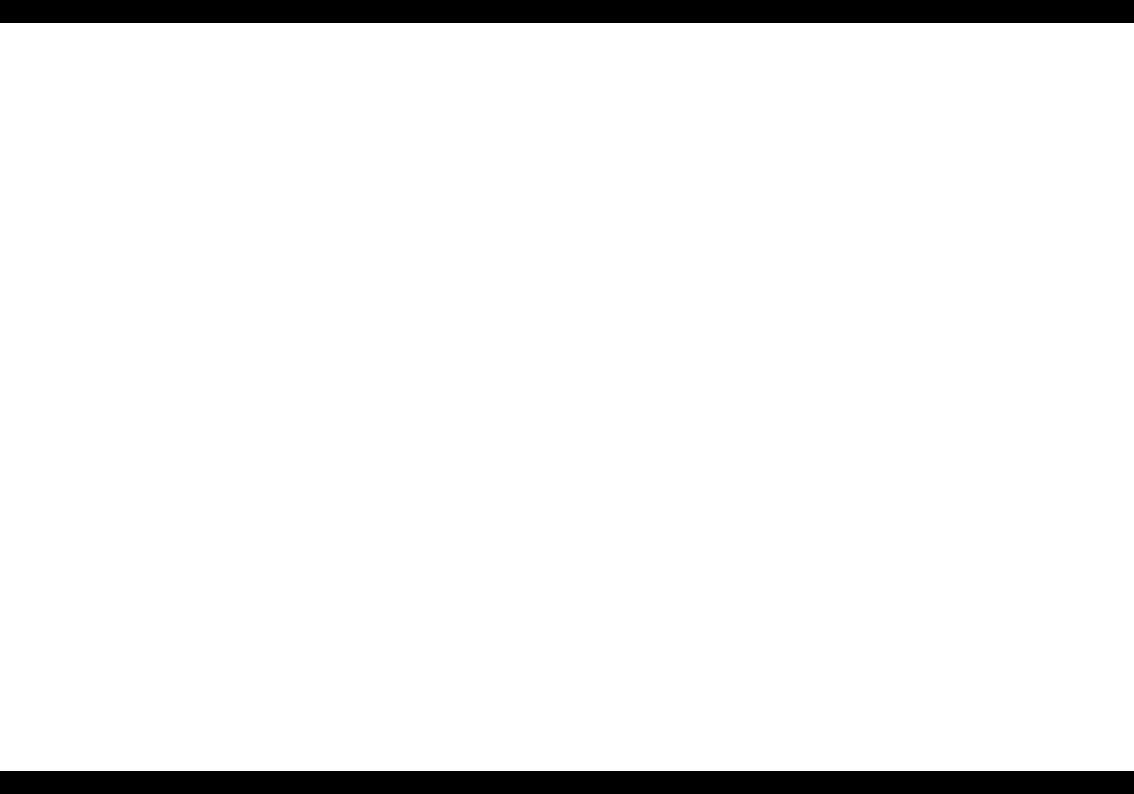}
}
\caption{Optimization for site 3}
\end{figure}
%
%
\begin{figure}[h]

\subfloat[Isometry]{
\includegraphics[width=0.3\columnwidth]{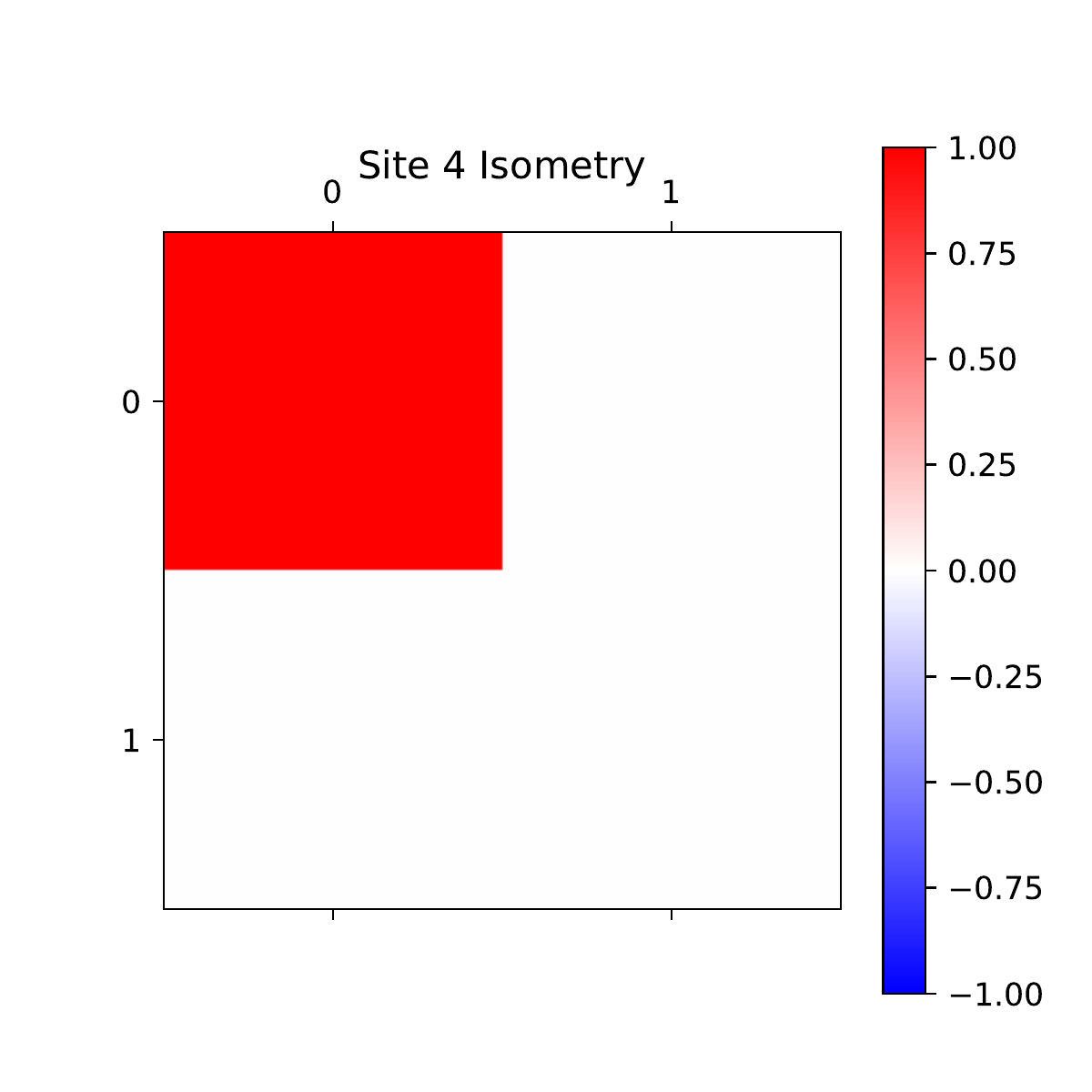}
}
\subfloat[Optimized gate]{
\includegraphics[width=0.3\columnwidth]{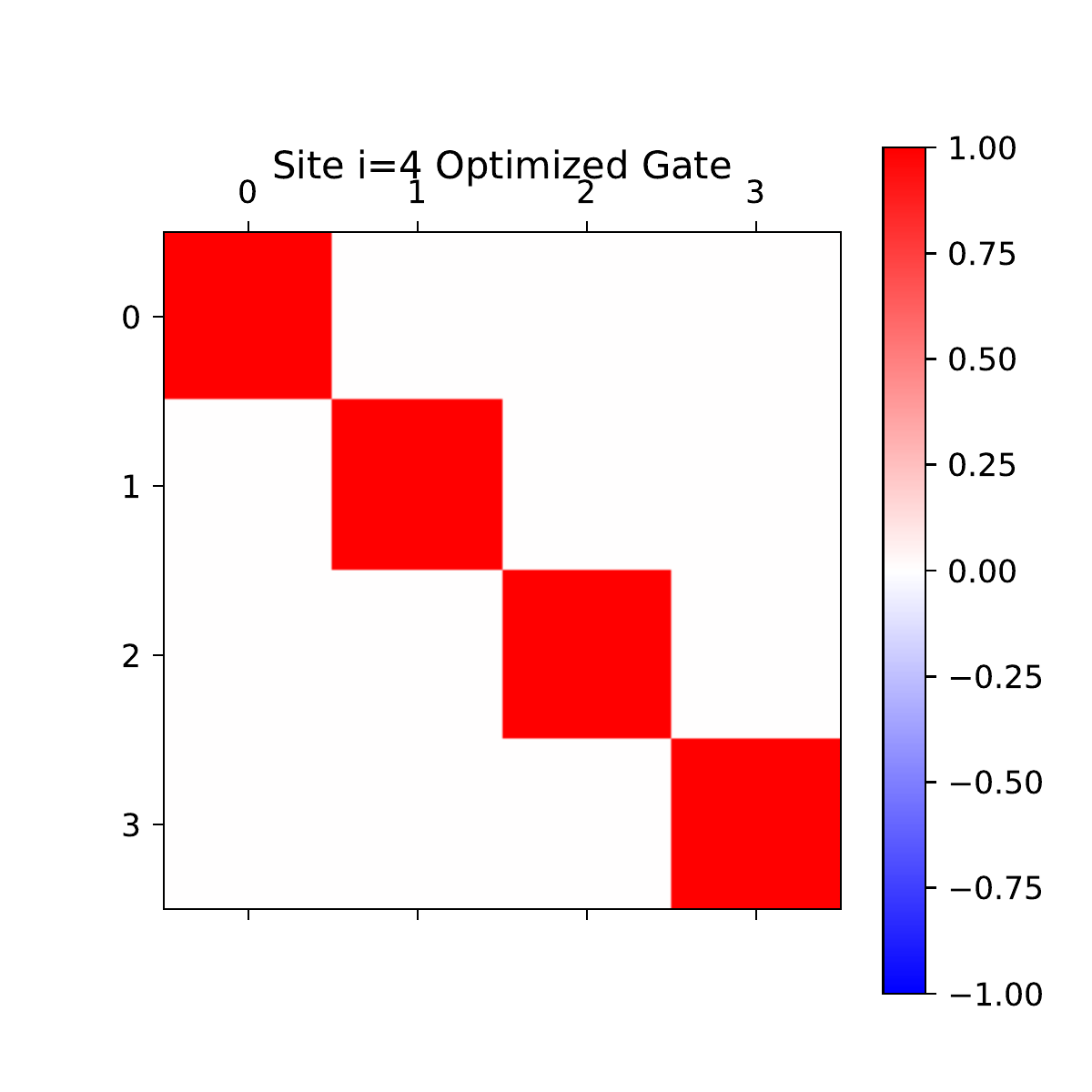}
}
\subfloat[Circuit from optimization]{
\includegraphics[width=0.3\columnwidth]{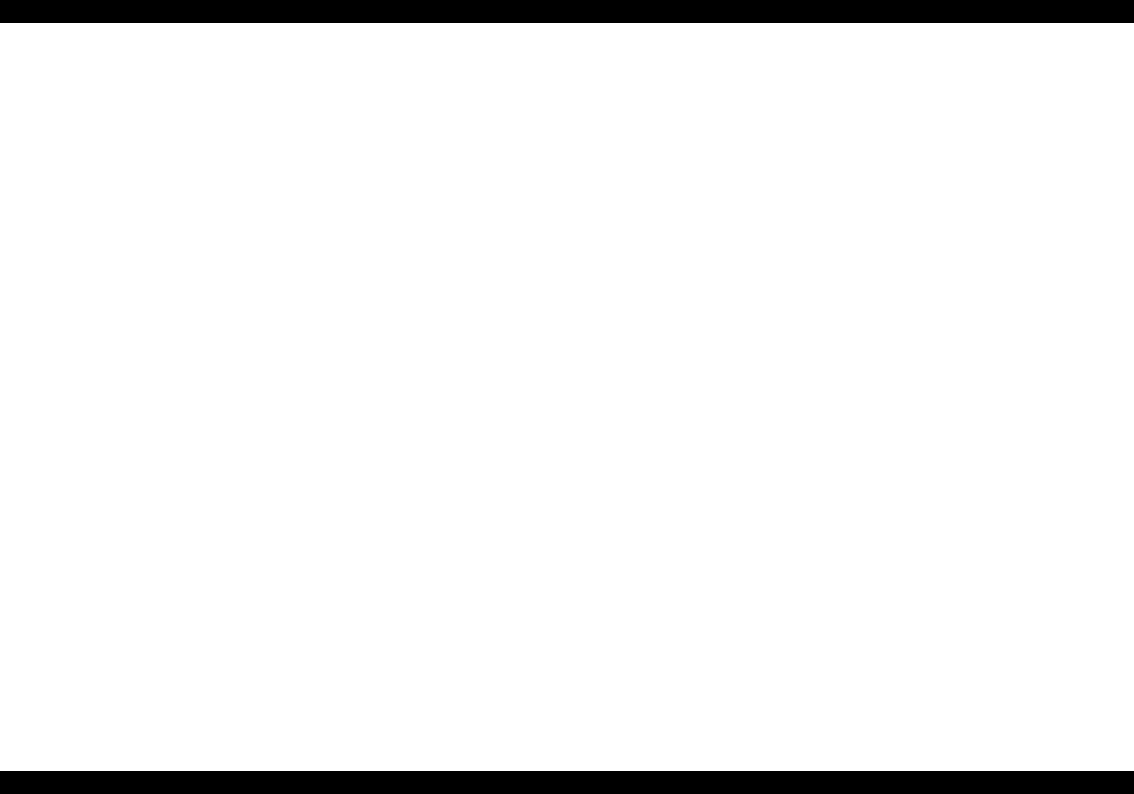}
}
\caption{Optimization for site 4}
\end{figure}
%
%
%
\begin{figure}[h]
%

\subfloat[Isometry]{
\includegraphics[width=0.3\columnwidth]{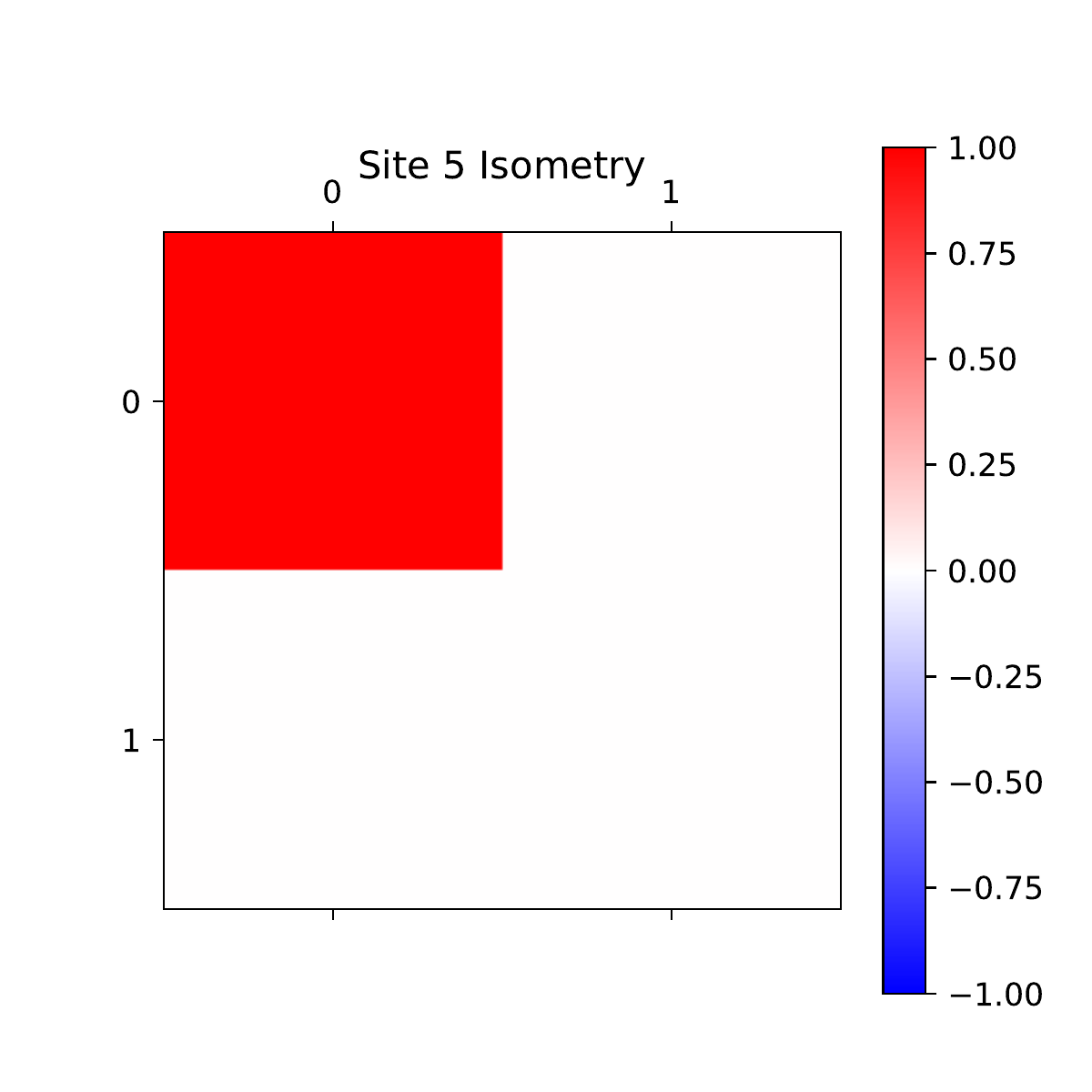}
}
\subfloat[Optimized gate]{
\includegraphics[width=0.3\columnwidth]{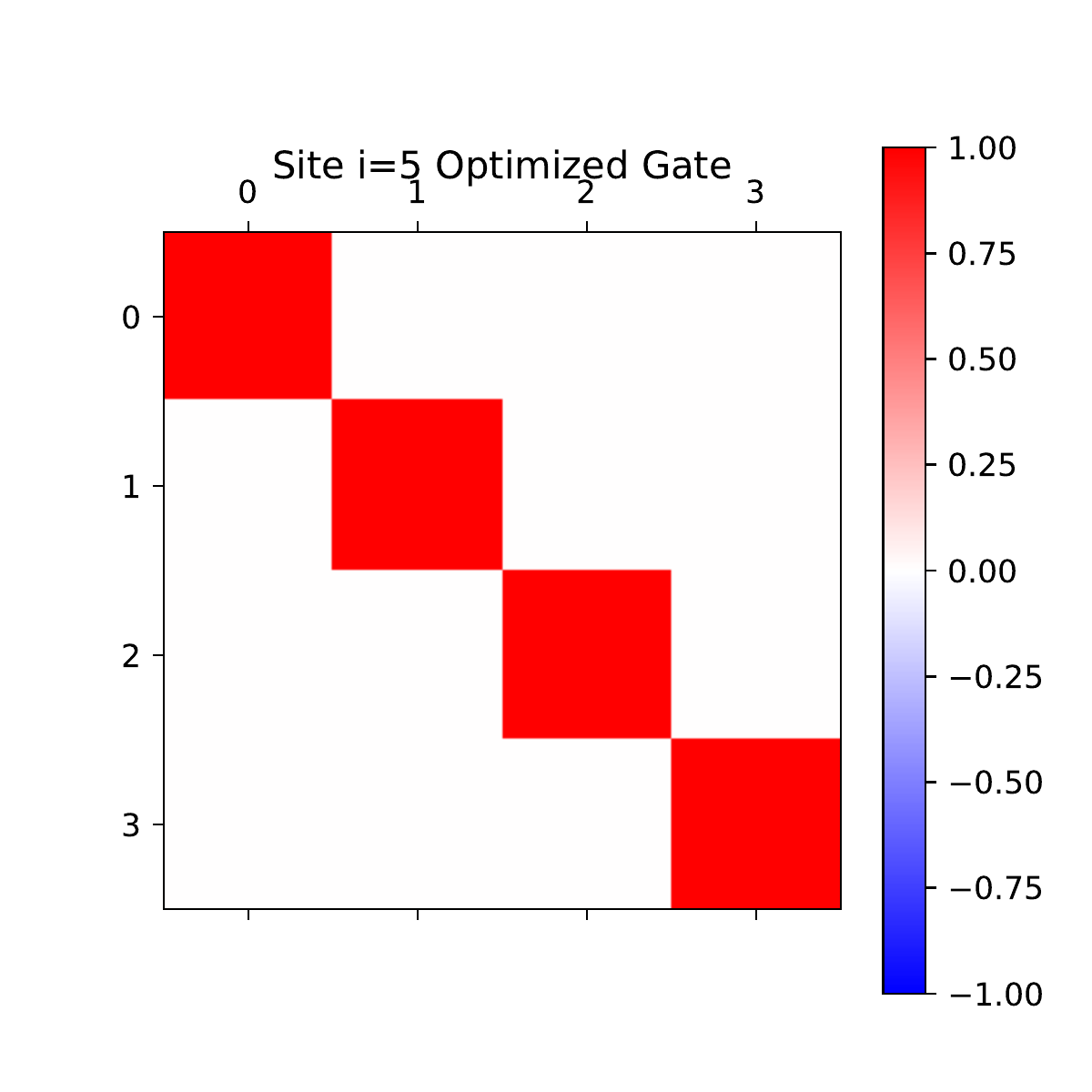}
}
\subfloat[Circuit from optimization]{
\includegraphics[width=0.3\columnwidth]{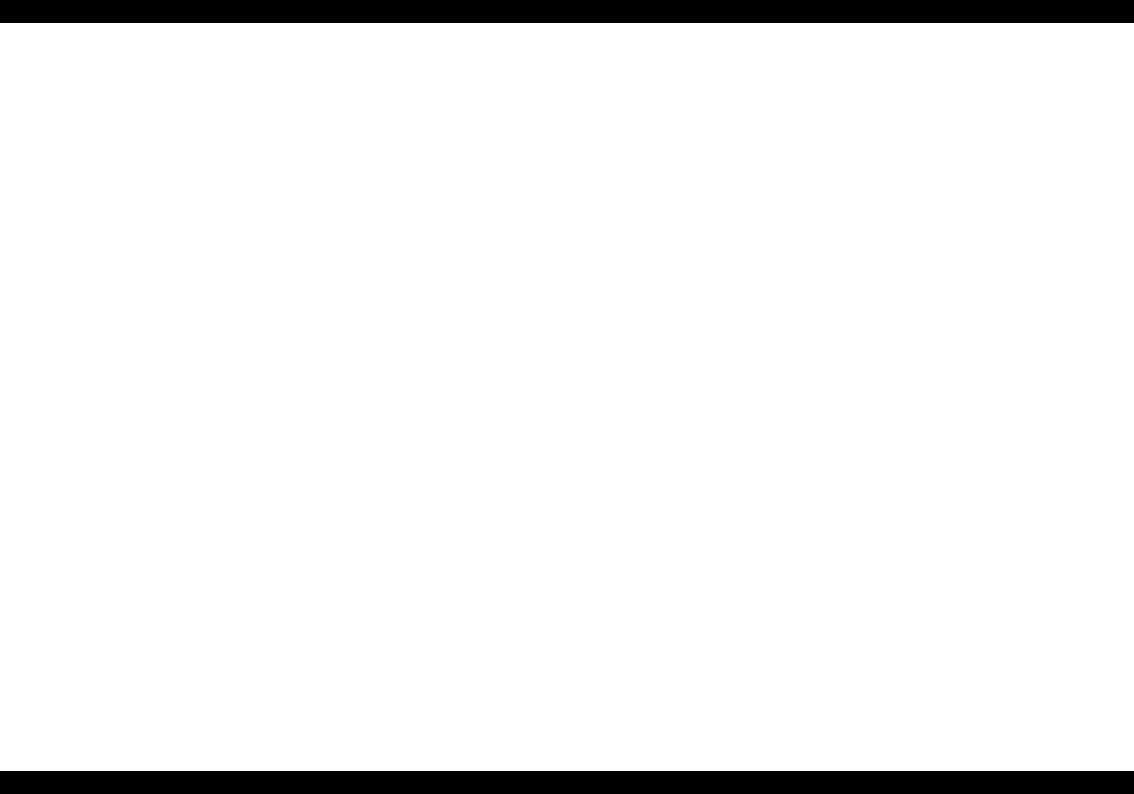}
}
\caption{Optimization for site 5}
\end{figure}
%
%
\begin{figure}[h]
%

\subfloat[Isometry]{
\includegraphics[width=0.3\columnwidth]{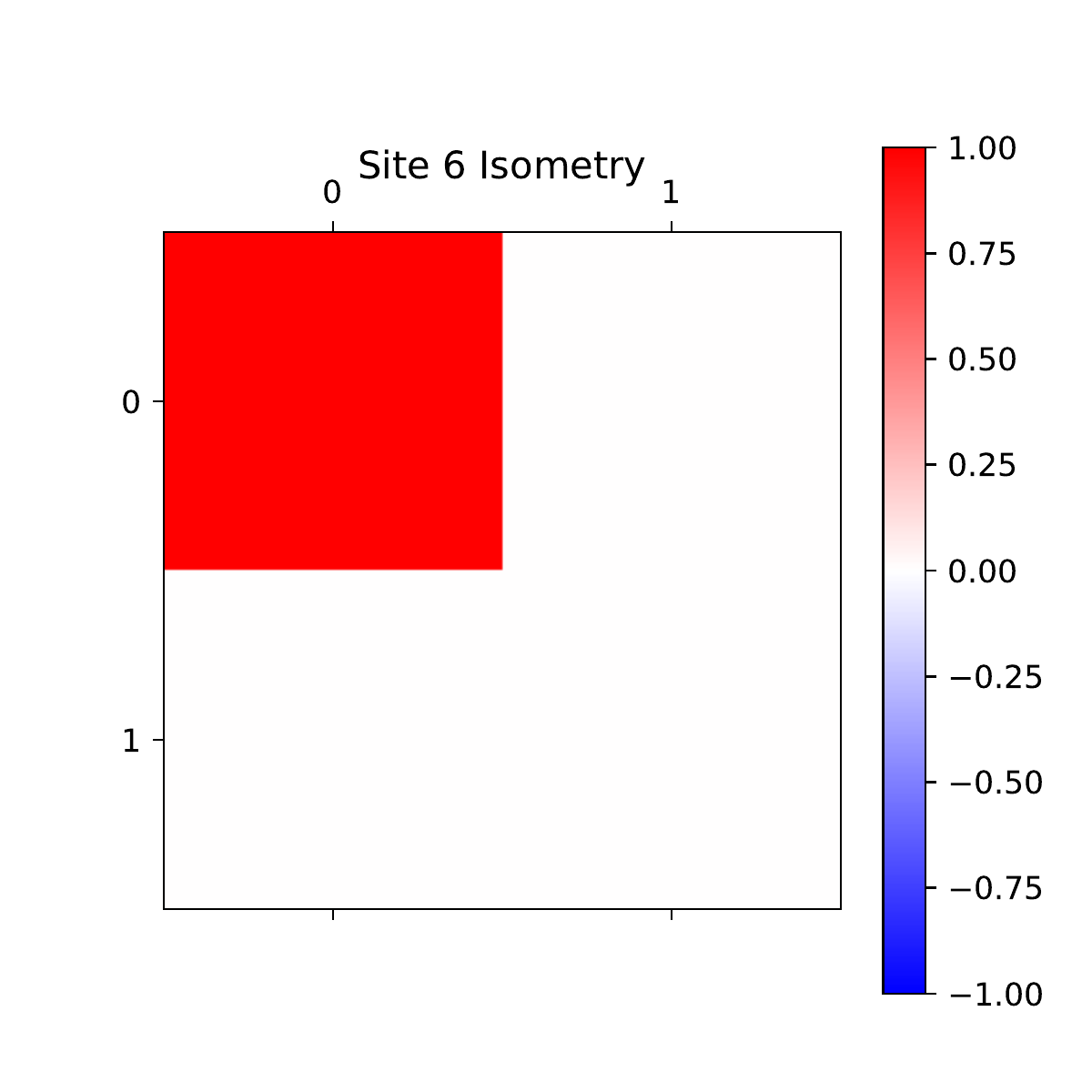}
}
\subfloat[Optimized gate]{
\includegraphics[width=0.3\columnwidth]{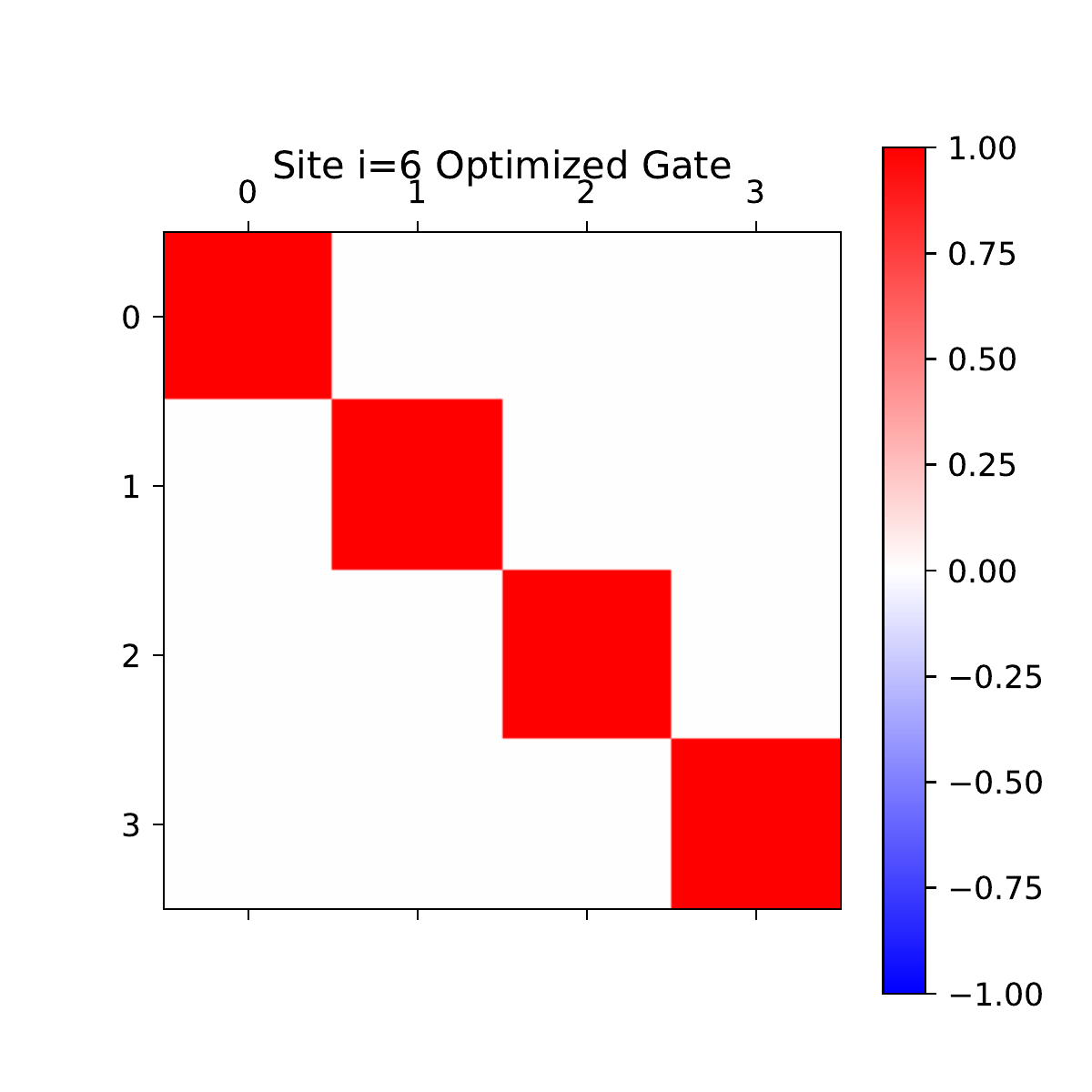}
}
\subfloat[Circuit from optimization]{
\includegraphics[width=0.3\columnwidth]{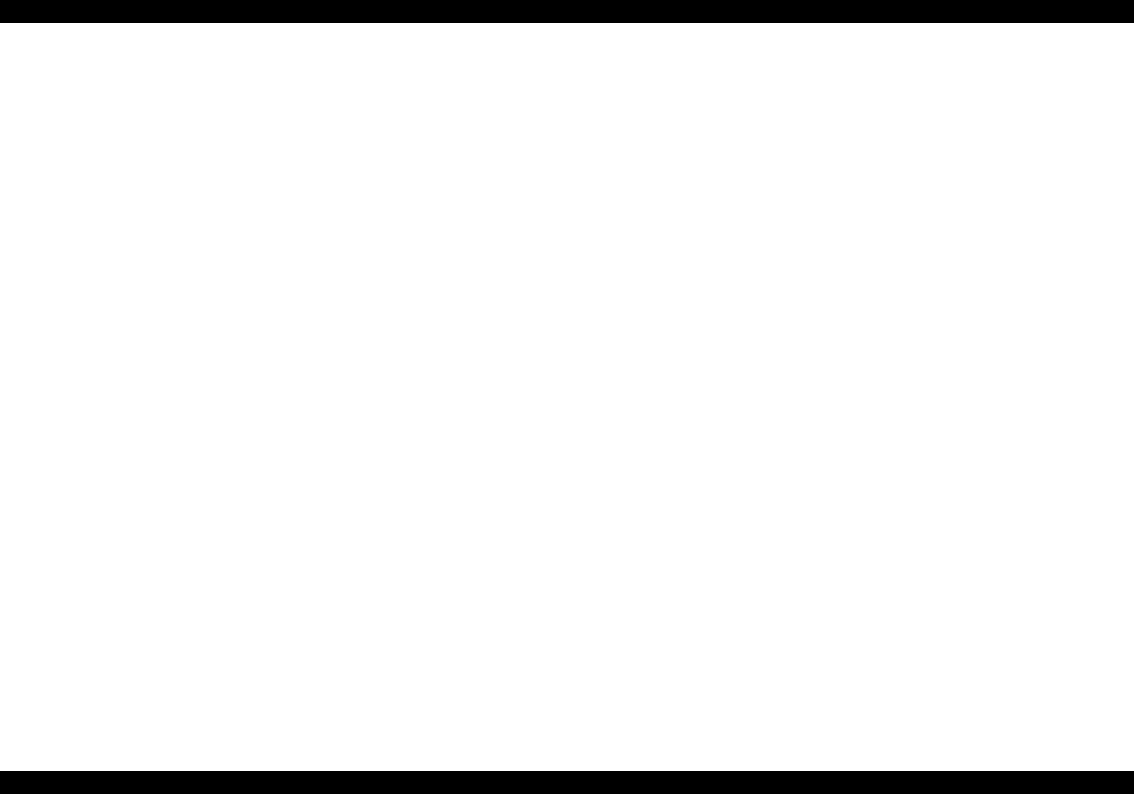}
}
\caption{Optimization for site 6}
\end{figure}
%
%
\begin{figure}[h]
%

\subfloat[Isometry]{
\includegraphics[width=0.3\columnwidth]{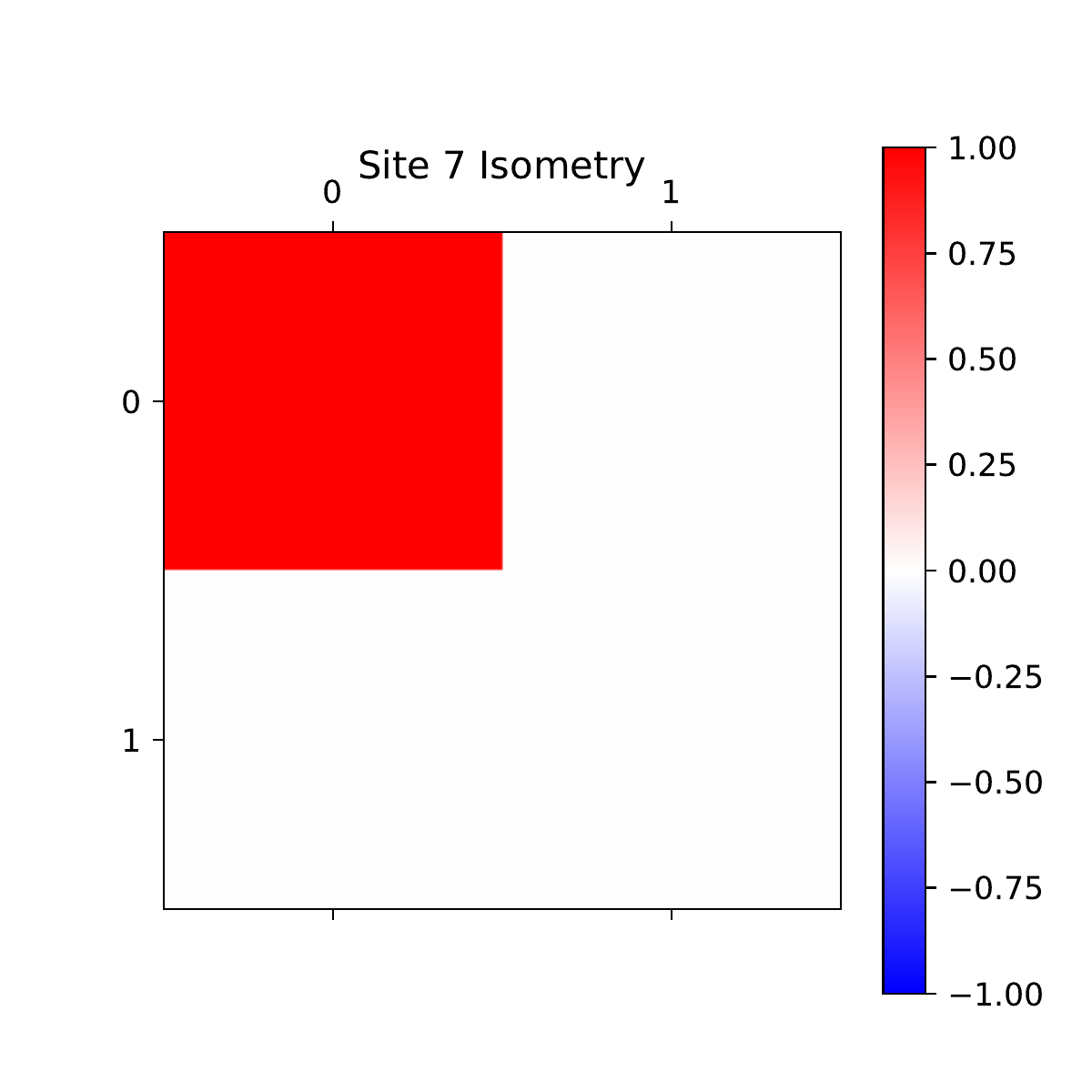}
}
\subfloat[Optimized gate]{
\includegraphics[width=0.3\columnwidth]{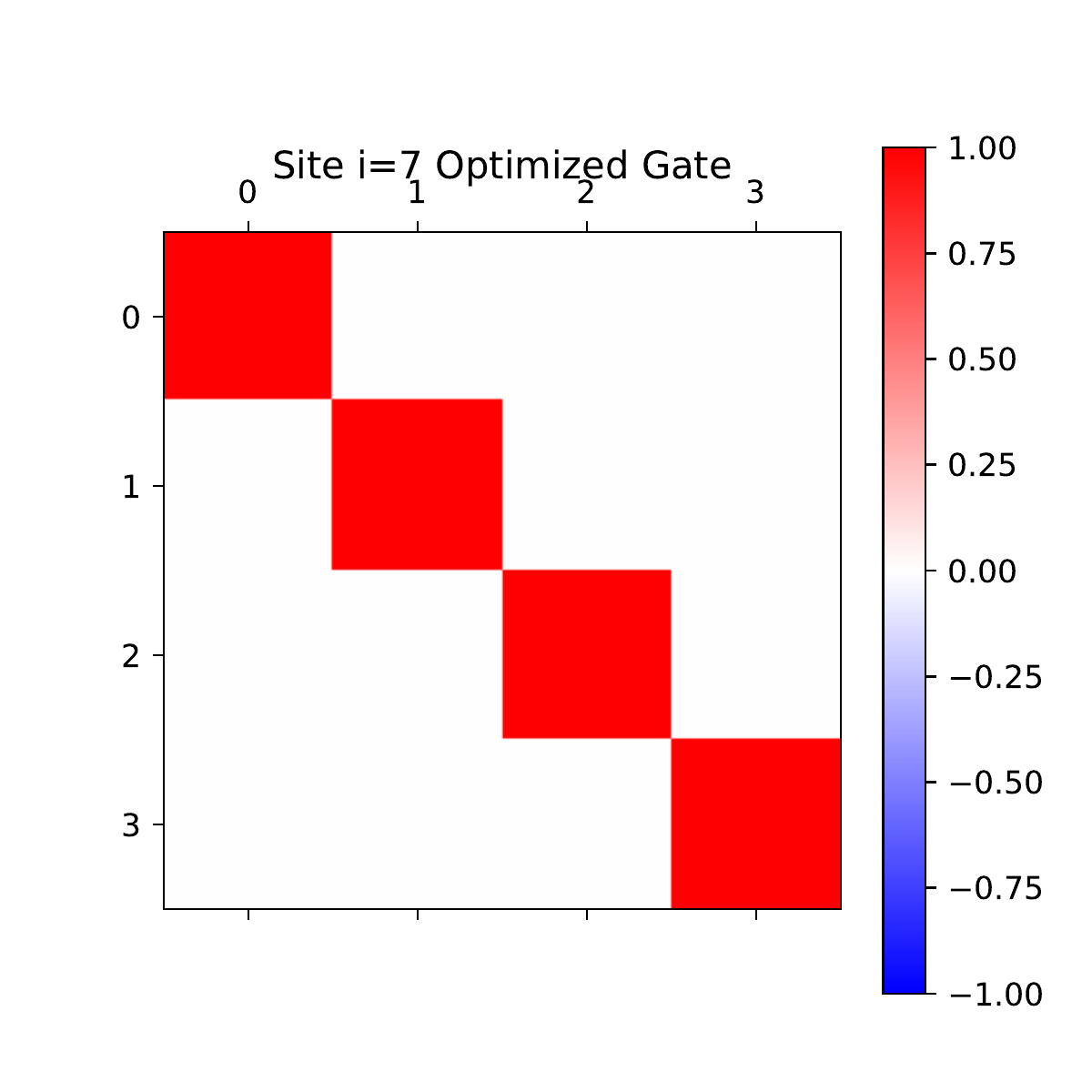}
}
\subfloat[Circuit from optimization]{
\includegraphics[width=0.3\columnwidth]{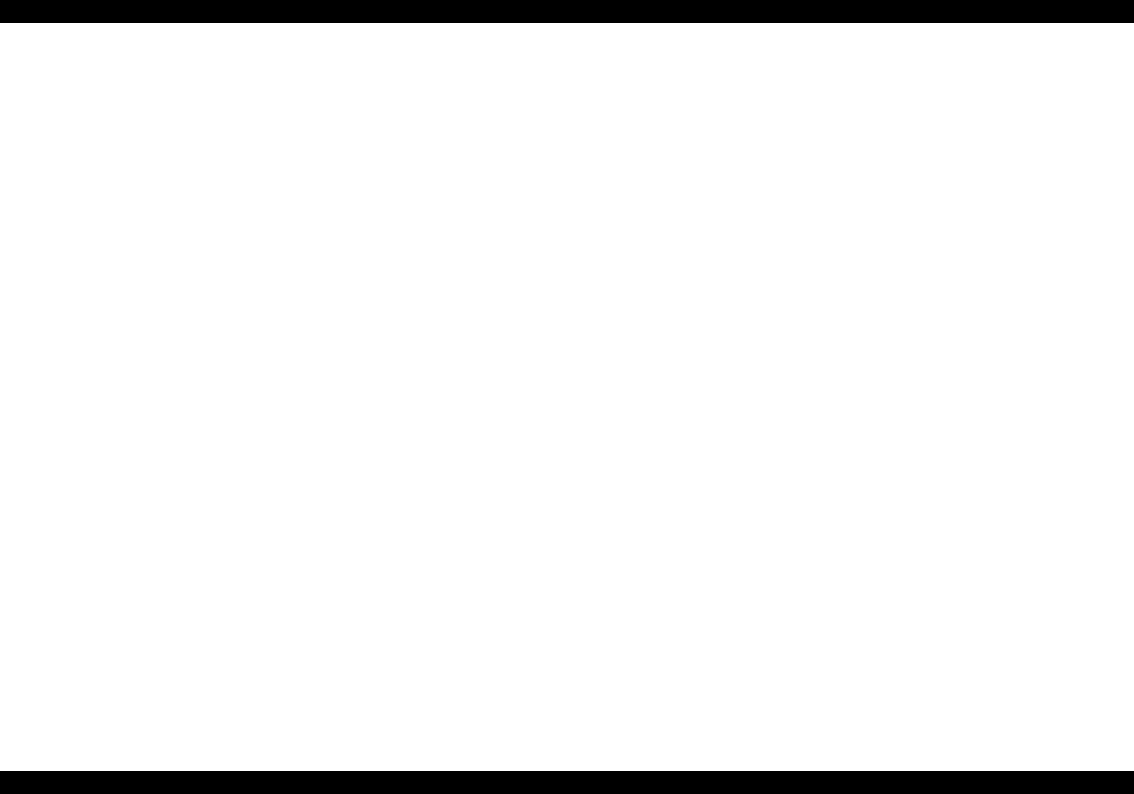}
}
\caption{Optimization for site 7}
\end{figure}
%
%
\begin{figure}[h]

\subfloat[Isometry]{
\includegraphics[width=0.45\columnwidth]{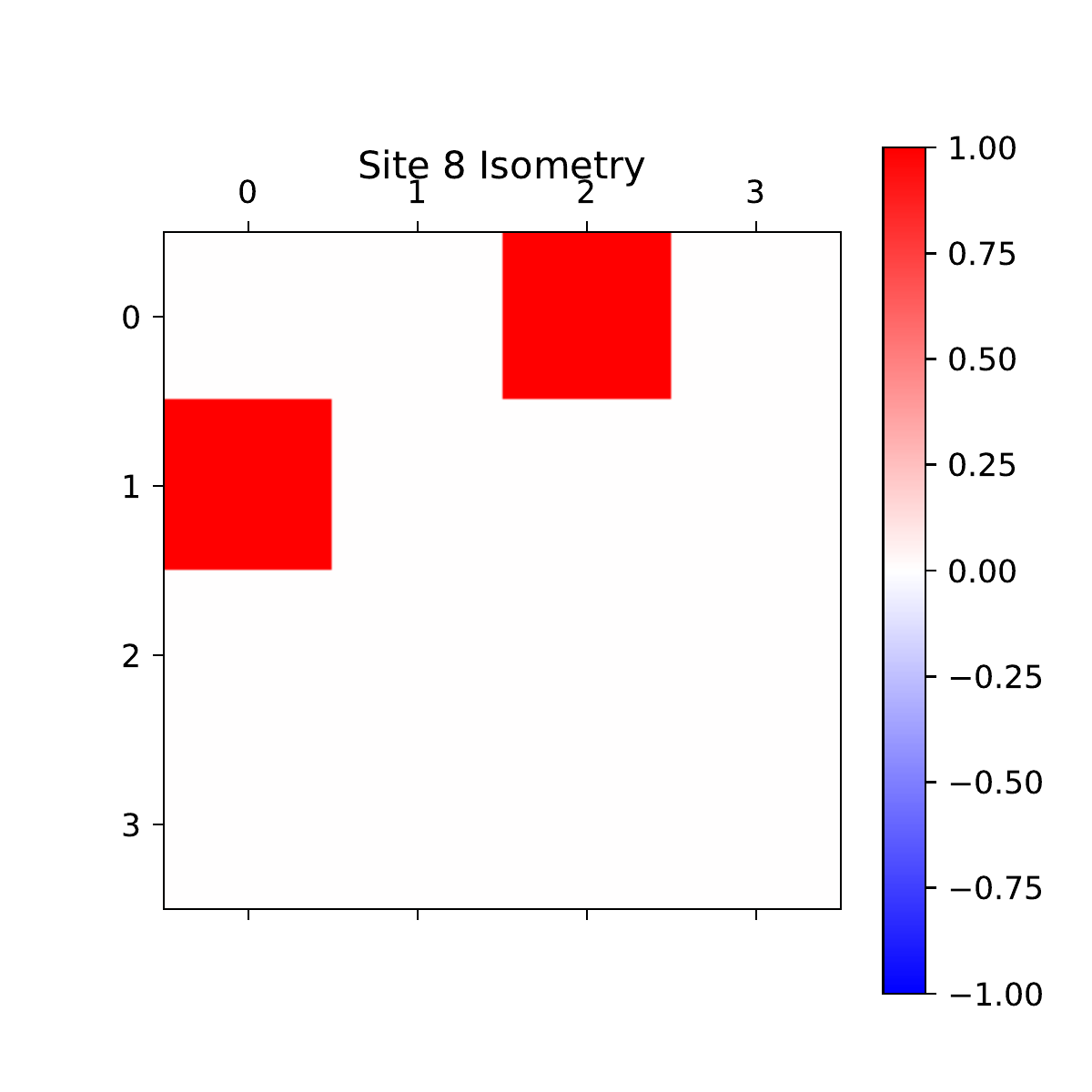}
}
\subfloat[Optimized gate]{
\includegraphics[width=0.45\columnwidth]{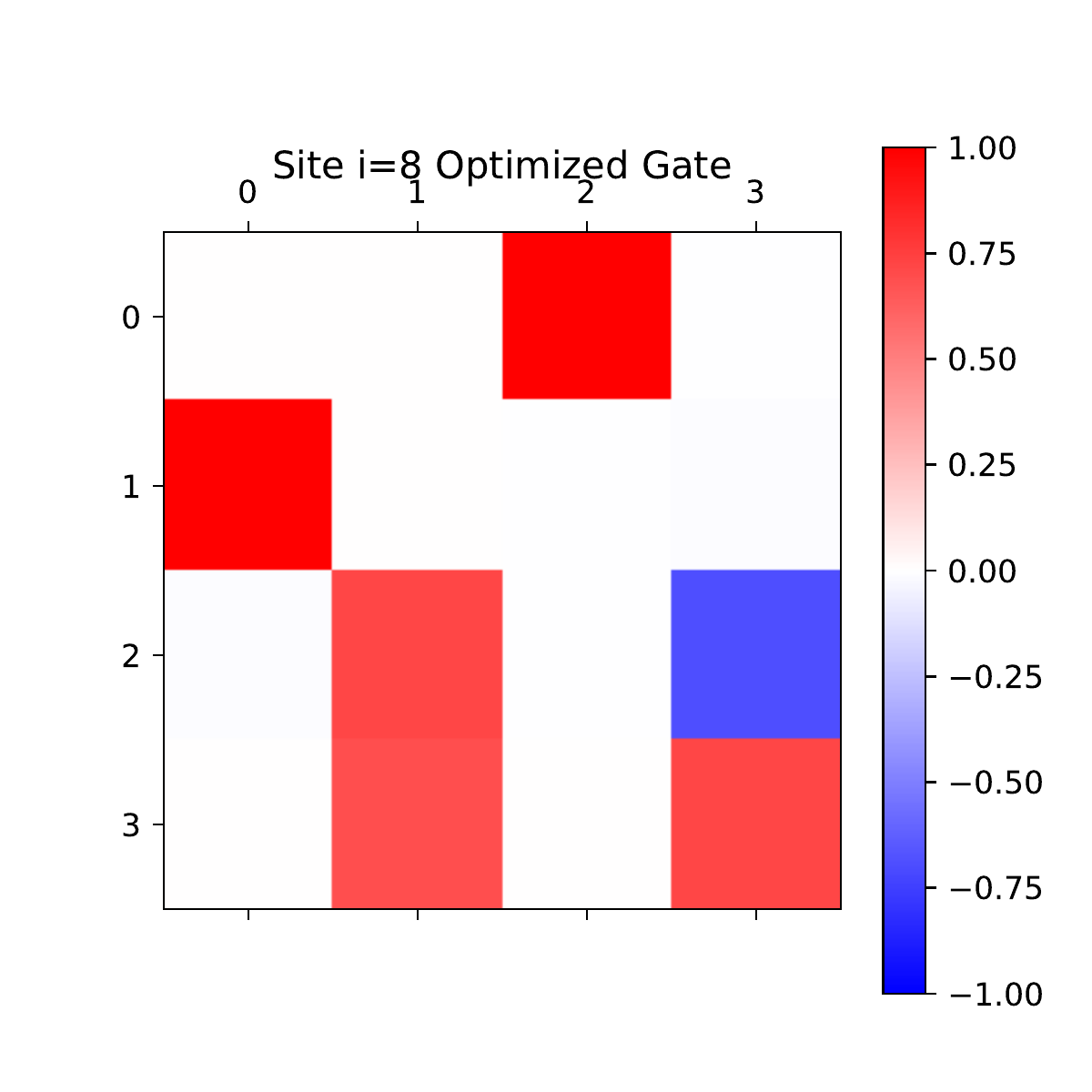}
}\\
\subfloat[Raw circuit from optimization]{
\includegraphics[width=0.9\columnwidth]{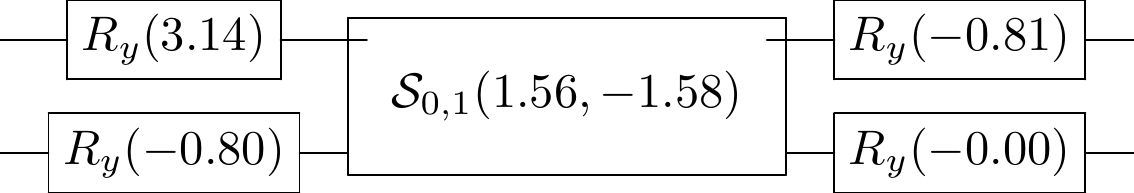}
}\\
\subfloat[Expanded and cleaned circuit from optimization]{
\includegraphics[width=0.9\columnwidth]{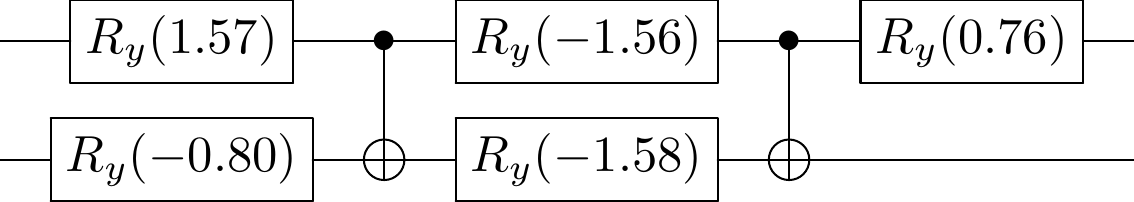}
}
\caption{Optimization for site 8}
\end{figure}
%
%
\begin{figure}[h]

\subfloat[Isometry]{
\includegraphics[width=0.45\columnwidth]{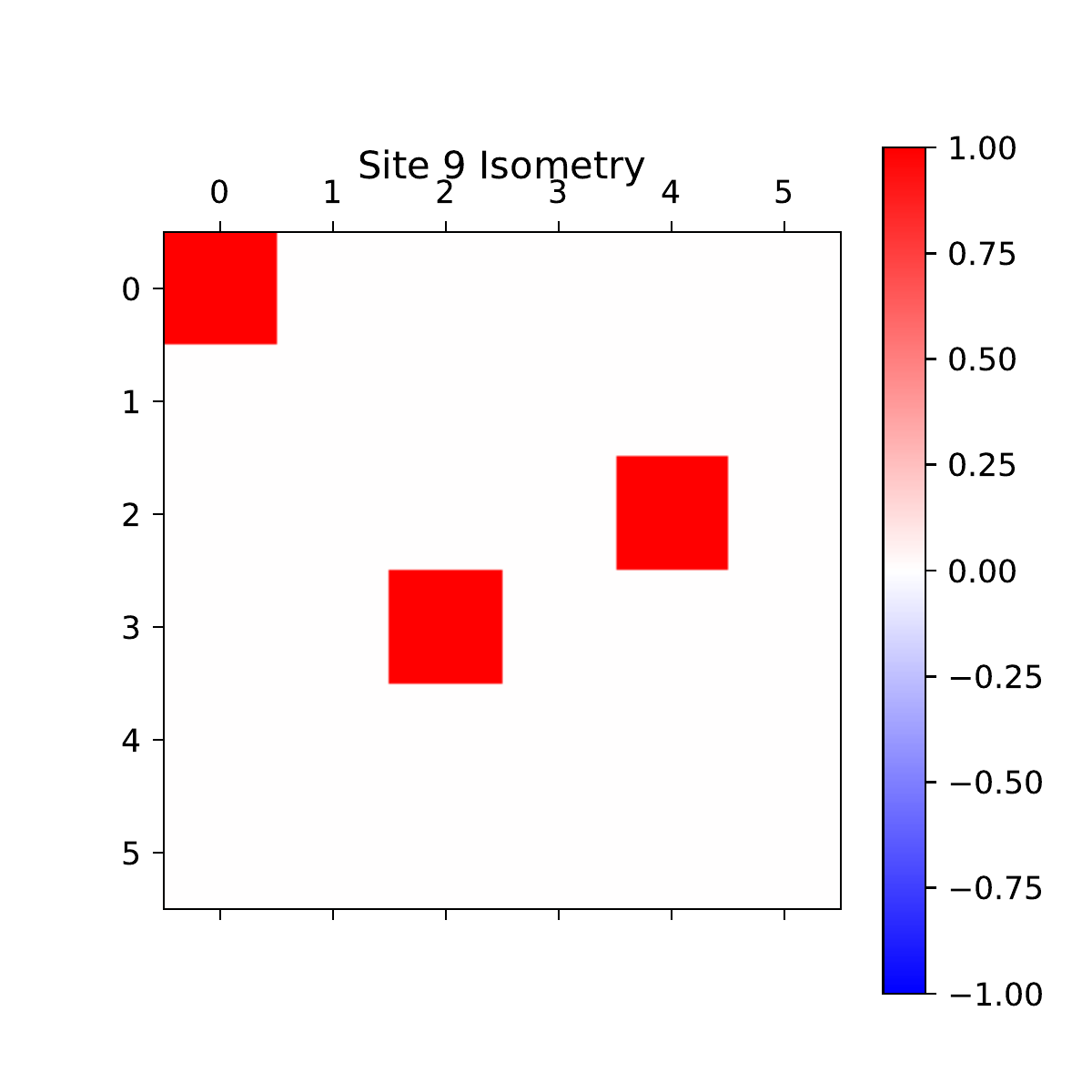}
}
\subfloat[Optimized gate]{
\includegraphics[width=0.45\columnwidth]{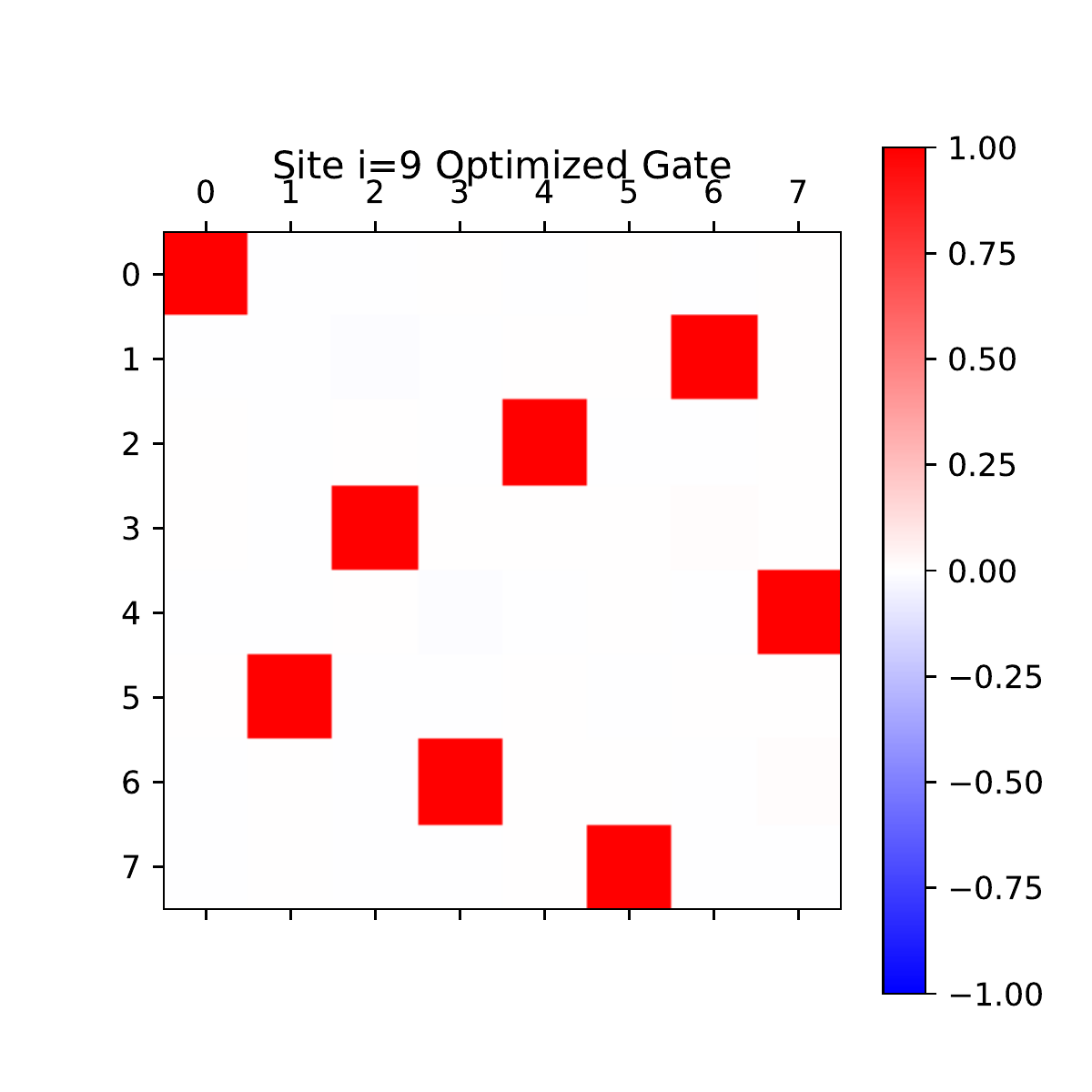}
}\\
\subfloat[Raw circuit from optimization]{
\includegraphics[width=0.9\columnwidth]{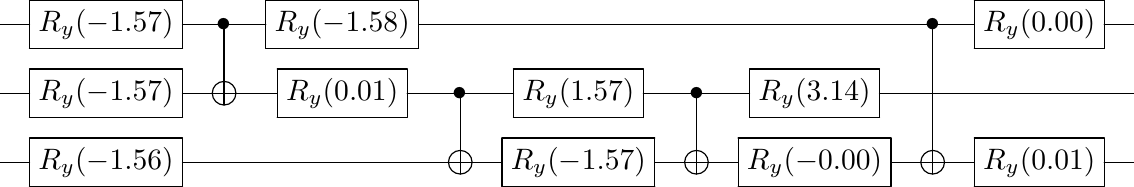}
}\\
\subfloat[Expanded and cleaned circuit from optimization]{
\includegraphics[width=0.9\columnwidth]{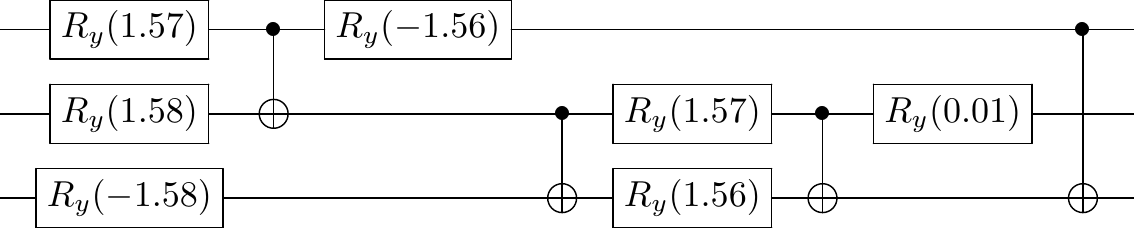}
}
\caption{Optimization for site 9}
\end{figure}
%
%
\begin{figure}[h]
\centering
\subfloat[Isometry]{
\includegraphics[width=0.45\columnwidth]{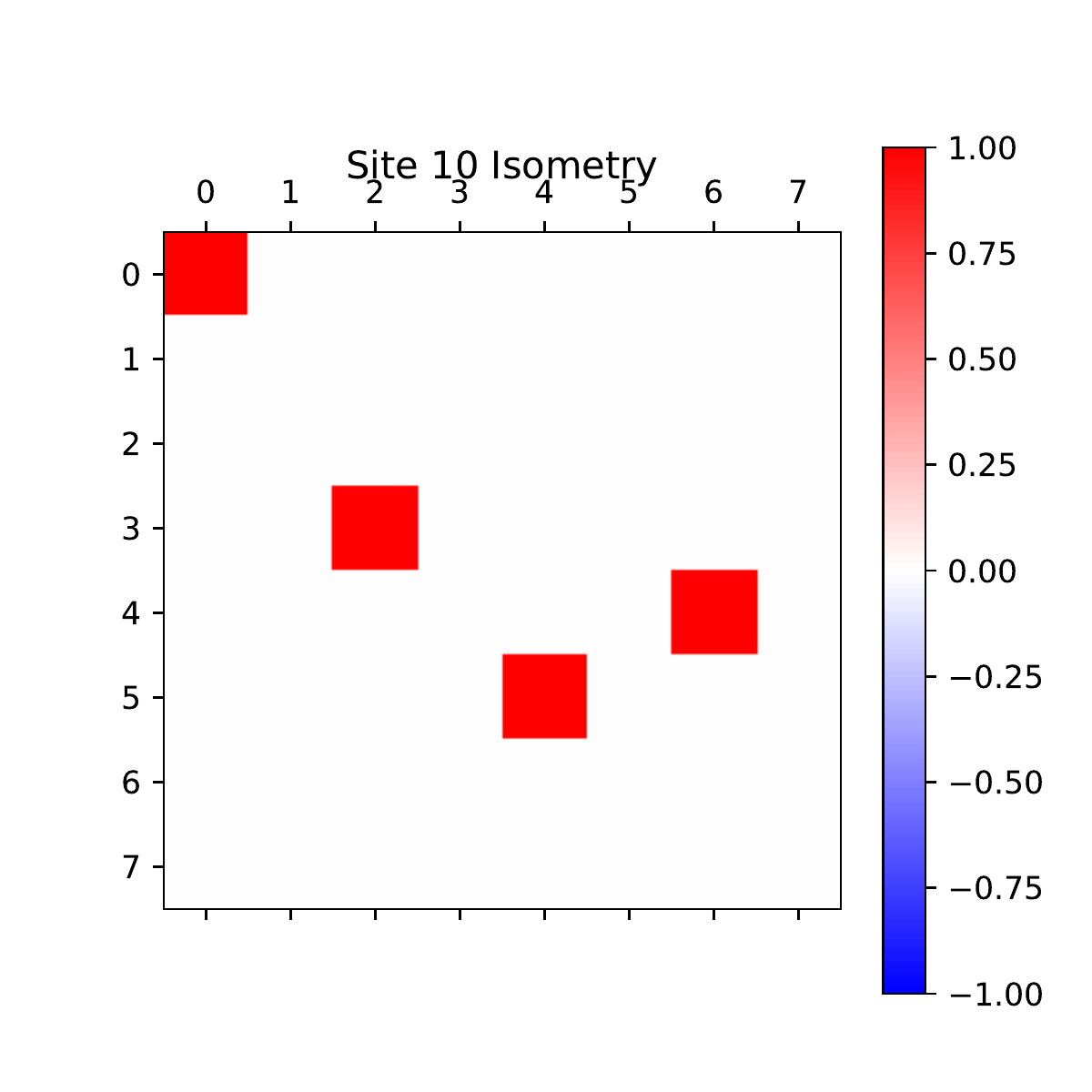}
}
\subfloat[Optimized gate]{
\includegraphics[width=0.45\columnwidth]{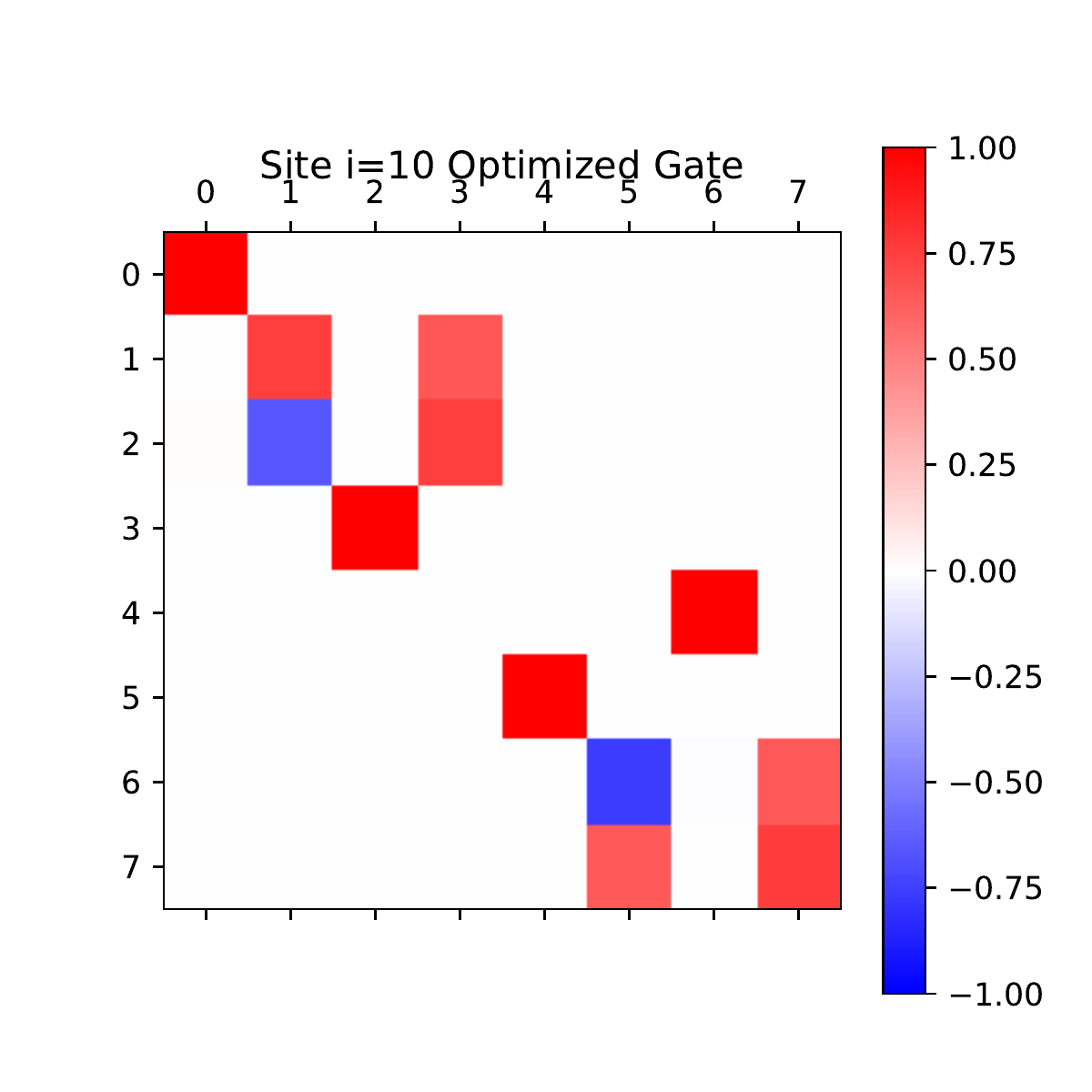}
}\\
\subfloat[Raw circuit from optimization]{
\includegraphics[width=0.9\columnwidth]{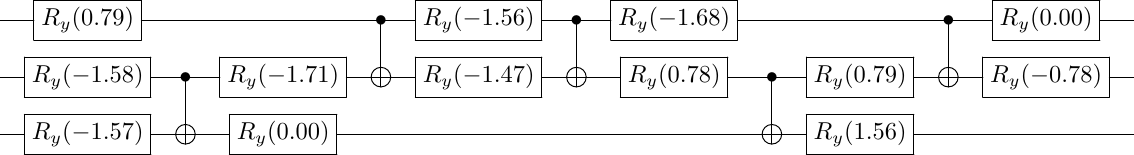}
}\\
\subfloat[Expanded and cleaned circuit from optimization]{
\includegraphics[width=0.9\columnwidth]{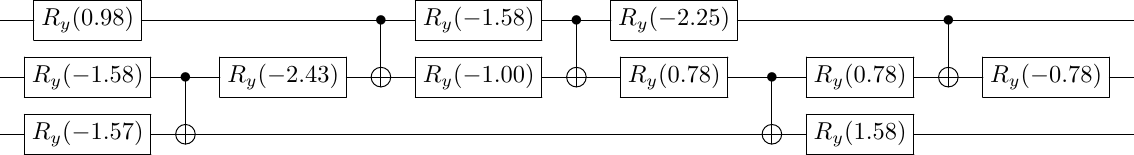}
}
\caption{Optimization for site 10}
\end{figure}
%
%
\begin{figure}[h]

\subfloat[Isometry]{
\includegraphics[width=0.45\columnwidth]{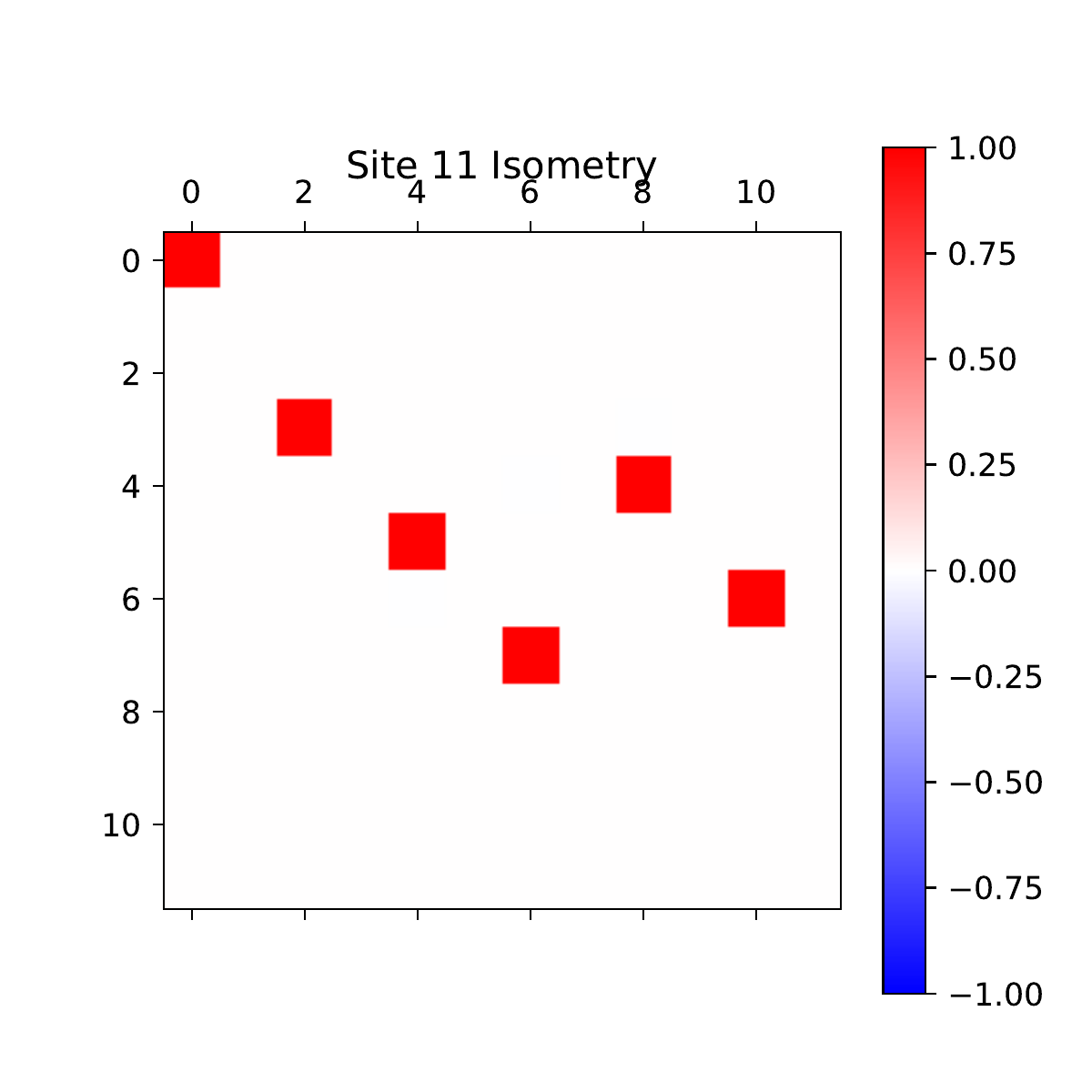}
}
\subfloat[Optimized gate]{
\includegraphics[width=0.45\columnwidth]{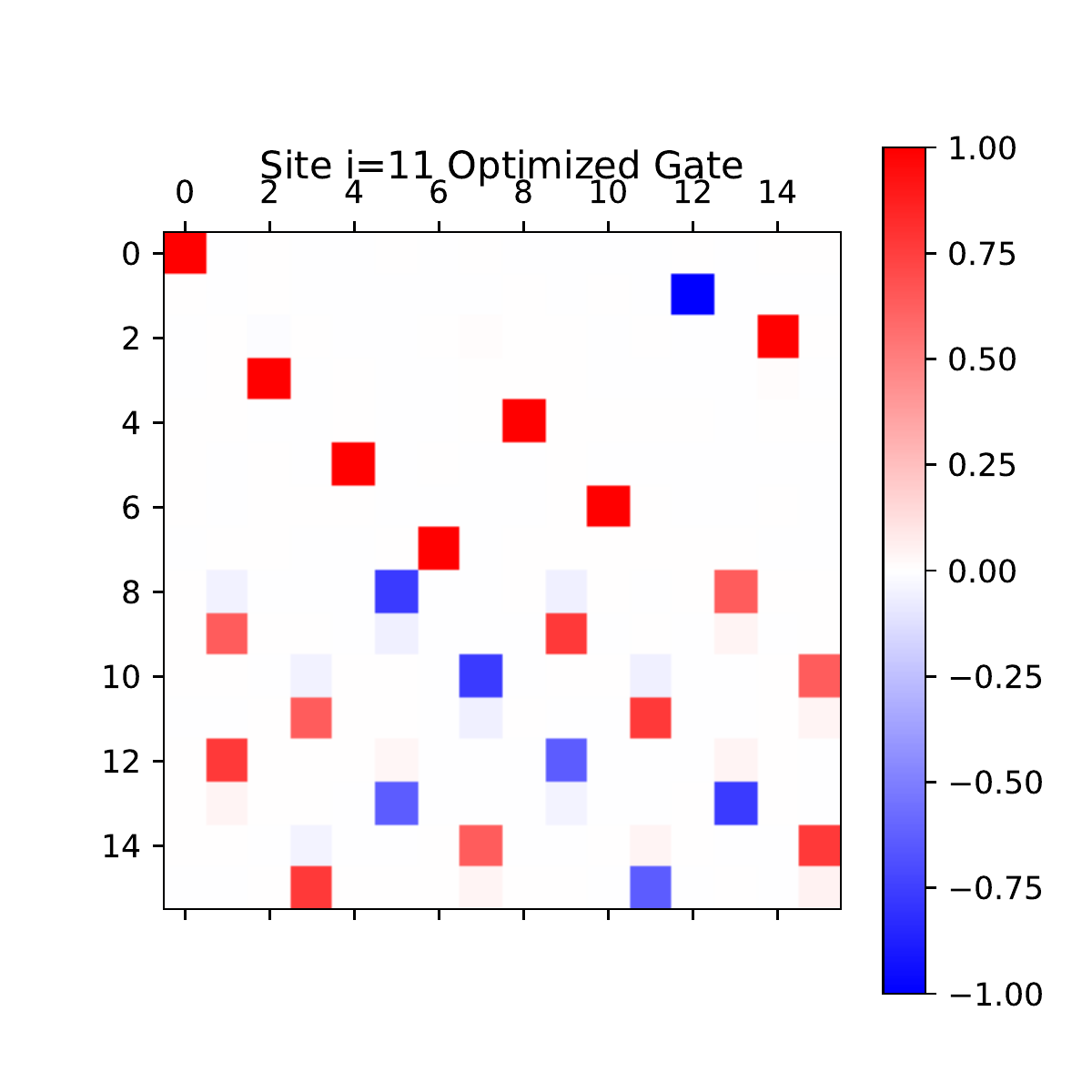}
}\\
\subfloat[Raw circuit from optimization]{
\includegraphics[width=0.9\columnwidth]{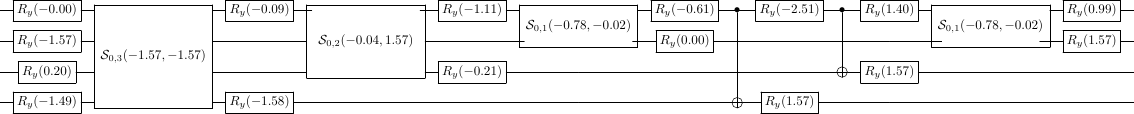}
}\\
\subfloat[Expanded and cleaned circuit from optimization]{
\includegraphics[width=0.9\columnwidth]{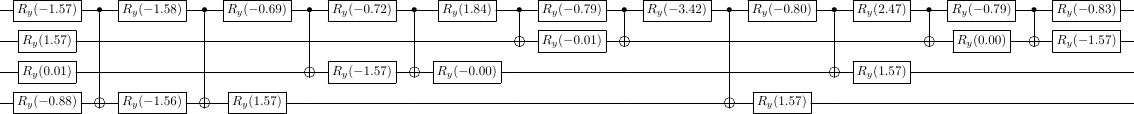}
}
\caption{Optimization for site 11}
\end{figure}
%
%
\begin{figure}[h]

\subfloat[Isometry]{
\includegraphics[width=0.45\columnwidth]{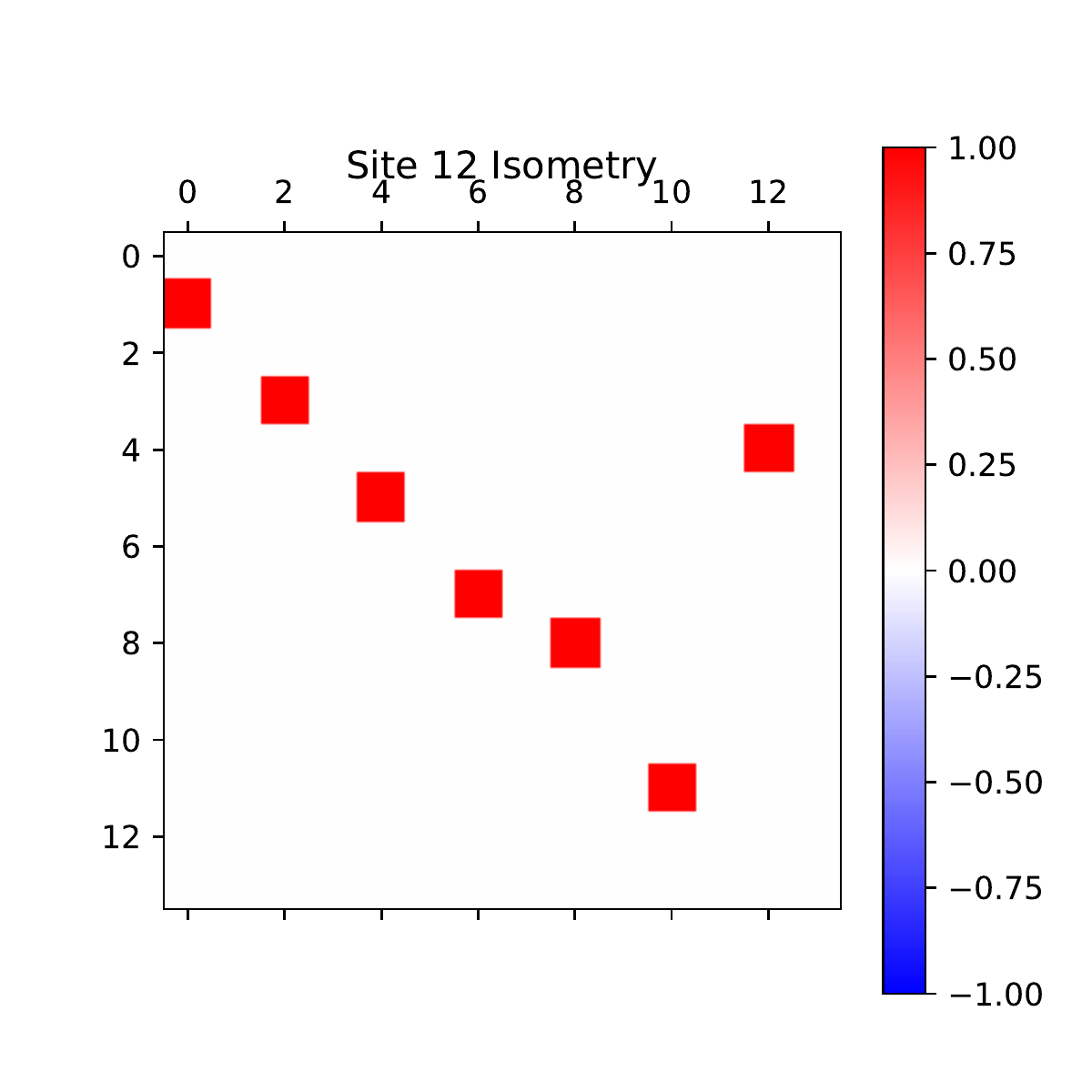}
}
\subfloat[Optimized gate]{
\includegraphics[width=0.45\columnwidth]{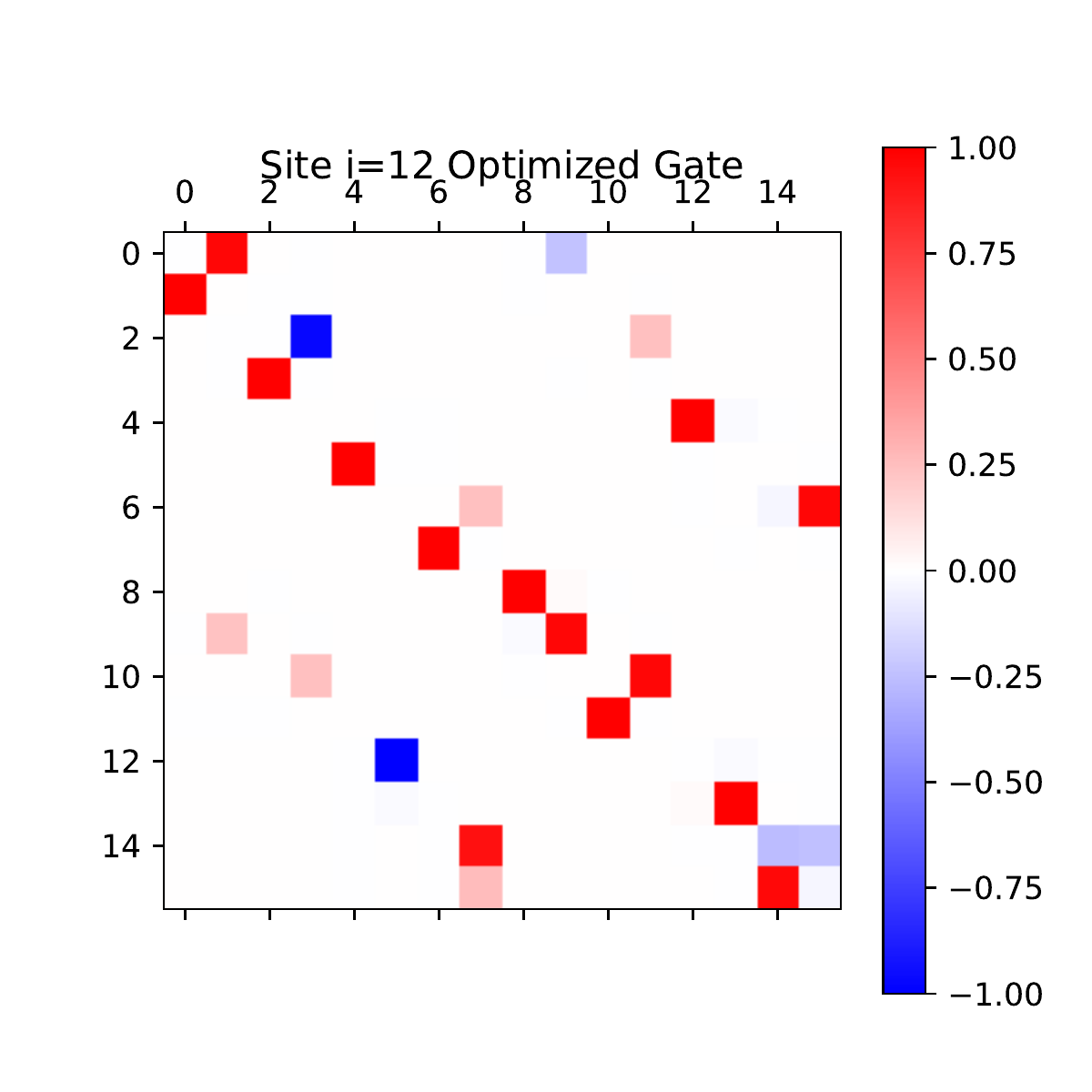}
}\\
\subfloat[Raw circuit from optimization]{
\includegraphics[width=0.9\columnwidth]{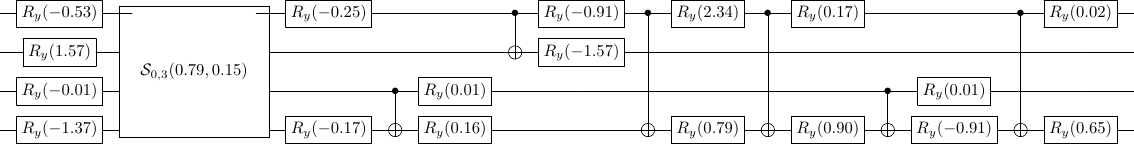}
}\\
\subfloat[Expanded and cleaned circuit from optimization]{
\includegraphics[width=0.9\columnwidth]{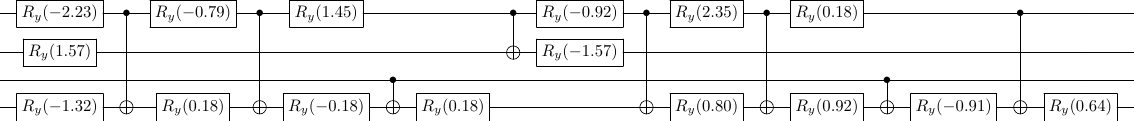}
}
\caption{Optimization for site 12}
\end{figure}
%
%
\begin{figure}[h]

\subfloat[Isometry]{
\includegraphics[width=0.3\columnwidth]{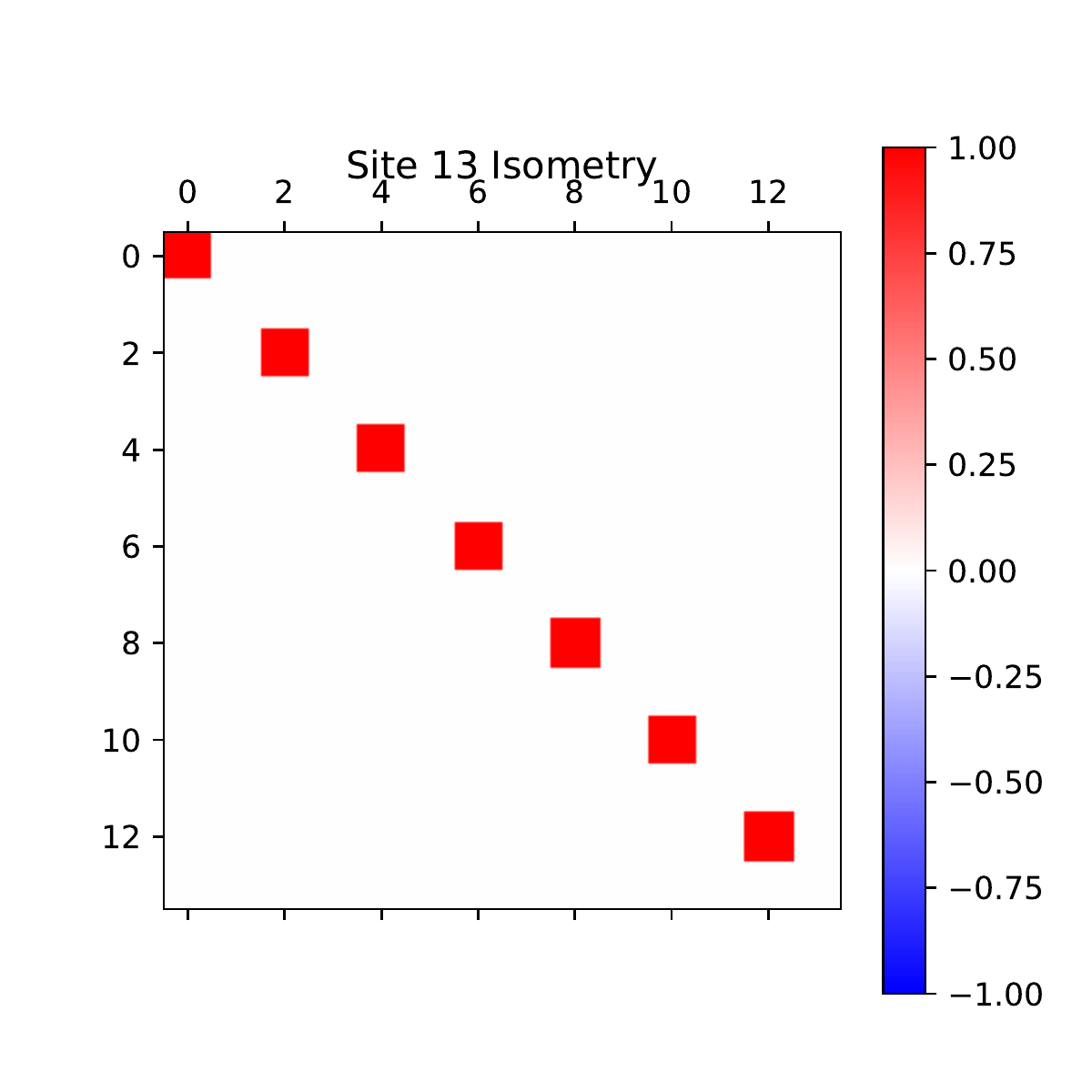}
}
\subfloat[Optimized gate]{
\includegraphics[width=0.3\columnwidth]{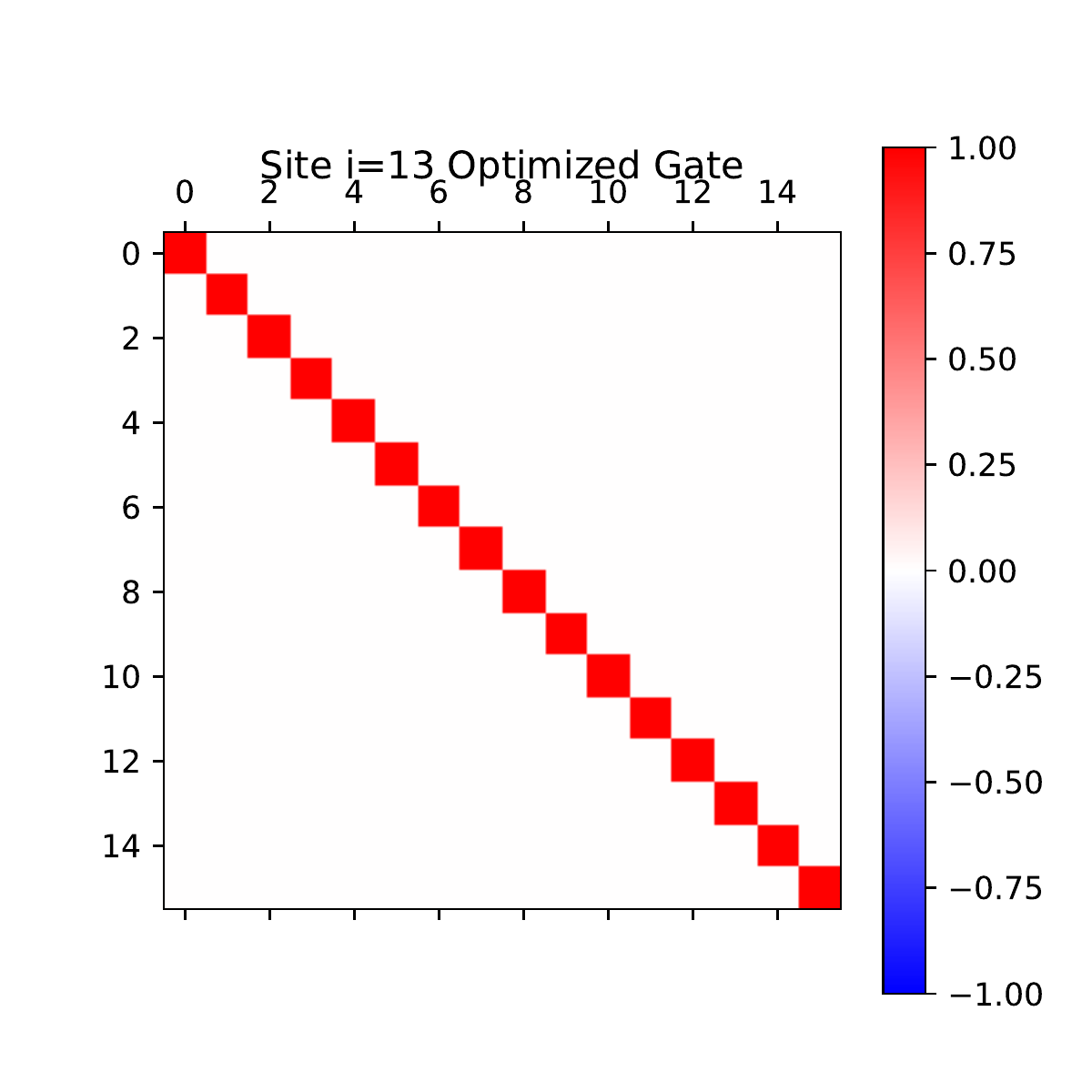}
}
\subfloat[Circuit from optimization]{
\includegraphics[width=0.2\columnwidth]{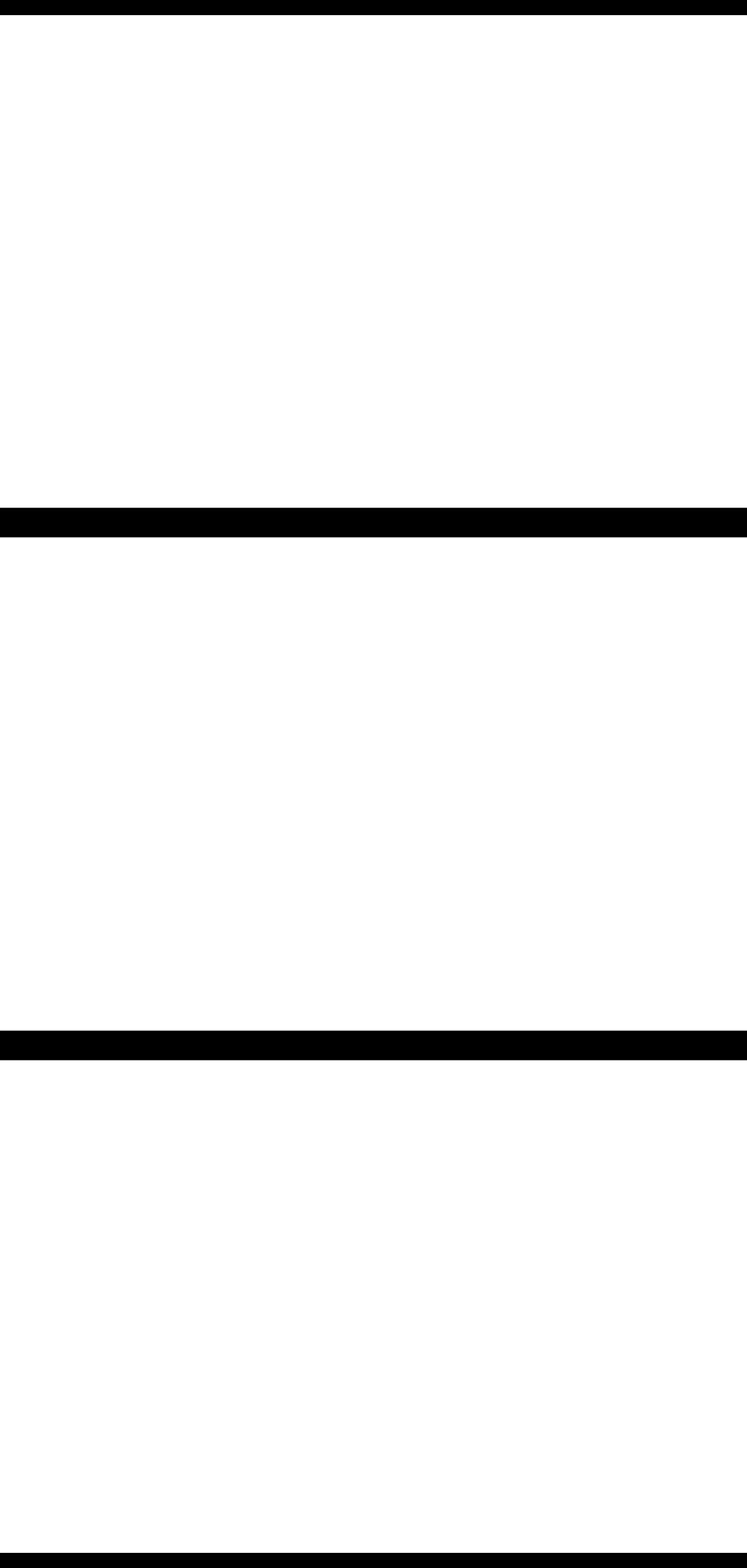}
}
\caption{Optimization for site 13}
\end{figure}
%
\begin{figure}[h]

\subfloat[Isometry]{
\includegraphics[width=0.3\columnwidth]{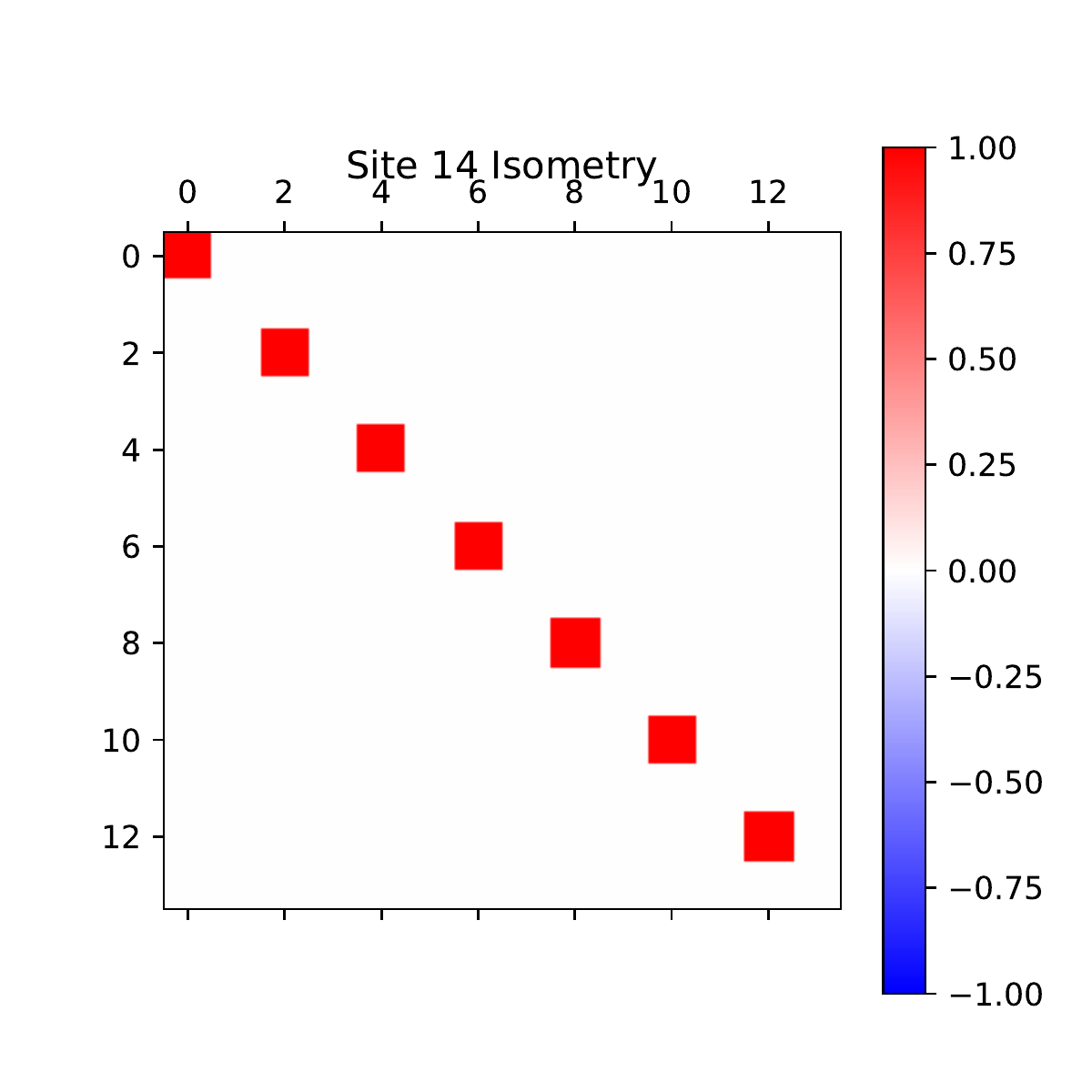}
}
\subfloat[Optimized gate]{
\includegraphics[width=0.3\columnwidth]{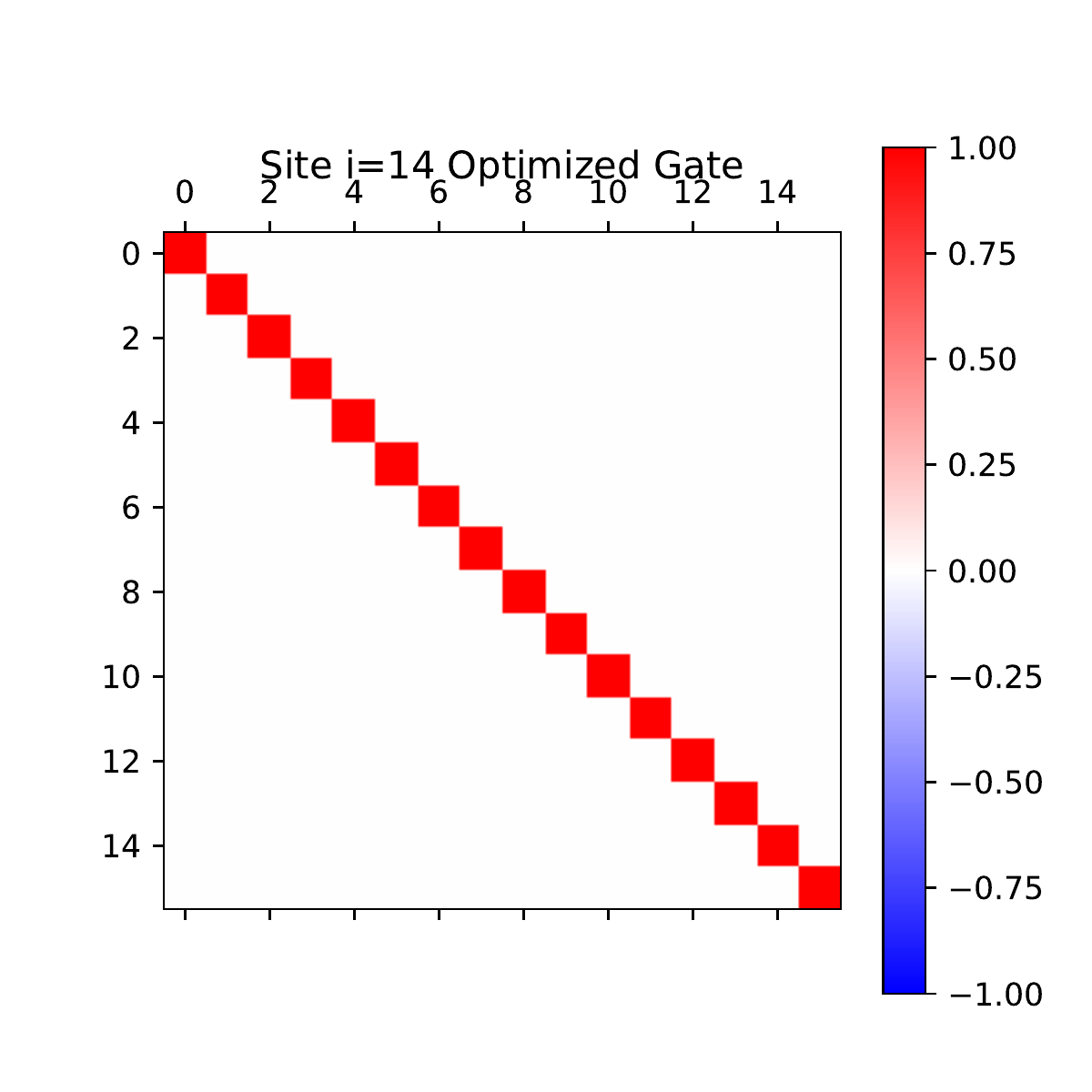}
}
\subfloat[Circuit from optimization]{
\includegraphics[width=0.2\columnwidth]{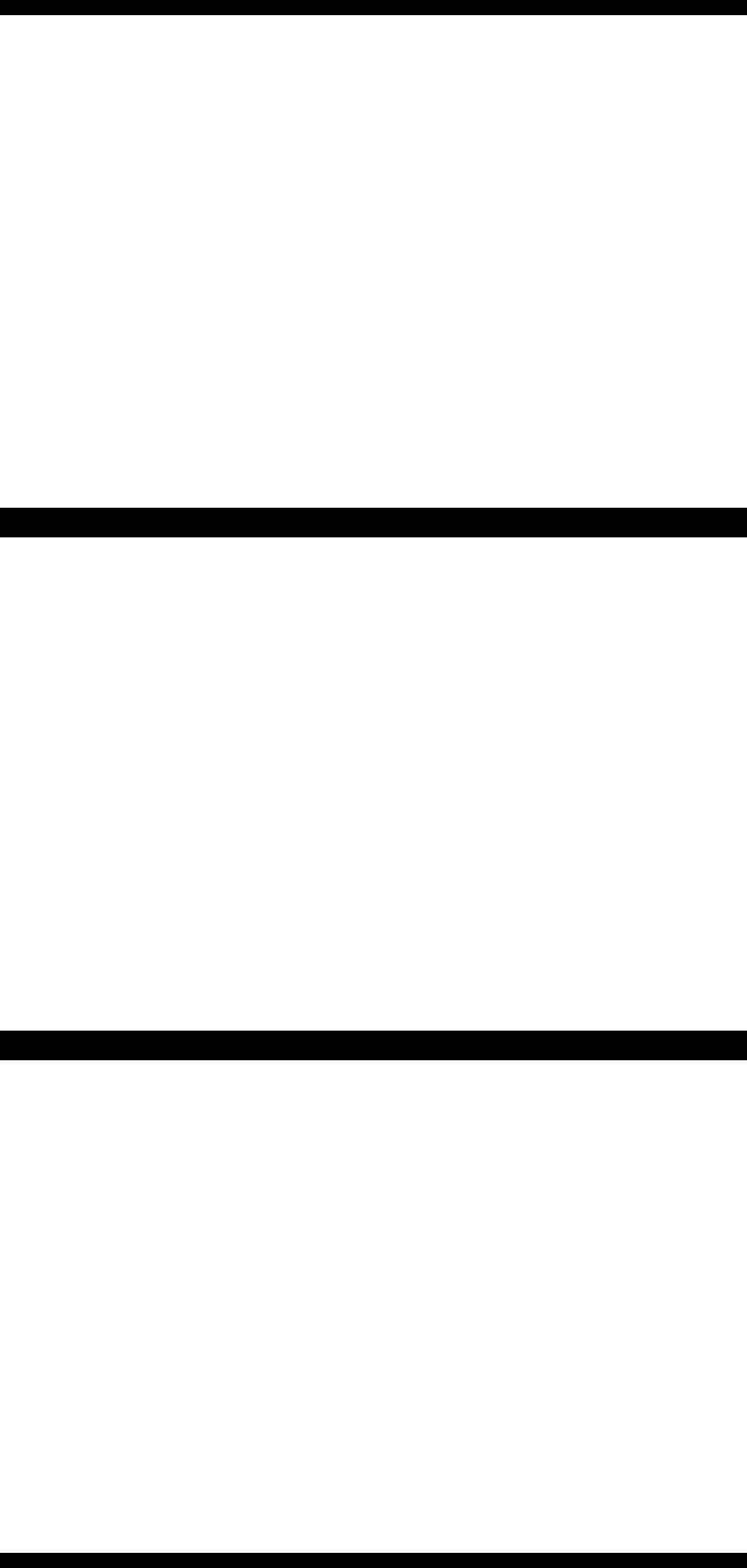}
}
\caption{Optimization for site 14}
\end{figure}
%
%
\begin{figure}[h]

\subfloat[Isometry]{
\includegraphics[width=0.45\columnwidth]{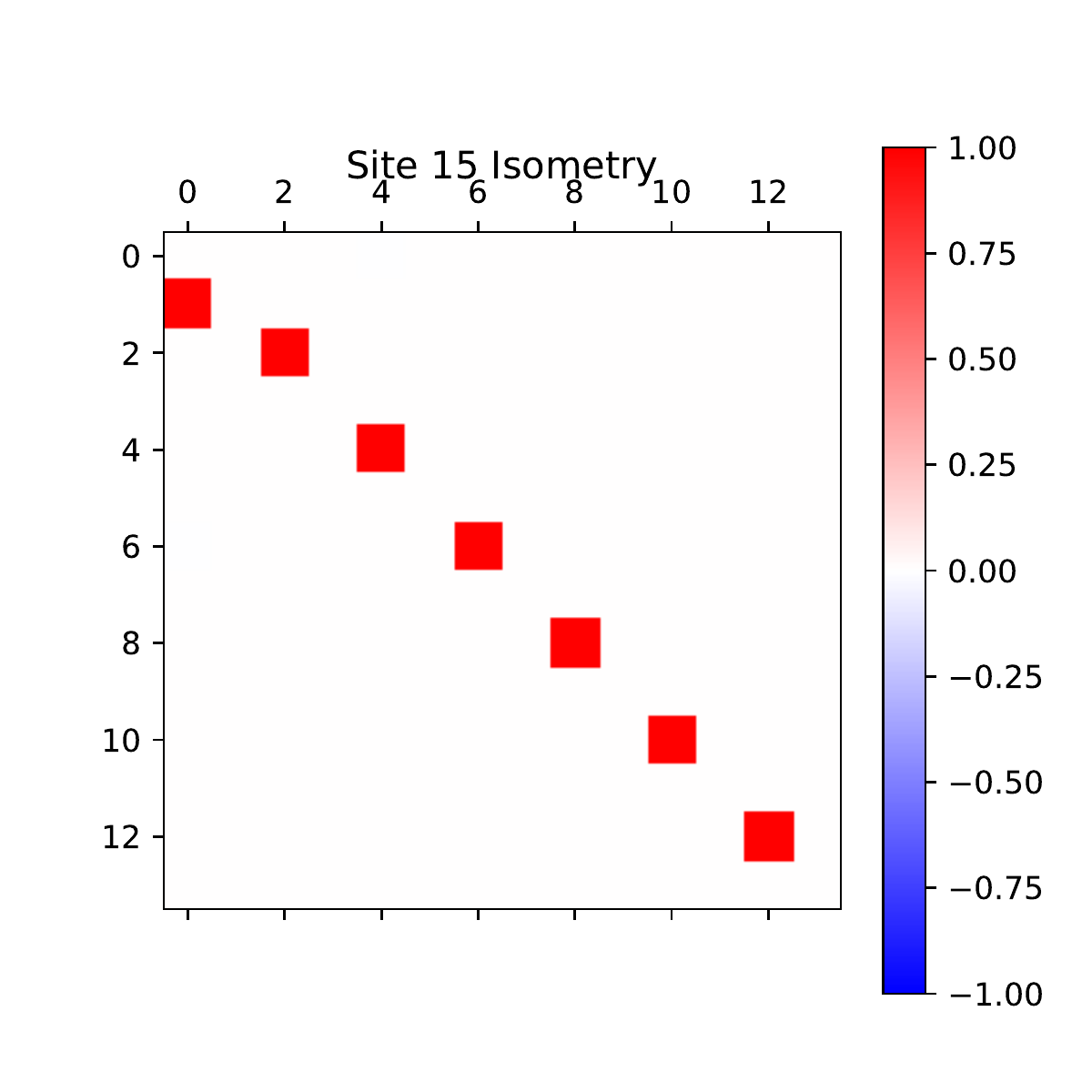}
}
\subfloat[Optimized gate]{
\includegraphics[width=0.45\columnwidth]{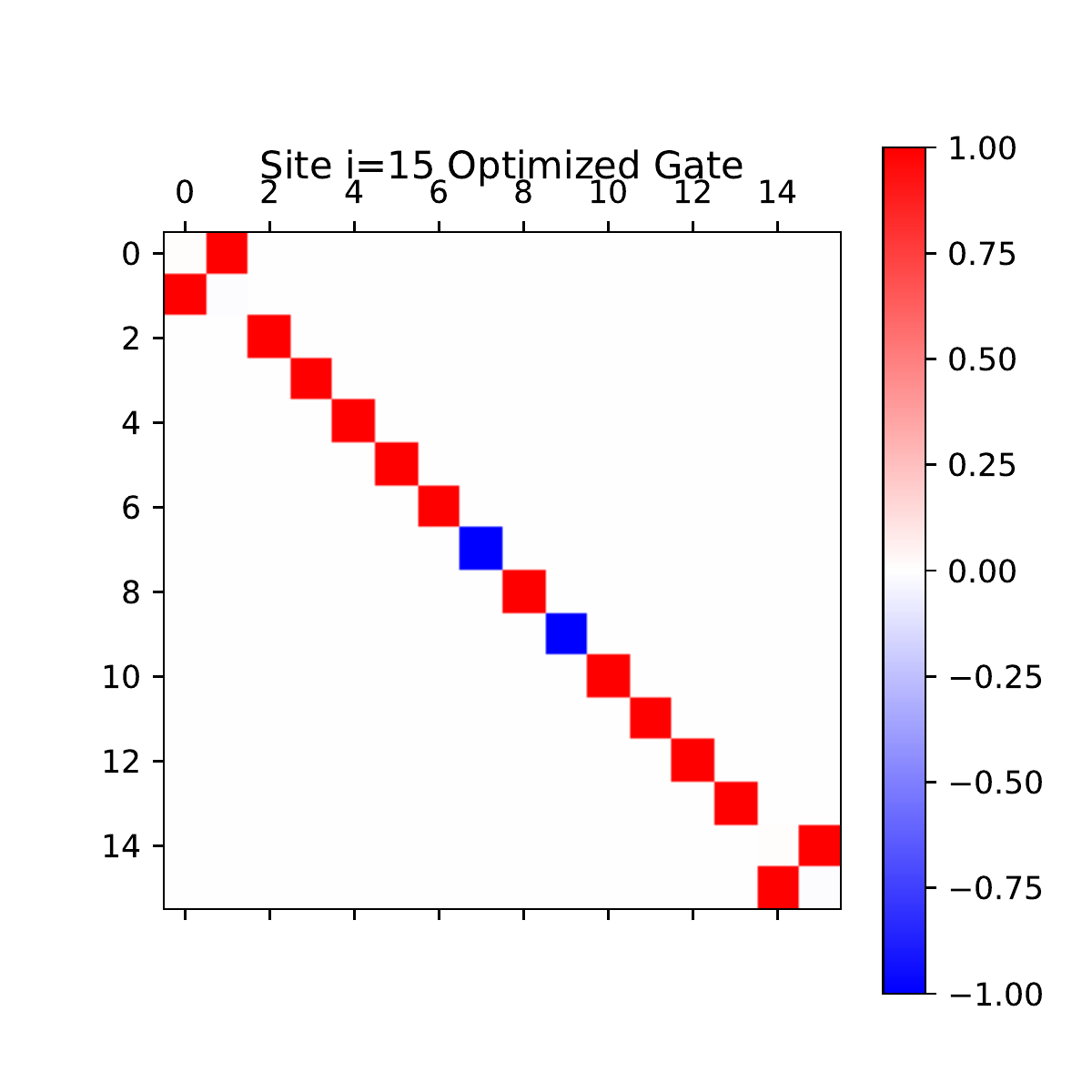}
}\\
\subfloat[Raw circuit from optimization]{
\includegraphics[width=0.9\columnwidth]{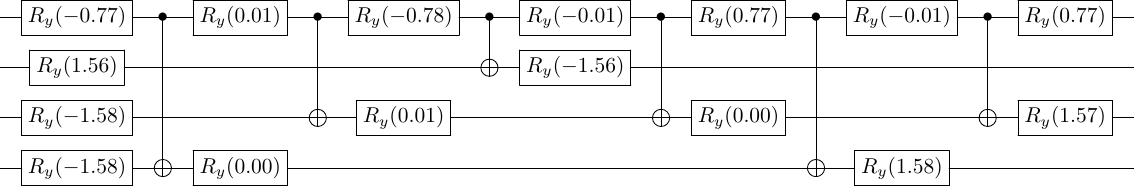}
}\\
\subfloat[Expanded and cleaned circuit from optimization]{
\includegraphics[width=0.9\columnwidth]{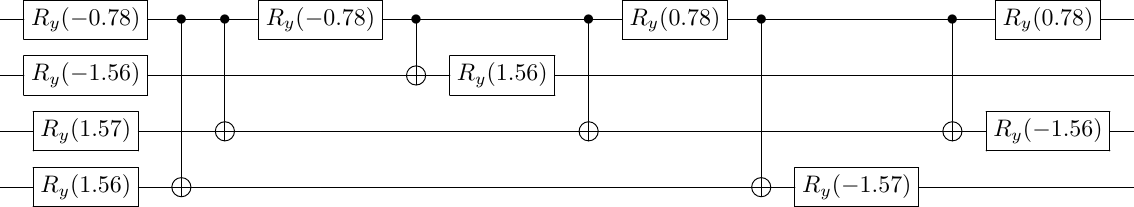}
}
\caption{\label{fig:site15} Optimization for site 15}
\end{figure}
%
%
\begin{figure}[h]

\subfloat[Isometry]{
\includegraphics[width=0.45\columnwidth]{"MNISTPlots/Site16_Isometry"}
}
\subfloat[Optimized gate]{
\includegraphics[width=0.45\columnwidth]{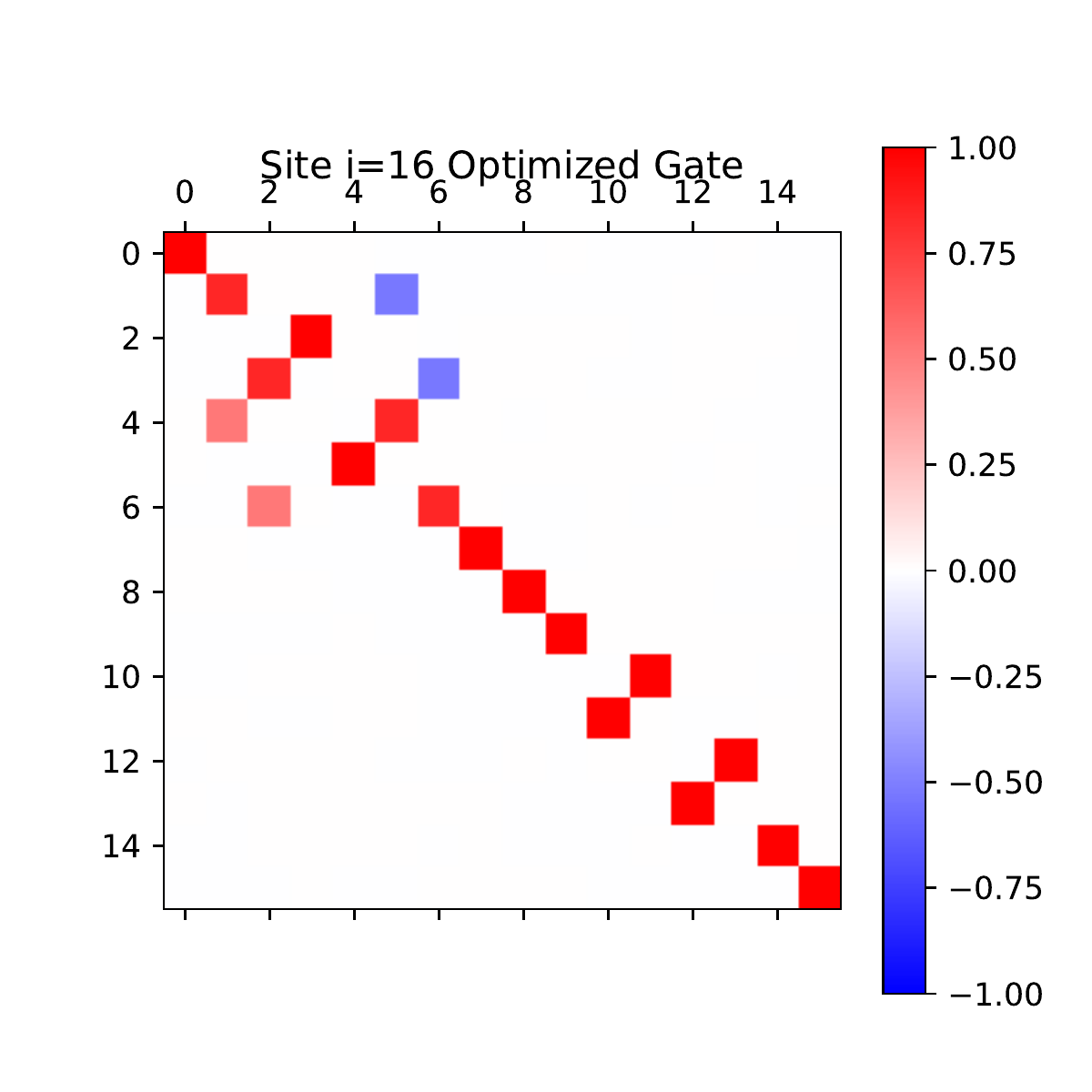}
}\\
\subfloat[Raw circuit from optimization]{
\includegraphics[width=0.9\columnwidth]{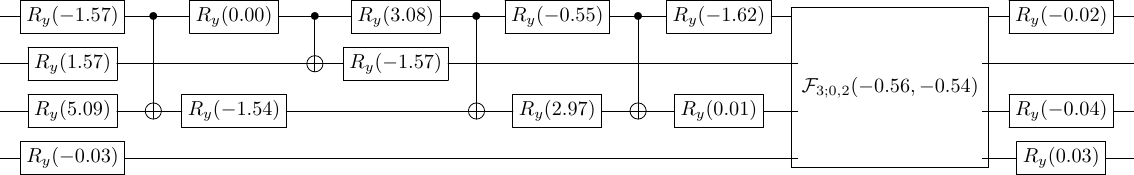}
}\\
\subfloat[Expanded and cleaned circuit from optimization]{
\includegraphics[width=0.9\columnwidth]{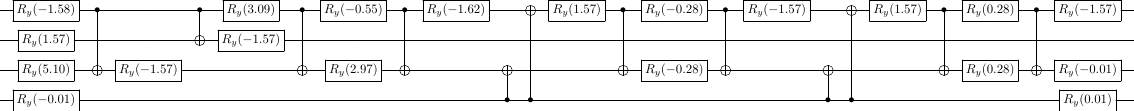}
}
\caption{\label{fig:site16} Optimization for site 16}
\end{figure}
%
%
\begin{figure}[h]

\subfloat[Isometry]{
\includegraphics[width=0.45\columnwidth]{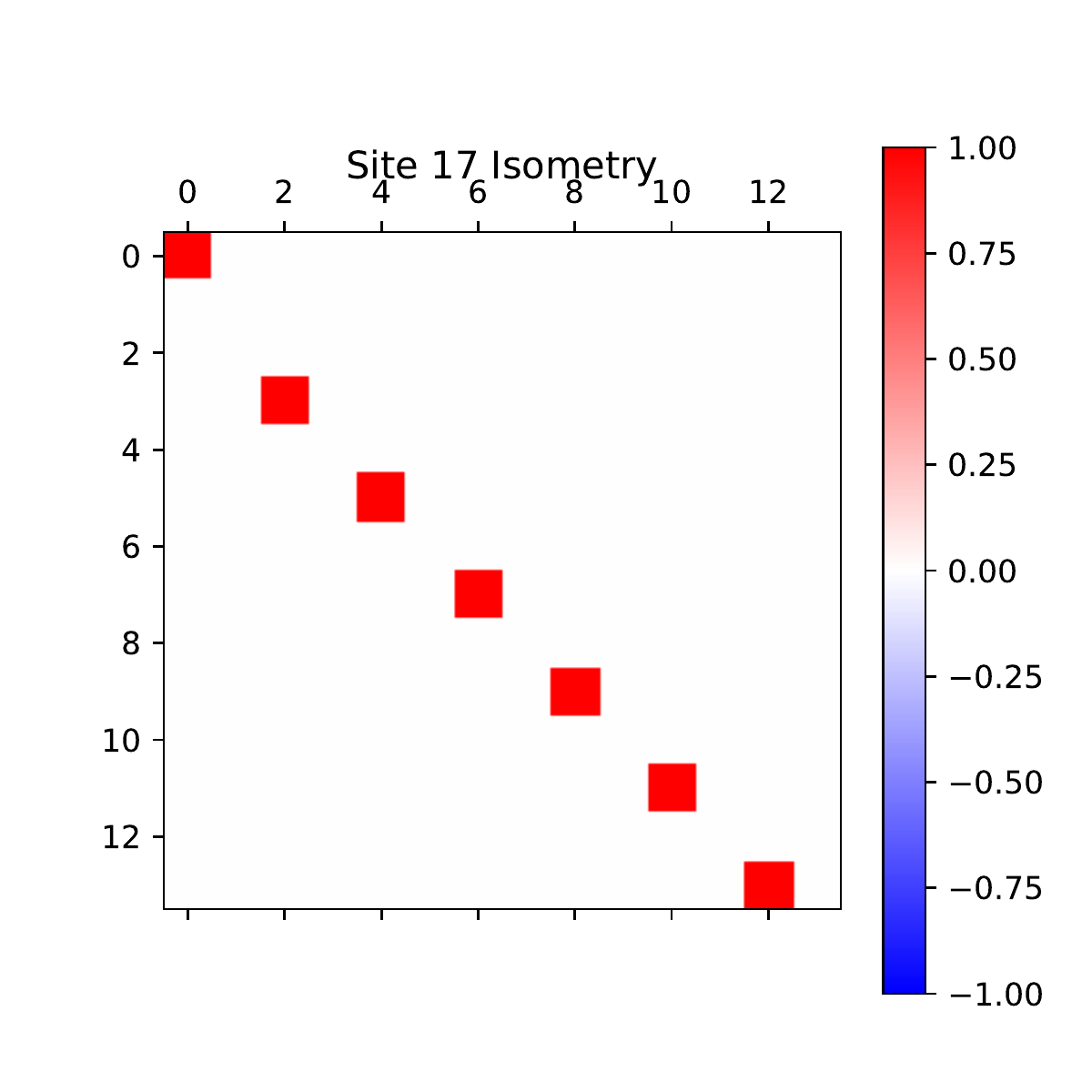}
}
\subfloat[Optimized gate]{
\includegraphics[width=0.45\columnwidth]{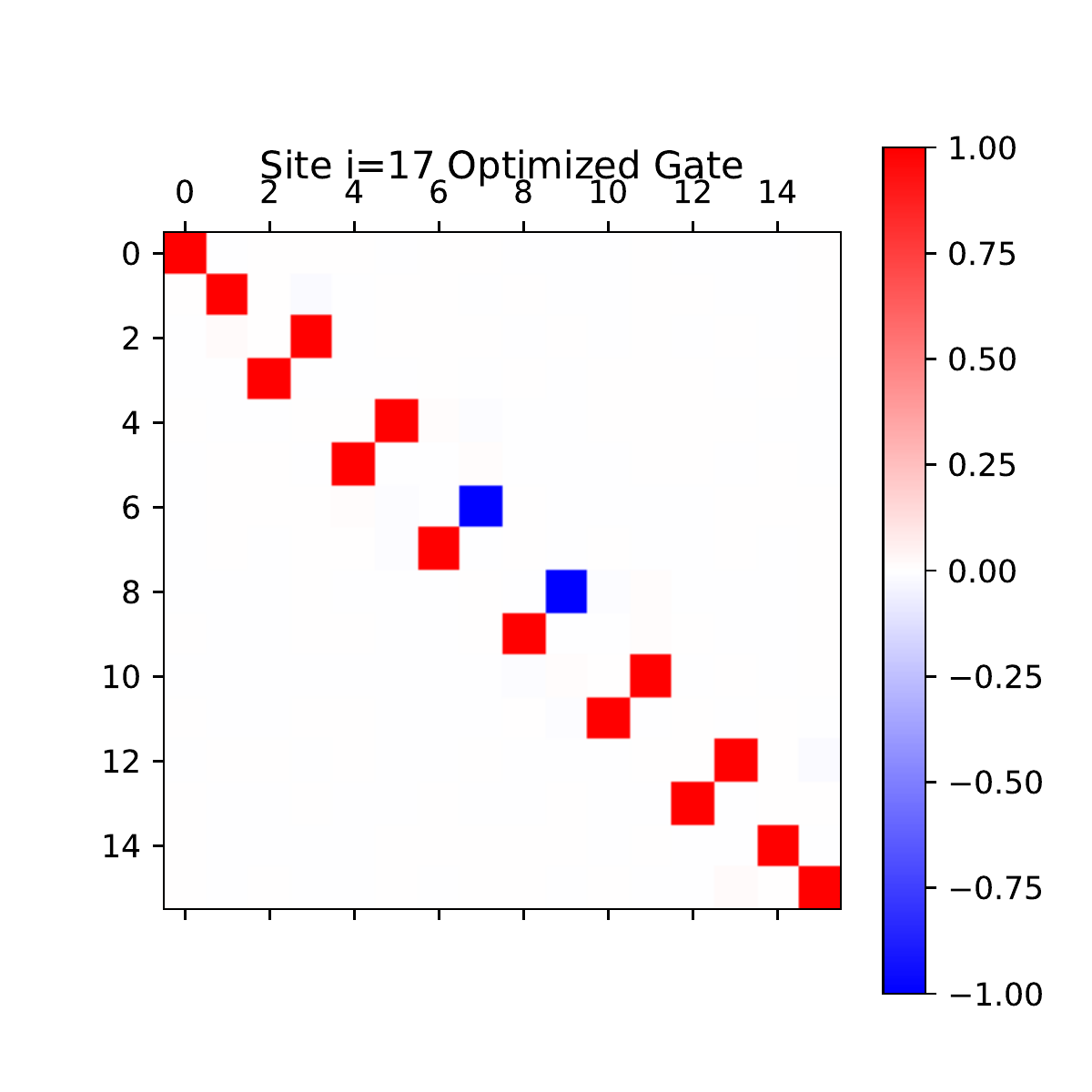}
}\\
\subfloat[Raw circuit from optimization]{
\includegraphics[width=0.9\columnwidth]{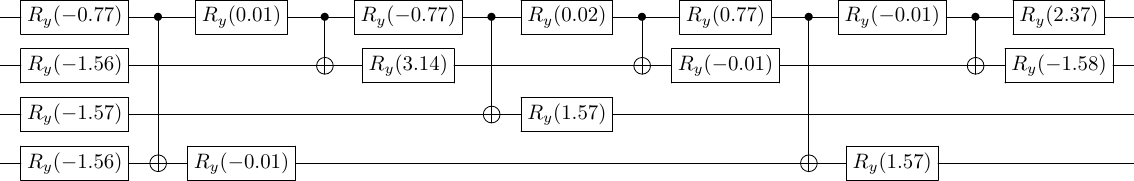}
}\\
\subfloat[Expanded and cleaned circuit from optimization]{
\includegraphics[width=0.9\columnwidth]{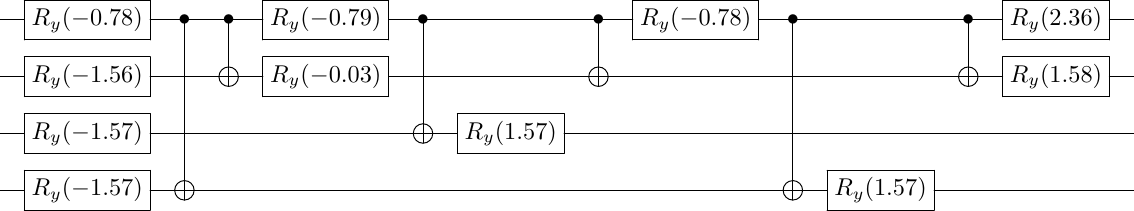}
}
\caption{Optimization for site 17}
\end{figure}
%
%
\begin{figure}[h]

\subfloat[Isometry]{
\includegraphics[width=0.45\columnwidth]{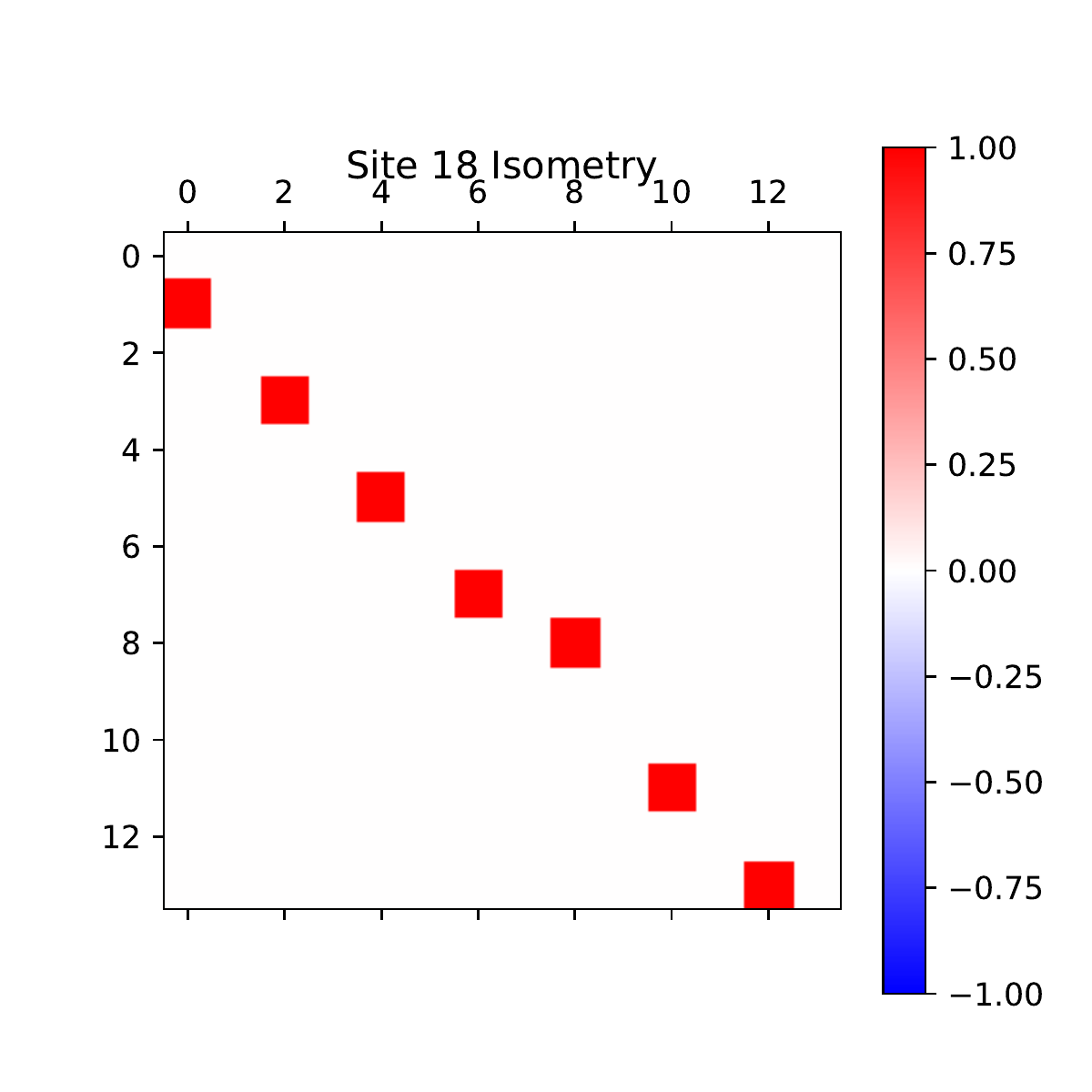}
}
\subfloat[Optimized gate]{
\includegraphics[width=0.45\columnwidth]{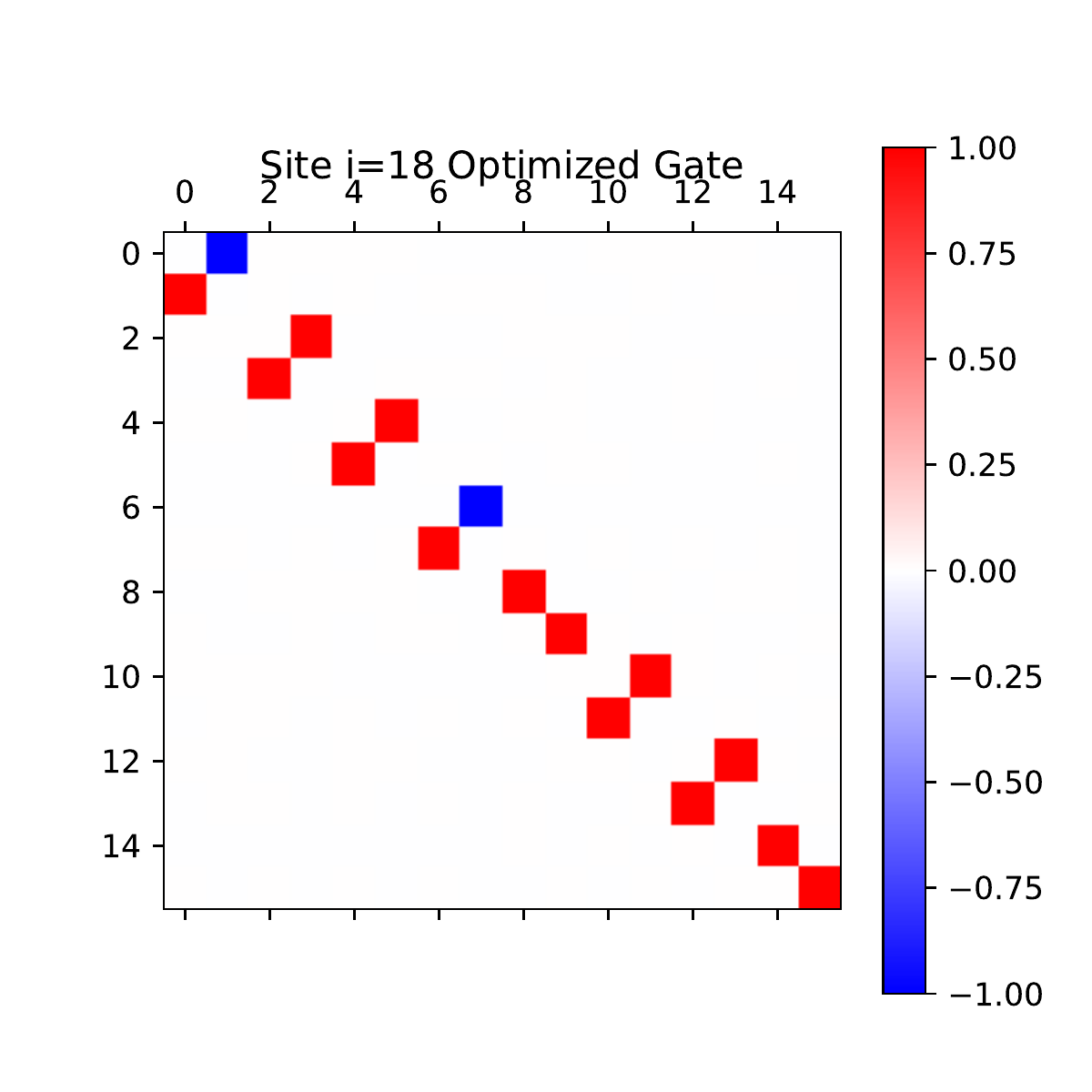}
}\\
\subfloat[Raw circuit from optimization]{
\includegraphics[width=0.9\columnwidth]{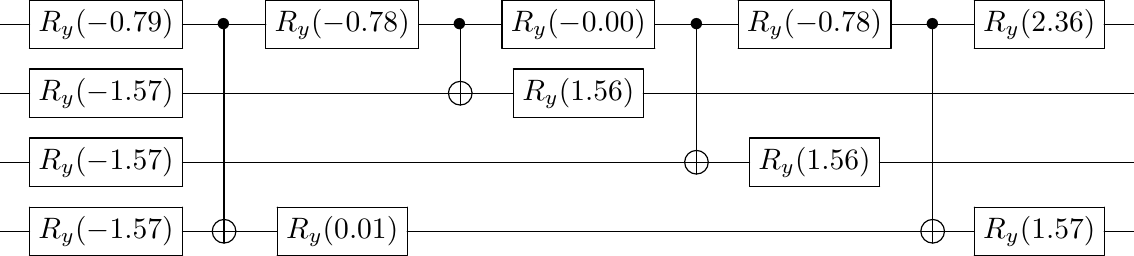}
}\\
\subfloat[Expanded and cleaned circuit from optimization]{
\includegraphics[width=0.9\columnwidth]{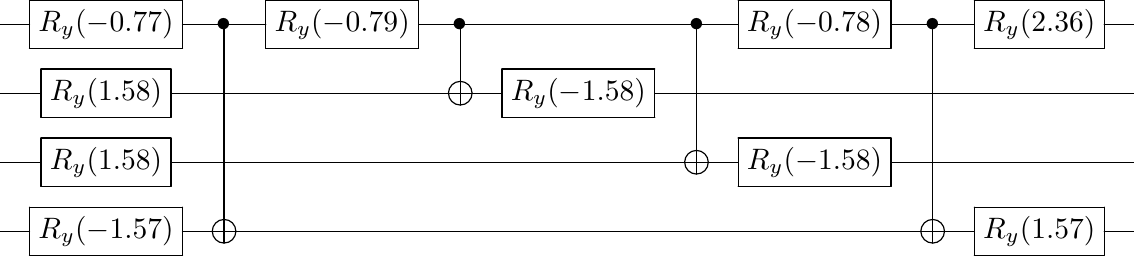}
}
\caption{Optimization for site 18}
\end{figure}
%
%
\begin{figure}[h]

\subfloat[Isometry]{
\includegraphics[width=0.45\columnwidth]{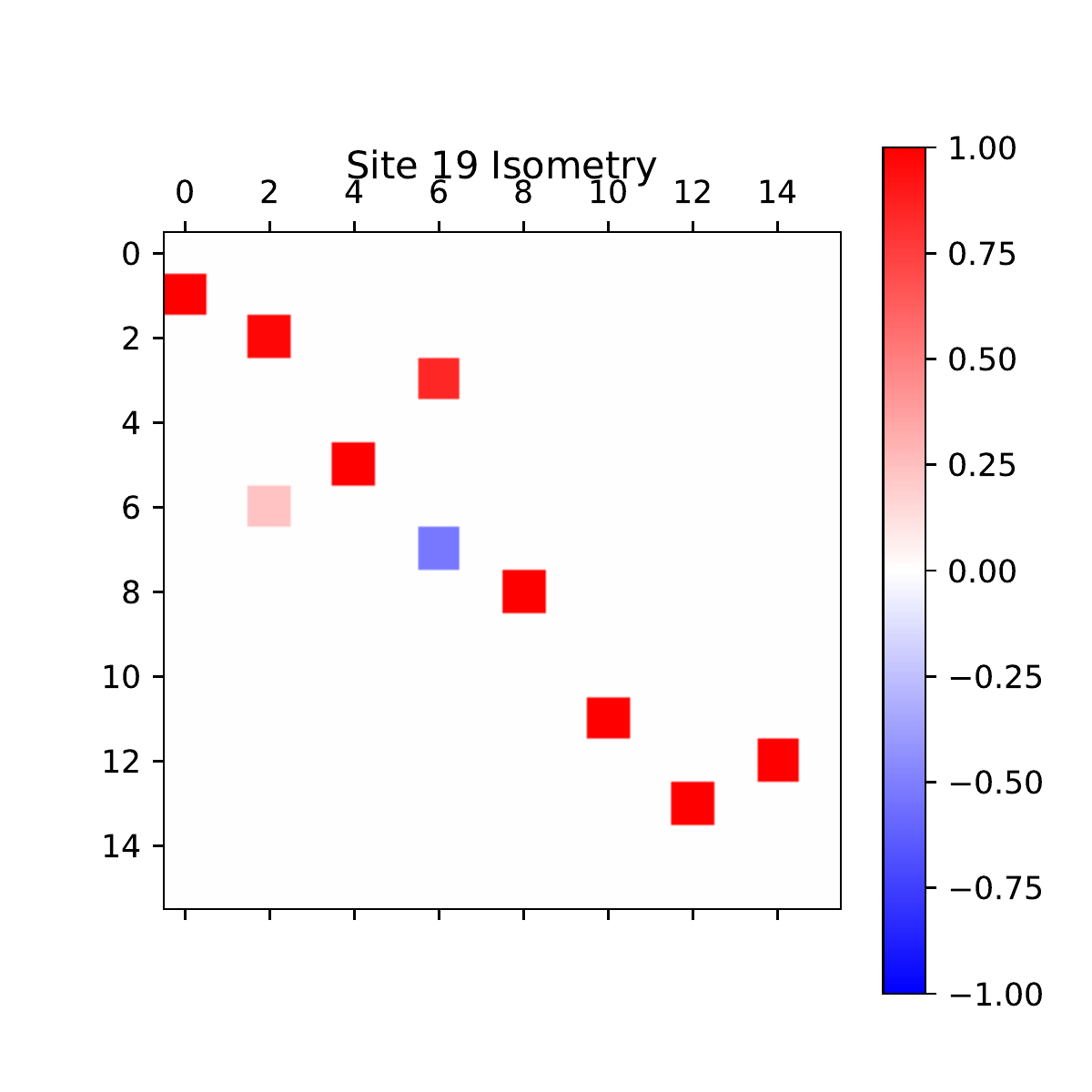}
}
\subfloat[Optimized gate]{
\includegraphics[width=0.45\columnwidth]{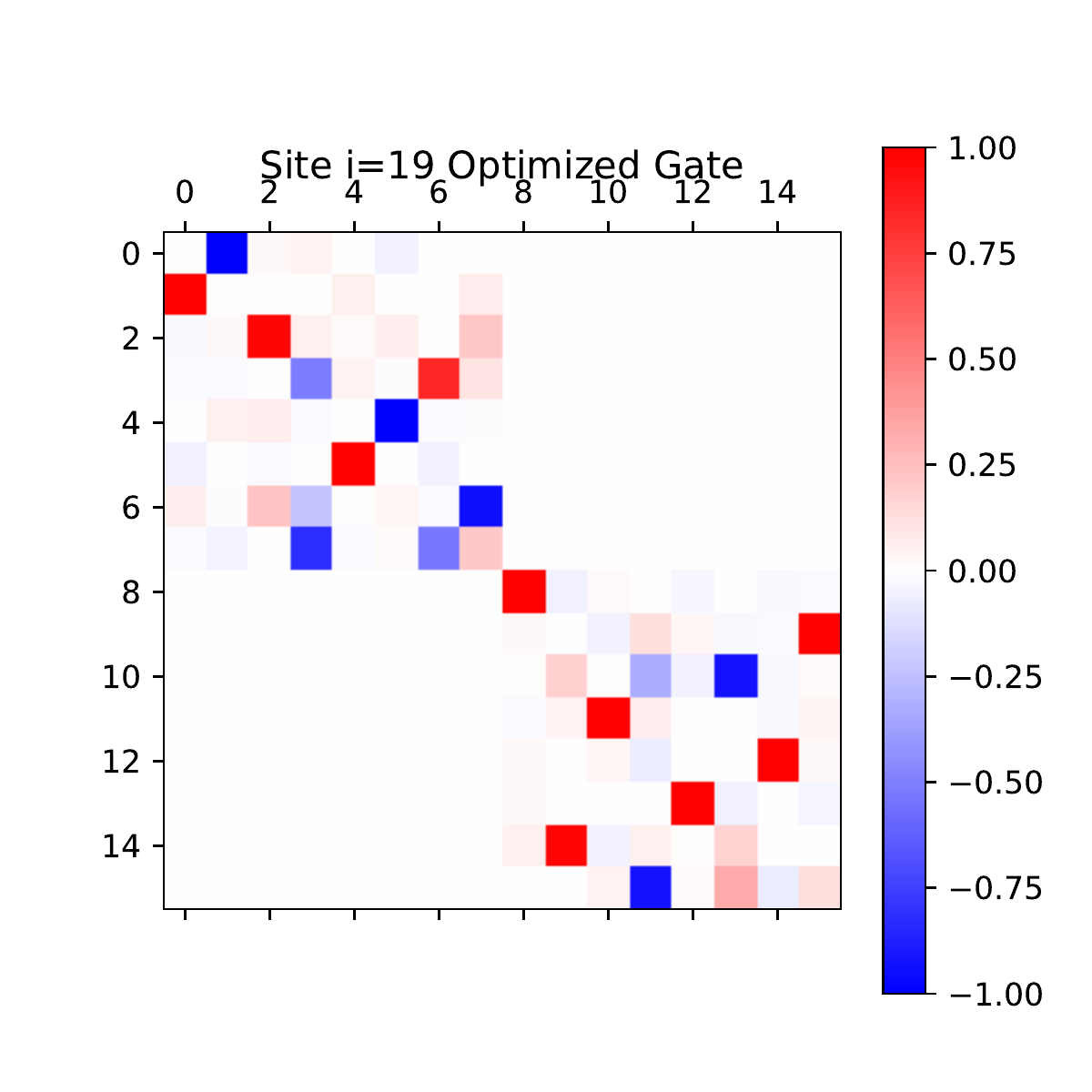}
}\\
\subfloat[Raw circuit from optimization]{
\includegraphics[width=0.9\columnwidth]{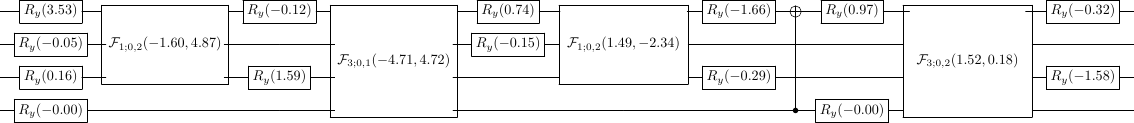}
}\\
\subfloat[Expanded and cleaned circuit from optimization]{
\includegraphics[width=0.9\columnwidth]{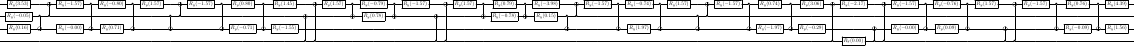}
}
\caption{Optimization for site 19}
\end{figure}
%
%
\begin{figure}[h]

\subfloat[Isometry]{
\includegraphics[width=0.3\columnwidth]{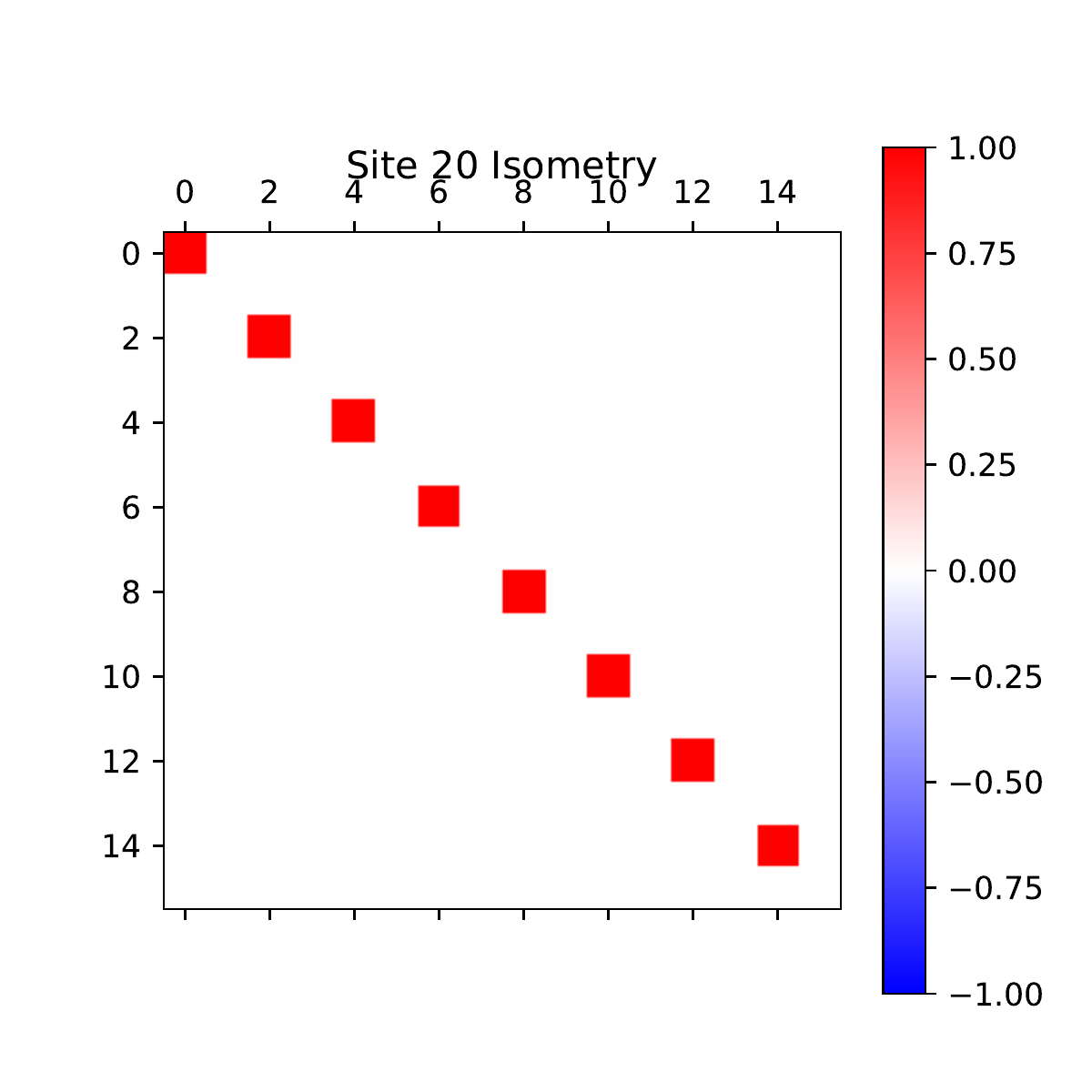}
}
\subfloat[Optimized gate]{
\includegraphics[width=0.3\columnwidth]{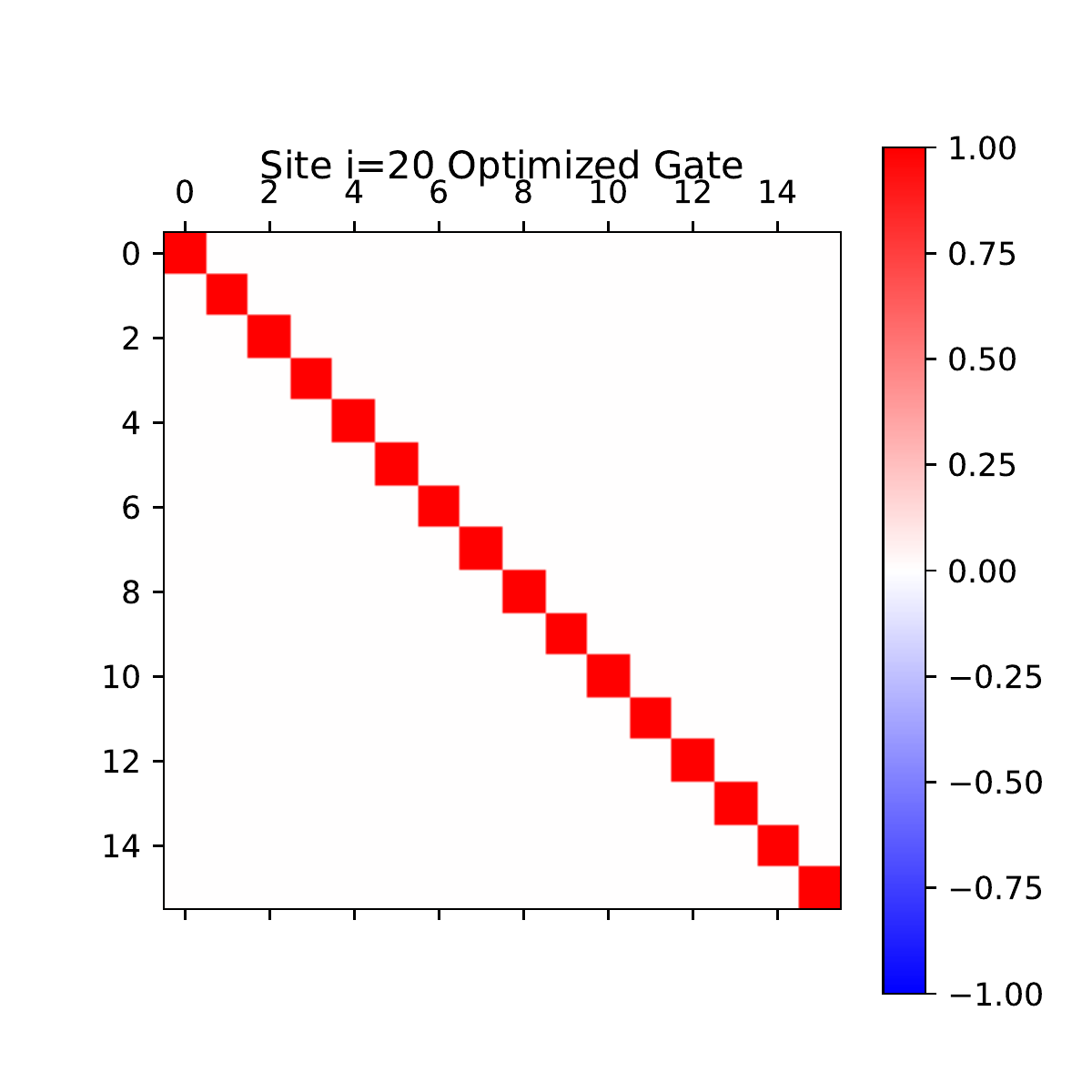}
}
\subfloat[Circuit from optimization]{
\includegraphics[width=0.2\columnwidth]{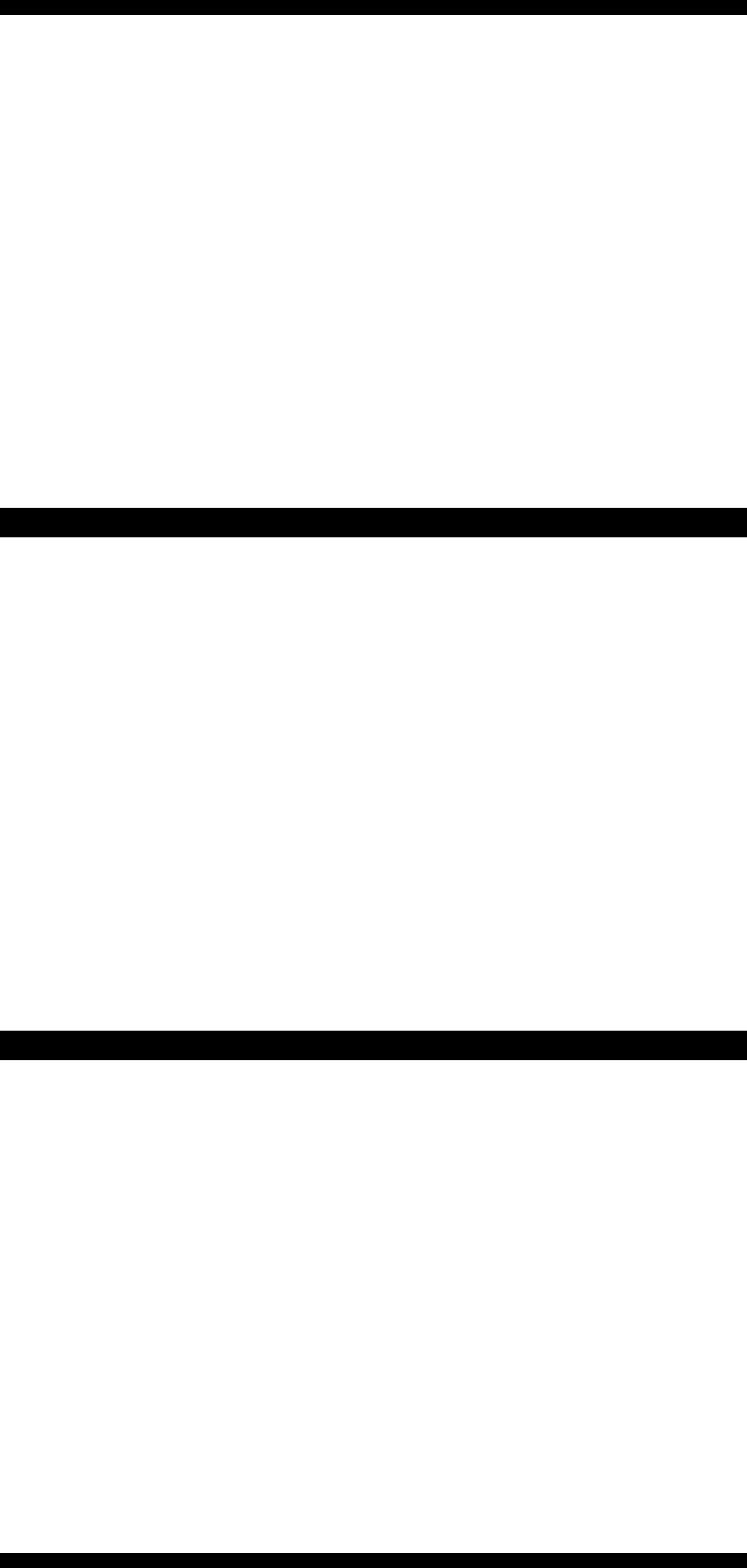}
}
\caption{Optimization for site 20}
\end{figure}
%
%
\begin{figure}[h]

\subfloat[Isometry]{
\includegraphics[width=0.45\columnwidth]{"MNISTPlots/Site21_Isometry"}
}
\subfloat[Optimized gate]{
\includegraphics[width=0.45\columnwidth]{"MNISTPlots/Site21_OptimizedGate"}
}\\
\subfloat[Raw circuit from optimization]{
\includegraphics[width=0.9\columnwidth]{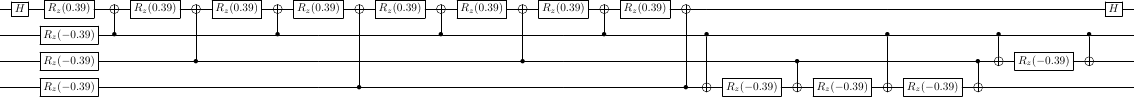}
}\\
\subfloat[Expanded and cleaned circuit from optimization]{
\includegraphics[width=0.9\columnwidth]{"MNISTPlots/Site21gate"}
}
\caption{Optimization for site 21}
\end{figure}
%
%
\begin{figure}[h]

\subfloat[Isometry]{
\includegraphics[width=0.45\columnwidth]{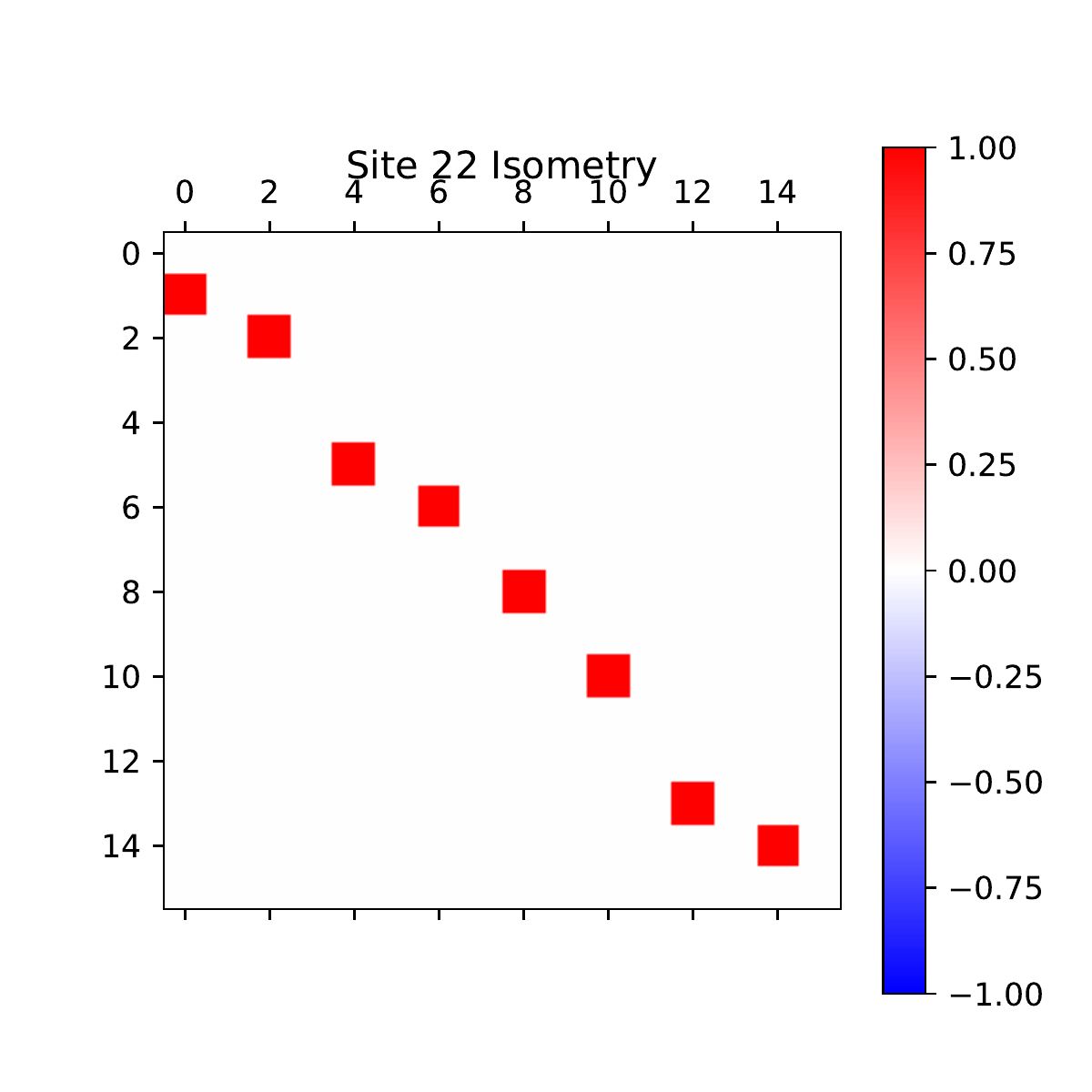}
}
\subfloat[Optimized gate]{
\includegraphics[width=0.45\columnwidth]{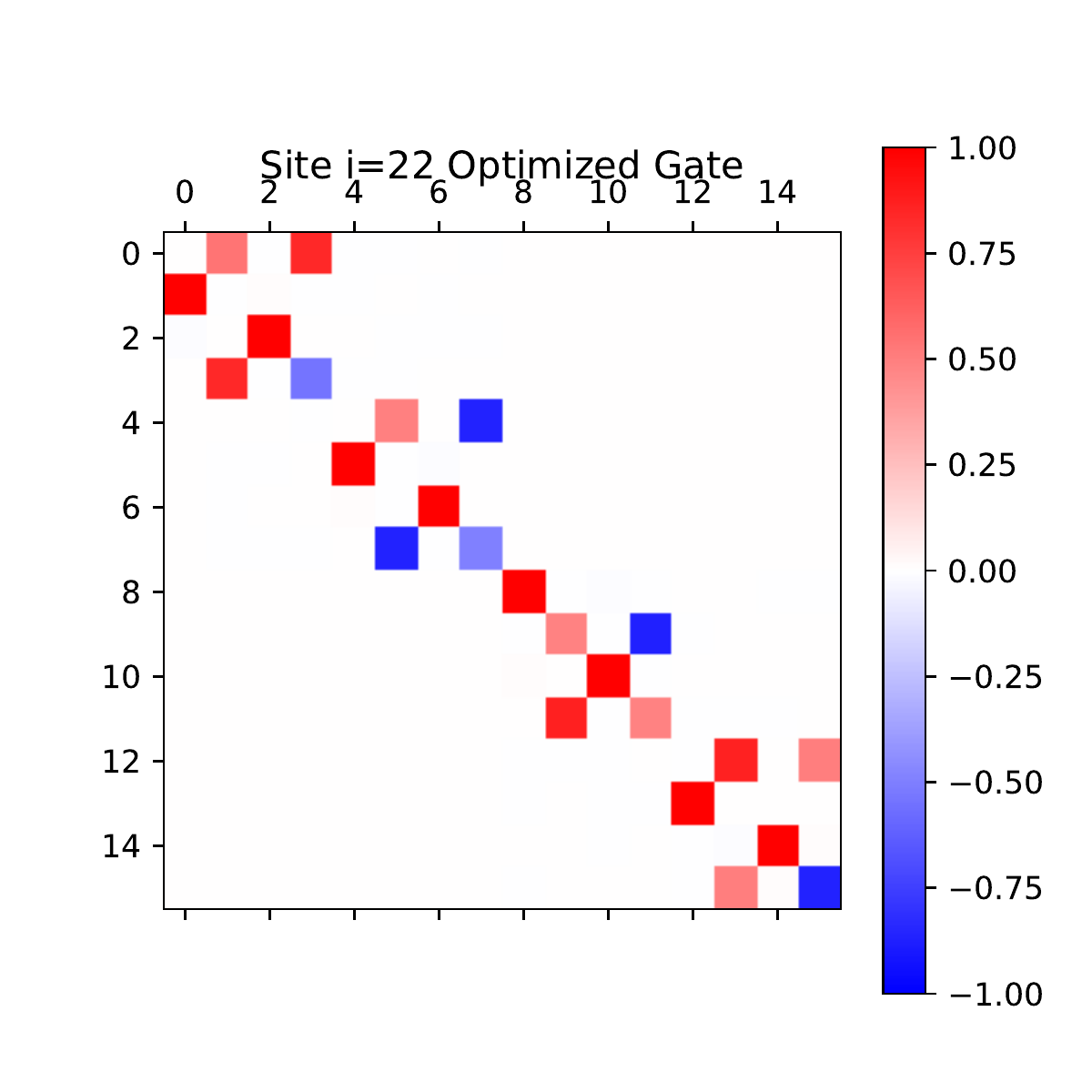}
}\\
\subfloat[Raw circuit from optimization]{
\includegraphics[width=0.9\columnwidth]{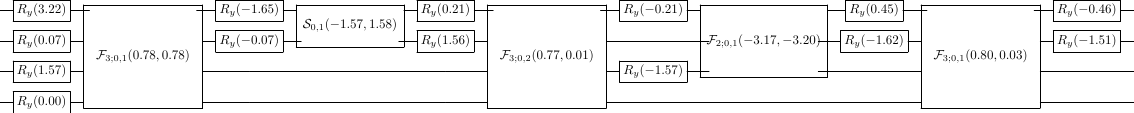}
}\\
\subfloat[Expanded and cleaned circuit from optimization]{
\includegraphics[width=0.9\columnwidth]{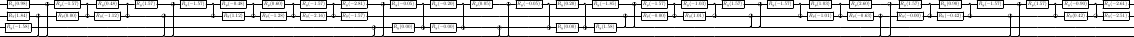}
}
\caption{Optimization for site 22}
\end{figure}
%
%
\begin{figure}[h]

\subfloat[Isometry]{
\includegraphics[width=0.45\columnwidth]{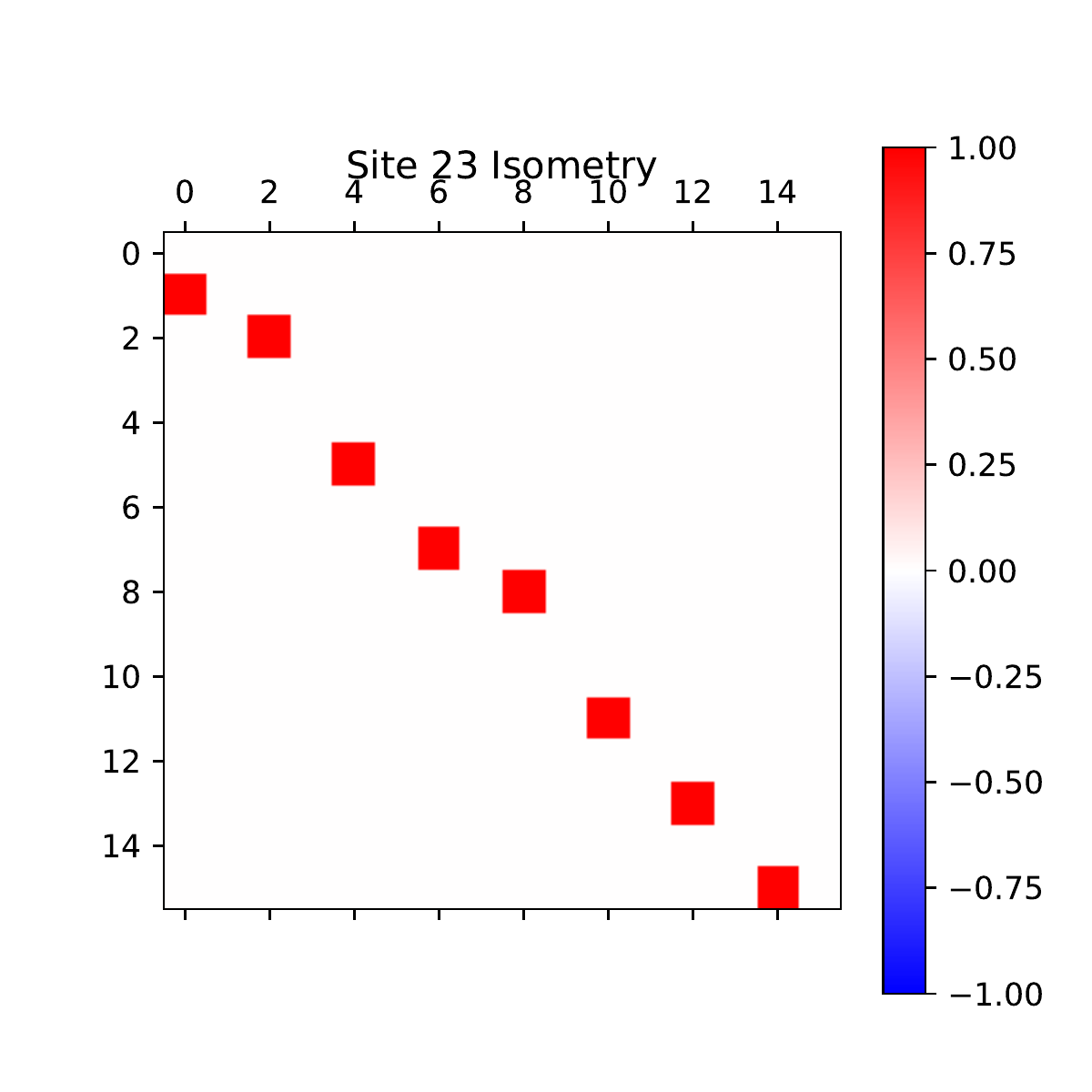}
}
\subfloat[Optimized gate]{
\includegraphics[width=0.45\columnwidth]{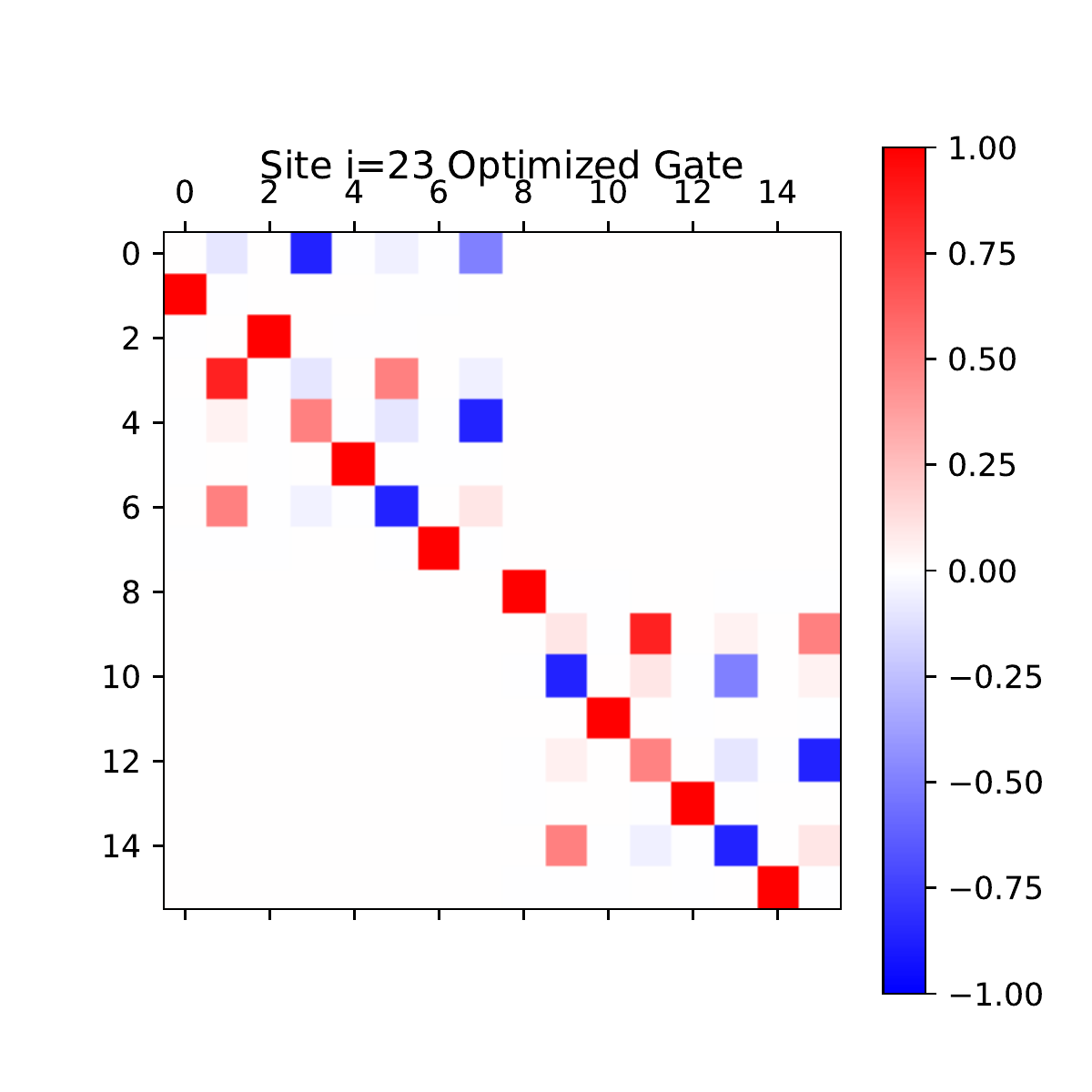}
}\\
\subfloat[Raw circuit from optimization]{
\includegraphics[width=0.9\columnwidth]{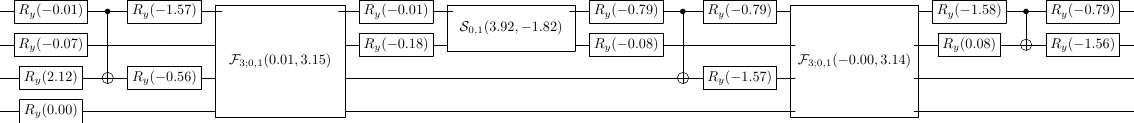}
}\\
\subfloat[Expanded and cleaned circuit from optimization]{
\includegraphics[width=0.9\columnwidth]{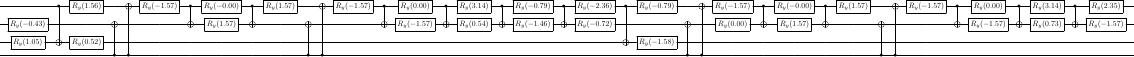}
}
\caption{Optimization for site 23}
\end{figure}
%
%
\begin{figure}[h]

\subfloat[Isometry]{
\includegraphics[width=0.45\columnwidth]{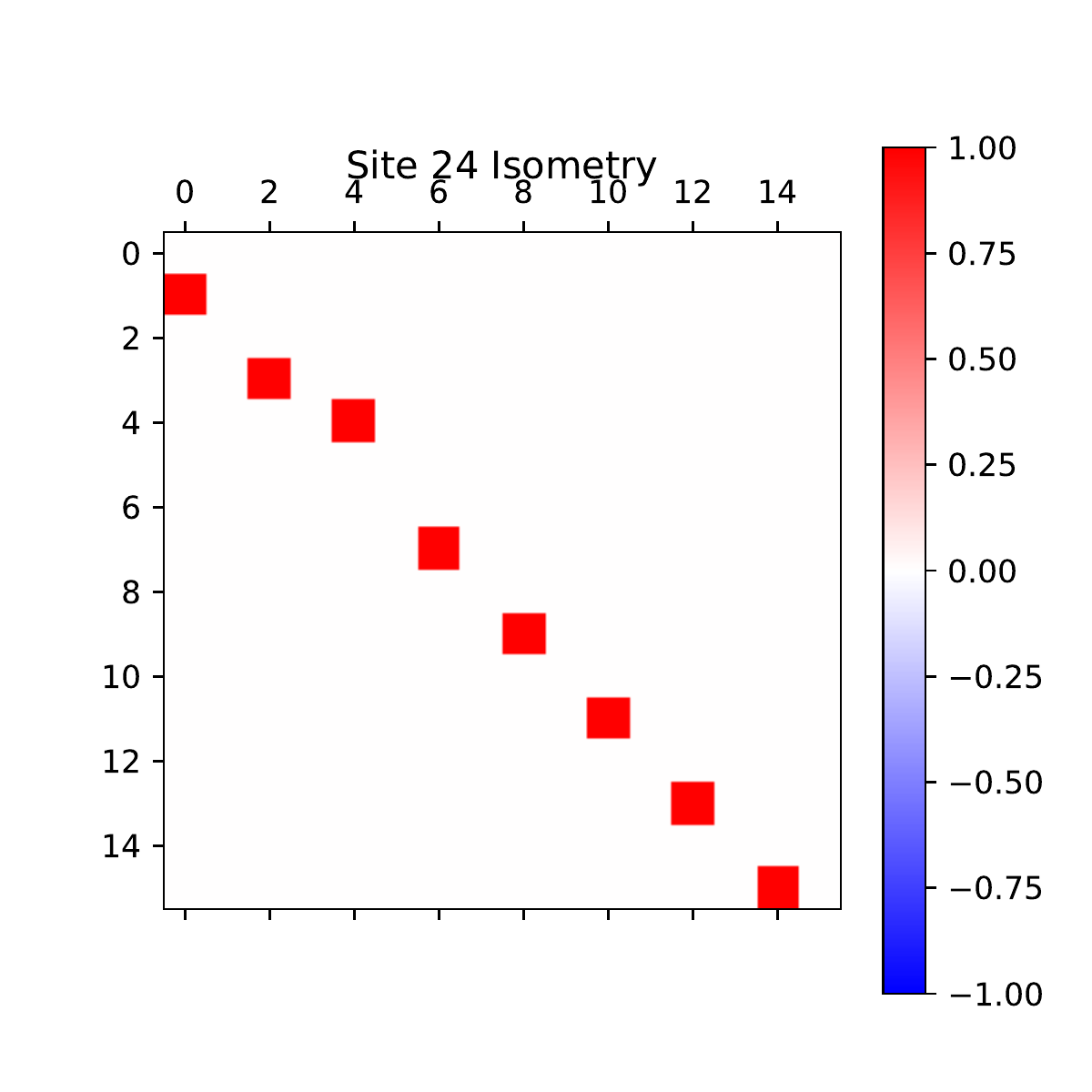}
}
\subfloat[Optimized gate]{
\includegraphics[width=0.45\columnwidth]{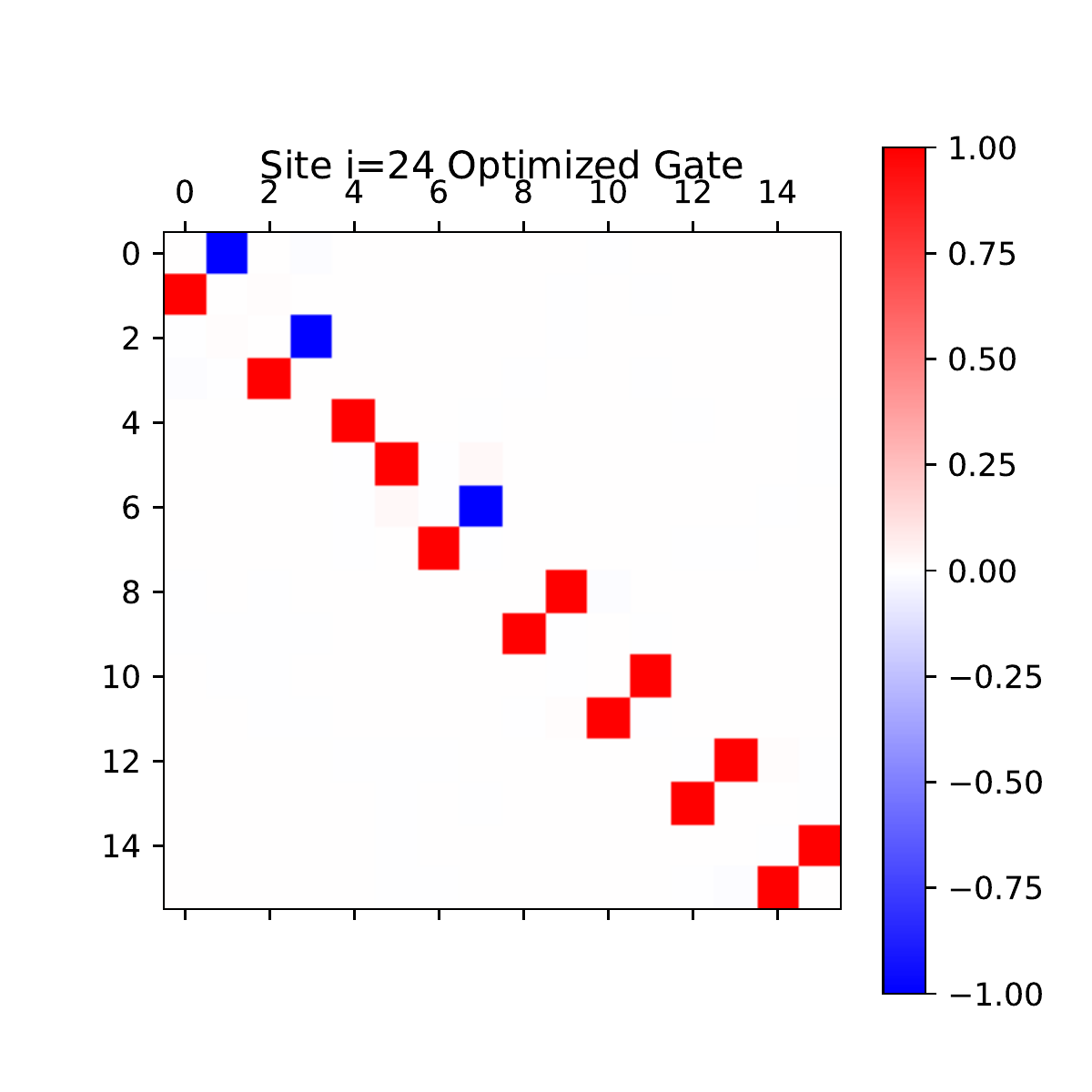}
}\\
\subfloat[Raw circuit from optimization]{
\includegraphics[width=0.9\columnwidth]{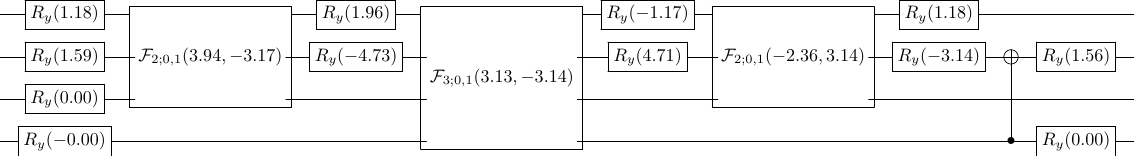}
}\\
\subfloat[Expanded and cleaned circuit from optimization]{
\includegraphics[width=0.9\columnwidth]{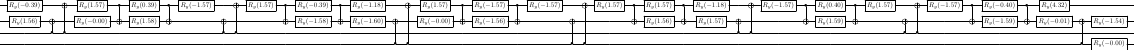}
}
\caption{Optimization for site 24}
\end{figure}
%
%
\begin{figure}[h]

\subfloat[Isometry]{
\includegraphics[width=0.45\columnwidth]{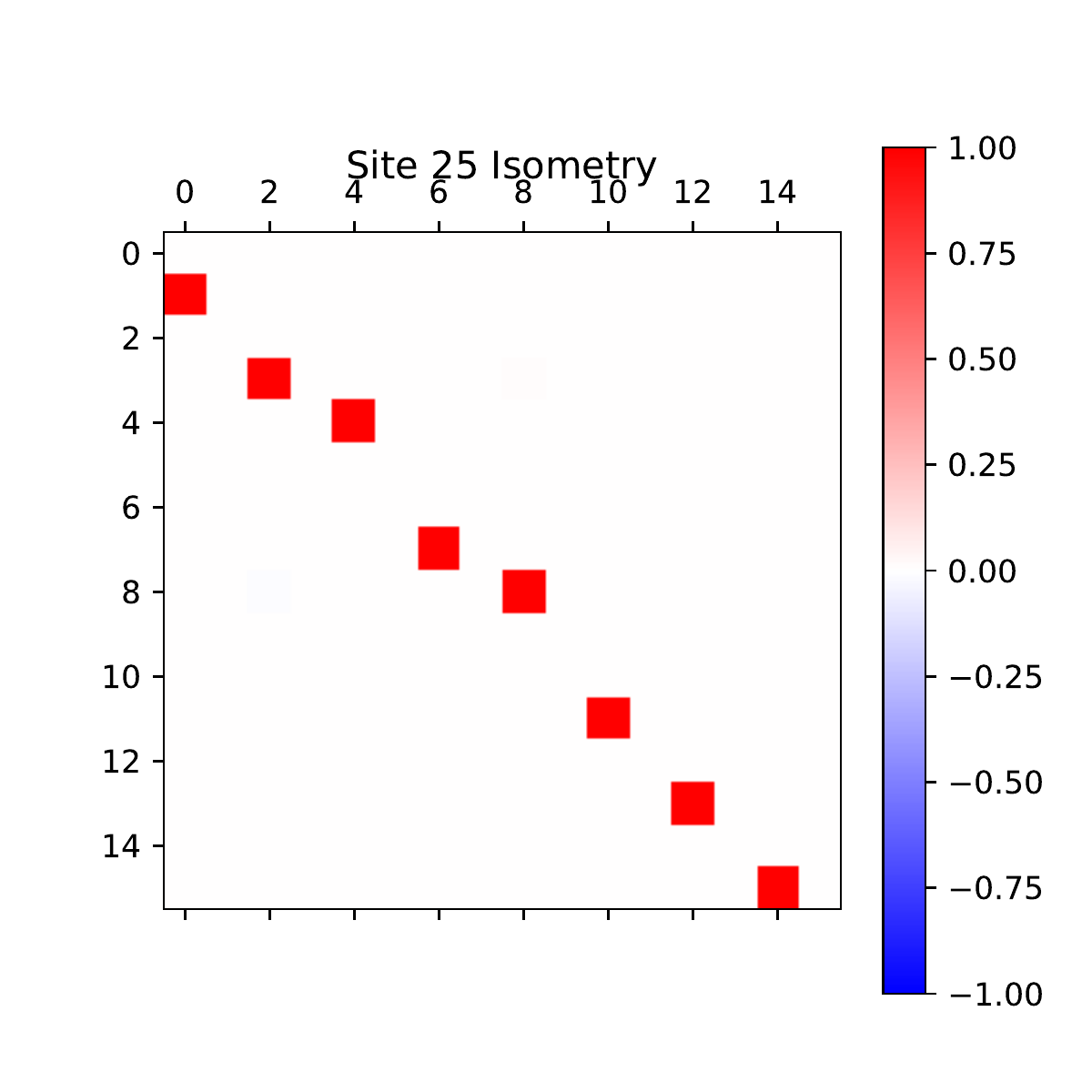}
}
\subfloat[Optimized gate]{
\includegraphics[width=0.45\columnwidth]{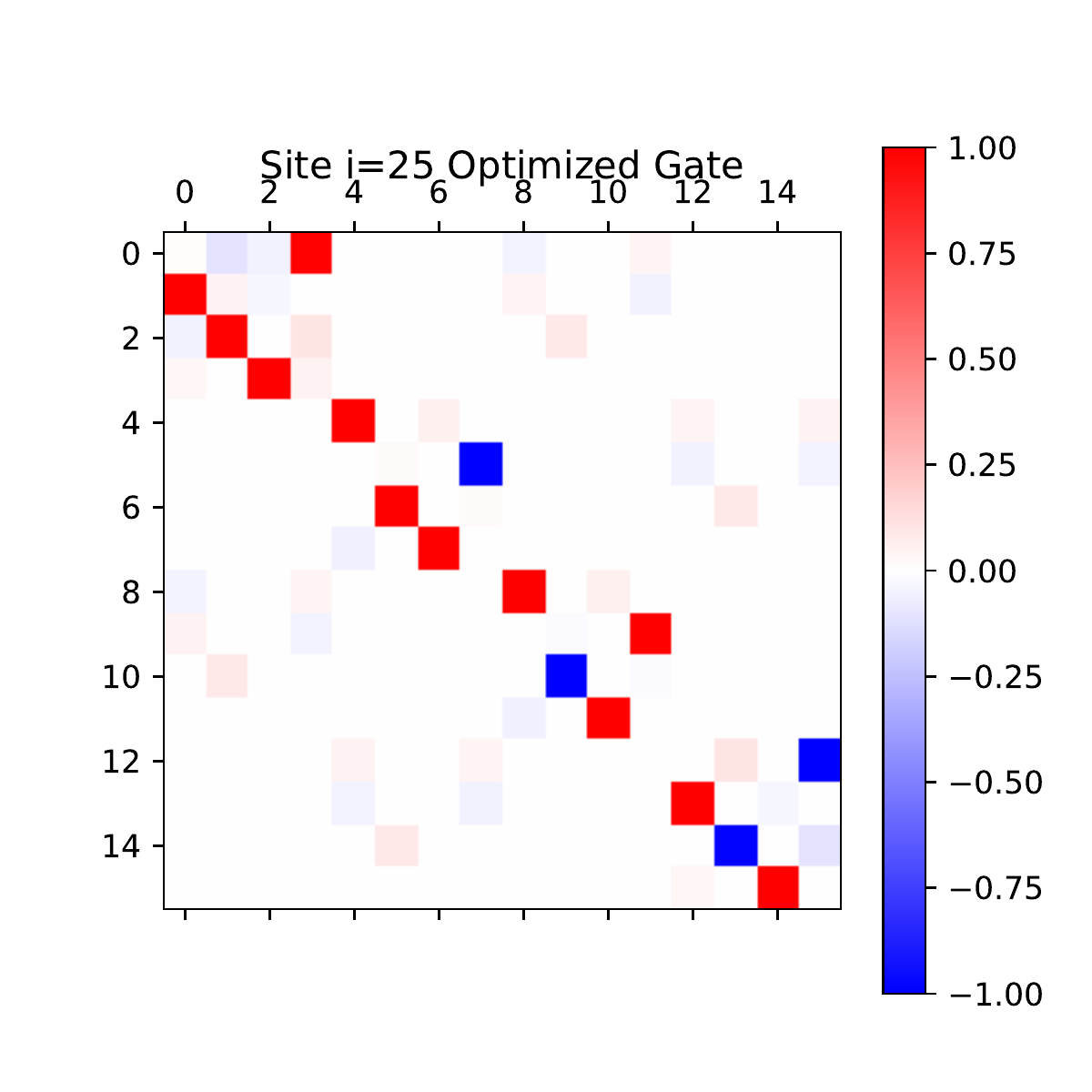}
}\\
\subfloat[Raw circuit from optimization]{
\includegraphics[width=0.9\columnwidth]{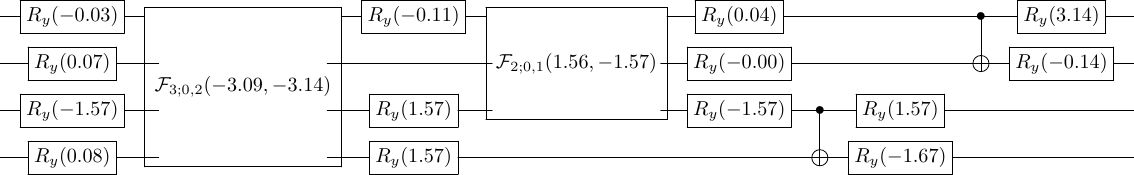}
}\\
\subfloat[Expanded and cleaned circuit from optimization]{
\includegraphics[width=0.9\columnwidth]{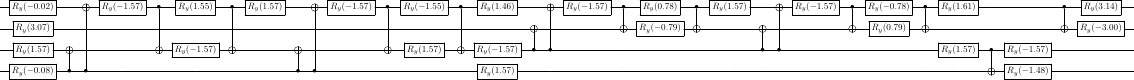}
}
\caption{Optimization for site 25}
\end{figure}
%
%
\begin{figure}[h]

\subfloat[Isometry]{
\includegraphics[width=0.45\columnwidth]{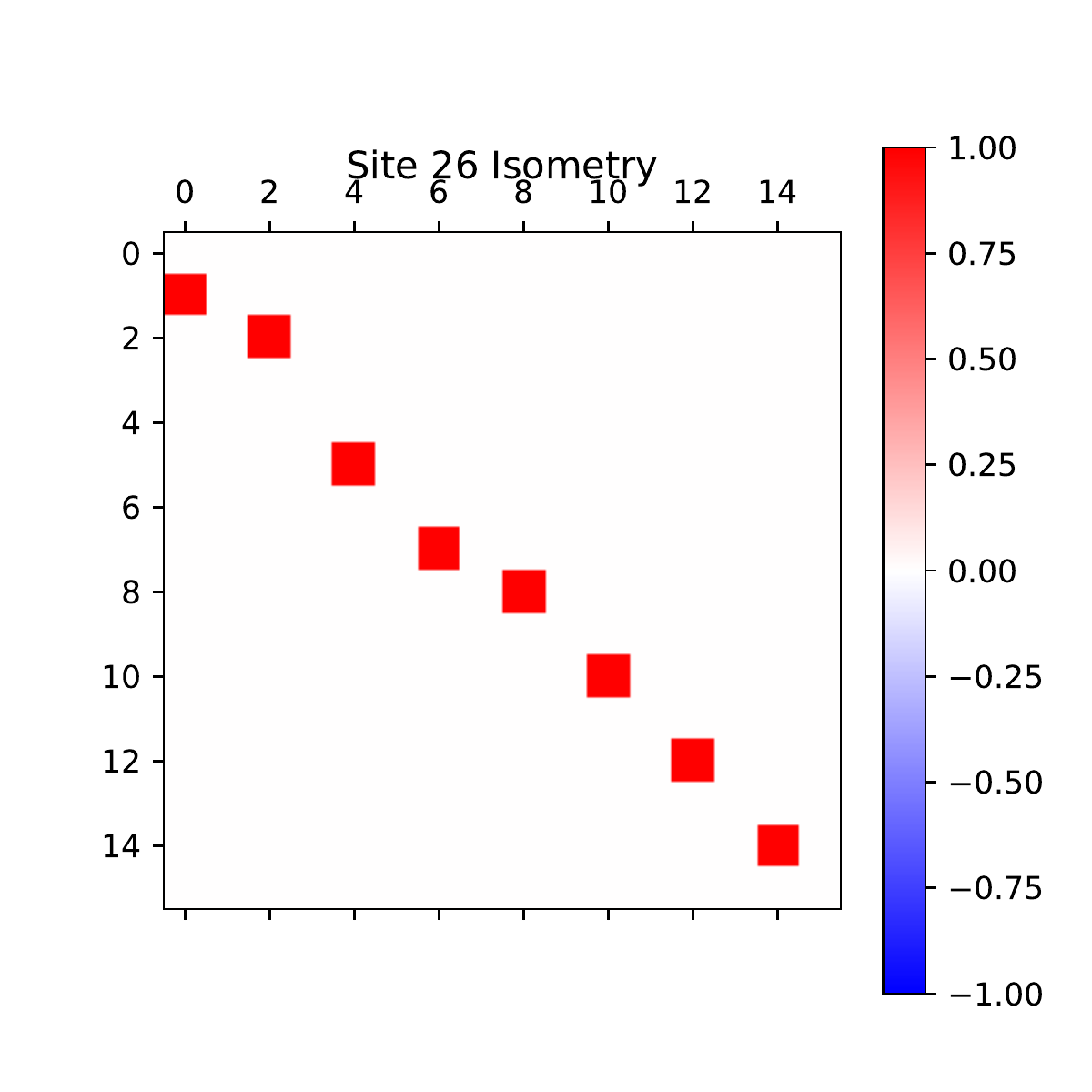}
}
\subfloat[Optimized gate]{
\includegraphics[width=0.45\columnwidth]{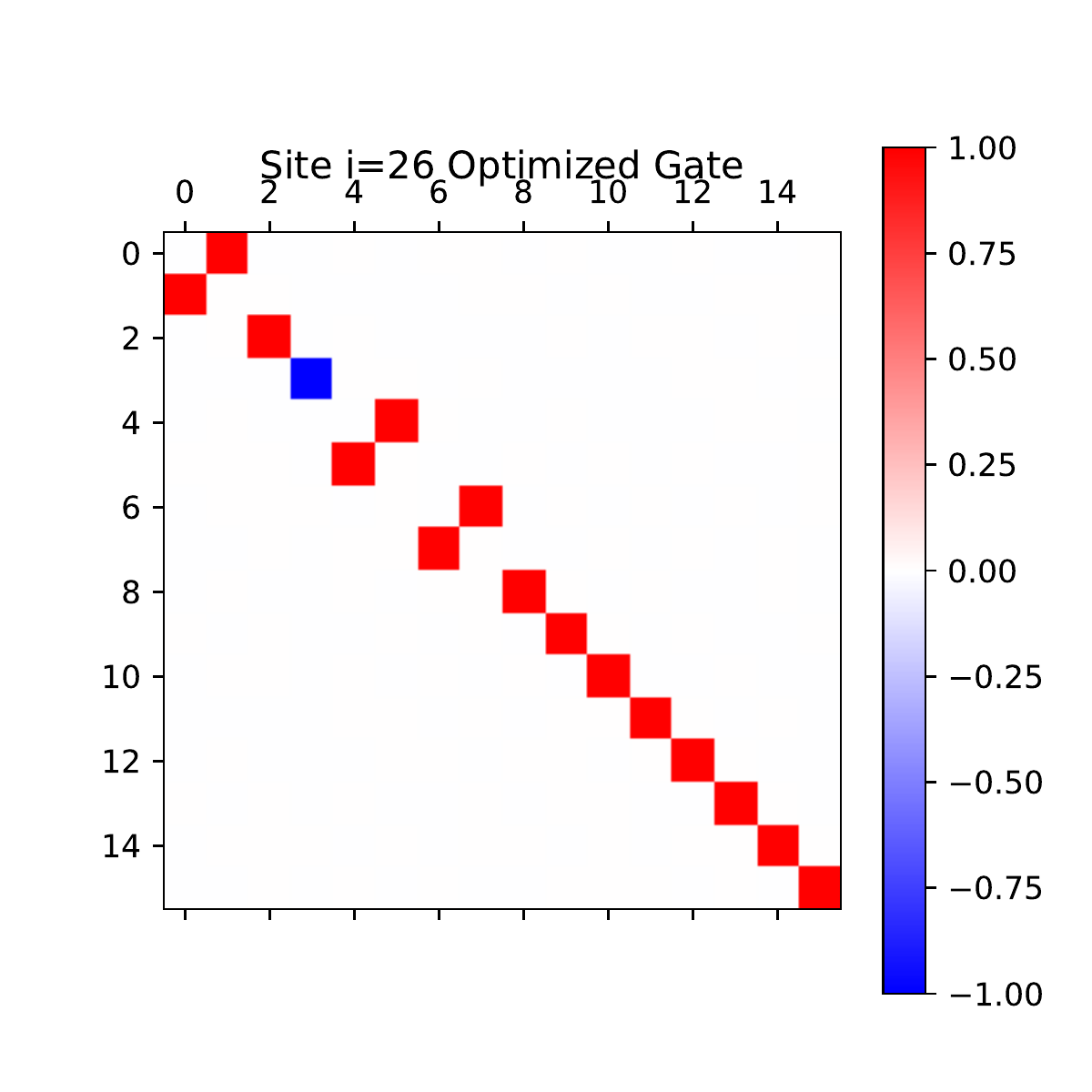}
}\\
\subfloat[Raw circuit from optimization]{
\includegraphics[width=0.9\columnwidth]{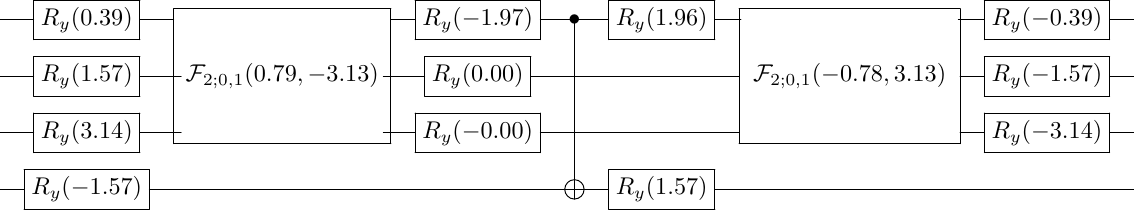}
}\\
\subfloat[Expanded and cleaned circuit from optimization]{
\includegraphics[width=0.9\columnwidth]{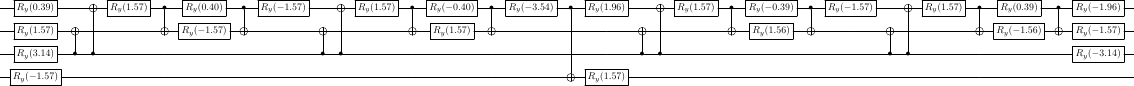}
}
\caption{Optimization for site 26}
\end{figure}
%
%
\begin{figure}[h]

\subfloat[Isometry]{
\includegraphics[width=0.3\columnwidth]{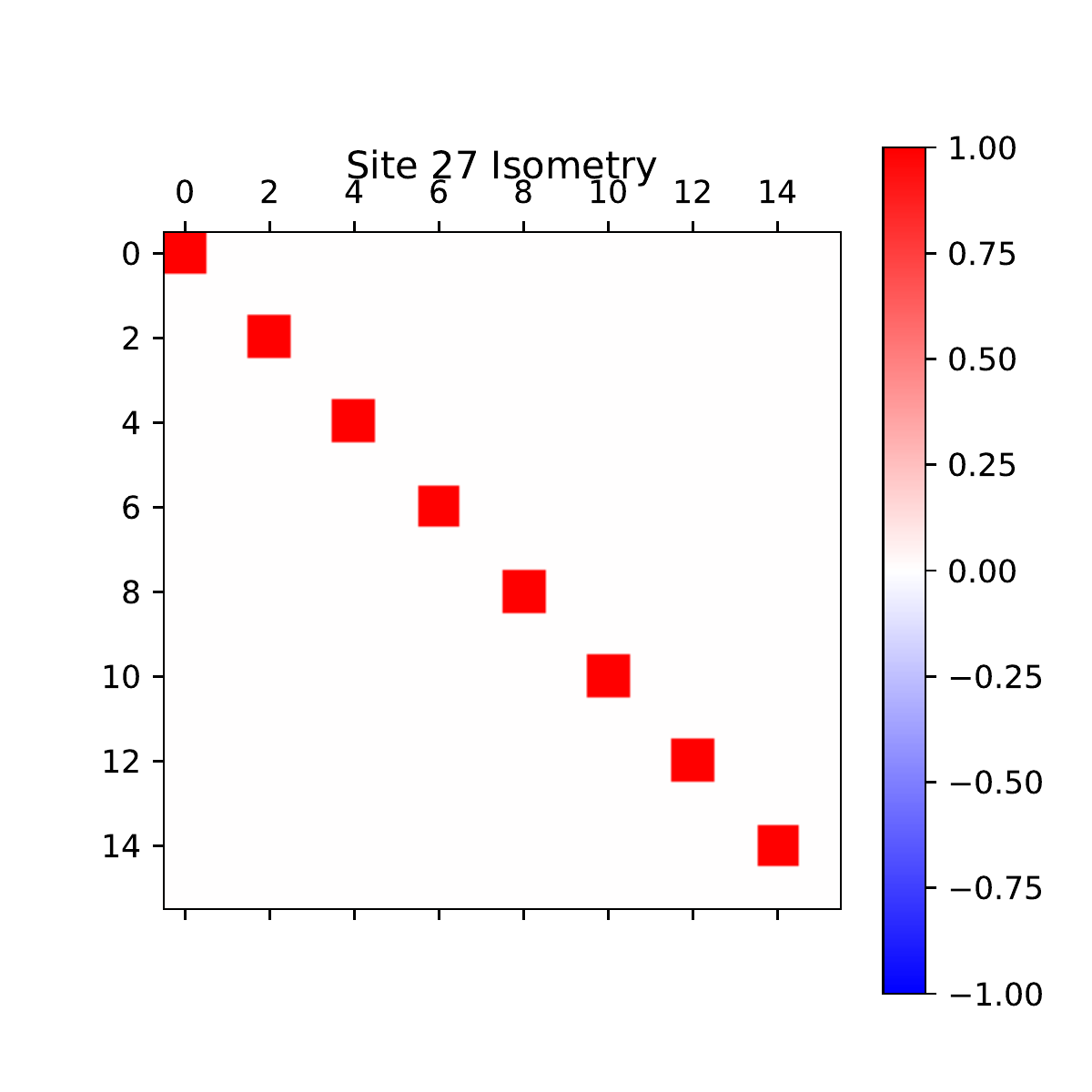}
}
\subfloat[Optimized gate]{
\includegraphics[width=0.3\columnwidth]{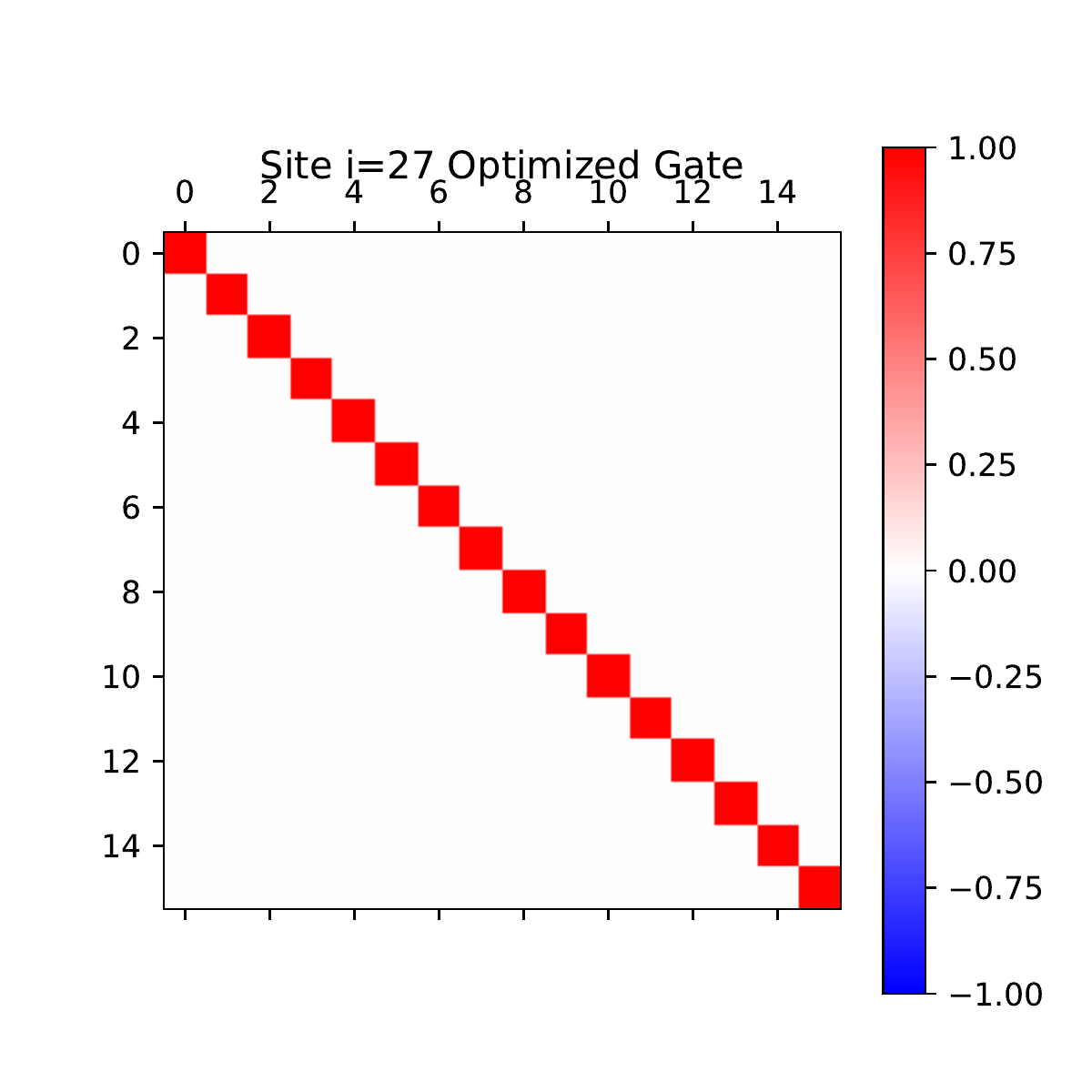}
}
\subfloat[Circuit from optimization]{
\includegraphics[width=0.2\columnwidth]{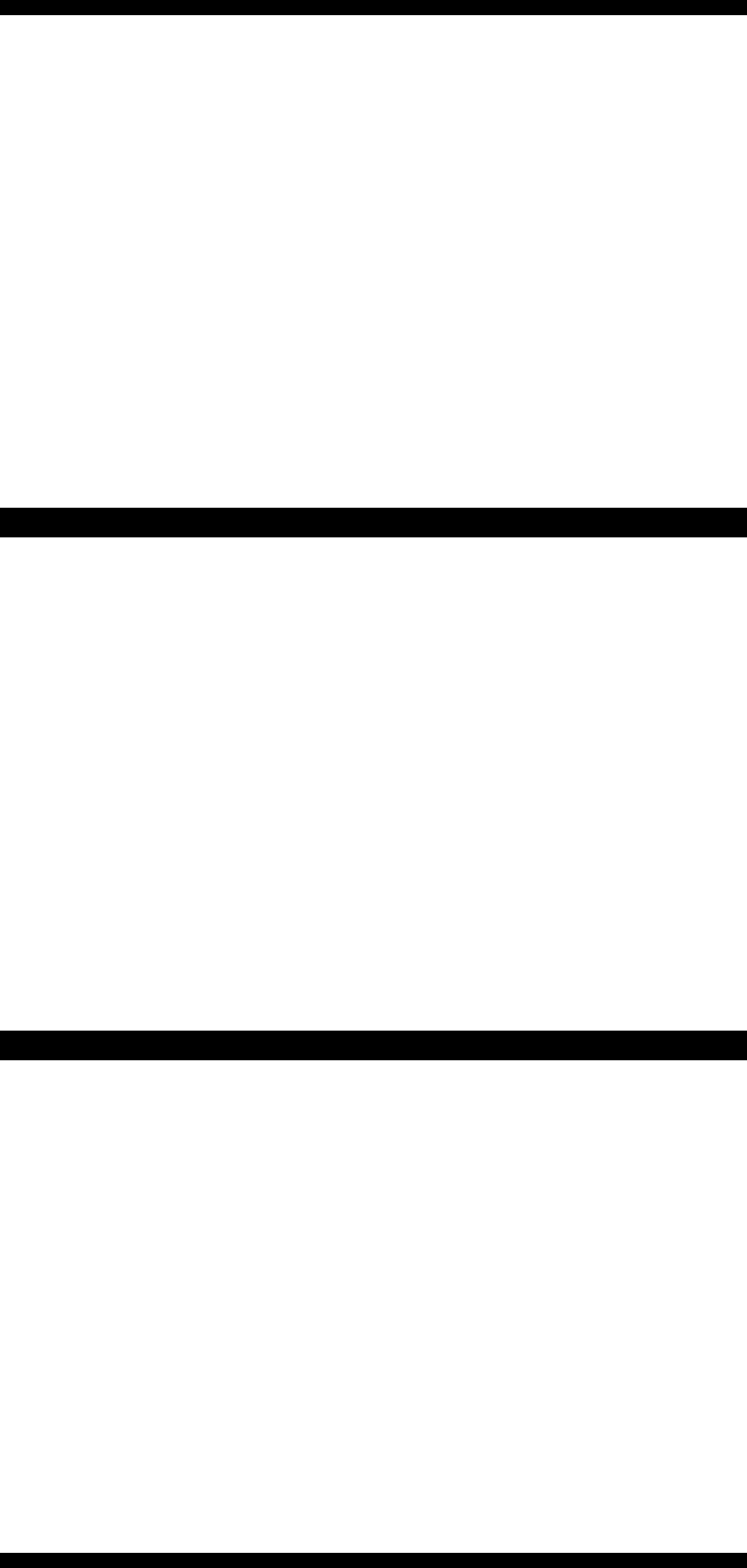}
}
\caption{Optimization for site 27}
\end{figure}
%
%
\begin{figure}[h]

\subfloat[Isometry]{
\includegraphics[width=0.3\columnwidth]{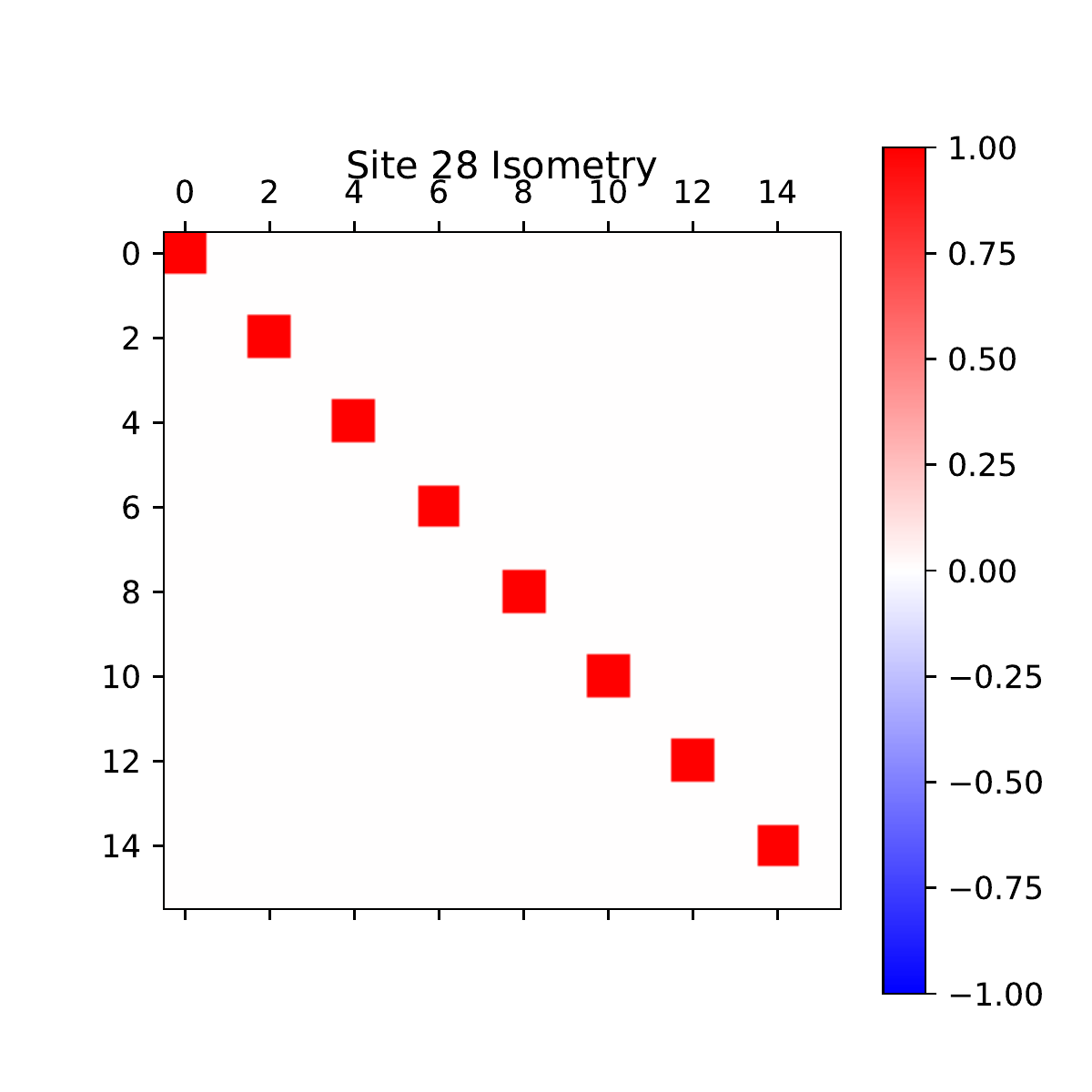}
}
\subfloat[Optimized gate]{
\includegraphics[width=0.3\columnwidth]{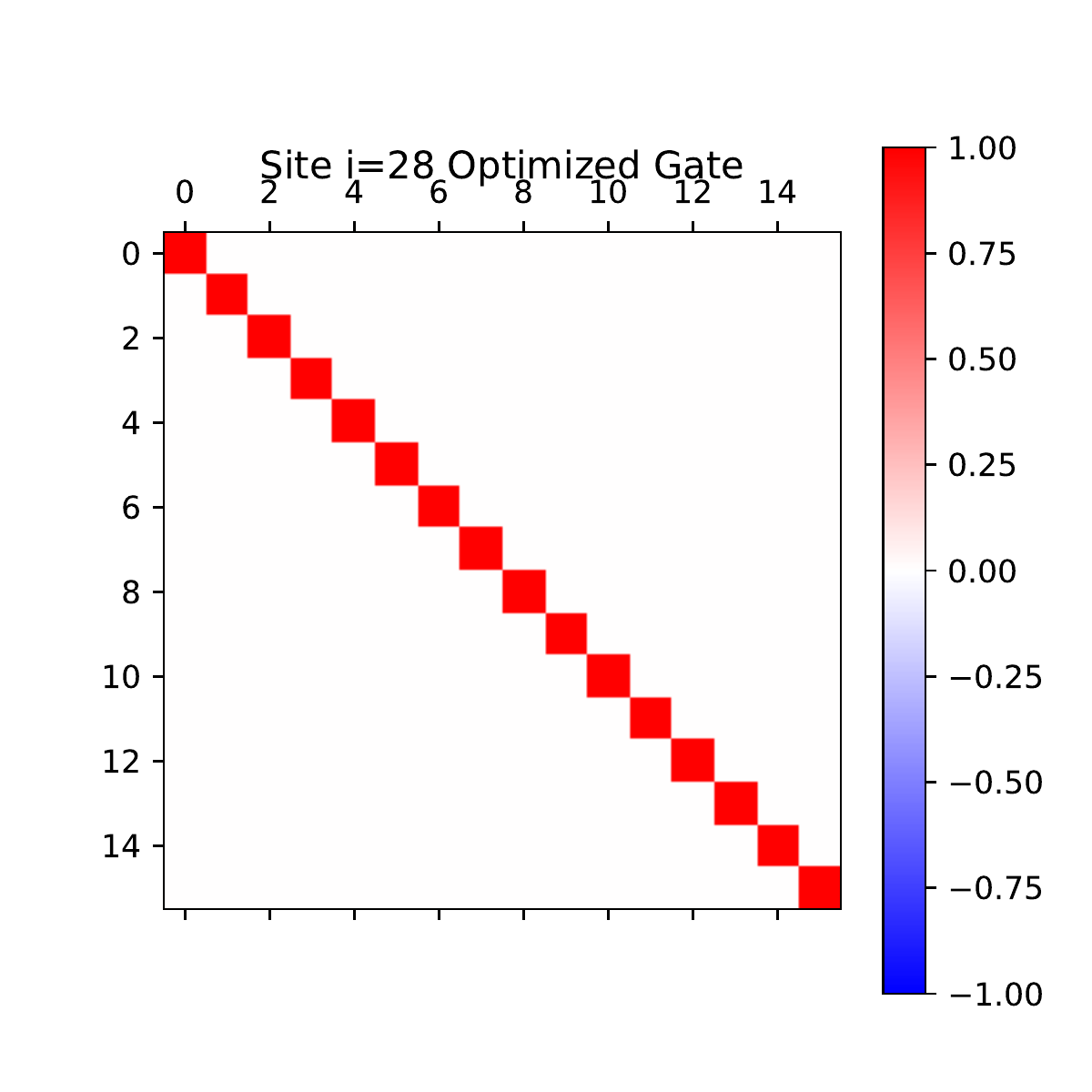}
}
\subfloat[Circuit from optimization]{
\includegraphics[width=0.2\columnwidth]{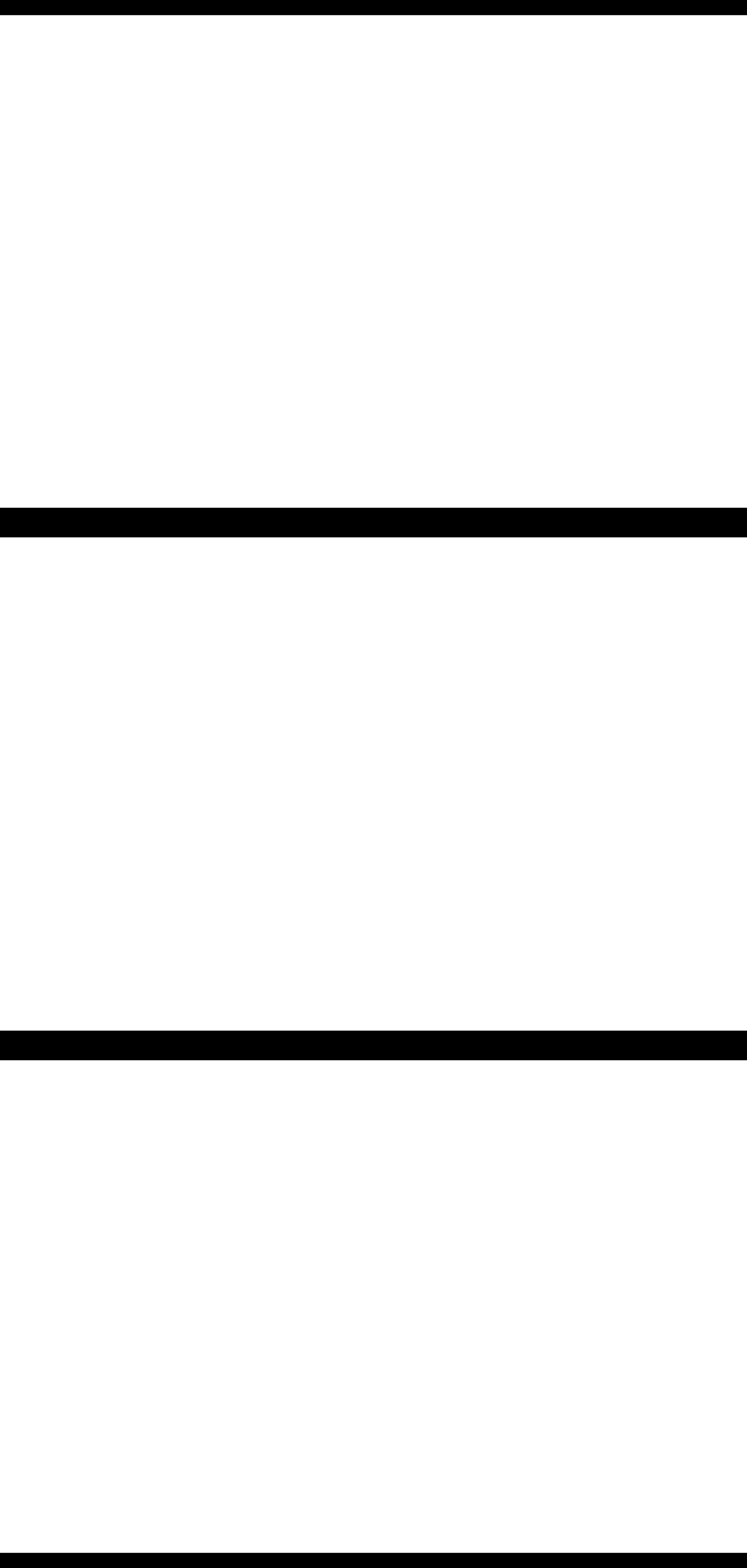}
}
\caption{Optimization for site 28}
\end{figure}
%
%
\begin{figure}[h]

\subfloat[Isometry]{
\includegraphics[width=0.45\columnwidth]{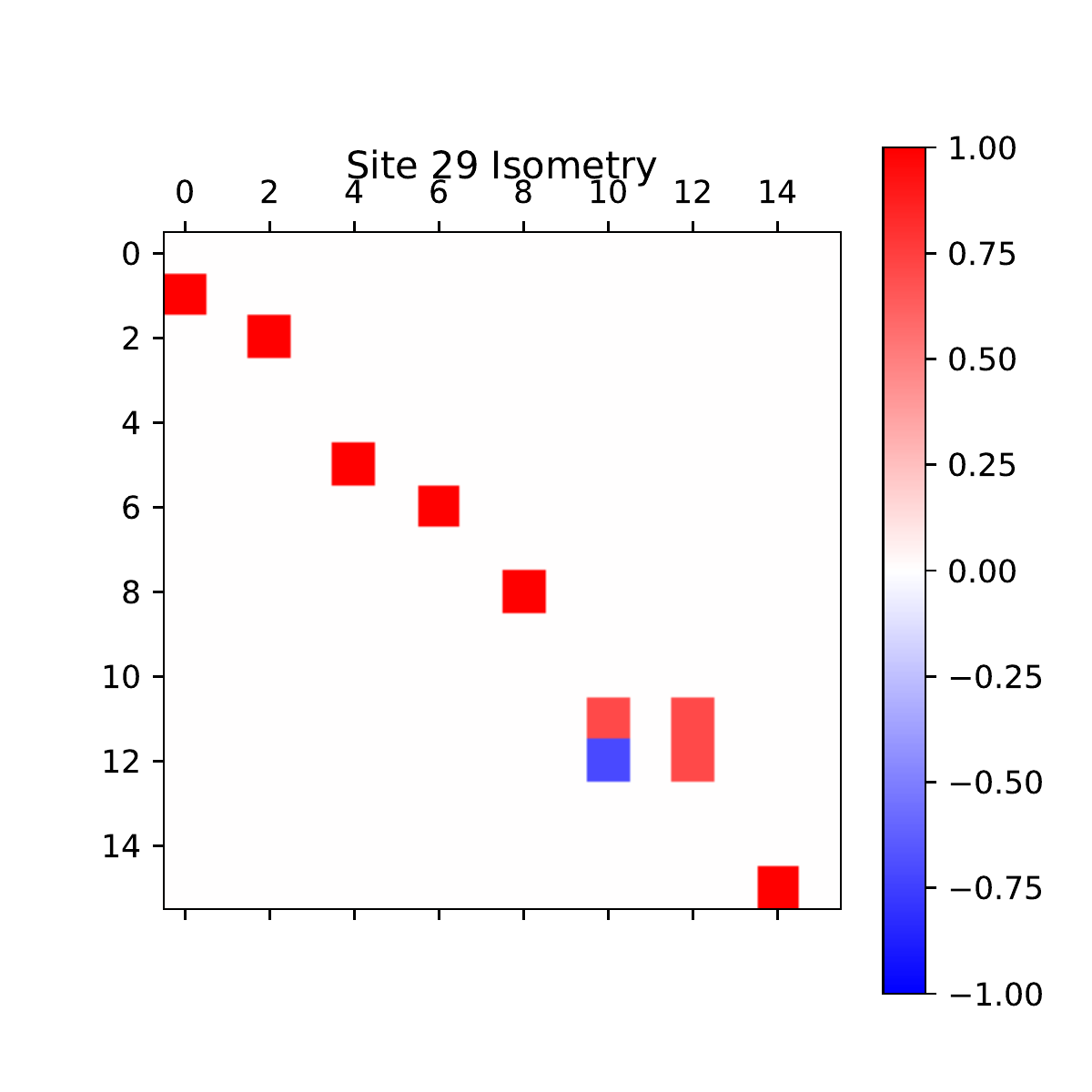}
}
\subfloat[Optimized gate]{
\includegraphics[width=0.45\columnwidth]{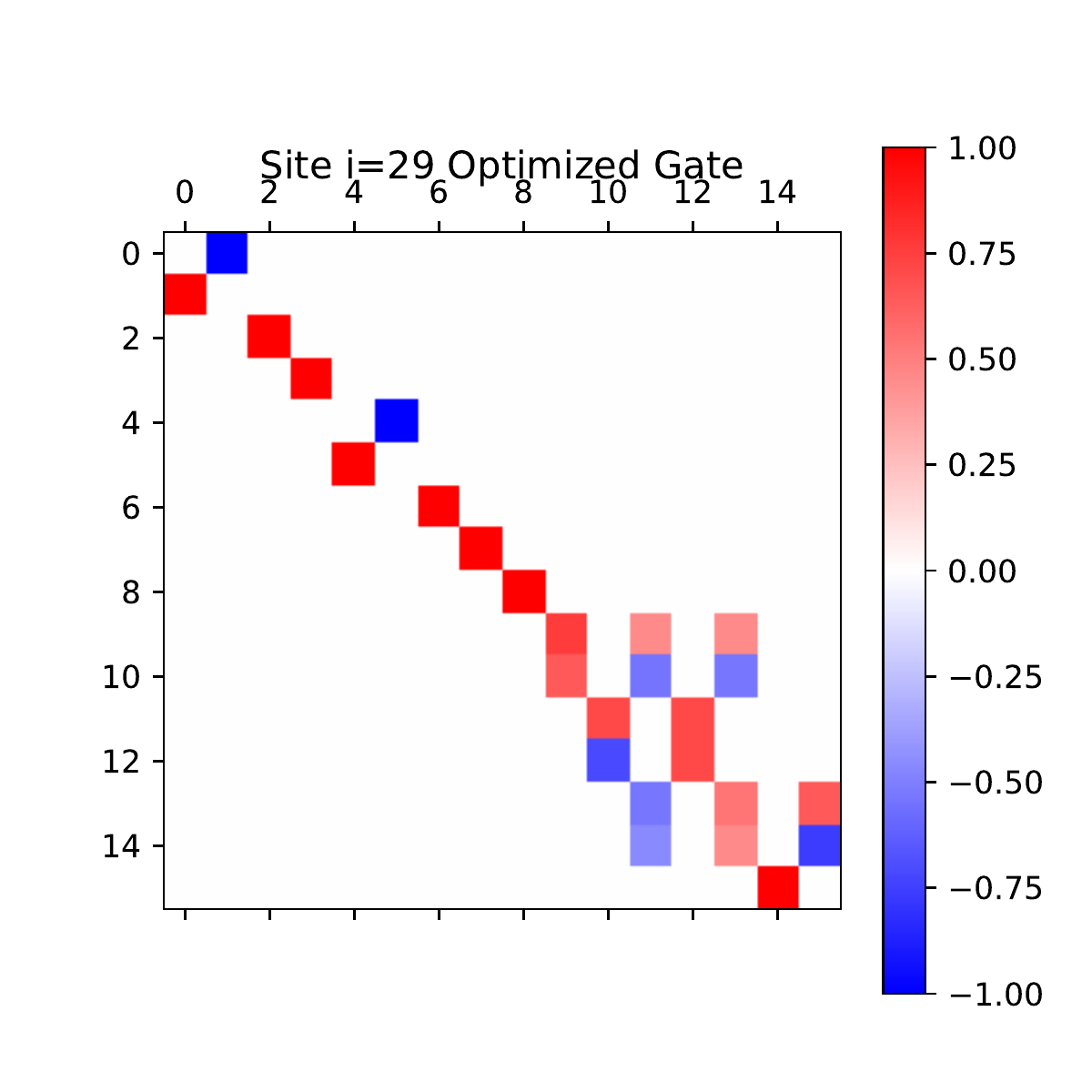}
}\\
\subfloat[Raw circuit from optimization]{
\includegraphics[width=0.9\columnwidth]{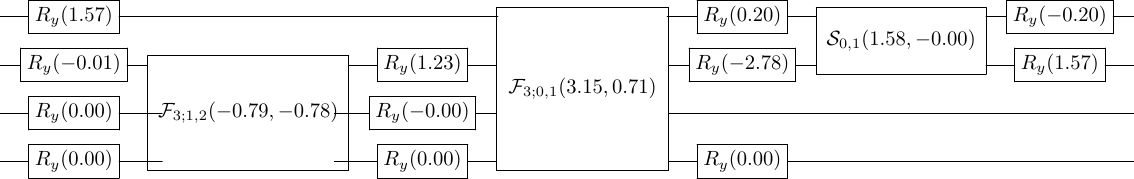}
}\\
\subfloat[Expanded and cleaned circuit from optimization]{
\includegraphics[width=0.9\columnwidth]{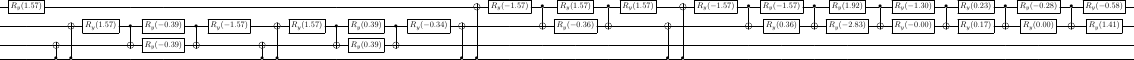}
}
\caption{Optimization for site 29}
\end{figure}
%
\begin{figure}[h]

\subfloat[Isometry]{
\includegraphics[width=0.45\columnwidth]{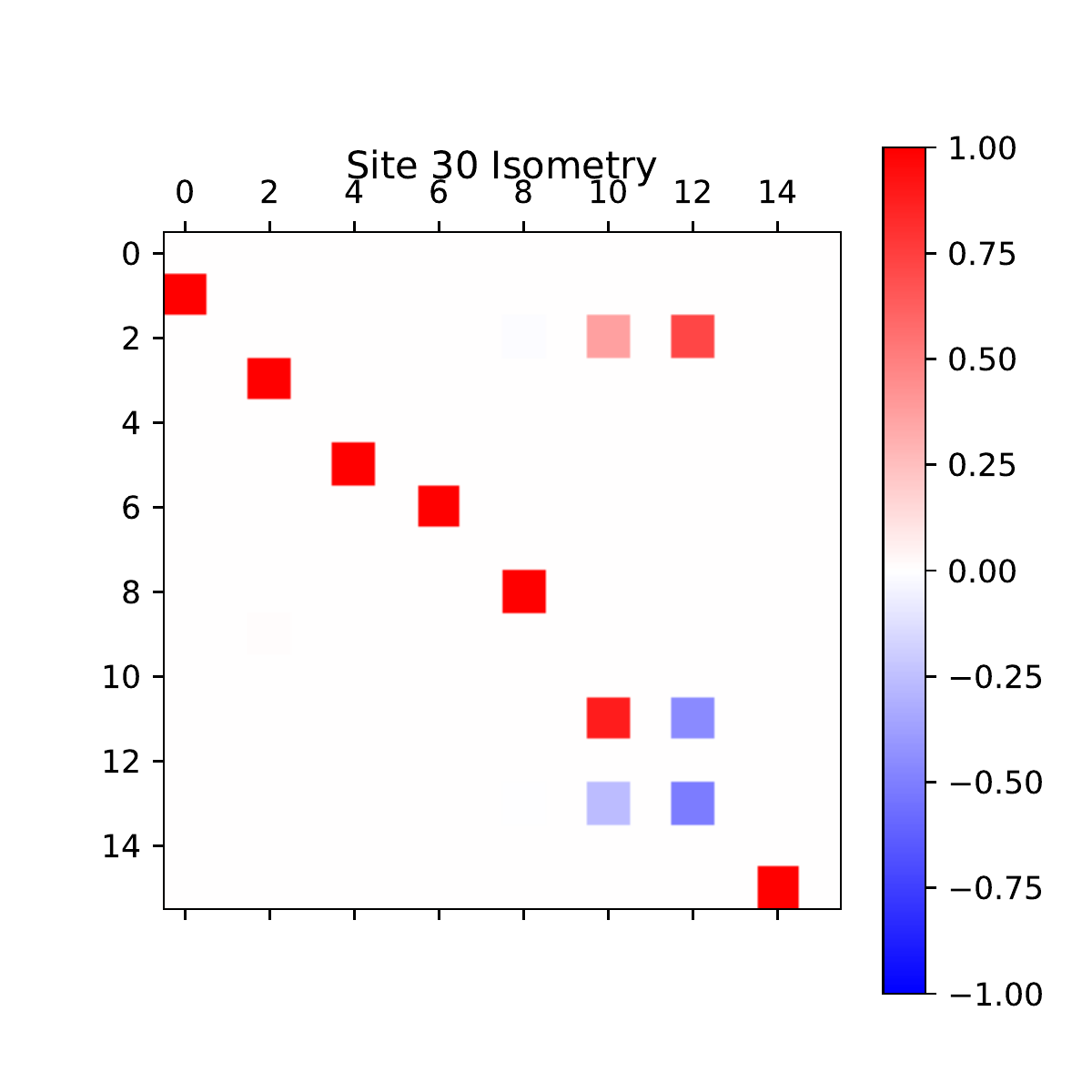}
}
\subfloat[Optimized gate]{
\includegraphics[width=0.45\columnwidth]{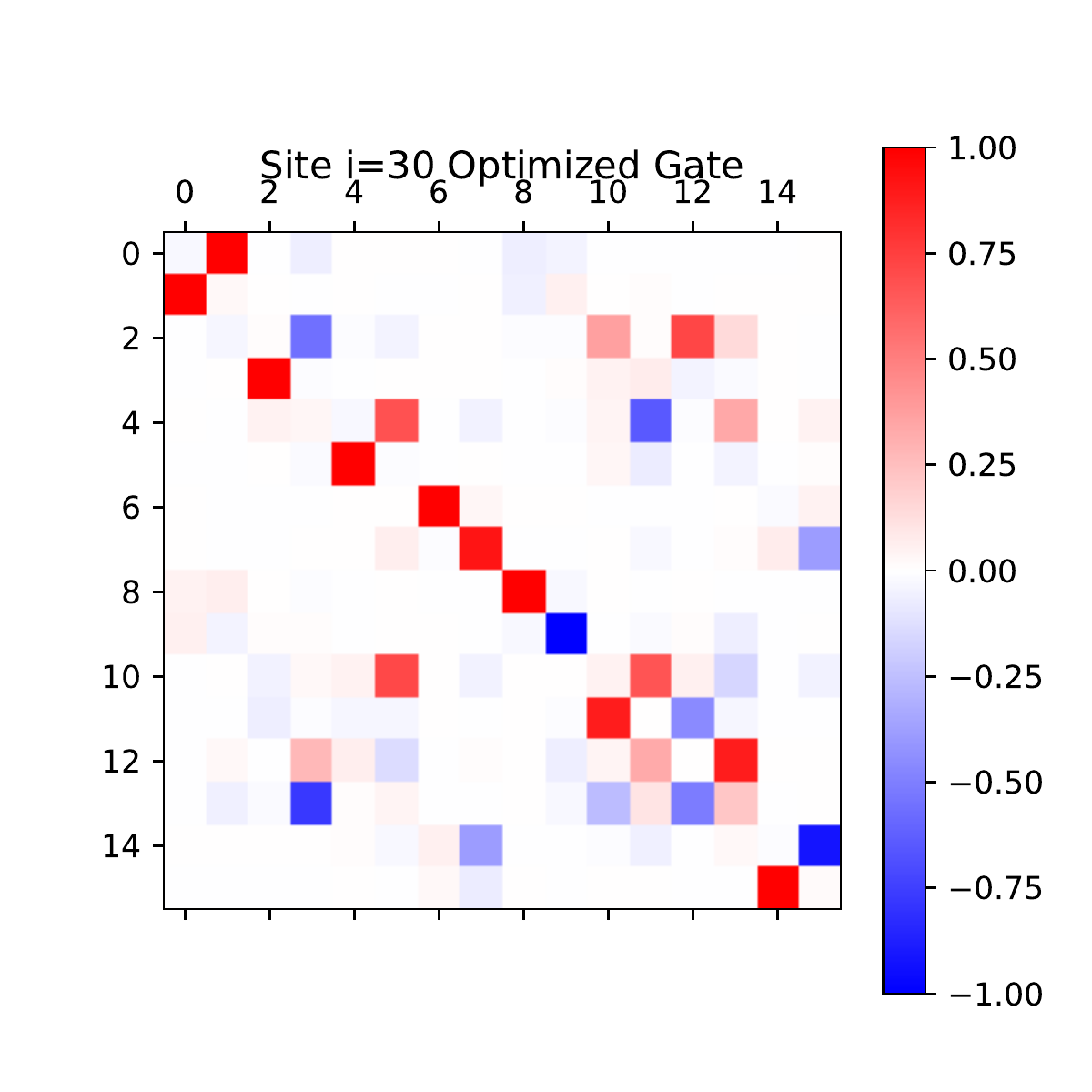}
}\\
\subfloat[Raw circuit from optimization]{
\includegraphics[width=0.9\columnwidth]{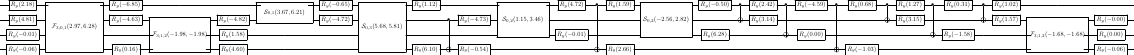}
}\\
\subfloat[Expanded and cleaned circuit from optimization]{
\includegraphics[width=0.9\columnwidth]{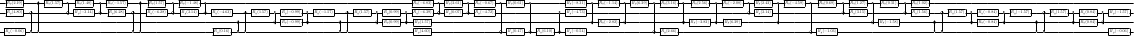}
}
\caption{Optimization for site 30}
\end{figure}
%
\begin{figure}[h]

\subfloat[Isometry]{
\includegraphics[width=0.45\columnwidth]{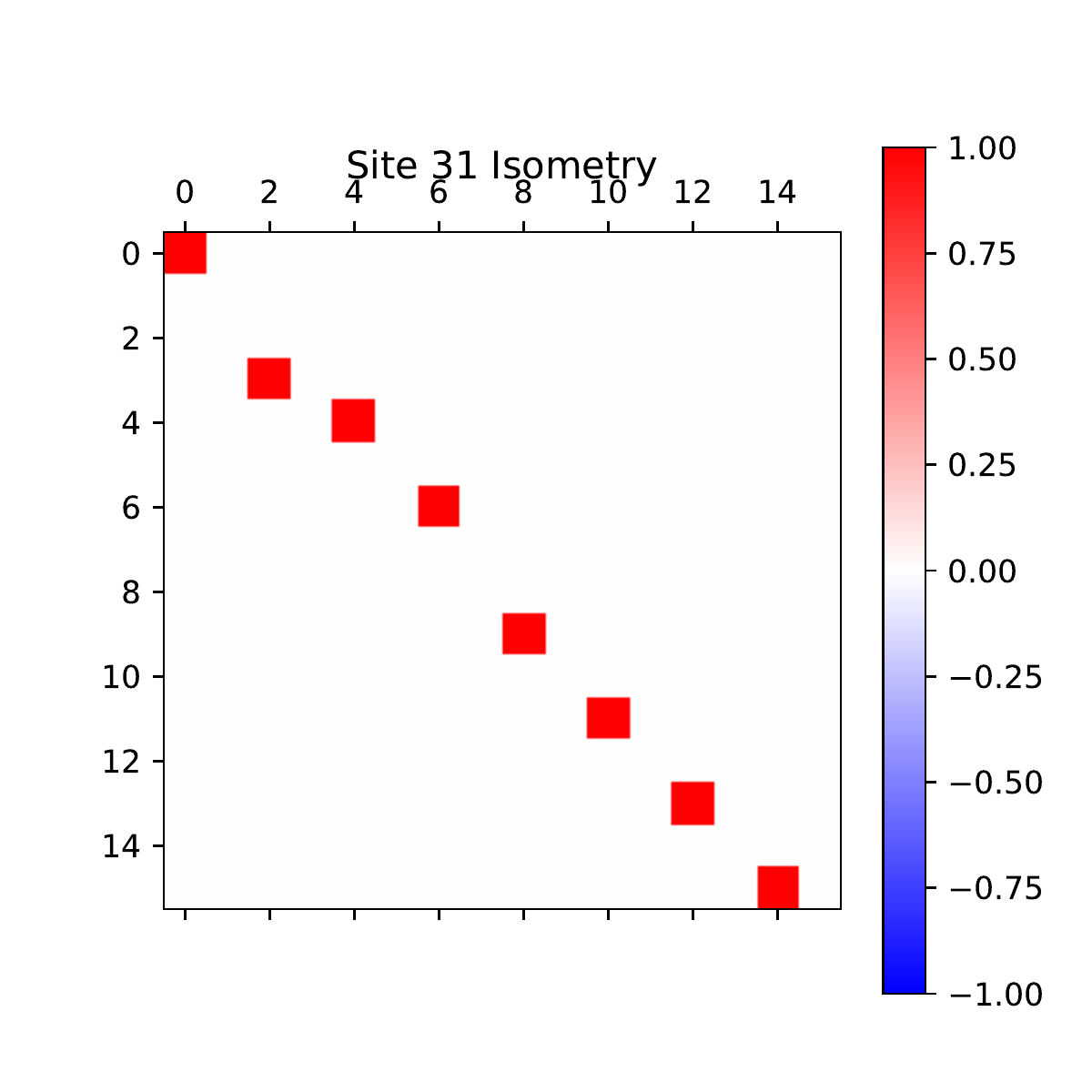}
}
\subfloat[Optimized gate]{
\includegraphics[width=0.45\columnwidth]{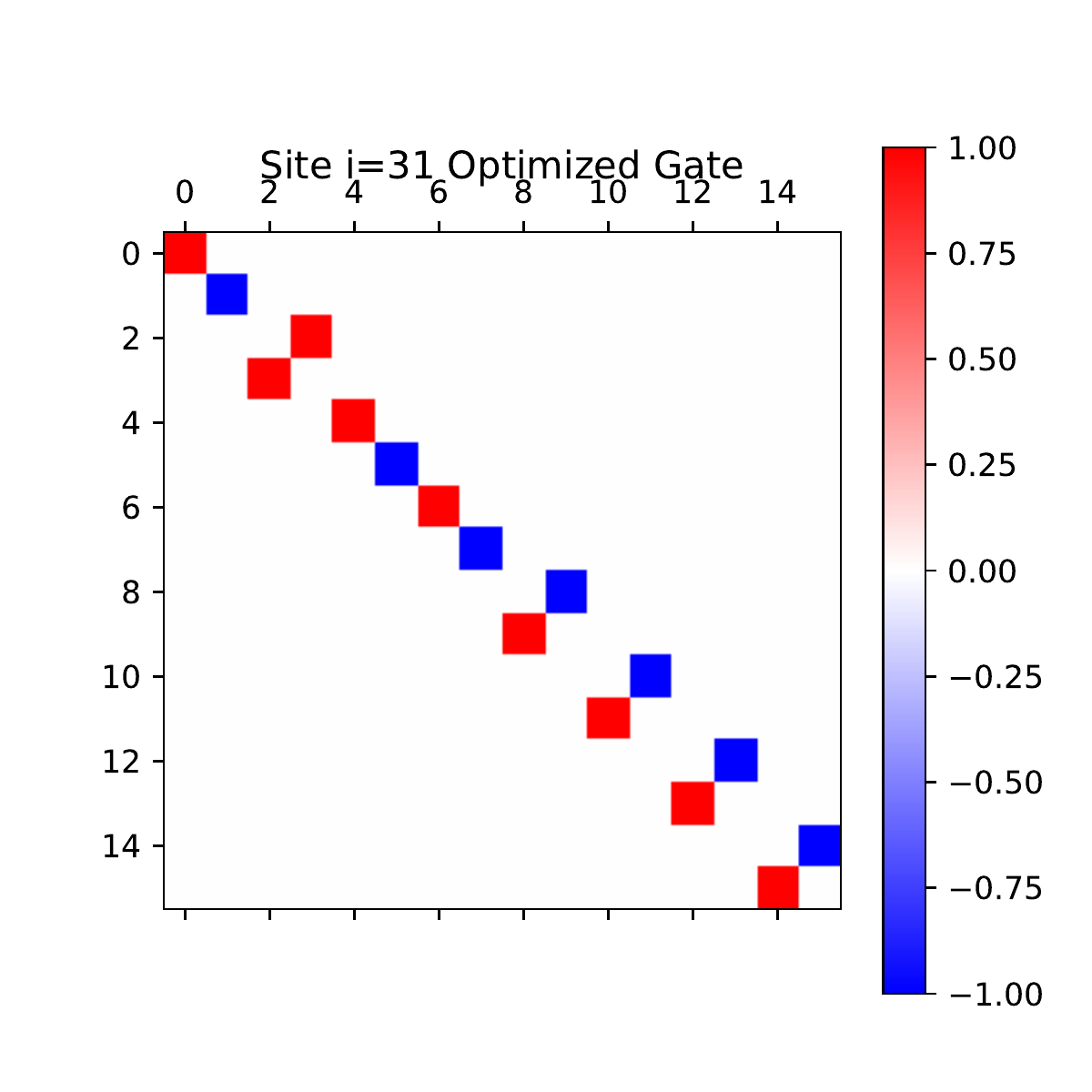}
}\\
\subfloat[Raw circuit from optimization]{
\includegraphics[width=0.9\columnwidth]{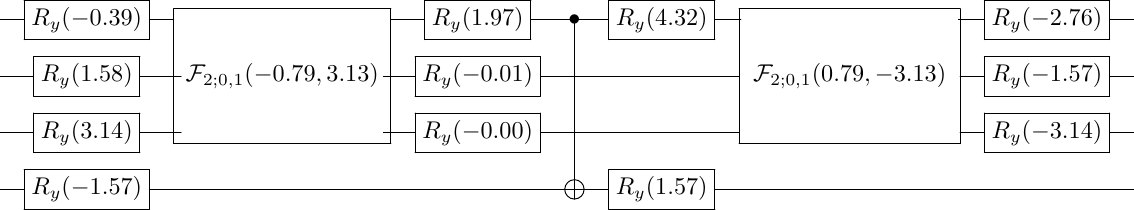}
}\\
\subfloat[Expanded and cleaned circuit from optimization]{
\includegraphics[width=0.9\columnwidth]{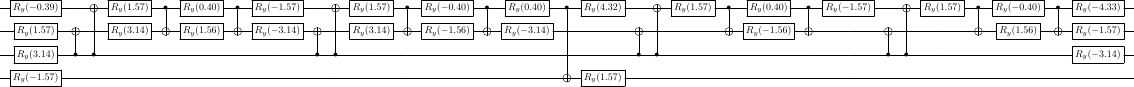}
}
\caption{Optimization for site 31}
\end{figure}
%
%
\begin{figure}[h]

\subfloat[Isometry]{
\includegraphics[width=0.45\columnwidth]{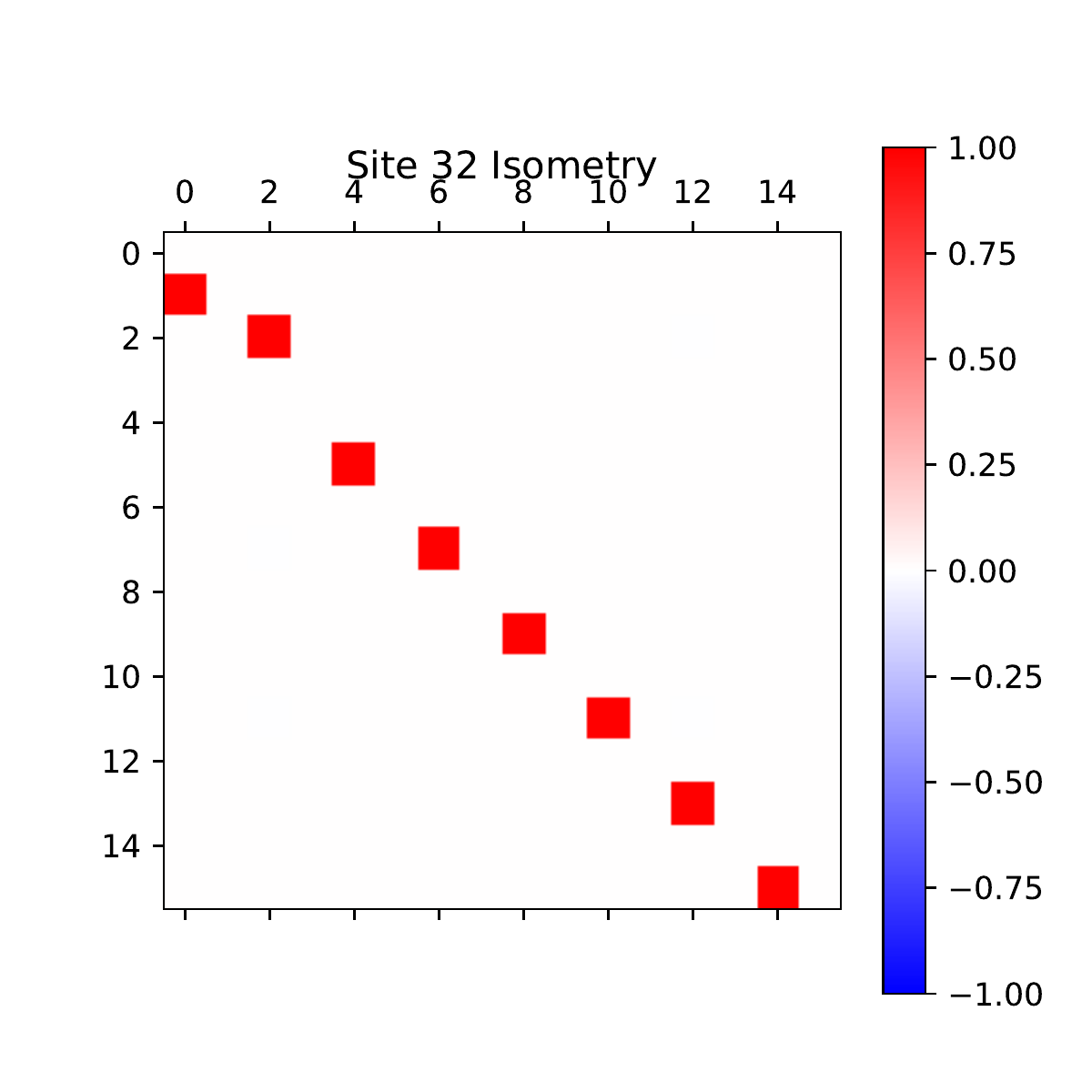}
}
\subfloat[Optimized gate]{
\includegraphics[width=0.45\columnwidth]{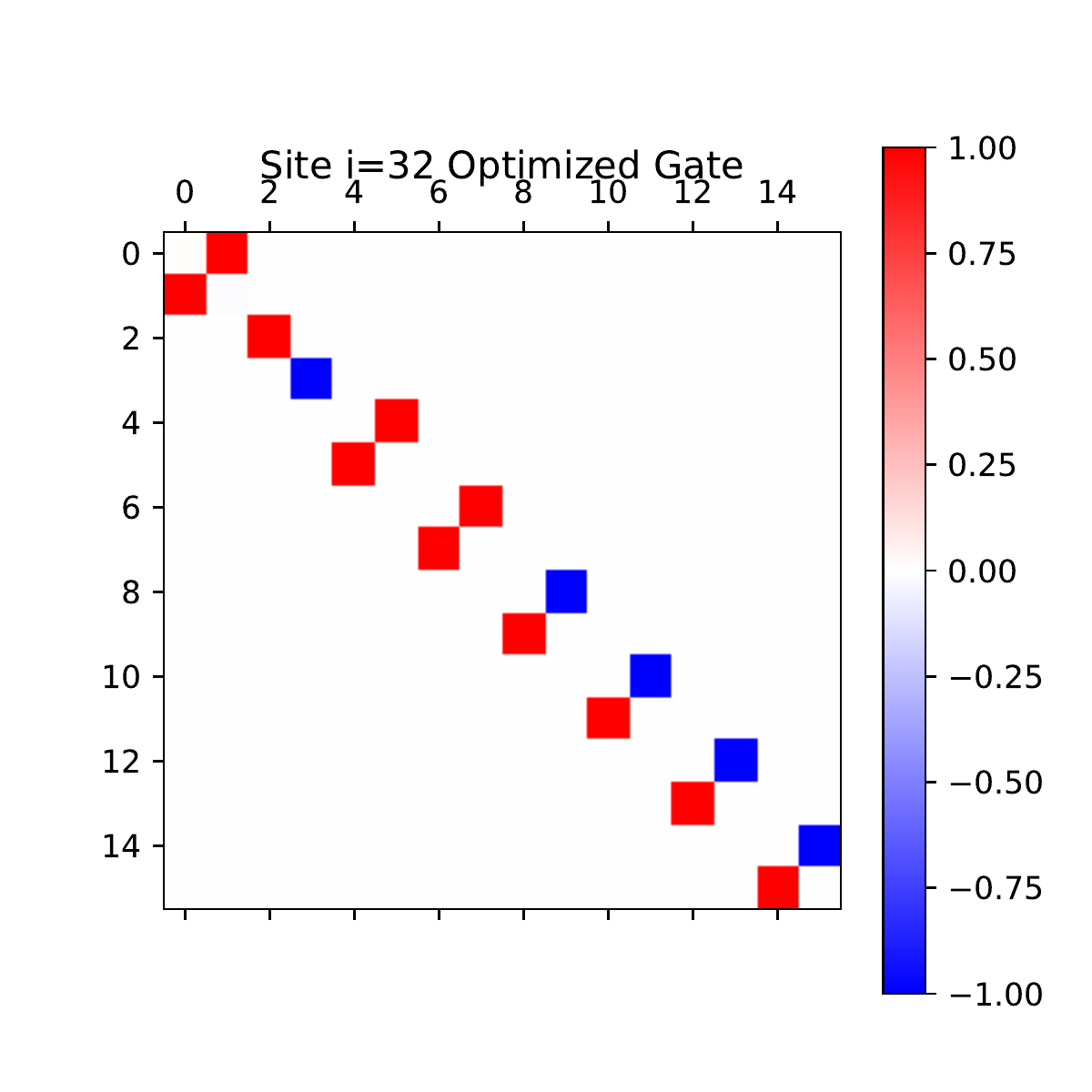}
}\\
\subfloat[Raw circuit from optimization]{
\includegraphics[width=0.9\columnwidth]{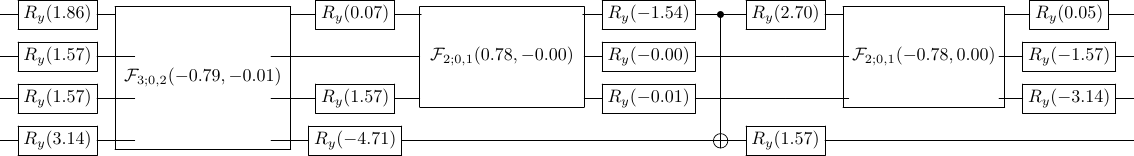}
}\\
\subfloat[Expanded and cleaned circuit from optimization]{
\includegraphics[width=0.9\columnwidth]{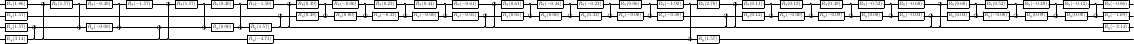}
}
\caption{Optimization for site 32}
\end{figure}
%
%
\begin{figure}[h]

\subfloat[Isometry]{
\includegraphics[width=0.45\columnwidth]{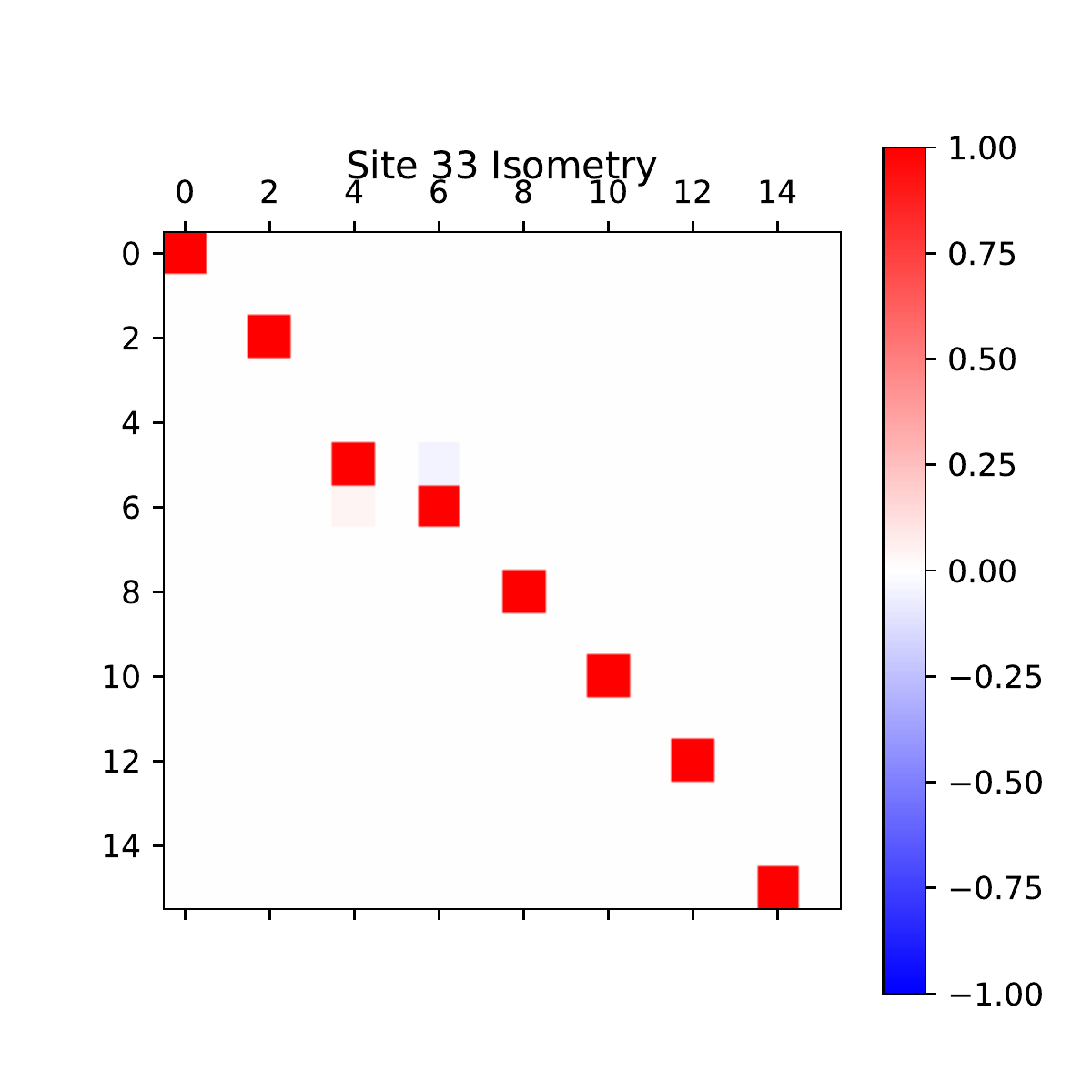}
}
\subfloat[Optimized gate]{
\includegraphics[width=0.45\columnwidth]{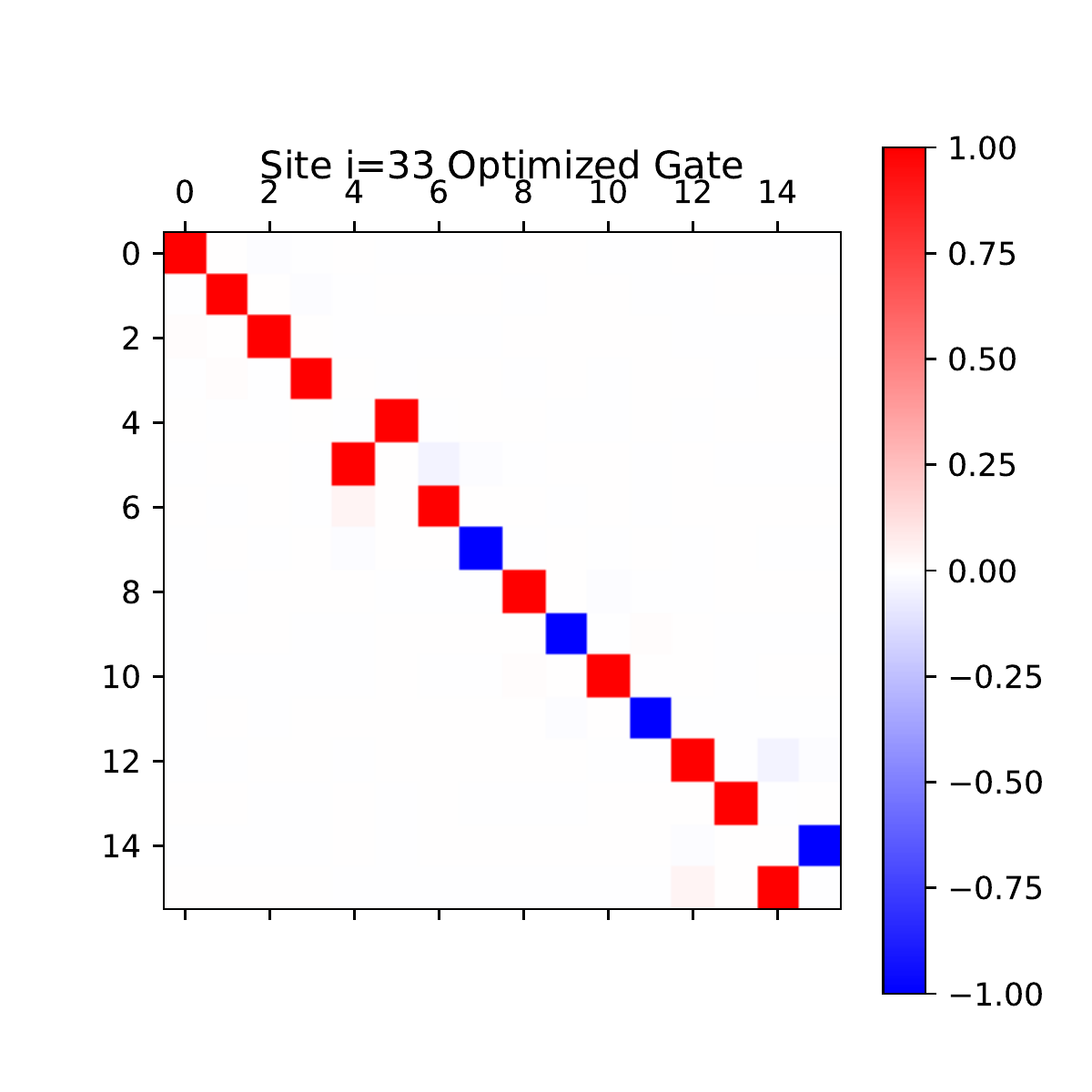}
}\\
\subfloat[Raw circuit from optimization]{
\includegraphics[width=0.9\columnwidth]{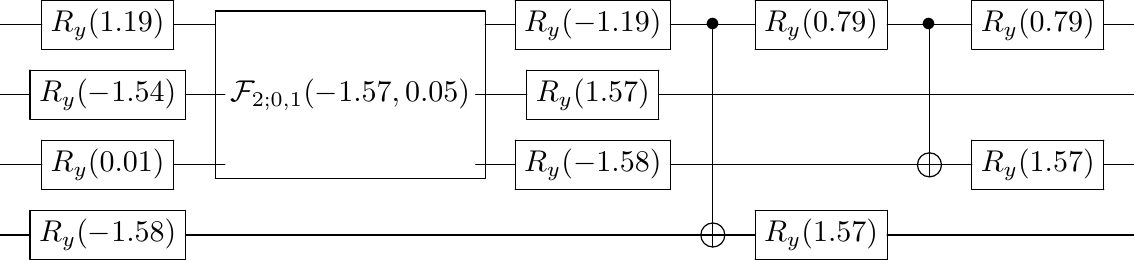}
}\\
\subfloat[Expanded and cleaned circuit from optimization]{
\includegraphics[width=0.9\columnwidth]{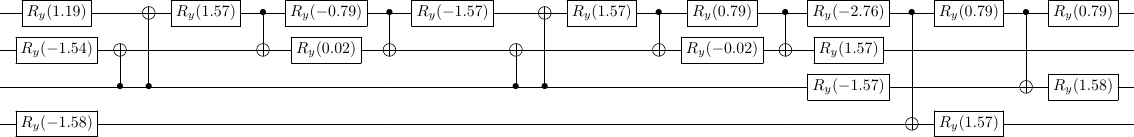}
}
\caption{Optimization for site 33}
\end{figure}
%
\begin{figure}[h]

\subfloat[Isometry]{
\includegraphics[width=0.45\columnwidth]{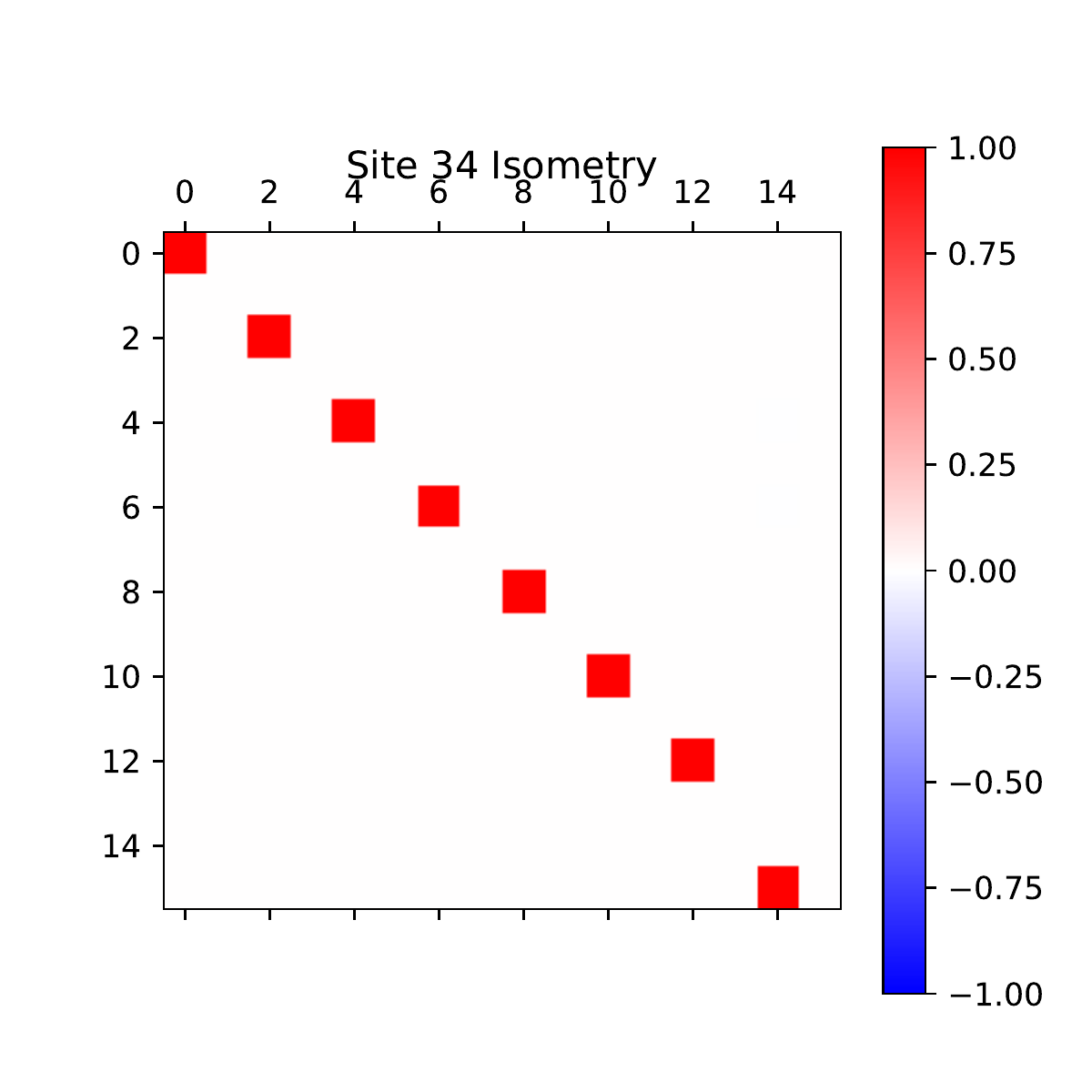}
}
\subfloat[Optimized gate]{
\includegraphics[width=0.45\columnwidth]{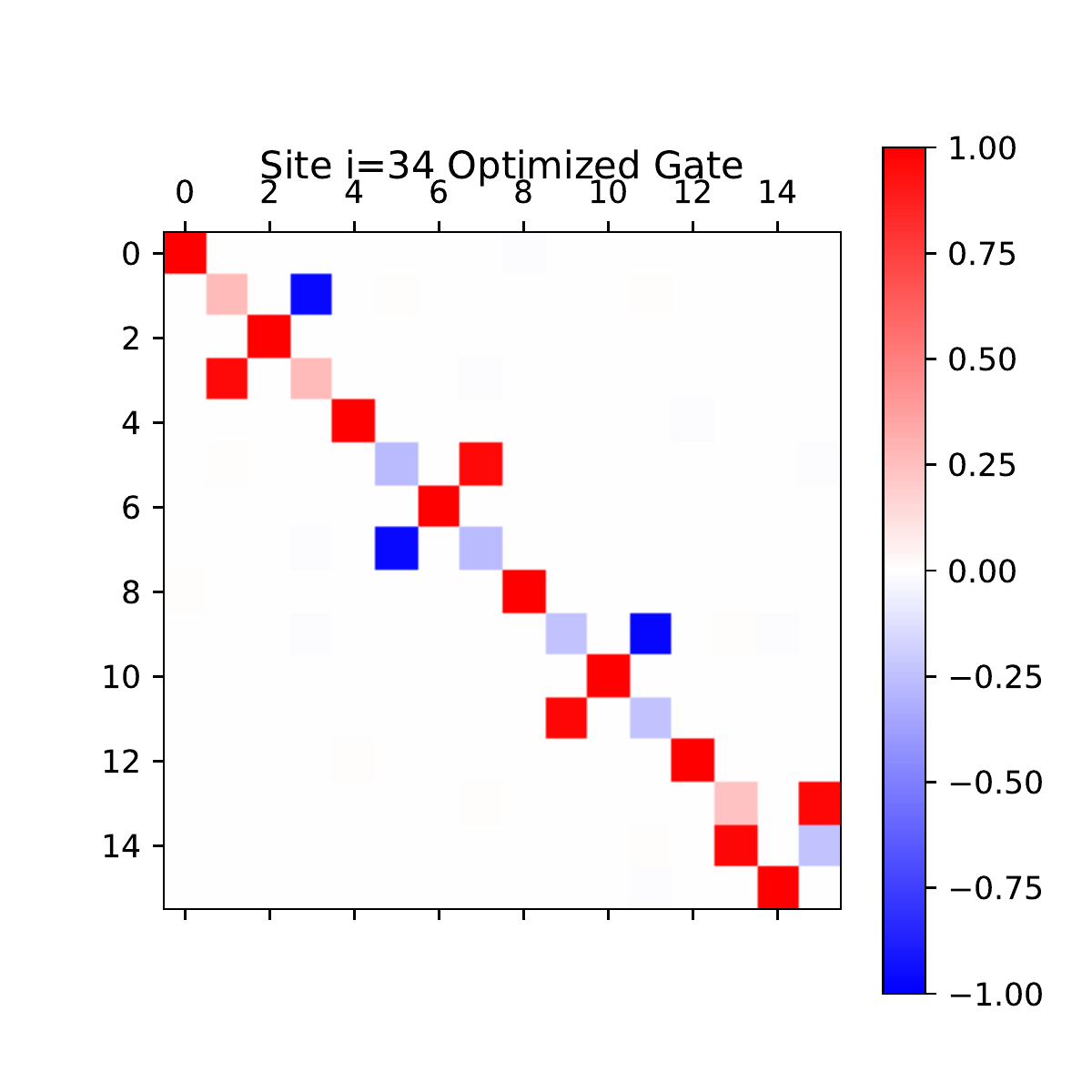}
}\\
\subfloat[Raw circuit from optimization]{
\includegraphics[width=0.9\columnwidth]{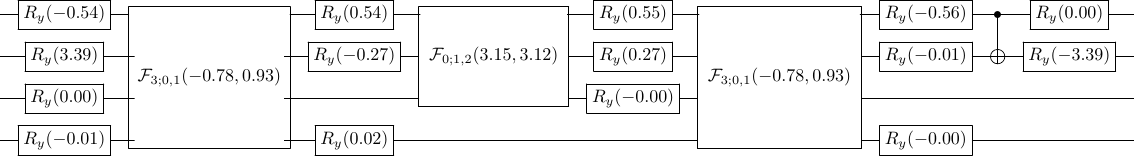}
}\\
\subfloat[Expanded and cleaned circuit from optimization]{
\includegraphics[width=0.9\columnwidth]{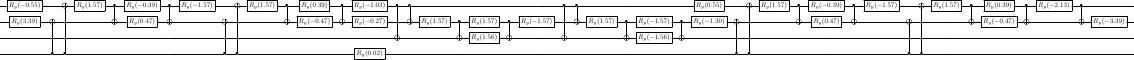}
}
\caption{Optimization for site 34}
\end{figure}
%
\begin{figure}[h]

\subfloat[Isometry]{
\includegraphics[width=0.3\columnwidth]{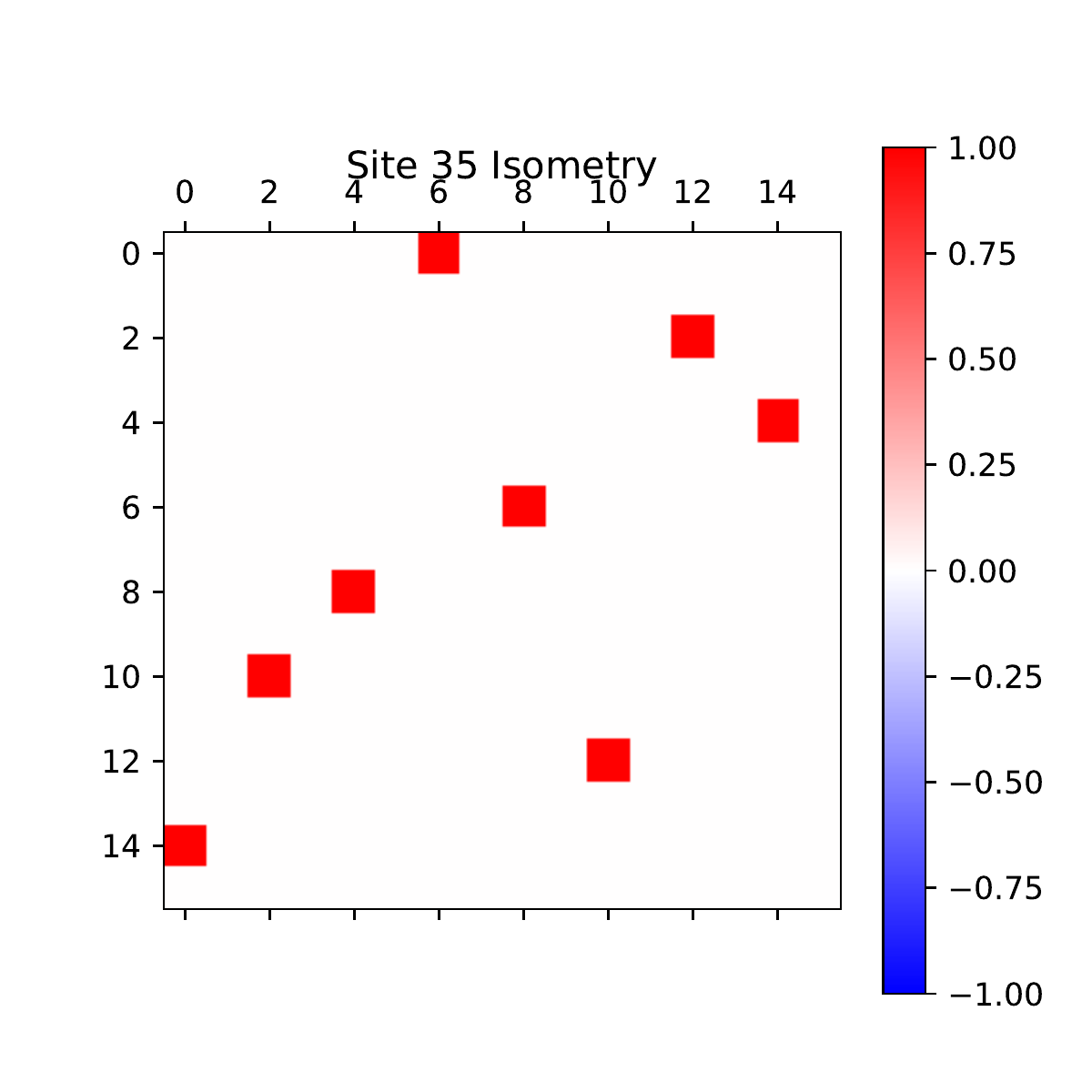}
}
\subfloat[Optimized gate]{
\includegraphics[width=0.3\columnwidth]{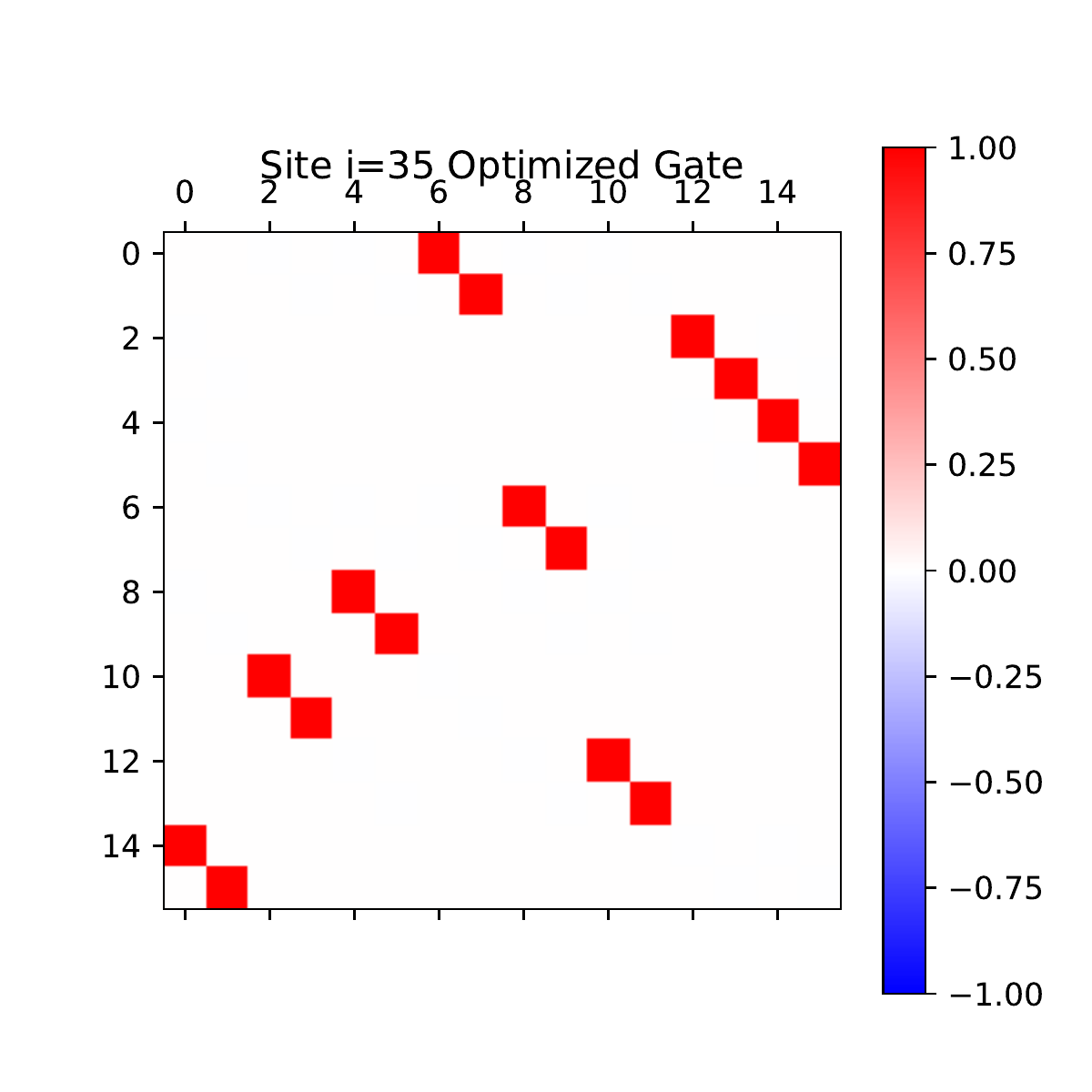}
}\\
\subfloat[Circuit from optimization]{
\includegraphics[width=0.9\columnwidth]{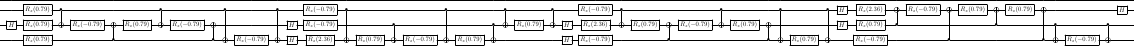}
}
\caption{Optimization for site 35}
\end{figure}
%
\begin{figure}[h]

\subfloat[Isometry]{
\includegraphics[width=0.45\columnwidth]{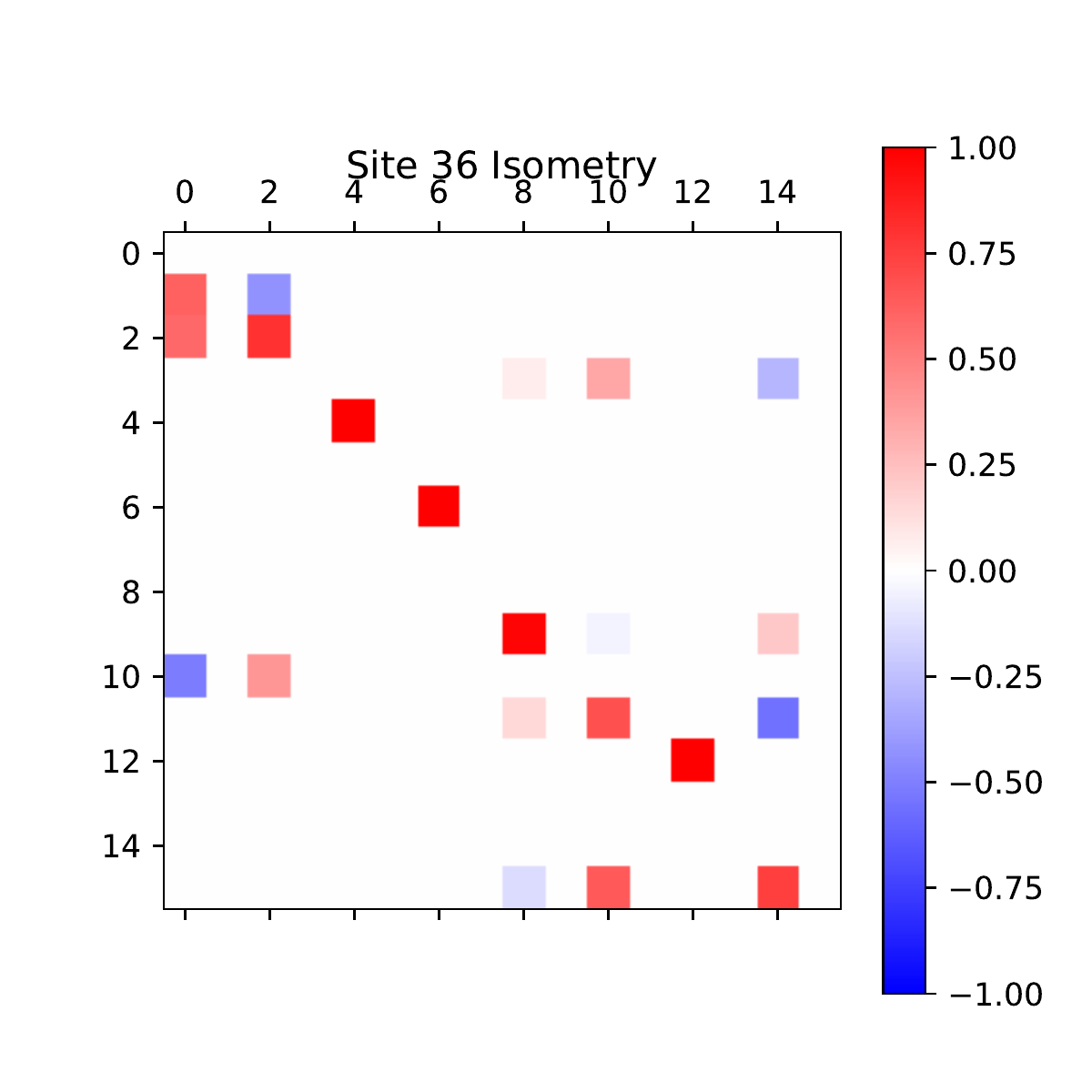}
}
\subfloat[Optimized gate]{
\includegraphics[width=0.45\columnwidth]{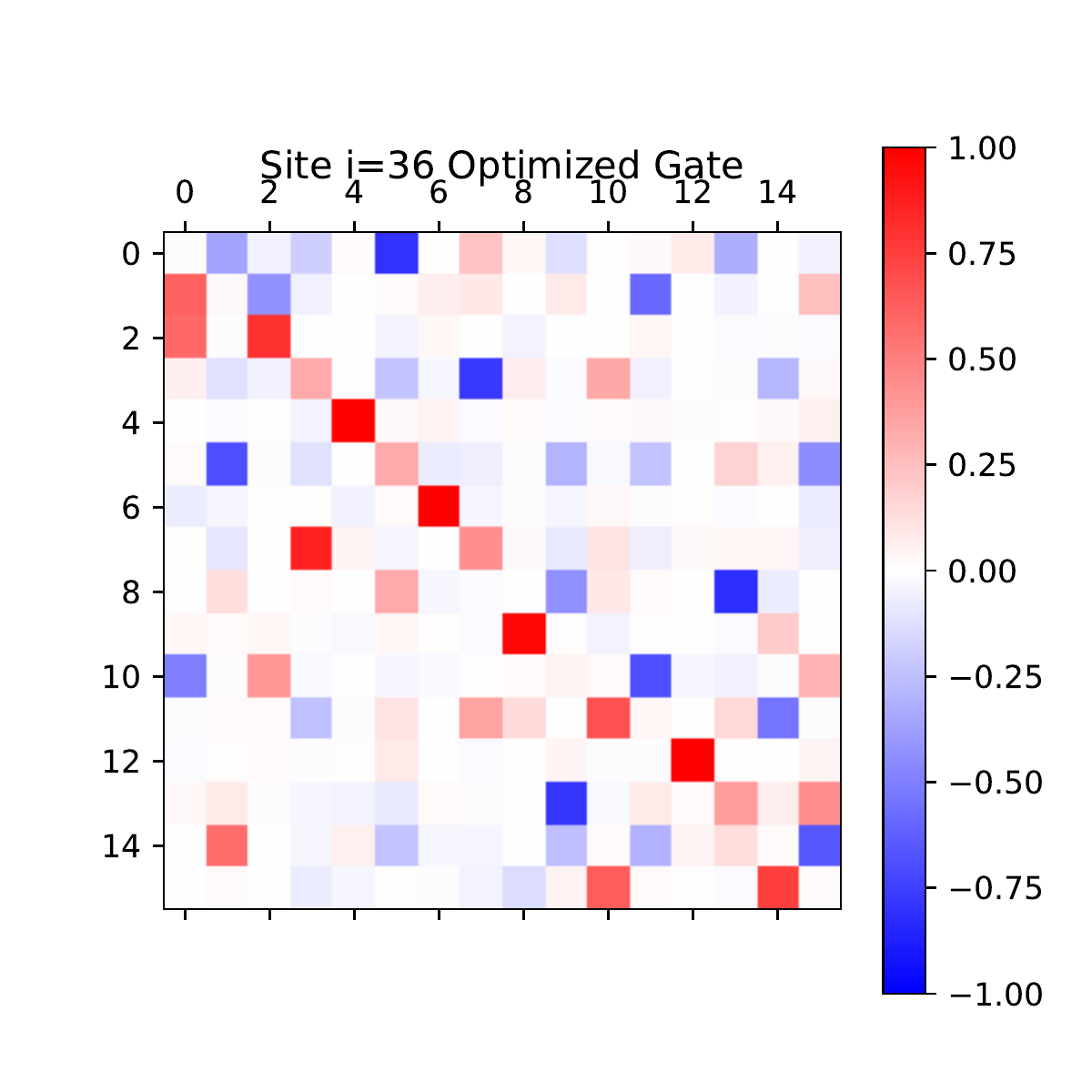}
}\\
\subfloat[Raw circuit from optimization]{
\includegraphics[width=0.9\columnwidth]{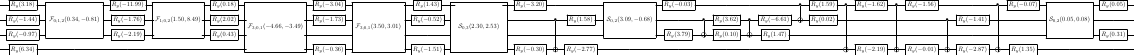}
}\\
\subfloat[Expanded and cleaned circuit from optimization]{
\includegraphics[width=0.9\columnwidth]{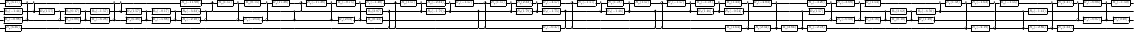}
}
\caption{Optimization for site 36}
\end{figure}
%
\begin{figure}[h]

\subfloat[Isometry]{
\includegraphics[width=0.45\columnwidth]{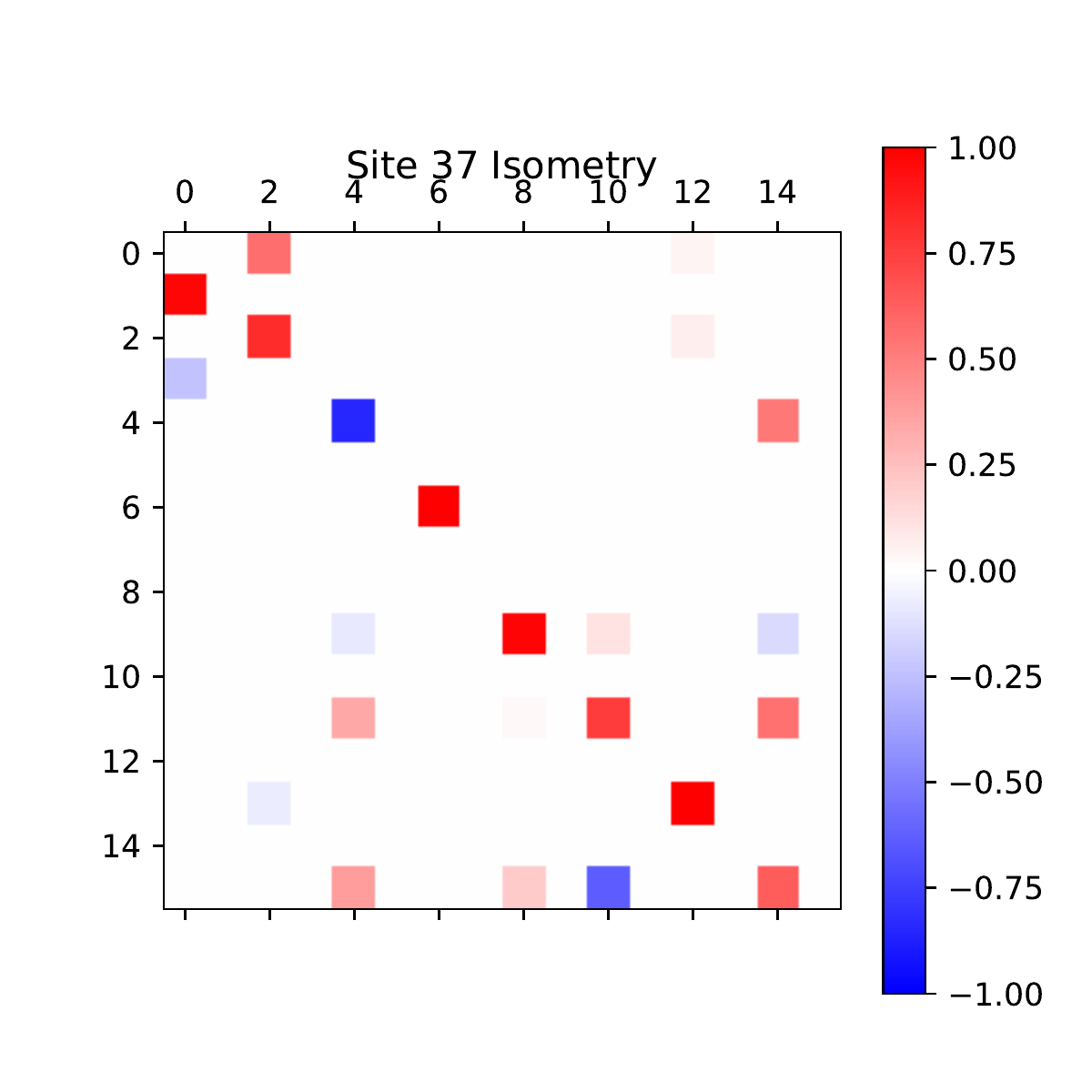}
}
\subfloat[Optimized gate]{
\includegraphics[width=0.45\columnwidth]{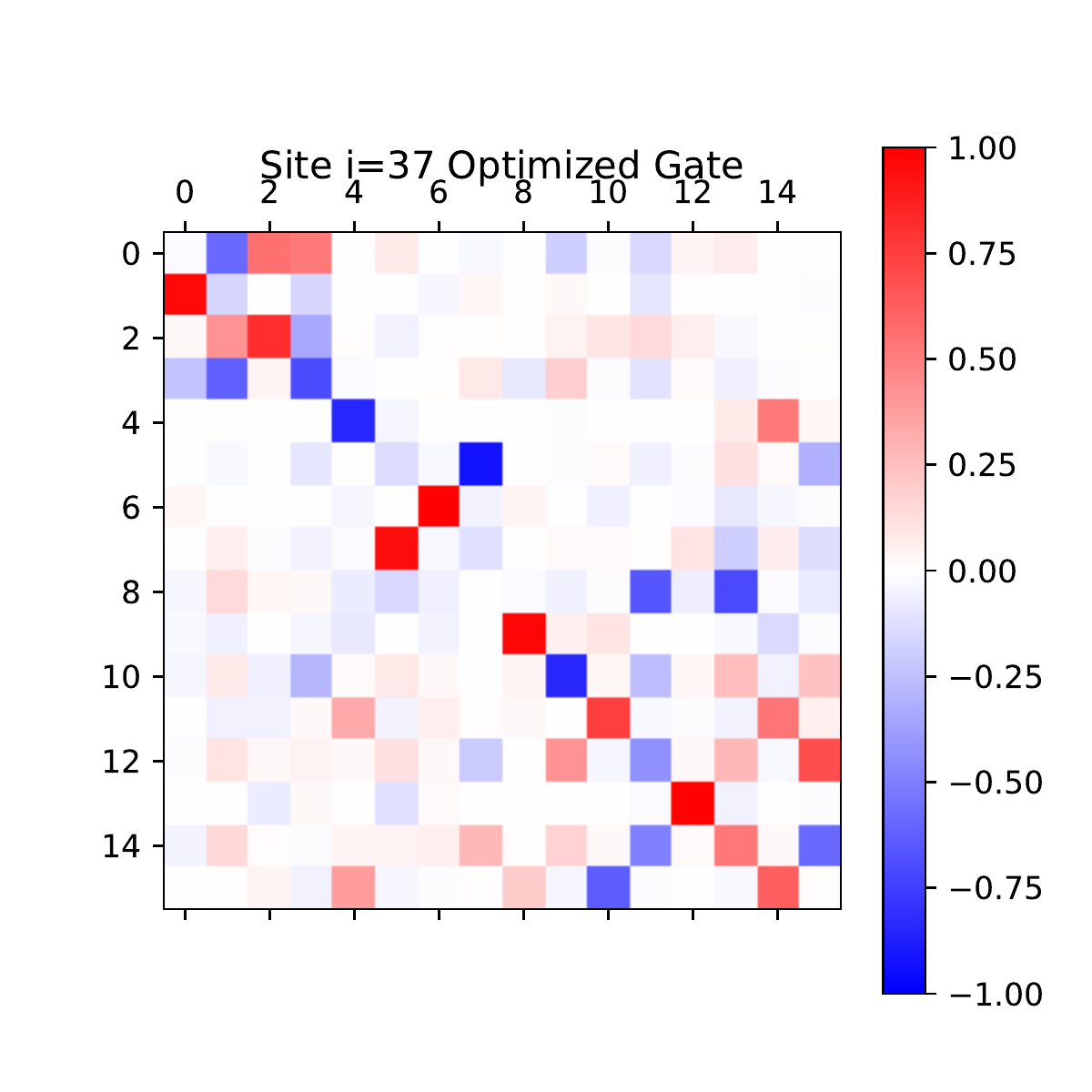}
}\\
\subfloat[Raw circuit from optimization]{
\includegraphics[width=0.9\columnwidth]{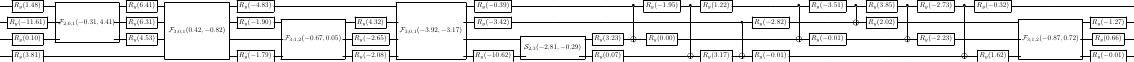}
}\\
\subfloat[Expanded and cleaned circuit from optimization]{
\includegraphics[width=0.9\columnwidth]{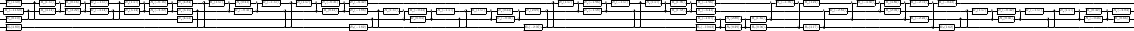}
}
\caption{Optimization for site 37}
\end{figure}
%
\begin{figure}[h]

\subfloat[Isometry]{
\includegraphics[width=0.45\columnwidth]{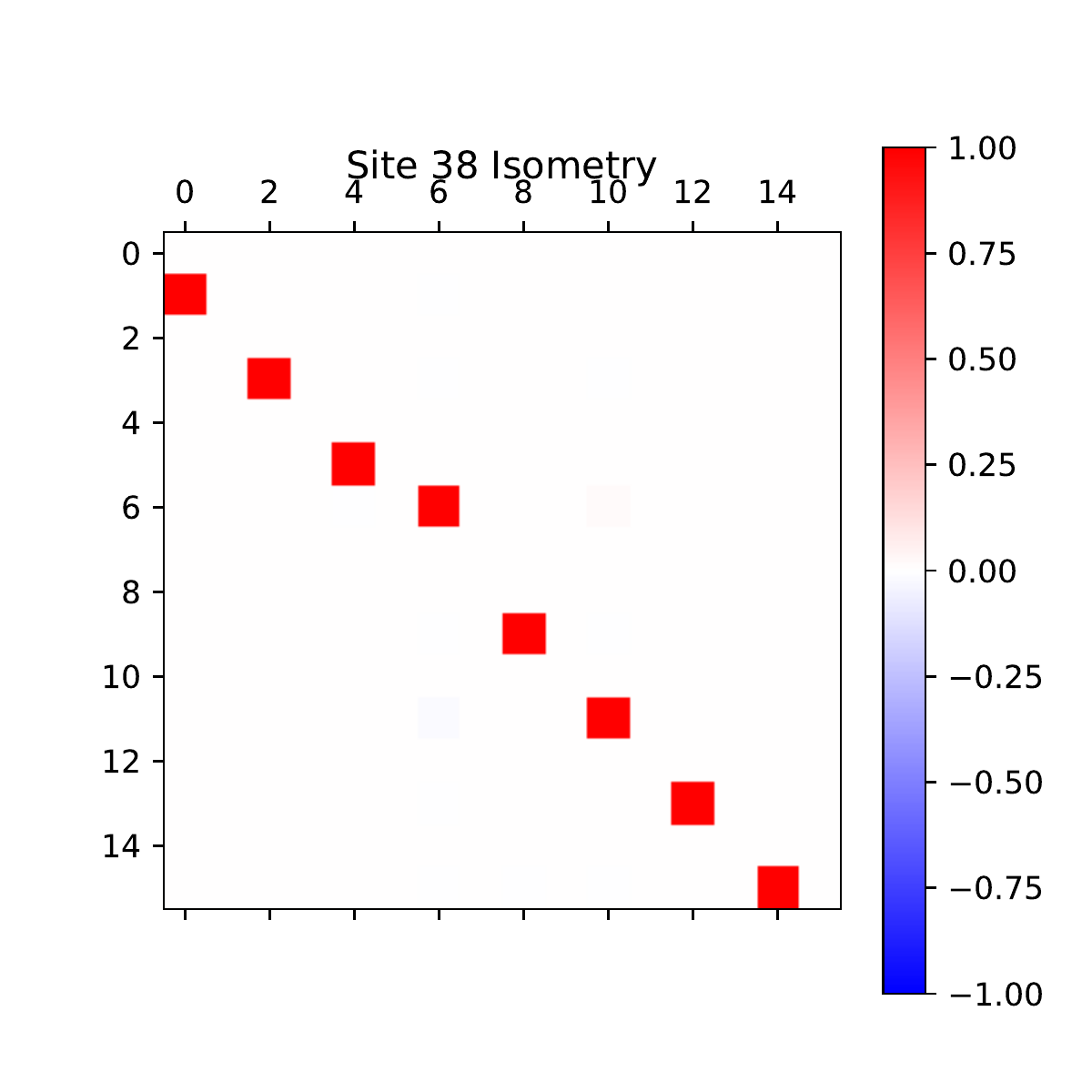}
}
\subfloat[Optimized gate]{
\includegraphics[width=0.45\columnwidth]{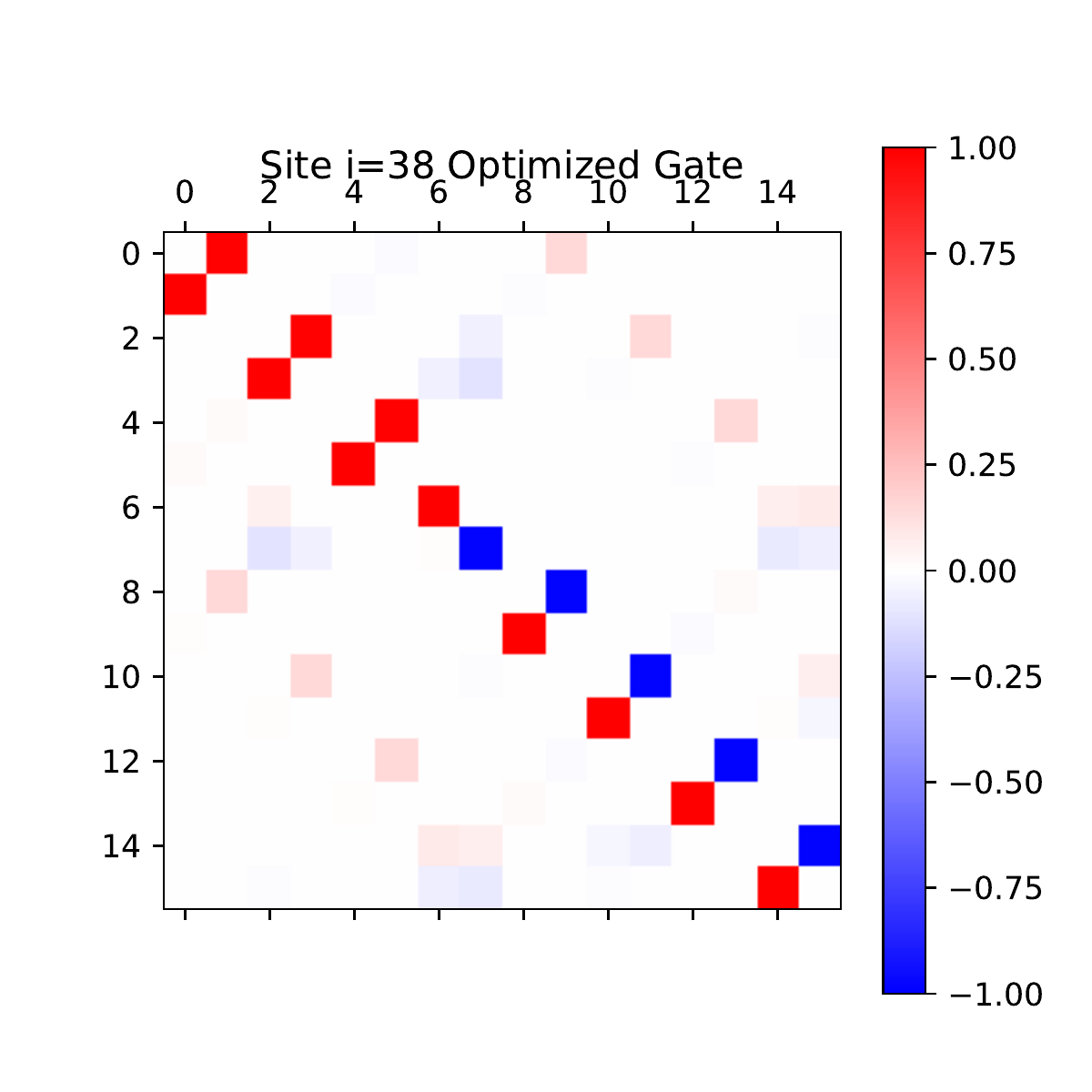}
}\\
\subfloat[Raw circuit from optimization]{
\includegraphics[width=0.9\columnwidth]{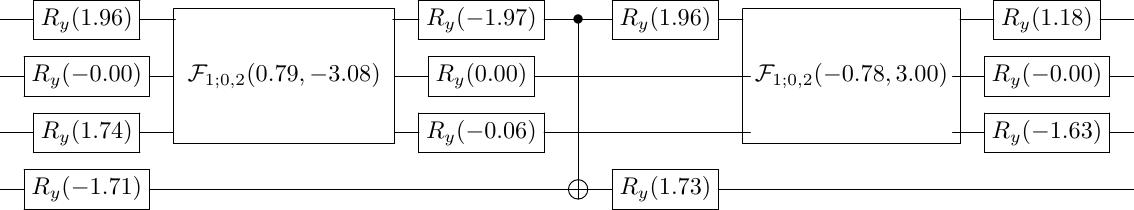}
}\\
\subfloat[Expanded and cleaned circuit from optimization]{
\includegraphics[width=0.9\columnwidth]{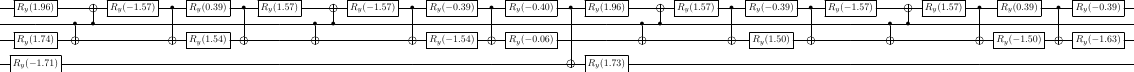}
}
\caption{Optimization for site 38}
\end{figure}
%
\begin{figure}[h]

\subfloat[Isometry]{
\includegraphics[width=0.45\columnwidth]{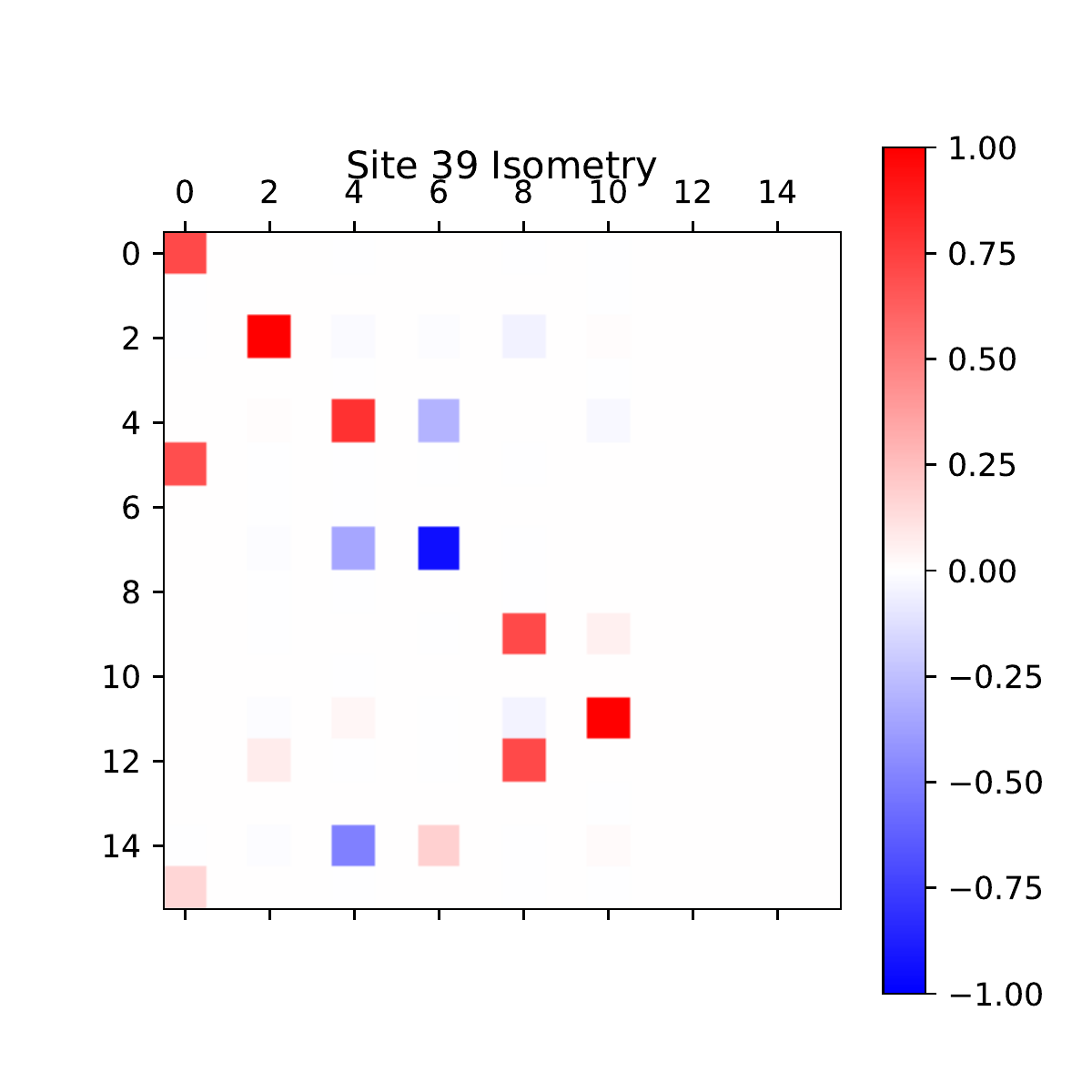}
}
\subfloat[Optimized gate]{
\includegraphics[width=0.45\columnwidth]{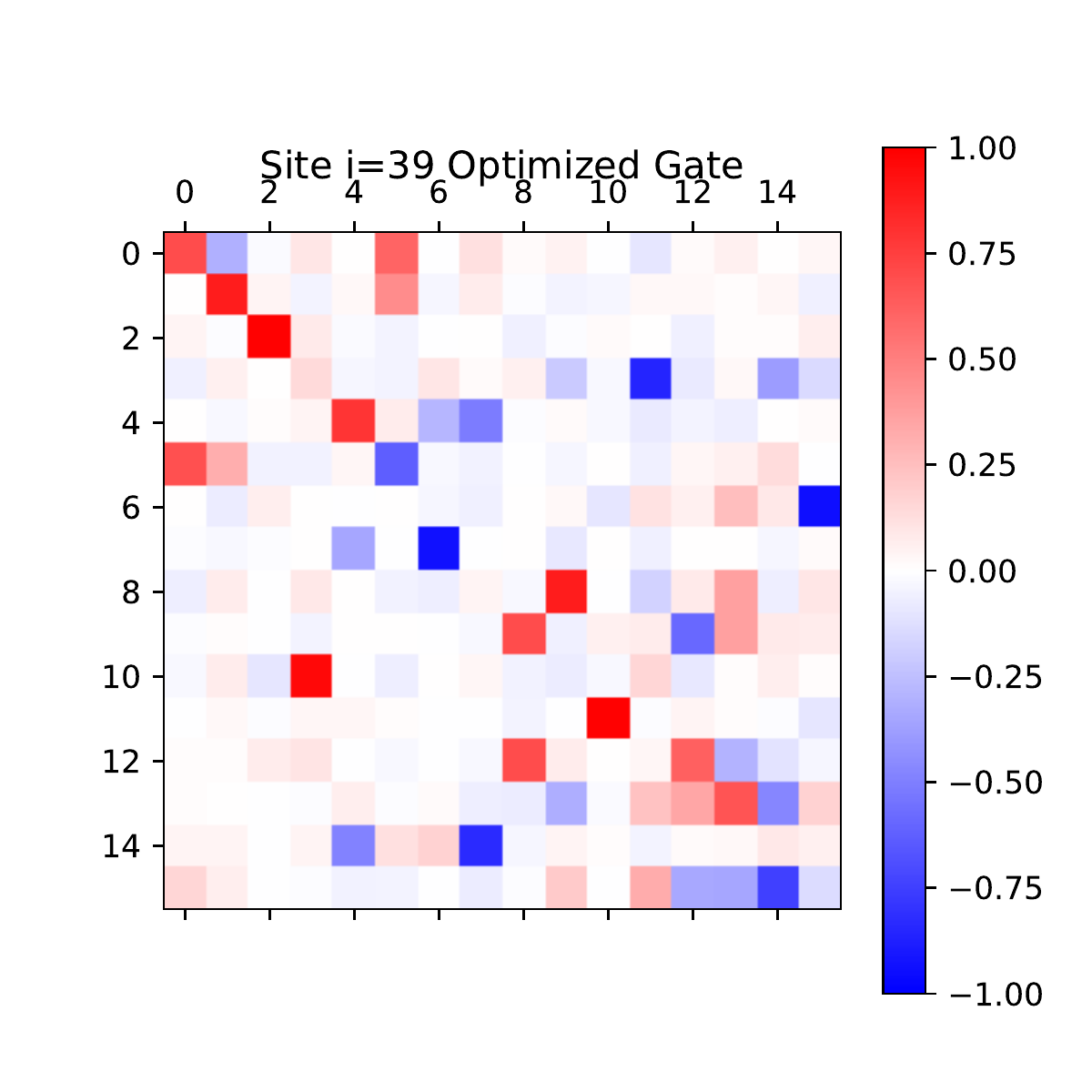}
}\\
\subfloat[Raw circuit from optimization]{
\includegraphics[width=0.9\columnwidth]{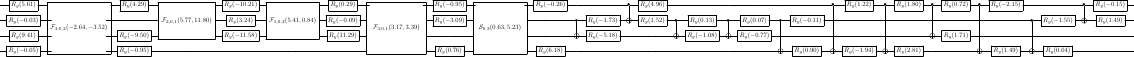}
}\\
\subfloat[Expanded and cleaned circuit from optimization]{
\includegraphics[width=0.9\columnwidth]{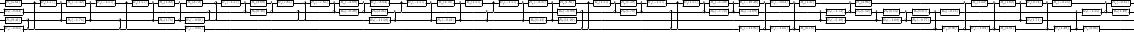}
}
\caption{Optimization for site 39}
\end{figure}
%
%
\begin{figure}[h]

\subfloat[Isometry]{
\includegraphics[width=0.3\columnwidth]{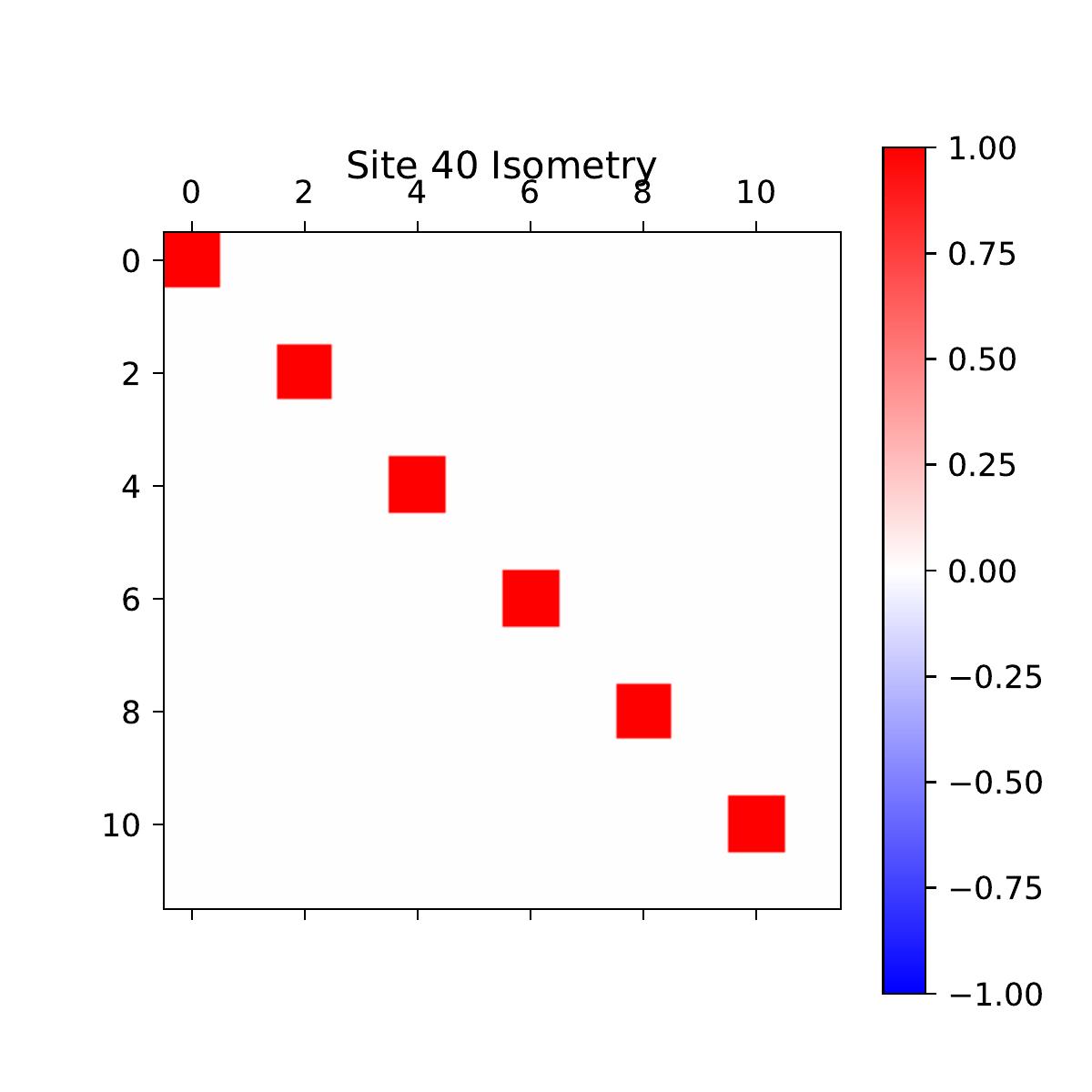}
}
\subfloat[Optimized gate]{
\includegraphics[width=0.3\columnwidth]{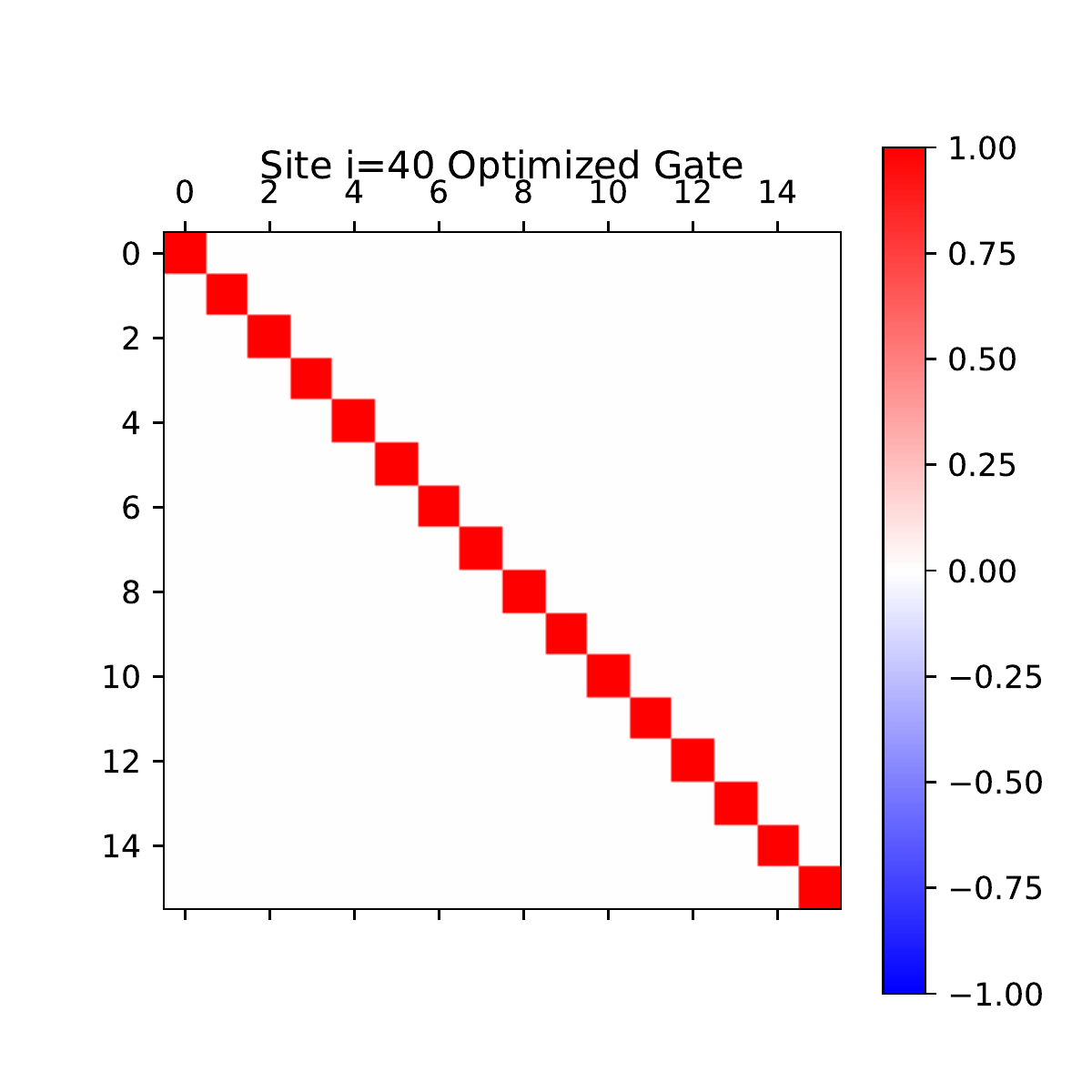}
}
\subfloat[Circuit from optimization]{
\includegraphics[width=0.2\columnwidth]{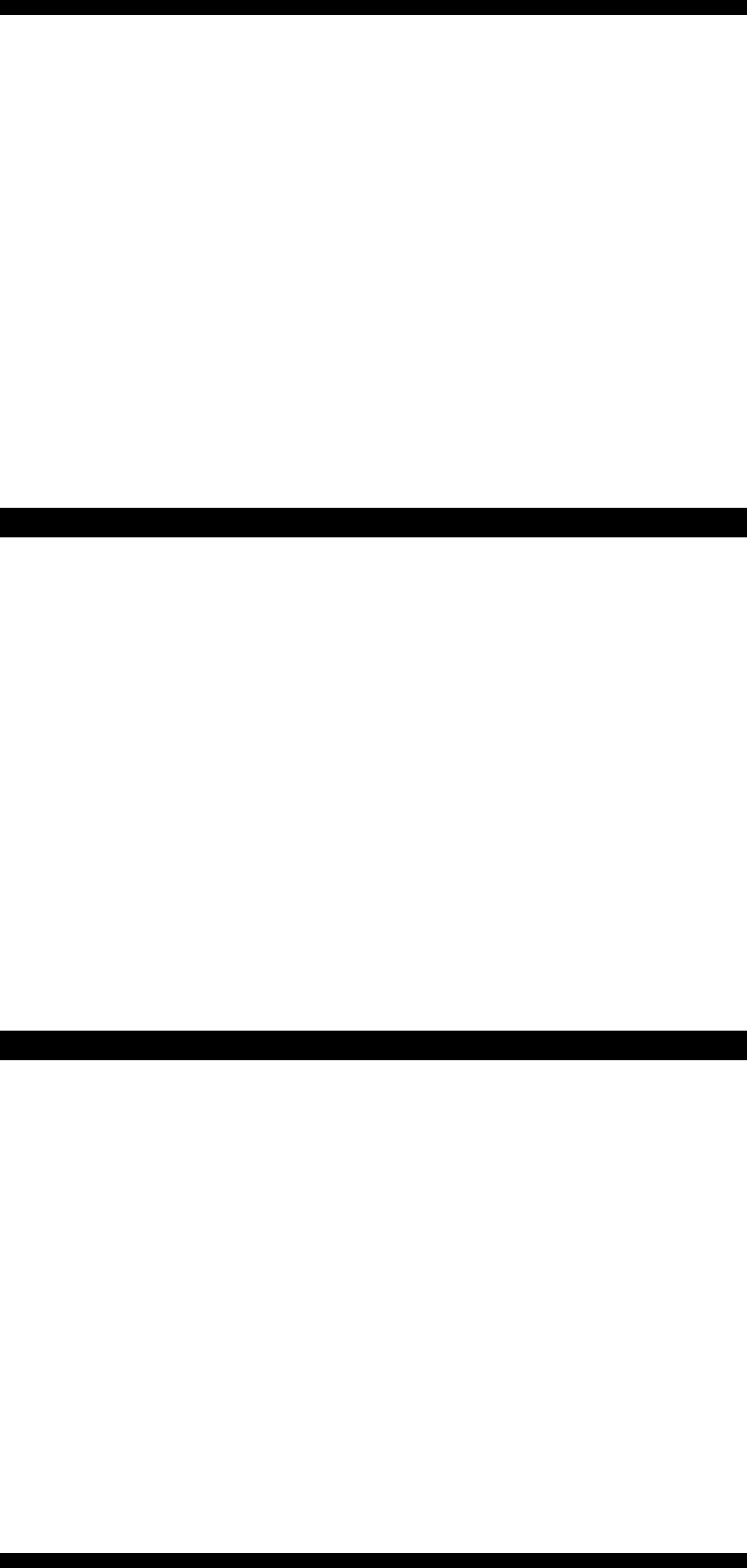}
}
\caption{Optimization for site 40}
\end{figure}
%
\begin{figure}[h]

\subfloat[Isometry]{
\includegraphics[width=0.3\columnwidth]{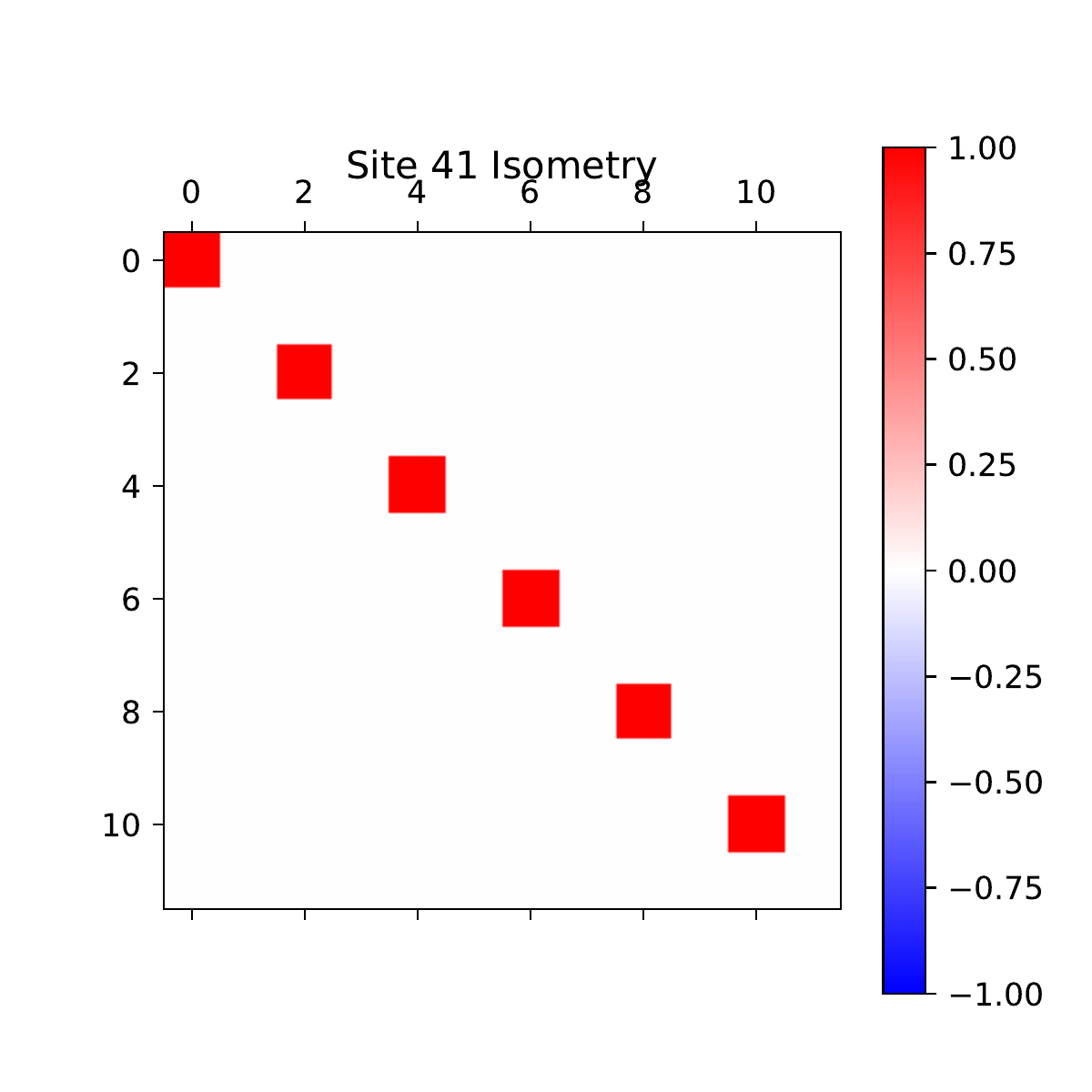}
}
\subfloat[Optimized gate]{
\includegraphics[width=0.3\columnwidth]{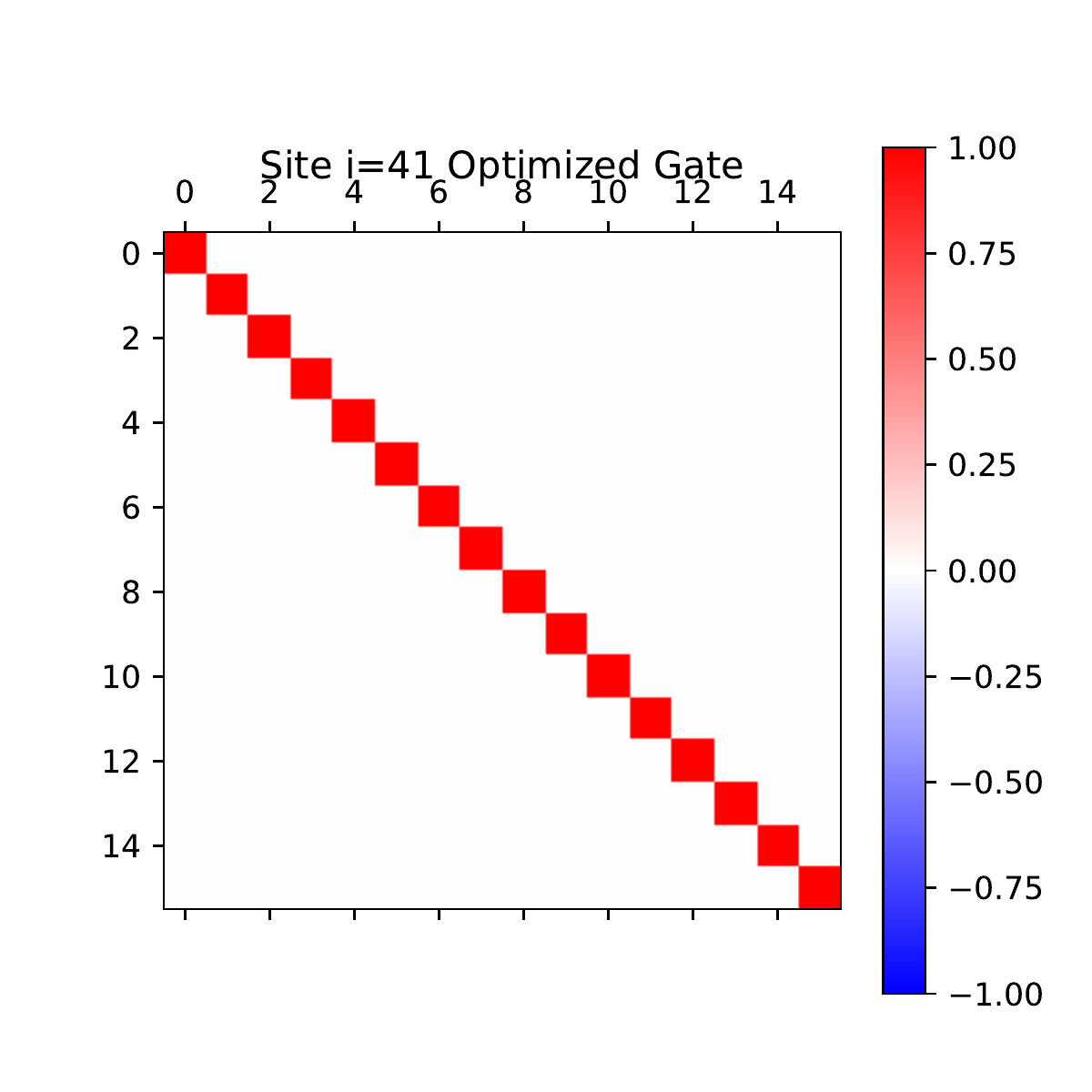}
}
\subfloat[Circuit from optimization]{
\includegraphics[width=0.2\columnwidth]{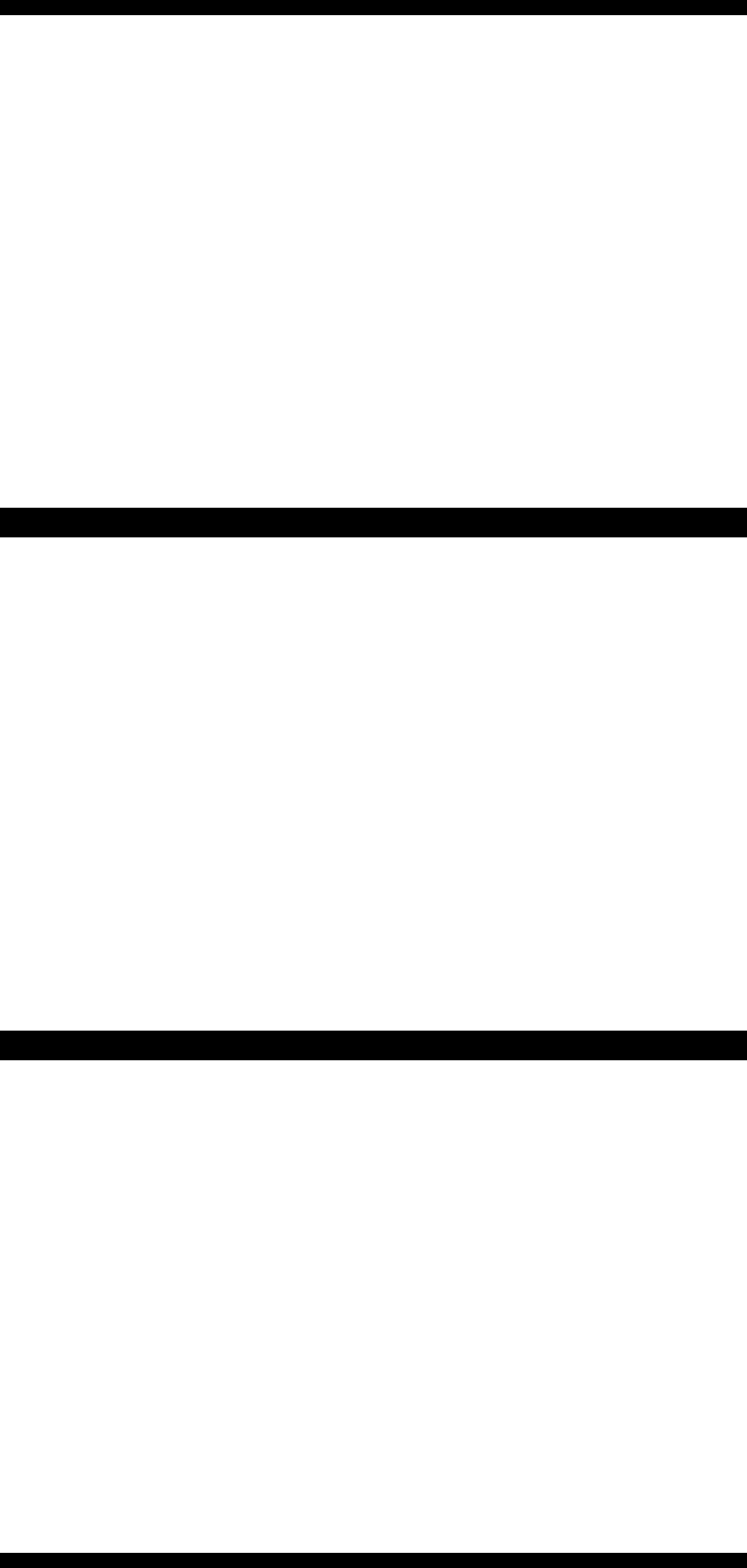}
}
\caption{Optimization for site 41}
\end{figure}
%
%
\begin{figure}[h]

\subfloat[Isometry]{
\includegraphics[width=0.3\columnwidth]{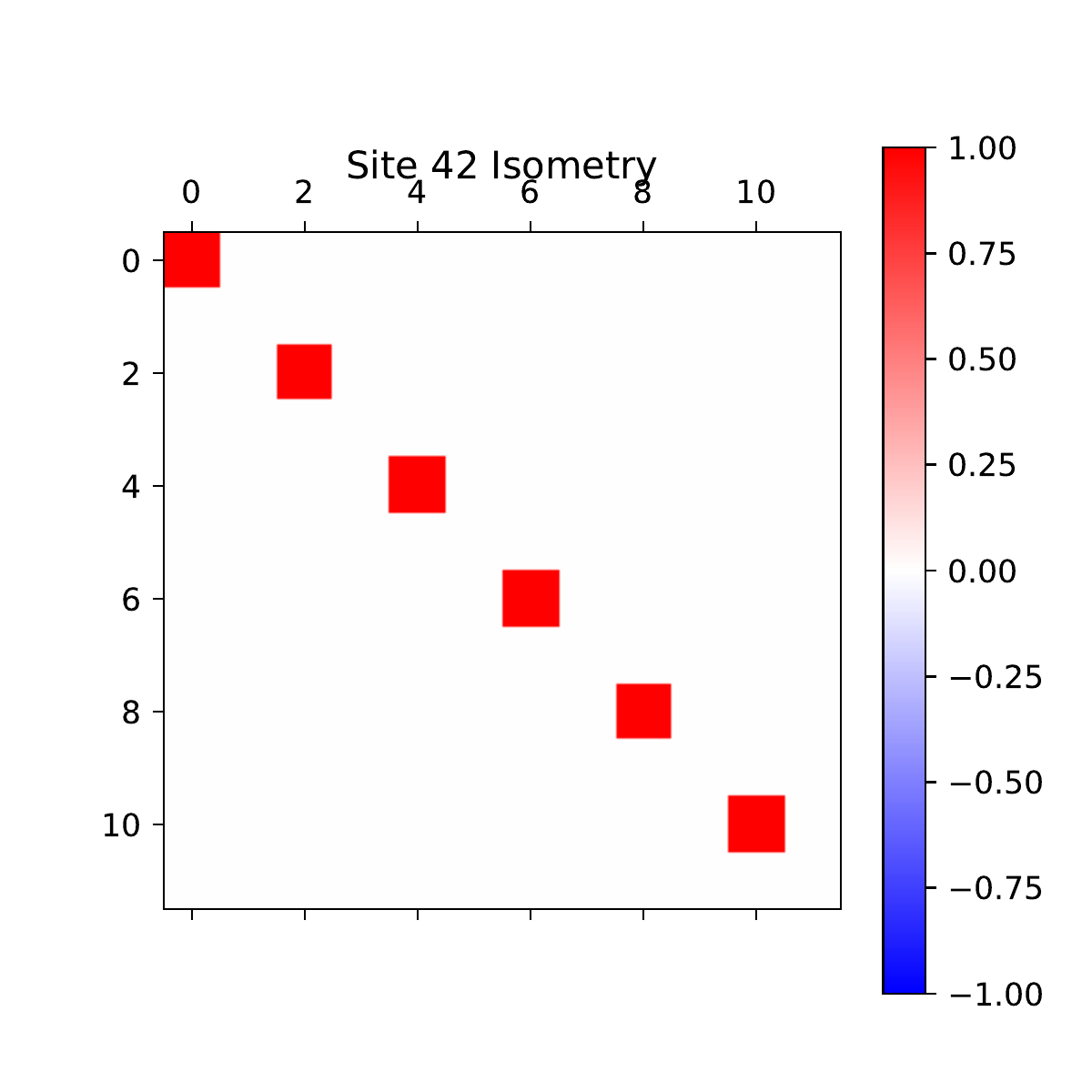}
}
\subfloat[Optimized gate]{
\includegraphics[width=0.3\columnwidth]{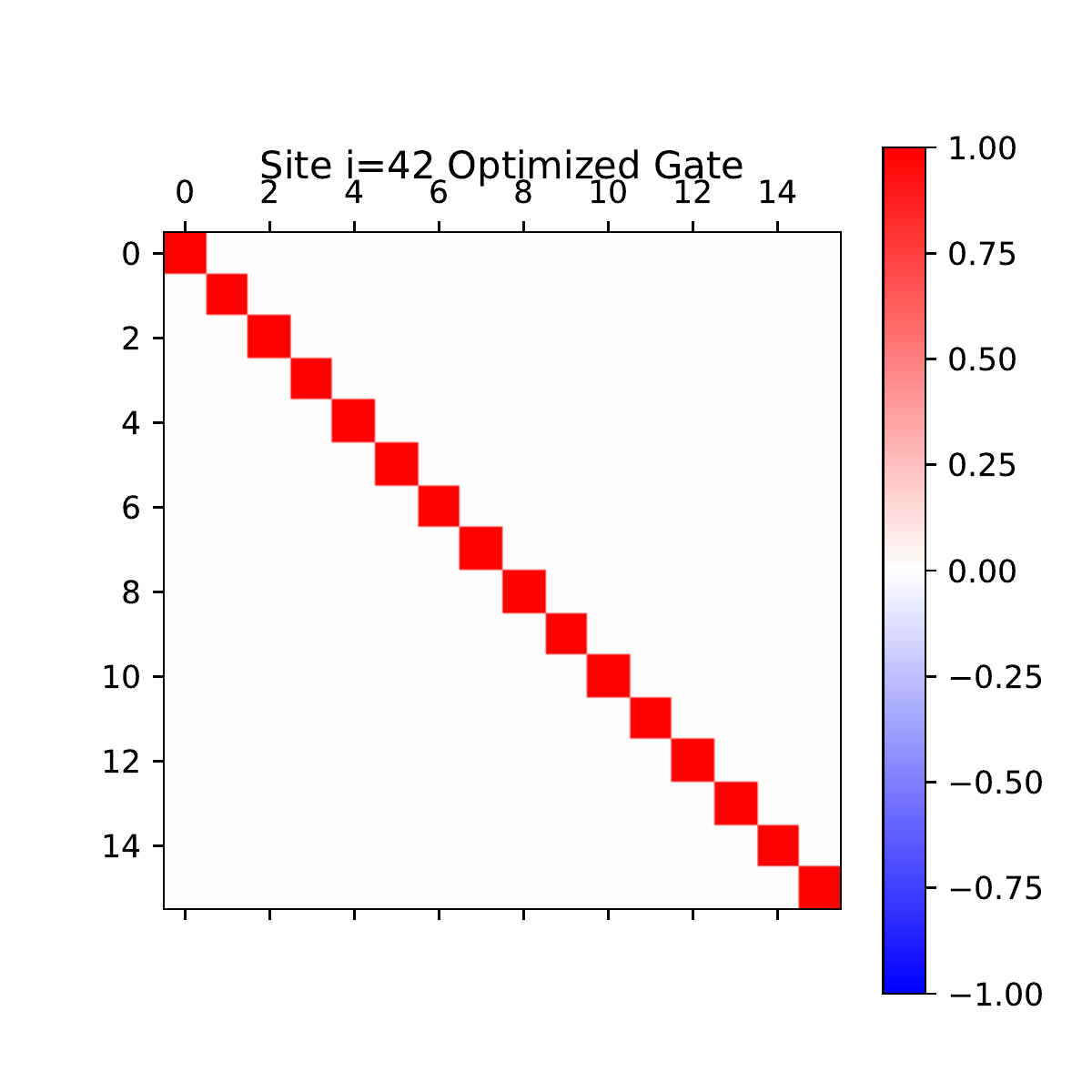}
}
\subfloat[Circuit from optimization]{
\includegraphics[width=0.2\columnwidth]{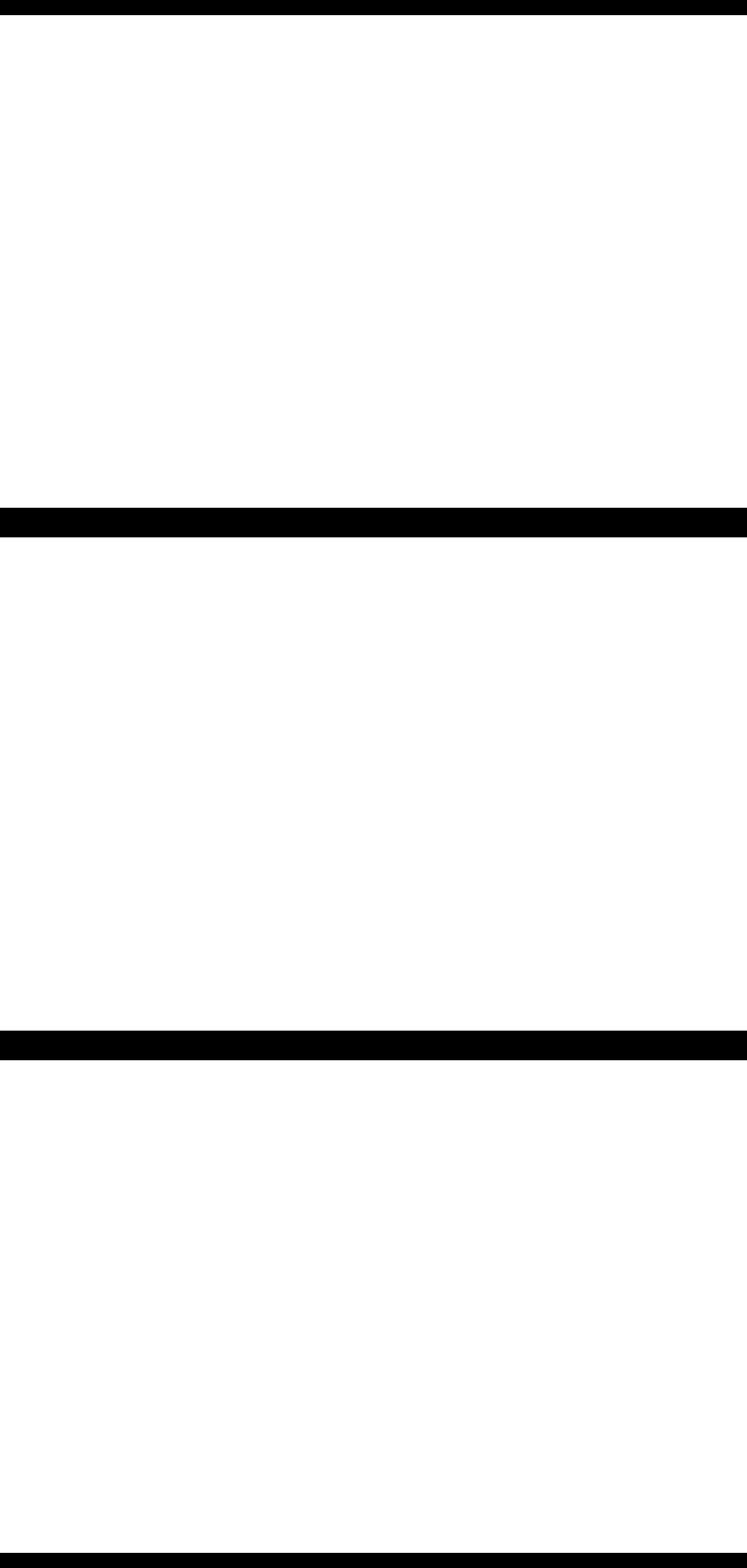}
}
\caption{Optimization for site 42}
\end{figure}
%
%
\begin{figure}[h]

\subfloat[Isometry]{
\includegraphics[width=0.45\columnwidth]{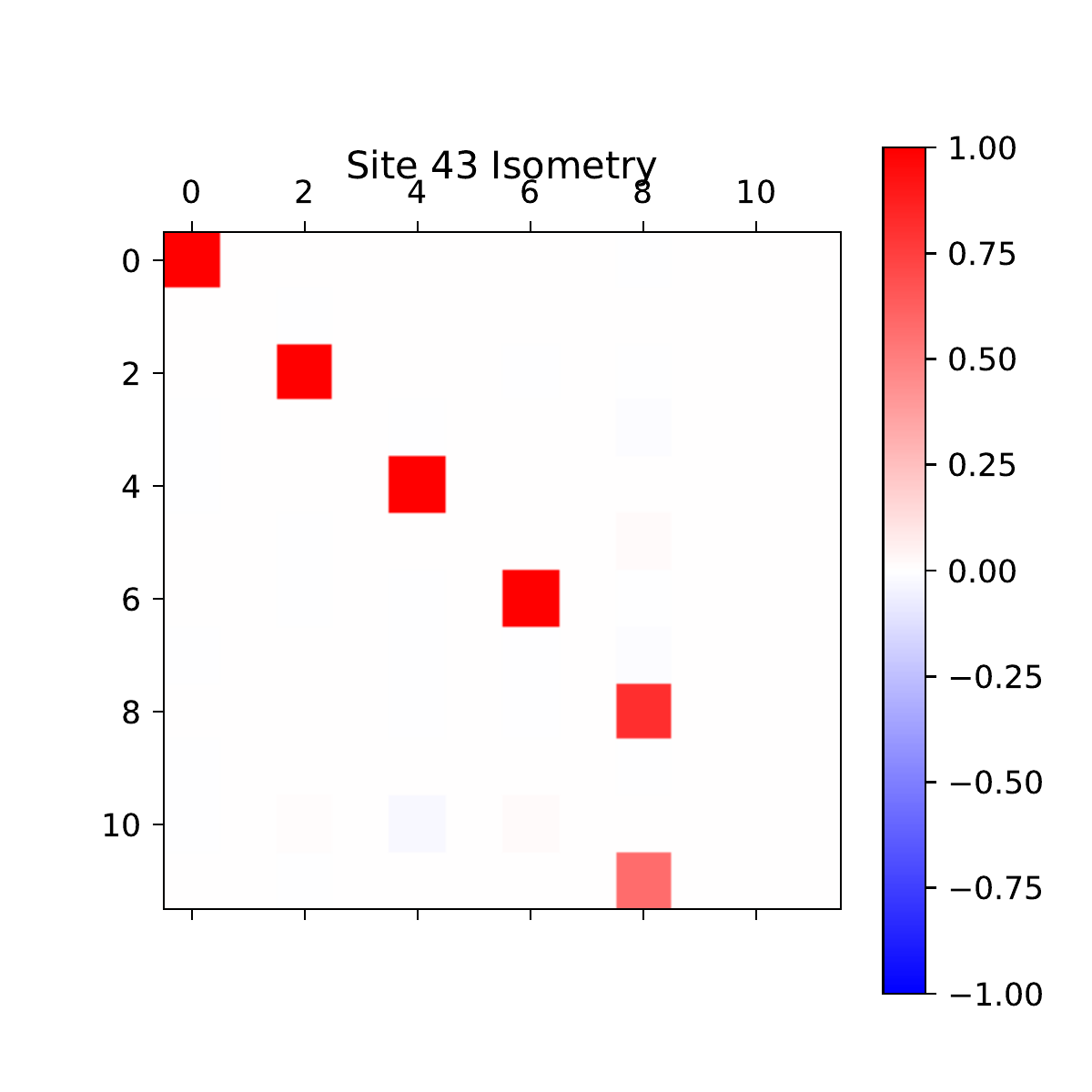}
}
\subfloat[Optimized gate]{
\includegraphics[width=0.45\columnwidth]{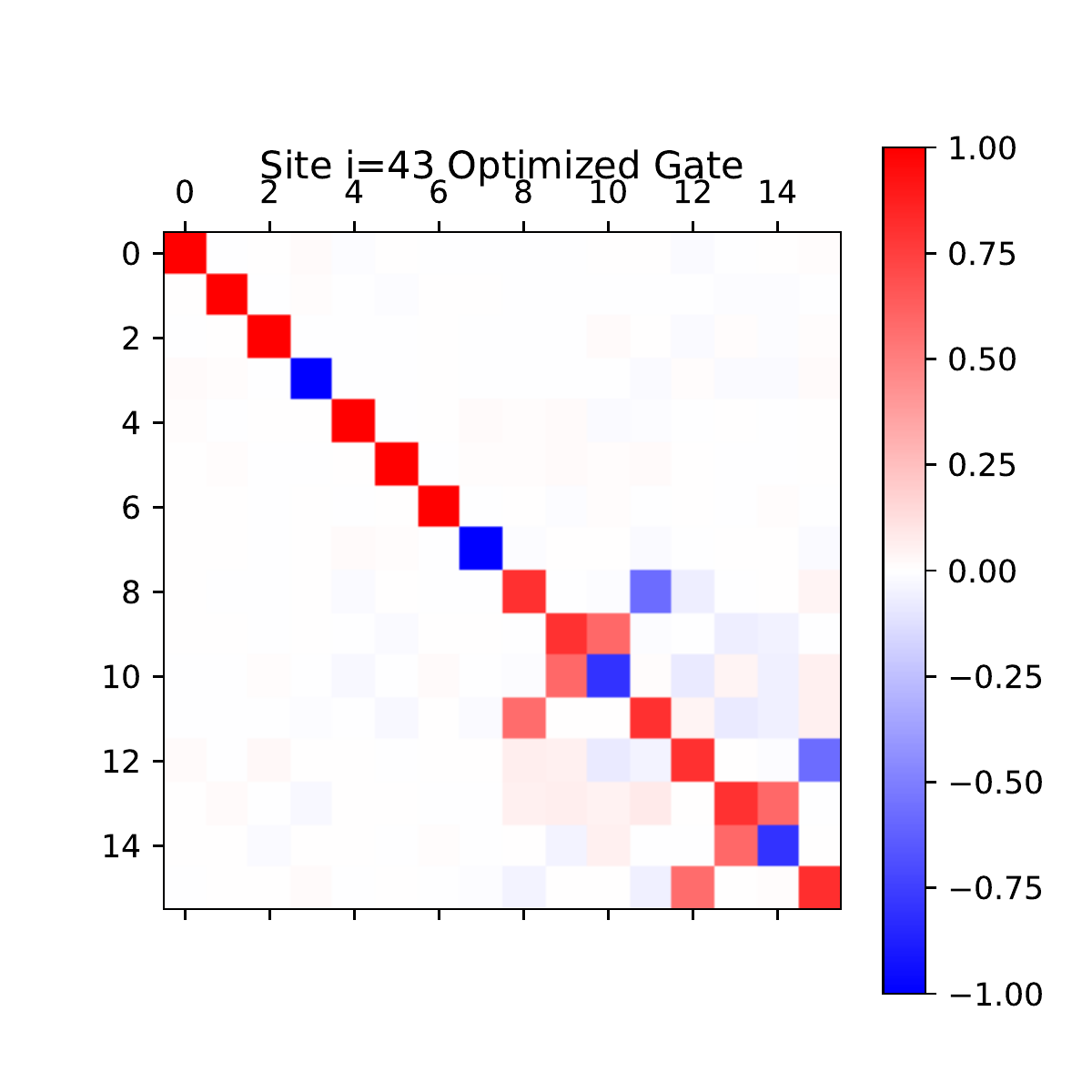}
}\\
\subfloat[Raw circuit from optimization]{
\includegraphics[width=0.9\columnwidth]{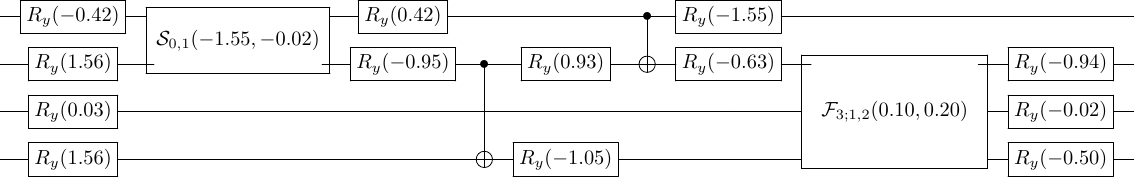}
}\\
\subfloat[Expanded and cleaned circuit from optimization]{
\includegraphics[width=0.9\columnwidth]{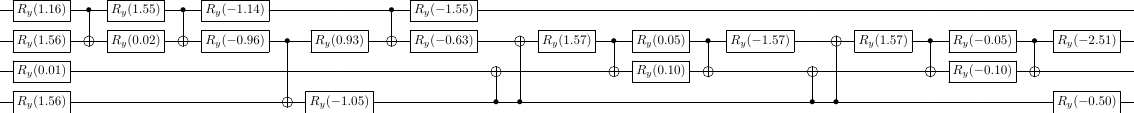}
}
\caption{Optimization for site 43}
\end{figure}
%
\begin{figure}[h]

\subfloat[Isometry]{
\includegraphics[width=0.45\columnwidth]{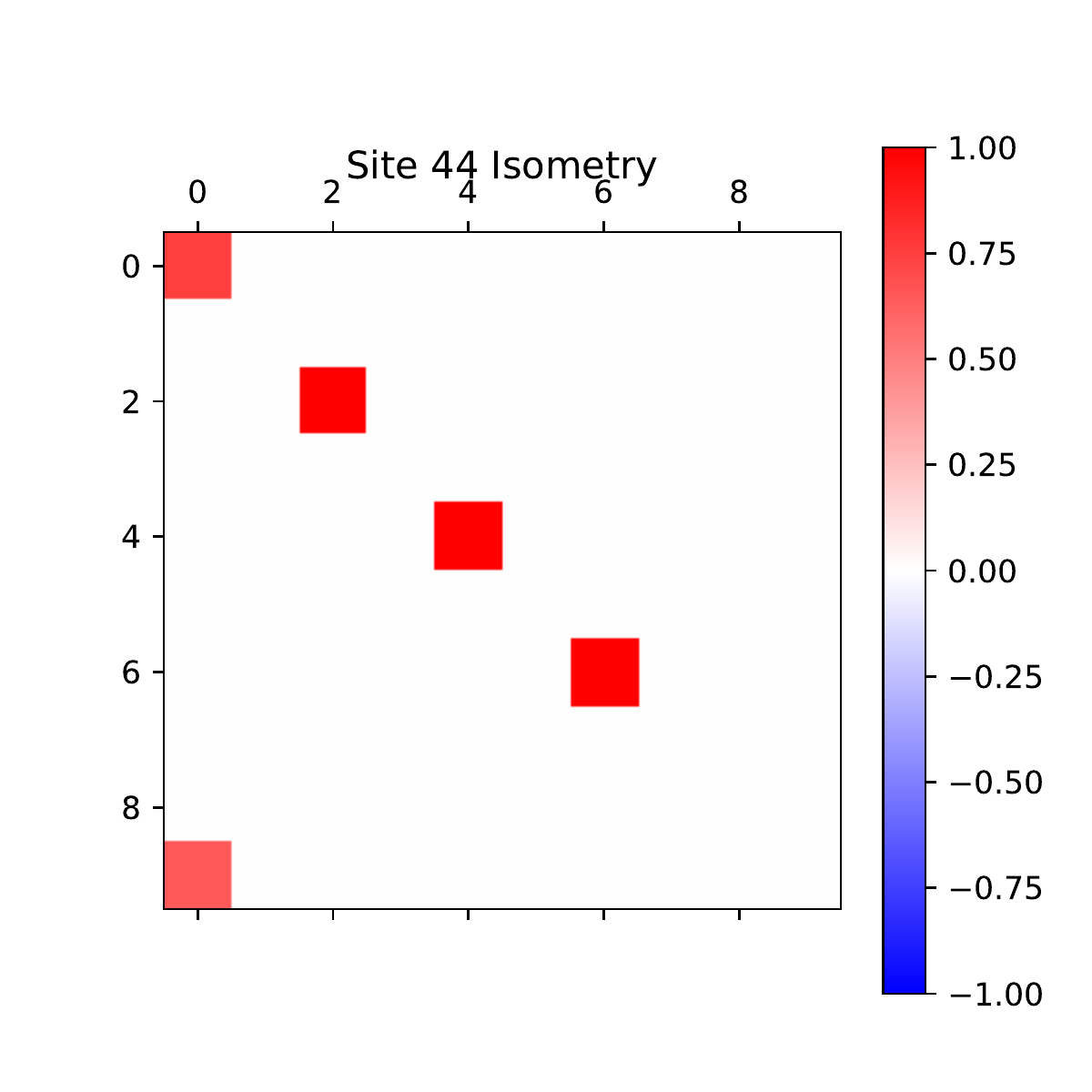}
}
\subfloat[Optimized gate]{
\includegraphics[width=0.45\columnwidth]{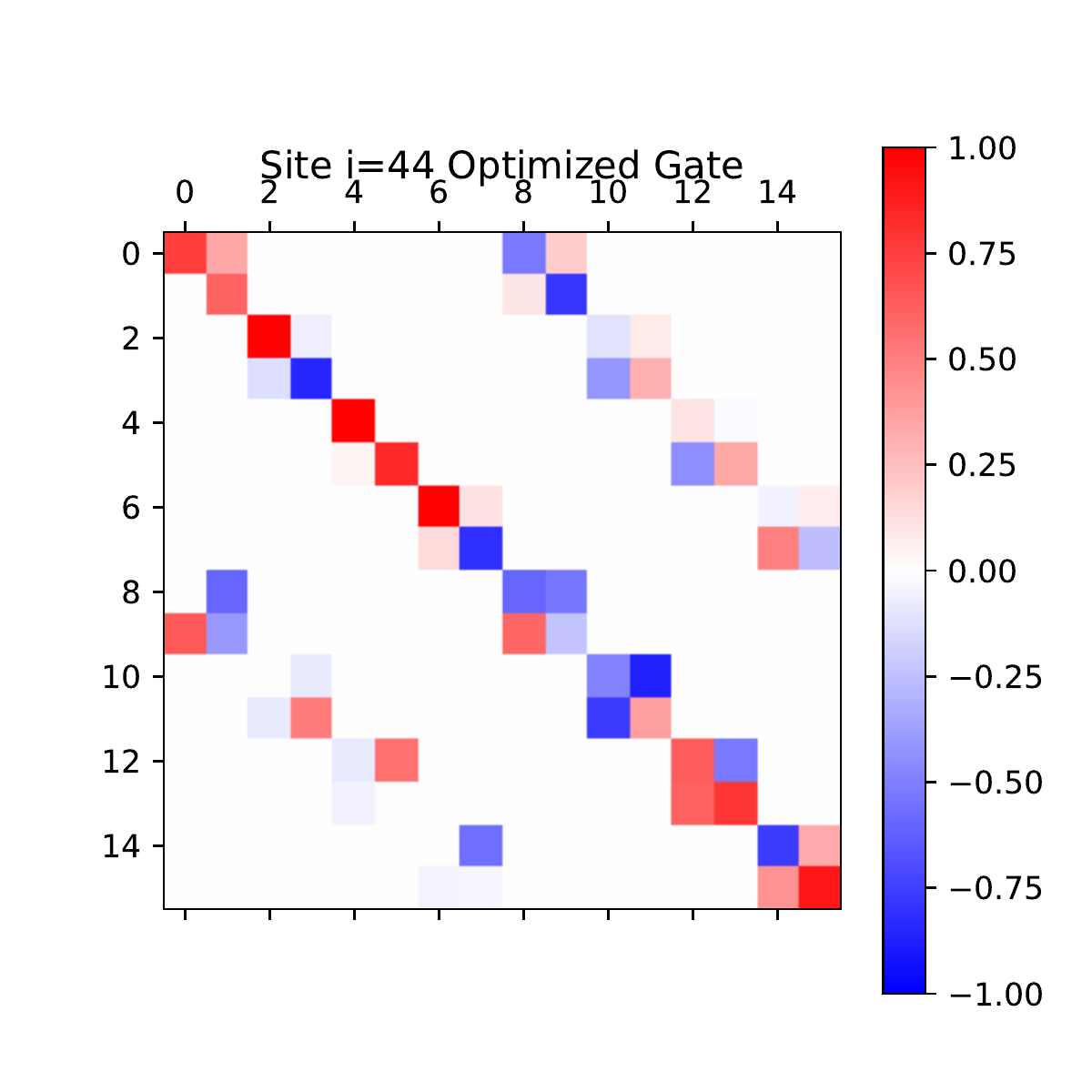}
}\\
\subfloat[Raw circuit from optimization]{
\includegraphics[width=0.9\columnwidth]{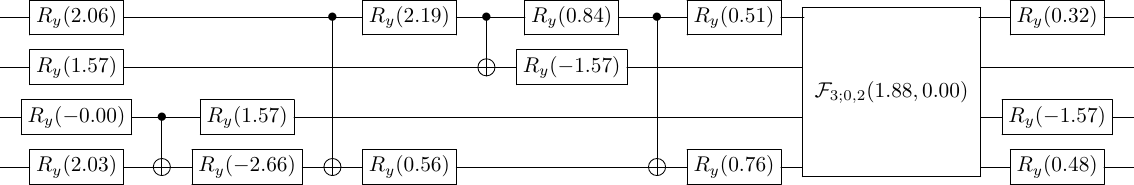}
}\\
\subfloat[Expanded and cleaned circuit from optimization]{
\includegraphics[width=0.9\columnwidth]{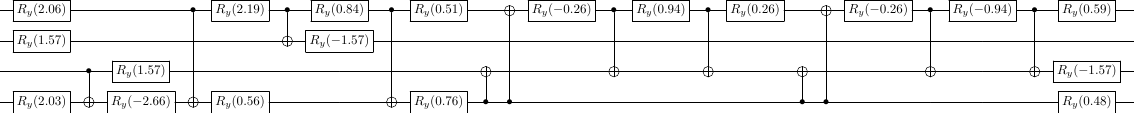}
}
\caption{Optimization for site 44}
\end{figure}
%
%
\begin{figure}[h]

\subfloat[Isometry]{
\includegraphics[width=0.45\columnwidth]{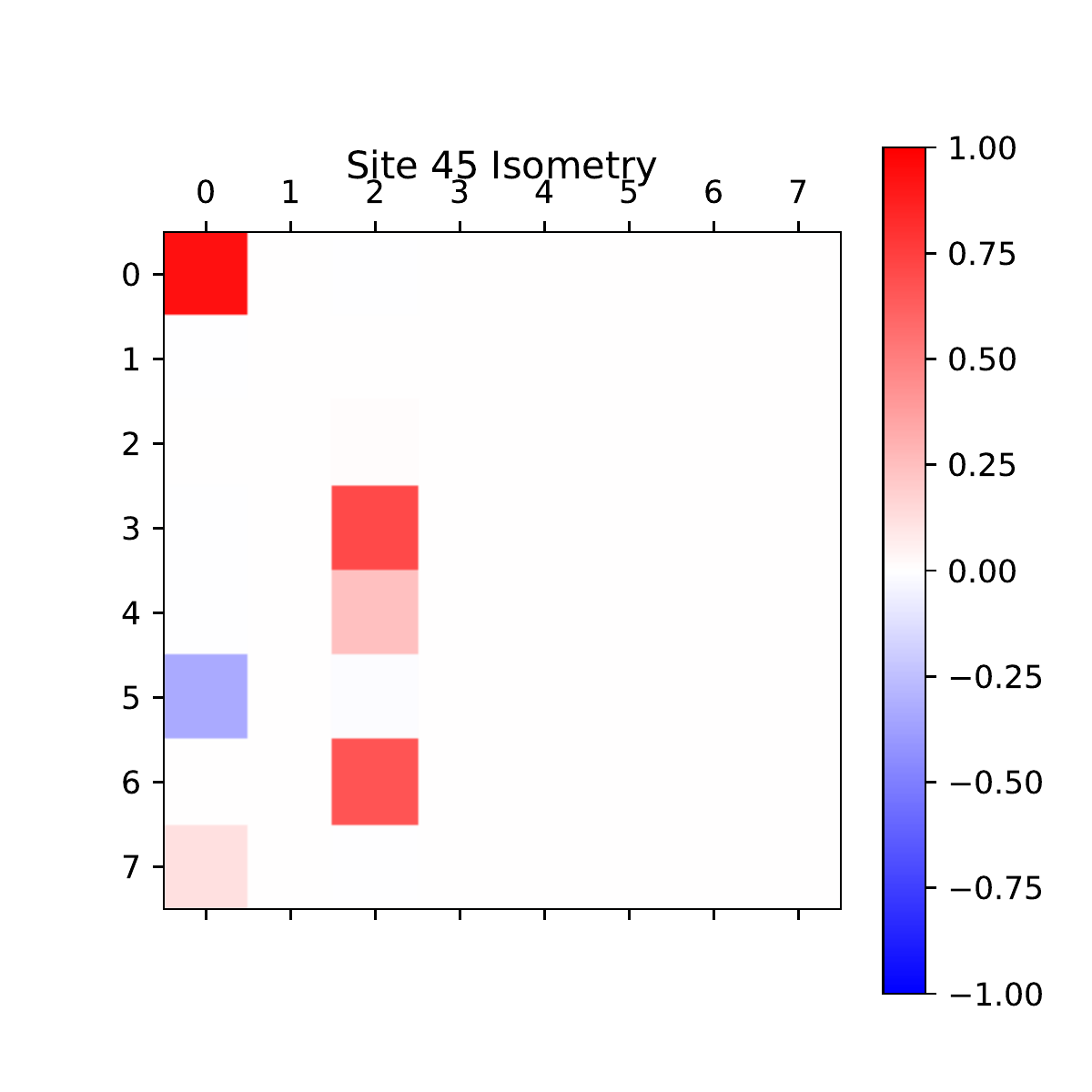}
}
\subfloat[Optimized gate]{
\includegraphics[width=0.45\columnwidth]{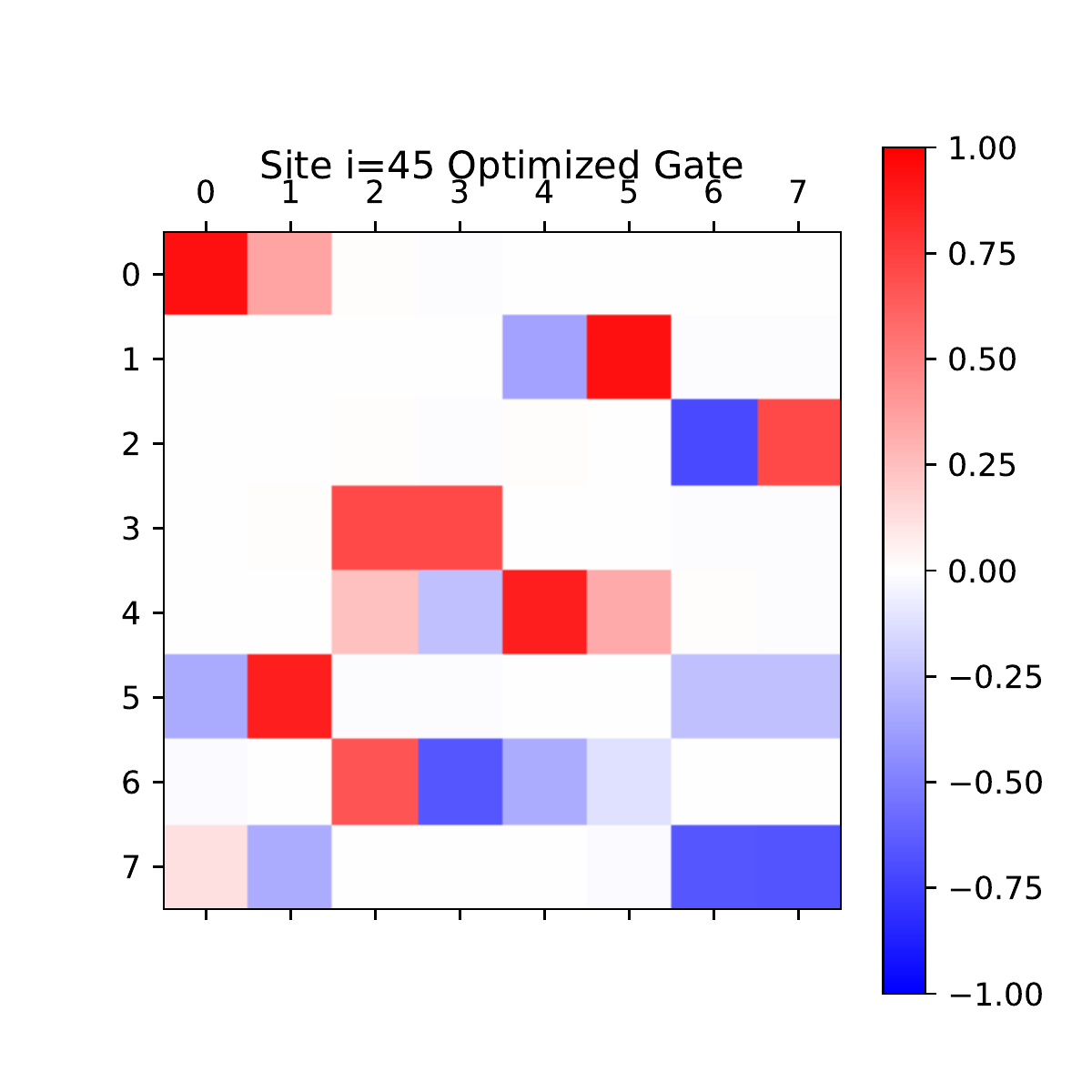}
}\\
\subfloat[Raw circuit from optimization]{
\includegraphics[width=0.9\columnwidth]{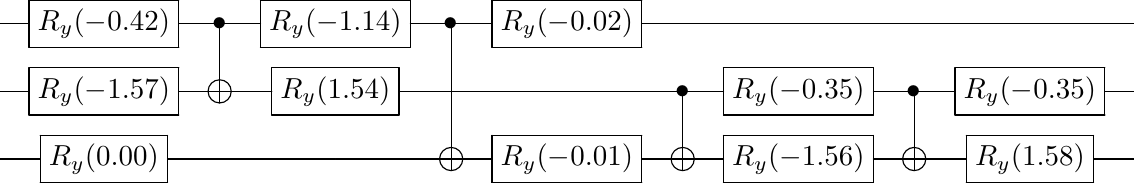}
}\\
\subfloat[Expanded and cleaned circuit from optimization]{
\includegraphics[width=0.9\columnwidth]{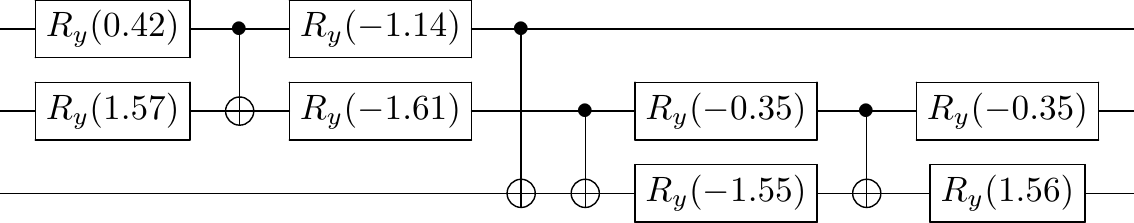}
}
\caption{Optimization for site 45}
\end{figure}
%
\begin{figure}[h]

\subfloat[Isometry]{
\includegraphics[width=0.45\columnwidth]{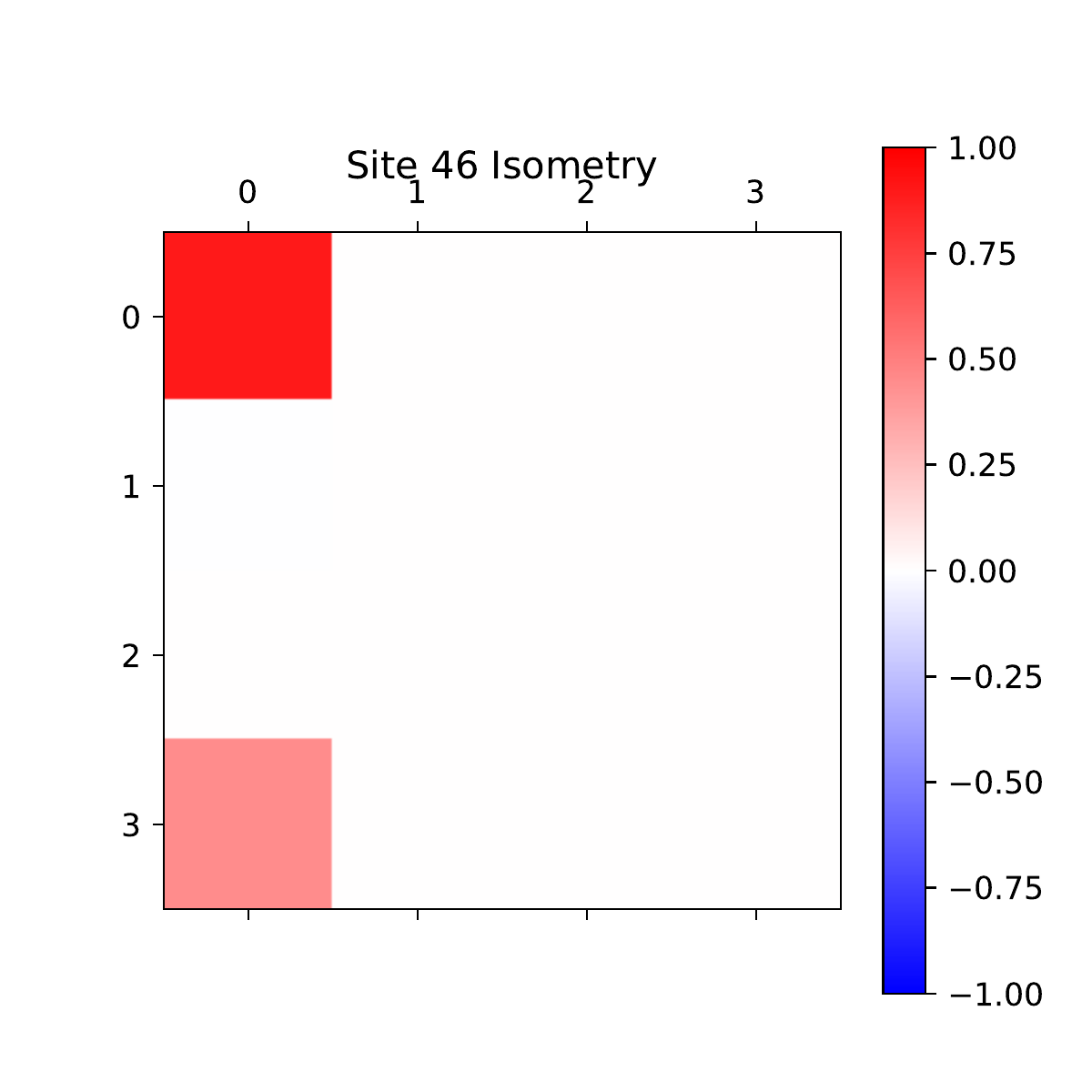}
}
\subfloat[Optimized gate]{
\includegraphics[width=0.45\columnwidth]{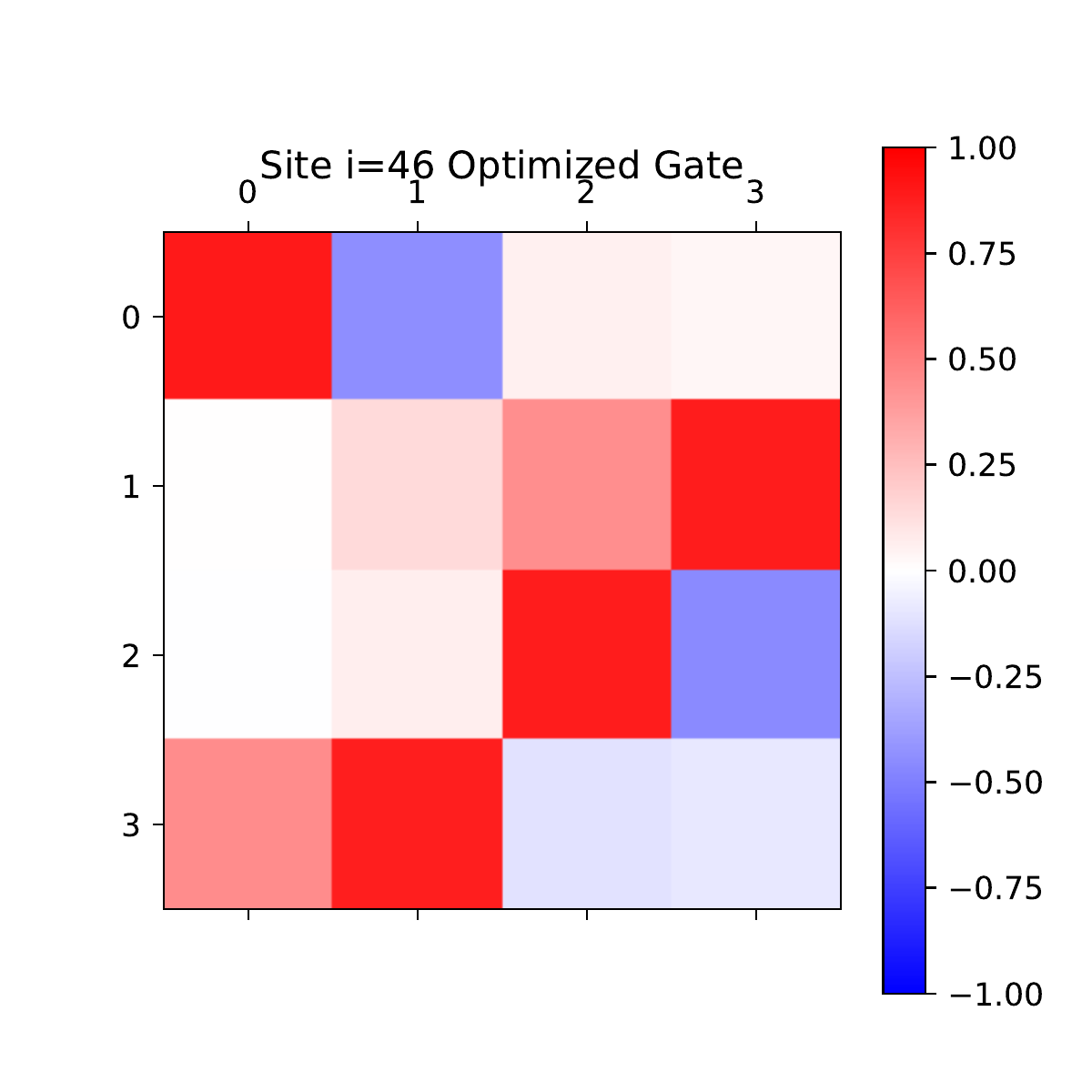}
}\\
\subfloat[Raw circuit from optimization]{
\includegraphics[width=0.9\columnwidth]{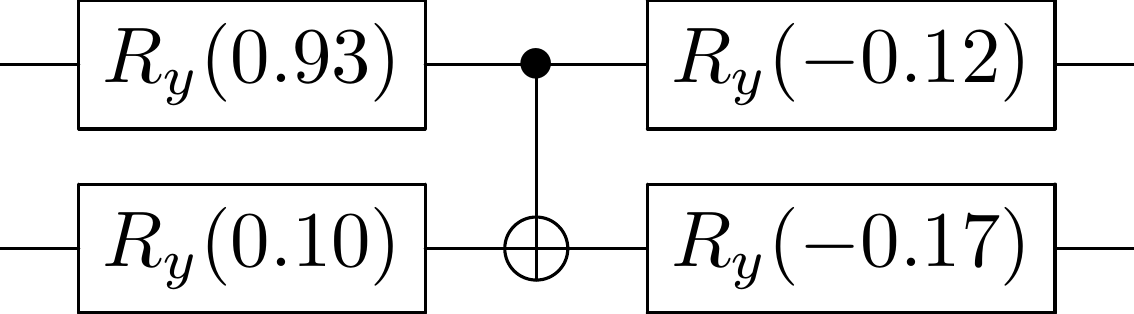}
}\\
\subfloat[Expanded and cleaned circuit from optimization]{
\includegraphics[width=0.9\columnwidth]{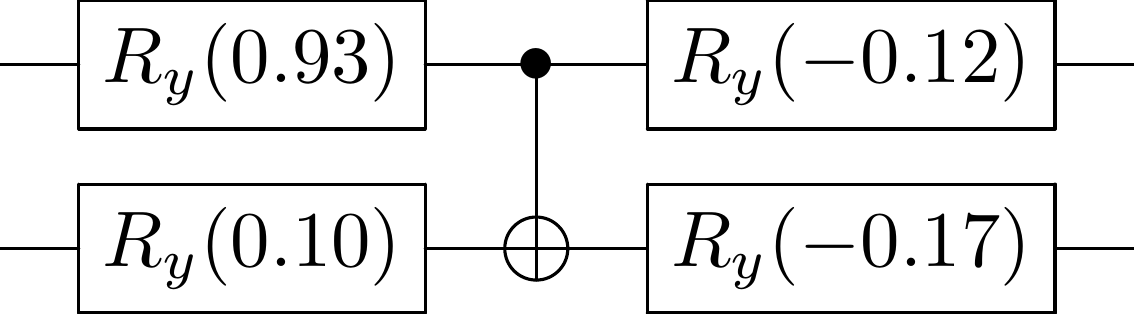}
}
\caption{Optimization for site 46}
\end{figure}
%
%
\begin{figure}[h]

\subfloat[Isometry]{
\includegraphics[width=0.3\columnwidth]{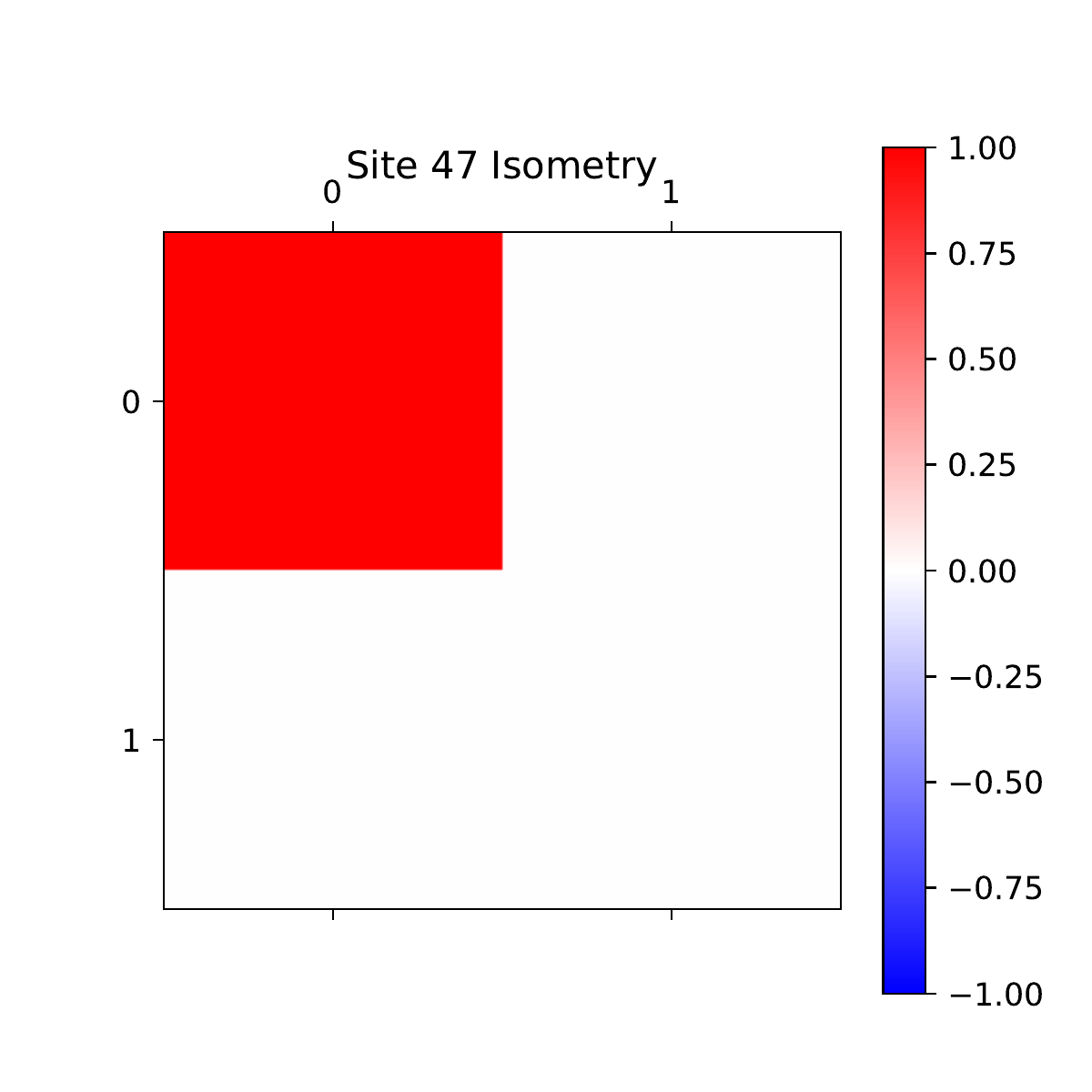}
}
\subfloat[Optimized gate]{
\includegraphics[width=0.3\columnwidth]{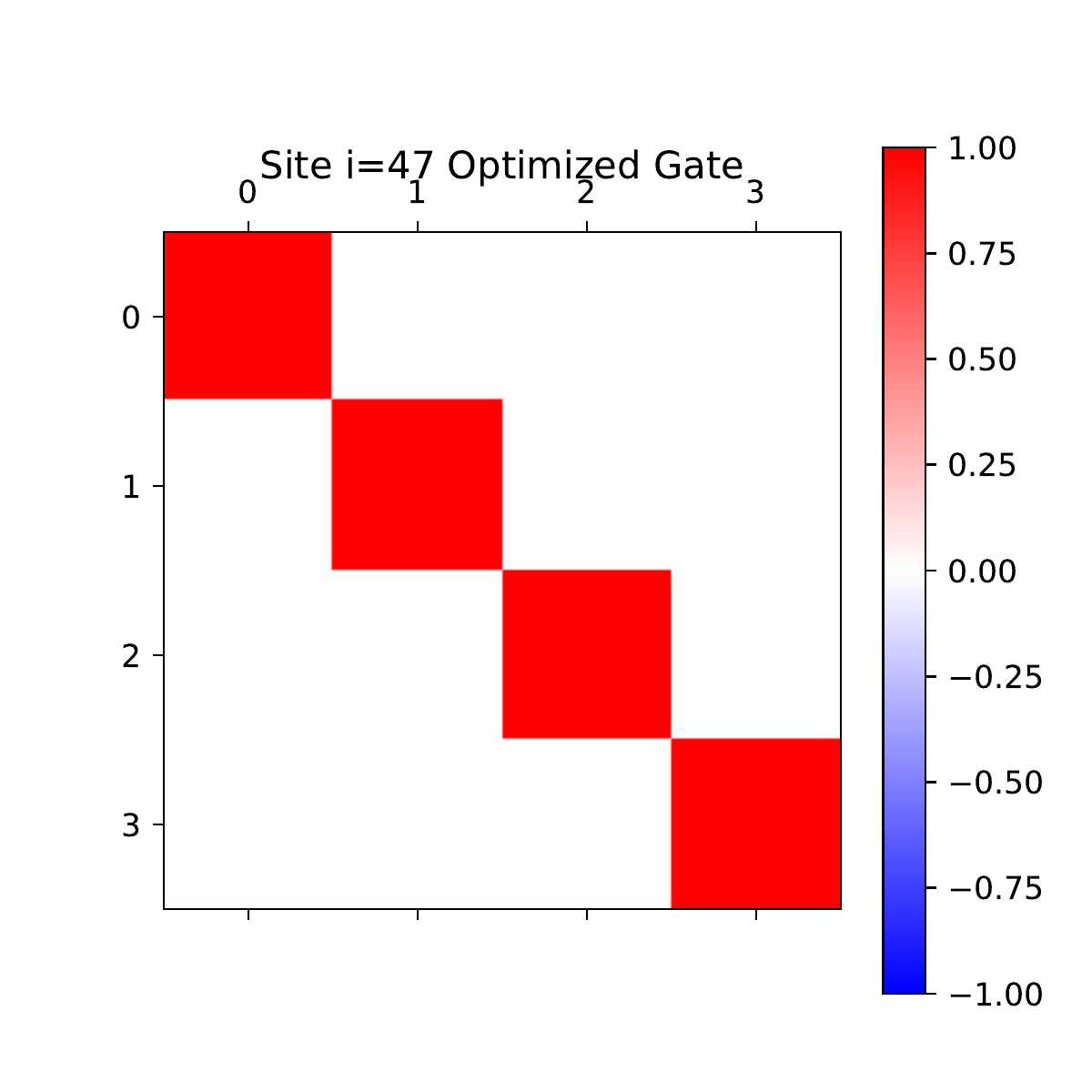}
}
\subfloat[Circuit from optimization]{
\includegraphics[width=0.3\columnwidth]{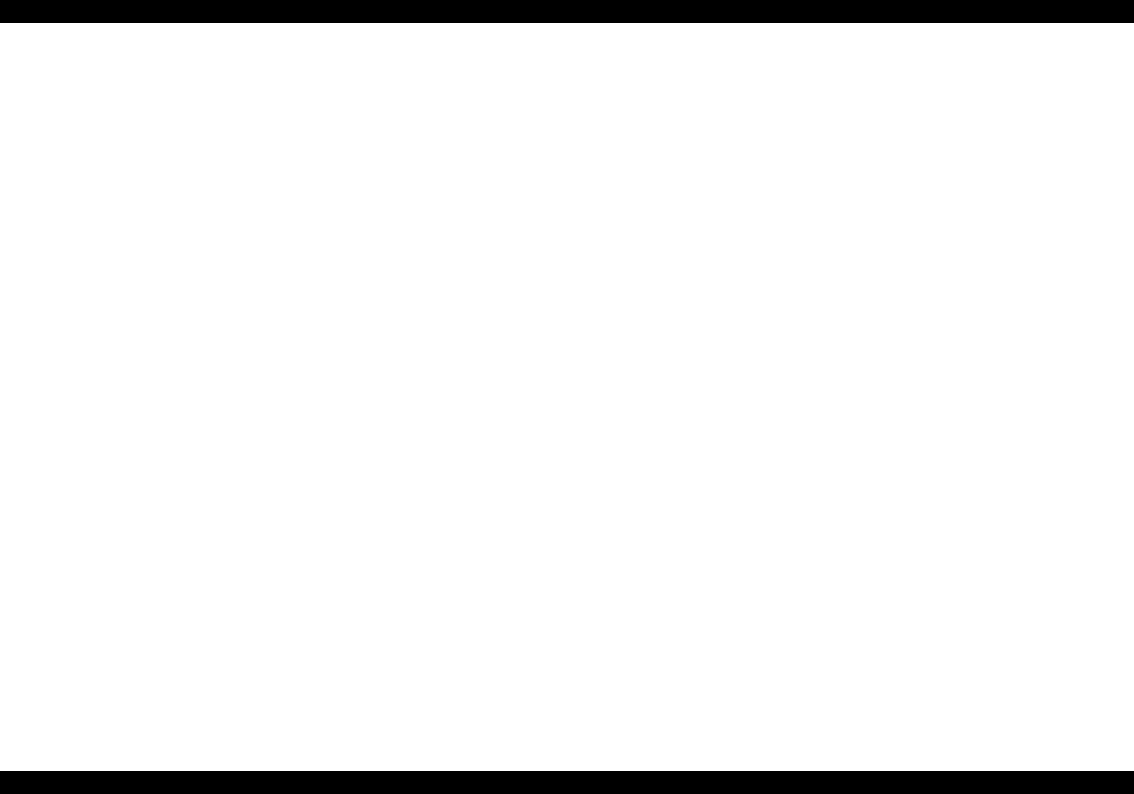}
}
\caption{Optimization for site 47}
\end{figure}
%
\begin{figure}[h]

\subfloat[Isometry]{
\includegraphics[width=0.3\columnwidth]{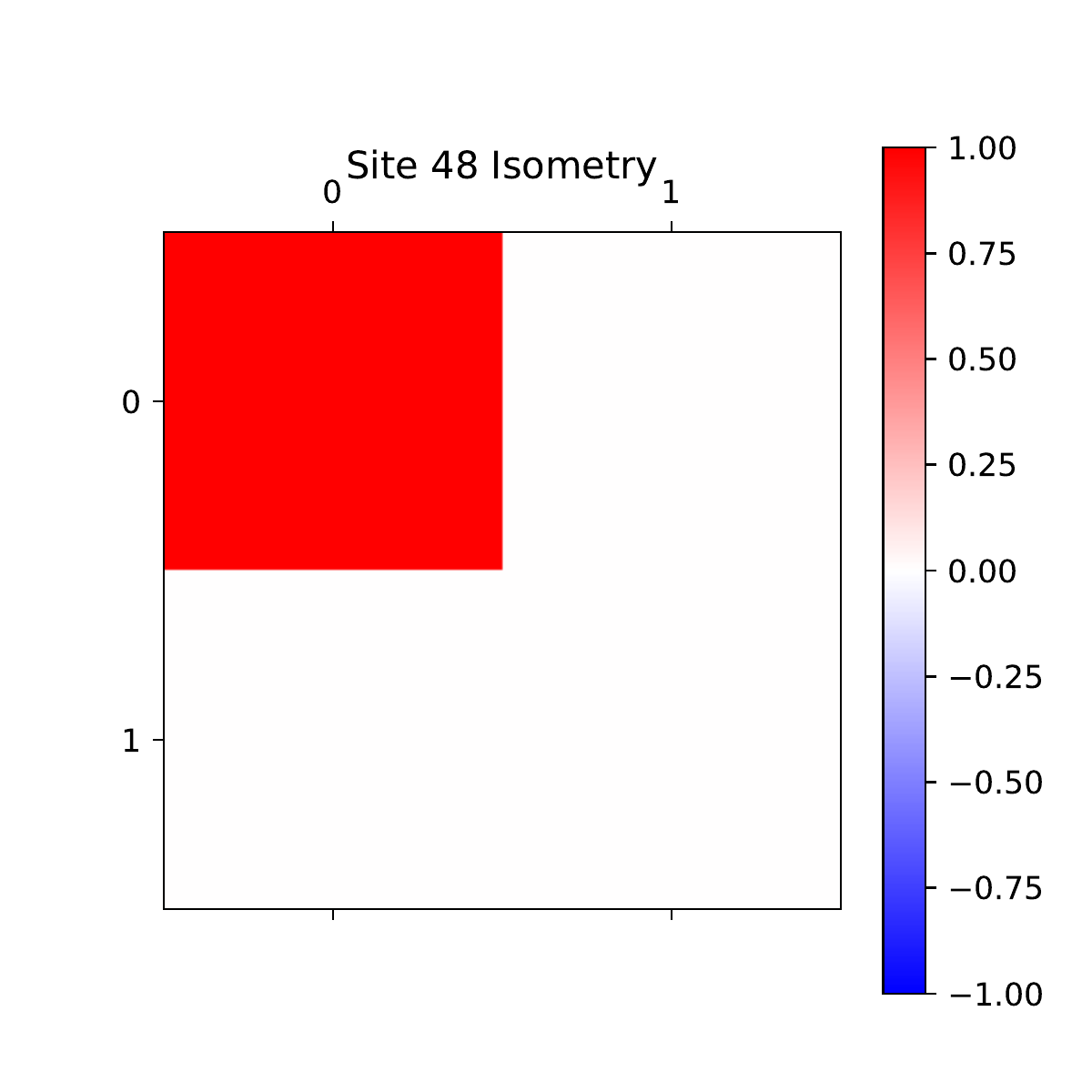}
}
\subfloat[Optimized gate]{
\includegraphics[width=0.3\columnwidth]{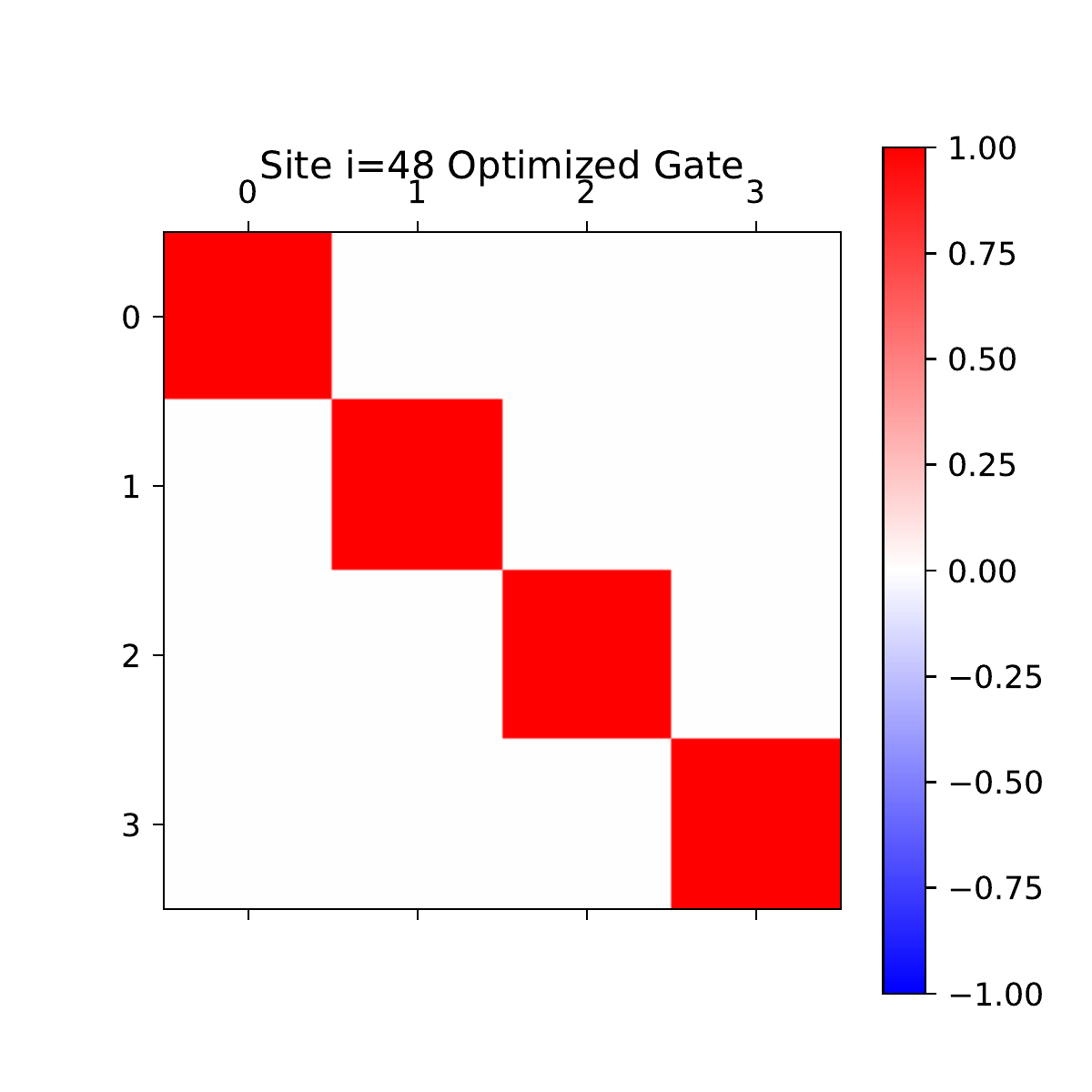}
}
\subfloat[Circuit from optimization]{
\includegraphics[width=0.3\columnwidth]{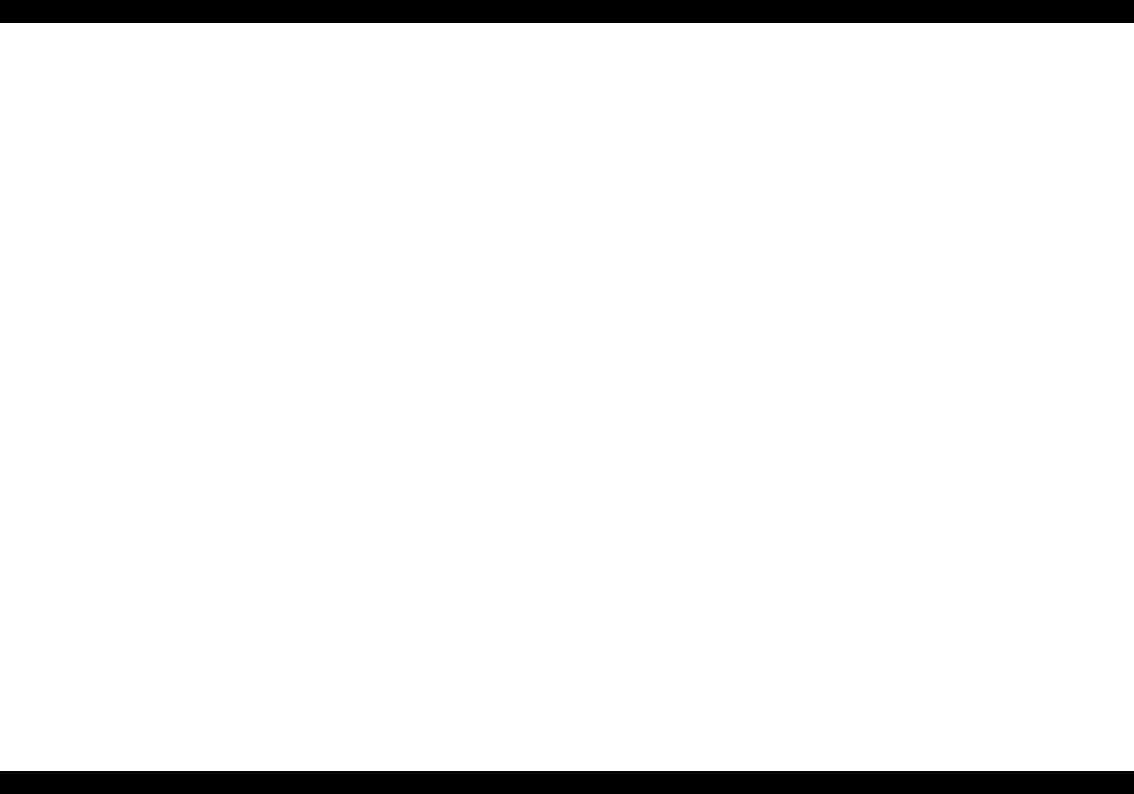}
}
\caption{\label{fig:site48} Optimization for site 48}
\end{figure}

\clearpage

\end{document}